%% file: elhatisari.tex
\documentclass[12pt,          
               phd,           
               doublespacing 
               ]{ncsuthesis}

\usepackage{booktabs}  
\usepackage{amsmath} 
\usepackage{textcomp}  
\usepackage{lipsum}    
\usepackage{subfig}    

\usepackage{mathtools}
\usepackage{slashbox}%
\usepackage{braket}
\usepackage{color}

\usepackage{amsmath,amssymb,amsfonts} 
\usepackage{graphicx,subfig}
\usepackage{dcolumn}
\usepackage{bm}
\usepackage{cancel}
\usepackage{verbatim}
\usepackage{ifthen}
\usepackage{url}
\usepackage{sectsty}
\usepackage{balance} 
\usepackage{caption}
\usepackage{graphicx} 
\usepackage{lastpage}
\usepackage[format=plain,justification=RaggedRight,singlelinecheck=false,font=small,labelfont=bf,labelsep=space]{caption} 
\usepackage{mathabx} 
\usepackage{slashed}
\usepackage{float}
\floatstyle{plaintop}
\restylefloat{table}

\newcommand{\ie}{\text{i.e.}}

\newcommand{\cf}{\text{cf.}}

\makeatletter

\newcommand{\Rmnum}[1]{\expandafter\@slowromancap\romannumeral #1@}
\makeatother

\usepackage{abbrevs}
\usepackage{etoolbox}
\robustify{\DateMark} 

\usepackage{units} 

\usepackage[sharp]{easylist} 

\usepackage[table]{xcolor}
\usepackage{array,booktabs}
\usepackage{colortbl}
\newcolumntype{L}{@{}>{\kern\tabcolsep}l<{\kern\tabcolsep}}

\usepackage{titlesec}

\titleformat{\paragraph}
{\normalfont\normalsize\bfseries}{\theparagraph}{1em}{}
\titlespacing*{\paragraph}
{0pt}{3.25ex plus 1ex minus .2ex}{1.5ex plus .2ex}

\usepackage{placeins}

\dispositionformat{\bfseries}            

\headingformat{\large\MakeUppercase}   

\include{optional}
\student{Serdar}{Elhatisari} 

\program{Physics}

\thesistitle{Low Energy Continuum and Lattice Effective Field Theories}

\newcommand{\eref}[1]{Eq.~(\ref{#1})}
\newcommand{\erefs}[2]{Eqs.~(\ref{#1})~\text{and}~(\ref{#2})}

\newcommand{\fref}[1]{Figure~\ref{#1}}
\newcommand{\tref}[1]{Table~\ref{#1}}
\newcommand{\sref}[1]{Section~\ref{#1}}
\newcommand{\aref}[1]{Appendix~\ref{#1}}
\newcommand{\prim}[1]{{#1}^{\prime}}
\newcommand{\pprim}[1]{{#1}^{\prime\prime}}

\graphicspath{{./Chapter-1/figs/}{./Chapter-2/figs/}{./Chapter-3/figs/}}

\usepackage{calc}
\newlength{\chaptercapitalheight}
\newlength{\chapterfootskip}
\newlength\graphht

\usepackage{pdflscape}
\usepackage{tikz}
\fancypagestyle{lscapedplain}{%
  \fancyhf{}
  \fancyfoot{%
    \tikz[remember picture,overlay]
      \node[outer sep=1cm,above,rotate=90] at (current page.east) {\thepage};}

}

\begin{document}
\pagestyle{plain}
\frontmatter

\include{front}

\mainmatter


\include{Chapter-1}
\include{Chapter-2}

\include{Chapter-3}
\include{Chapter-4}
\include{Chapter-5}
\include{Chapter-6}
\restoregeometry


\bibliography{elhatisari}{}
\bibliographystyle{plain}

\restoregeometry
\appendix


\include{Appendix-A}

\include{Appendix-B}
\include{Appendix-C}

\include{Appendix-D}

\restoregeometry

\backmatter

\end{document}

%% file: optional.tex
\usepackage[Lenny]{fncychap}

\usepackage[
  pdfauthor={Serdar Elhat{\i}sar{\i}},
  pdftitle={Low energy continuum and lattice effective field theories},
  pdfcreator={pdftex},
  pdfsubject={NC State ETD Thesis},
  pdfkeywords={causality, unitarity, lattice, adiabatic, monte-carlo, effective field theory},
  colorlinks=true,
  linkcolor=black,
  citecolor=black,
  filecolor=black,
  urlcolor=black,
]{hyperref}

\usepackage{microtype}

\usepackage[T1]{fontenc}
\usepackage{mathptmx}



%% file: front.tex
\maketitlepage

\makecopyrightpage

\begin{abstract}

In this thesis we investigate several constraints and their impacts on the short-range potentials in the low-energy limits of quantum mechanics. We also present lattice Monte Carlo calculations using the adiabatic projection method.

In the first part of the thesis we consider the constraints of causality and unitarity for particles interacting via strictly finite-range interactions. We generalize Wigner's
causality bound to the case of non-vanishing partial-wave mixing. Specifically we analyze the system of the low-energy interactions between protons and neutrons. We also analyze low-energy scattering for systems with arbitrary short-range interactions plus an attractive $1/r^{\alpha}$ tail for $\alpha\geq2$. In particular, we focus on the case of $\alpha=6$ and we derive the constraints of causality and unitarity also for these systems and find that the van der Waals length scale dominates over parameters characterizing the short-distance physics of the interaction.  This separation of scales suggests a separate universality class for physics characterizing interactions with an attractive $1/r^{6}$ tail.  We argue that a similar universality class exists for any attractive potential $1/r^{\alpha}$ for $\alpha\geq2$.

In the second part of the thesis we present lattice Monte Carlo calculations of fermion-dimer scattering in the limit of zero-range interactions using the adiabatic projection method.  The adiabatic projection method uses a set of initial cluster states and Euclidean time projection to  give a systematically improvable description of the low-lying scattering   cluster states in a finite volume.  We use L{\"u}scher's finite-volume relations to determine the $s$-wave, $p$-wave, and $d$-wave phase shifts.  For comparison, we also compute exact lattice results using Lanczos iteration and continuum results using the Skorniakov-Ter-Martirosian equation.  For our Monte Carlo calculations we use a new lattice algorithm called impurity lattice Monte Carlo.  This algorithm can be viewed as a hybrid technique which incorporates  elements of both worldline and auxiliary-field Monte Carlo simulations.

\end{abstract}

\begin{dedication}
 \centering \textit{Dedicated to my parents and siblings}.
\end{dedication}


\begin{acknowledgements}

It is with great pleasure that I take this opportunity to express my deepest appreciation to all of those who supported me in any respect during the completion of the thesis.

First, I am very grateful to my advisor Dean Lee for his continuous supports, encouragements, endless helps and being a constant source of intellectual stimulation in every step of my studies at NC State. I am very fortunate not only because I have had the opportunity to learn from his brilliant ideas but also because I have been involved in such a friendly scientific environment provided by his kindness.

I wish to thank the Department of Physics at NC State University for taking such a nice care of all students, and want to thank everyone in the department for establishing a very nice atmosphere. I am grateful to the Turkish Government Ministry of National Education for supporting me with a doctoral fellowship and the Department of Physics at NC State University for additional supports. I am also thankful to the North Carolina State University High Performance Computing center for extensive computer support and HPC Remedy Workgroup, specially Gary Howell, for valuable technical supports.

I would like to thank Michelle Pine, Sebastian K\"{o}nig and Shahin Bour who are members of the Lee Research Group. A special thank to Sebastian K\"{o}nig for his collaborations and useful discussions, and Michelle Pine for carefully reading and useful comments on the manuscript of my thesis. I also thank the physicist from the Nuclear Lattice Effective Field Theory collaboration for their valuable discussions on my work. Specially I am grateful to Gautam Rupak for extensive discussions on several aspects of the adiabatic projection method, and I thank Shahin Bour, Hans-Werner Hammer and Ulf-G. Mei{\ss}ner for discussions on the impurity Monte Carlo method.

As a member of an internationally recognized group, I have had several opportunities to travel to many places and meet pioneers in the fields. I would like to thank Ruhr-Universit{\"{a}}t Bochum for its hospitality during a couple of weeks in the summers of 2011, 2012 and 2013, and thank Evgeny Epelbaum and Hermann Krebs for useful discussions. I wish to thank the organizers of the program ``Light nuclei from first principles" in the Institute of Nuclear Theory at the University of Washington for inviting me during two weeks of the program and thank the INT for its hospitality. Also I would like to thank the organizer of the Third UiO-MSU-ORNL-UT School on Topics in Nuclear Physics held at the Oak Ridge Laboratory for the financial support to attend the winter school.

Last, but surely not least, I would like to thank all my friends for their unquestioning support of my goals and dreams. I also wish to thank those whom I have met in the Triangle area for making the last four years of my life more fun, special thanks to Ako{\u{g}}lu, Ay and G{\"o}k{\c{c}}e families, H. {\.{I}}brahim Aky{\i}ld{\i}z, Murat An,  Zubair Azad, Esra \"{O}zt{\"{u}}rk, {\.{I}}. \c{S}afak Bayram, Shaan Qamar, Enis {\"{U}}{\c{c}}{\"u}nc{\"u} and Hasan Y{\i}ld{\i}r{\i}m.

\end{acknowledgements}

\thesistableofcontents

\thesislistoftables

\thesislistoffigures

%% file: Chapter-1.tex
\chapter{Overview}

Nuclear structure and reaction studies are at the heart of low energy nuclear physics research. A broad goal of this research is to understand the basic interactions among fundamental particles. By fundamental particles we refer to the relevant degrees of freedom that can be probed by the energy scales of a given system. In nuclear physics physical phenomena are observed over a range of different energy scales. In low energy nuclear physics the relevant degrees of freedom are nucleons and light mesons which mediate forces between nucleons, and physical observables are insensitive to the details at high energies or equivalently short distances.  

Effective field theory (EFT) is a very general framework to understand physics. The general idea is that a simple effective description of the physics can describe the relevant features of a given system even if the details at short distances are disregarded. In low-energy nuclear effective field theory the detailed shape of the short-range nuclear forces are not important. Instead effective field theory organizes the nuclear interactions as an expansion in powers of momenta and other low energy scales such as the pion mass.

Lattice effective field theory is a powerful numerical method which is formulated in the framework of effective field theory. The method is quite economical when one uses pionless effective field theory with the nucleons interacting via only local contact interactions. Recent developments in lattice EFT allow one to study nuclear scattering and reactions. The adiabatic projection method is a general framework for calculating scattering and reactions on the lattice. This method uses a set of initial cluster states and Euclidean time projection to construct a low-energy effective field theory  for the cluster states.

In this thesis we summarize work done on a number of topics in the framework of effective field theories during my Ph.D. study. We start with a brief review of scattering processes with strictly finite-range interactions in Chapter~\ref{chap:Scattering-theory}. We also discuss the case where long-range forces are present in addition to the short-range interactions, and we specifically consider Coulomb interactions.

In Chapter~\ref{chap:LatticeEFT} after a brief introduction we define the basic continuum and lattice formulations of non-relativistic quantum mechanics.  We then take a short detour to derive the connection between the operator formalism in quantum mechanics and lattice Grassmann path integrals. We introduce the adiabatic projection method and describe the implementation of the method in lattice effective field theory. We discuss the details of and some mathematical tools for scattering phase shift calculations at finite volume. Then we close this chapter by reviewing Monte Carlo simulations employed to compute observables in finite volumes with periodic boundary conditions.

In Chapter~\ref{chap:Neutron-proton-scattering} we consider the Wigner causality constraint and unitarity for the low-energy interactions with strictly finite range. We derive the generalization of Wigner's causality bound to the case of non-vanishing partial-wave mixing. As an application in nuclear physics, we analyse the low-energy
interactions of protons and neutrons.  We investigate the constrains on the range of the interactions between neutrons and protons in effective field theory and universality in many-body
Fermi systems.

In Chapter~\ref{chap:van-der-Waals} we analyze low-energy scattering for arbitrary short-range interactions plus
any attractive potential $1/r^{\alpha}$
for $\alpha\geq2$. This type of long-range force plays an important role in low-energy atomic and nuclear physics. In particular, we consider the van der Waals interaction and derive the constraints of causality and
unitarity. We also briefly discuss multichannel systems near
a magnetic Feshbach resonance.

In Chapter~\ref{chap:fermion-dimer-scattering}
we present scattering of composite particles using lattice Monte Carlo simulations with the adiabatic projection method. For our Monte Carlo simulations we introduce the impurity lattice Monte Carlo algorithm. As an application we consider fermion-dimer elastic scattering in the limit of zero-range interactions. Then we present results for the scattering phase shifts in the continuum and infinite volume limits as well as in finite volumes.

%% file: Chapter-2.tex
\chapter{Scattering Theory}
\label{chap:Scattering-theory}

\section{Introduction}
\label{sec:ScattThey-Preliminaries}

The scattering of two spinless particles is described by an incident plane wave along the $z$ direction and spherical outgoing scattered wave with amplitude $f$ as
\begin{equation}
\psi(\vec{r}) \, \buildrel{r\to\infty}\over\longrightarrow\, \frac{1}{(2\pi)^{3/2}}
\left[
e^{i  p z} + f(p,\theta)
 \, \frac{e^{ipr}}{r}
\right]
\,,
\label{eqn:asymtoticWave-0001}
\end{equation}
where $\vec{p}$ is the outgoing relative momenta, $\vec{r}$ is the relative coordinates and $\theta$ is the angle between $\vec{p}$ and the $z$ axis. The scattering amplitude is a non-trivial physical quantity in which all quantitative information about the scattering process is contained. An important constraint on the scattering amplitude is unitarity.  Unitarity requires that the sum of all outcome probabilities is one. In other word  the normalization of the incoming wave must be preserved. 

In addition, the time evolution of any quantum mechanical system obeys \emph{causality} as well as unitarity. Causality requires that the cause of an event must occur before any resulting consequences are produced. In non-relativistic quantum mechanics the causality constraint means that the outgoing wave may depart only after the incoming wave reaches the scattering object. The constraints of causality in quantum mechanics with finite range interactions were first
studied by Wigner~\cite{Wigner1954}. Phillips and Cohen~\cite{Phillips1997} derived the causality bound on the scattering parameters of low energy effective field theories by considering the $s$-wave case and finite-range interactions in three
dimensions. Later, the causality bounds for arbitrary dimension $d$ or arbitrary angular
momentum $\ell$ were investigated in Refs.~\cite{Hammer-Lee9,HammerDean27}.

The scattering processes described above is an idealized system with the assumptions that particles are structureless. This simplified process is called \emph{single-channel} since there is only one possible final configuration which is the same as the initial one, $a(b,a)b$. In general, elementary or composite particles have spin structures; and accordingly, the interaction potentials become spin-dependent which makes the scattering processes more complicated. In these types of system different possible final outcomes exist~\cite{taylorNRQM}, and the scattering processes are called \textit{multi-channel} scattering. See Ref.~\cite{Helv.Phys.Acta31.661} for a detailed rigorous mathematical formulation of the system with short-range interactions, and the general multi-channel problems in the presence of long-range potentials were studied in Ref.~\cite{JMP:1704171,PhysRevA.26.2441}.

\section{The Schr\"{o}dinger wave equation}

In this and the following sections we will give brief reviews on several topics of the two-body elastic scattering problem.  Following Refs.~\cite{R.G.Newton} with slightly changed notation, we will discuss the scattering of particles within the framework of the formal scattering theory in configuration space.

\subsection{Single channel}

We start with an idealized system and consider elastic scattering of two spinless\footnote{By spinless particles we mean either exactly spinless fundamental particles or composite particles with zero-intrinsic angular momentum.} particles interacting via a spherically
symmetric potential in the center-of-mass frame. \ We use units where $\hbar=1$. The free radial Schr\"{o}dinger equation
with energy $E=p^{2}/(2\mu)$ is
\begin{equation}
\left[  -\frac{1}{r^{2}}\frac{d}{dr}\left(  r^{2}\frac{d}{dr}\right)
+\frac{\ell(\ell+1)}{r^{2}}\right]  R_{\ell}^{(p)}(r) = p^{2} \,  R_{\ell}^{(p)}(r).
\label{eqn:radialeqR}%
\end{equation}
where $\ell$ is the orbital angular momentum. It is convenient to use the rescaled radial function $u_{\ell}^{(p)}(r)$ given by
\begin{equation}
u_{\ell}(pr)= r \, R_{\ell}^{(p)}(r) \,,
\label{eqn:rescaled-radial}%
\end{equation}
then the Schr\"{o}dinger equation describes the system is alternatively 
\begin{equation}
 \left[-\frac{d^{2}}{dr^{2}}+\frac{\ell(\ell+1)}{r^{2}}\right] u_{\ell}(pr)
 =p^{2} \,  u_{\ell}(pr) \,,
 \label{eqn:FreeSchrodingerEq-0001}
\end{equation}
 Two linearly independent (regular and irregular) solutions of \eref{eqn:FreeSchrodingerEq-0001} are the Riccati-Bessel $S_{\ell}(pr)$ and Riccati-Neumann $C_{\ell}(pr)$ functions, respectively, which are defined in terms of Bessel and Neumann functions in \eref{eqn:append-SLpr001} and \eref{eqn:append-CLpr001}. The asymptotic form of these functions for large arguments are
\begin{align}
 S_{\ell}(\rho) \buildrel{|\rho|\to\infty}\over\sim  \sin\left(\rho -\frac{\ell \pi}{2}\right)  \,,
 \label{eqn:asymp-RiccatiBessel-0001}
 \\
 C_{\ell}(\rho) \buildrel{|\rho|\to\infty}\over\sim  \cos\left(\rho -\frac{\ell \pi}{2}\right)  \,.
 \label{eqn:asymp-RiccatiNewmann-0001}
\end{align}
The total wave function in the partial wave expansion form is
\begin{equation}
 \Psi^{\pm}=(p,\vec{r})
 =
\left(\frac{2\mu p}{\pi}\right)^{1/2} \,
\, \sum_{\ell}
    \sum_{m =-\ell}^{\ell} 
     i^{\ell} \,  \frac{\psi^{\pm}_{\ell}(pr)}{pr}
 \text{Y}_{\ell}^{m}(\hat{r})
 \, {\text{Y}_{\ell}^{m}}^{*}(\hat{z}) \,,
 \label{eqn:totalwave-Singlechannel-0001}
\end{equation}
where $\mu$ is the reduced mass, $\text{Y}_{\ell}^{m}(\hat{\rho})$
are the spherical harmonics, $\hat{r}$ denotes the polar angles $(\theta,\vartheta)$ of $\vec{r}$, $\psi^{\pm}_{\ell}(pr)$ is the radial part of the wave function. The solution with the superscript $+/-$ corresponds to the out/in asymptotic wave when we go to a time-dependent formalism. Using the Legendre polynomial,
\begin{equation}
 P_{\ell}\left(\cos\theta\right)
 =
\frac{4\pi}{2\ell+1}
   \sum_{m_{\ell} =-\ell}^{\ell}  
    \text{Y}_{\ell}^{m}(\hat{r})
    \, {\text{Y}_{\ell}^{m}}^{*}(\hat{z})\,,
\end{equation}
the total wave function is rewritten as
\begin{equation}
 \Psi^{\pm}(p,\vec{r})
 =
\left(\frac{2\mu}{\pi p}\right)^{1/2} \frac{1}{4\pi r} \,
\, \sum_{\ell}
     i^{\ell} \,  (2\ell+1) \, 
      P_{\ell}\left(\cos\theta\right) \, \psi^{\pm}_{\ell}(pr) \,.
\end{equation}
Now inserting the total wave function in the Lippmann-Schwinger integral equation, we get the following integral equation for the radial wave function, 
\begin{equation}
\psi^{\pm}_{\ell}(pr)
 = S_{\ell}(pr)
 +
 2\mu \int_{0}^{\infty}\, \int_{0}^{\infty} d\prim{r} d\pprim{r} \,
 G_{\ell}^{\pm}(p;r,\prim{r}) \,
 W(\prim{r},\pprim{r}) \,
 \, \psi^{\pm}_{\ell}(p\prim{r}) \,,
\label{eqn:radialwave-Singlechannel-0010}
\end{equation}
where $ W(\prim{r},\pprim{r})$ is the interaction potential assumed to be a rotationally invariant operator, and $G^{\pm}_{{\ell}}(p;r,\prim{r})$ is the partial wave Green's function which satisfies the differential equation
\begin{equation}
 \left[-\frac{d^{2}}{dr^{2}}+\frac{\ell(\ell+1)}{r^{2}} -p^{2}\right] G^{\pm}_{{\ell}}(p;r,\prim{r})
 =-\delta\left(r-\prim{r}\right) \,.
 \label{eqn:Greens-func-0001}
\end{equation}
The Green's function that satisfies \eref{eqn:Greens-func-0001} with suitable boundary conditions is defined in terms of the Riccati-Bessel and Riccati-Henkel functions as
\begin{equation}
G^{\pm}_{{\ell}}(p;r,\prim{r})
= - \frac{1}{p} S_{\ell}(pr_{<}) \, h^{\pm}_{\ell}(pr_{>})
\,,
 \label{eqn:Greens-func-0005}
\end{equation}
where $r_{<}$ signifies the smaller of $r$ and $\prim{r}$ and  $r_{>}$ is the larger, and $h_{\ell}^{\pm}$ are the Riccati-Henkel functions defined in terms of the Riccati-Bessel and Neumann functions
\begin{align}
h_{\ell}^{\pm}(\rho) = \pm i  S_{\ell}(\rho) +C_{\ell}(\rho)\,,
\end{align}
and their asymptotic forms for large arguments are 
\begin{equation}
h_{\ell}^{\pm}(\rho)  \buildrel{|\rho|\to\infty}\over\sim  e^{\pm i\left(\rho -\frac{\ell \pi}{2} \right)} \,.
\label{eqn:RiccatiHenkel}
\end{equation}
\eref{eqn:radialwave-Singlechannel-0010} is the solution of the following Schr\"{o}dinger wave equation,
\begin{equation}
 \left[-\frac{d^{2}}{dr^{2}}+\frac{\ell(\ell+1)}{r^{2}}\right] \psi^{+}_{\ell}(pr)
  +
  2\mu \int_{0}^{\infty}\, d\prim{r} \,
  W({r},\prim{r}) \,
  \, \psi^{+}_{\ell}(p\prim{r})
 =p^{2} \,  \psi^{+}_{\ell}(pr) \,.
 \label{eqn:SchrodingerEq-0001}
\end{equation}
At the moment we do not impose any condition on the potential and postpone the discussion till Section~\ref{chap1:interaction-potential}. At $r \to \infty$ the asymptotic form of \eref{eqn:radialwave-Singlechannel-0010} has a formal solution written in terms of the incident wave, the scattered wave and the scattering matrix, ${\rm{S}}_{\ell}(p)$, as
\begin{equation}
\psi^{+}_{\ell}(pr)
 = \frac{i}{2}\left[h^{-}_{\ell}(pr) -
{\rm{S}}_{\ell}(p) \, h^{+}_{\ell}(pr) \right] \,,
\label{eqn:radialwave-Singlechannel-0020}
\end{equation}
where
\begin{equation}
{\rm{S}}_{\ell}(p) =  1 - \frac{4i\mu}{p} \int_{0}^{\infty} \,\int_{0}^{\infty} d\prim{r} d\pprim{r} \,
  S_{\ell}(p\prim{r}) \,
 W(\prim{r},\pprim{r}) \,
 \, \psi^{+}_{\ell}(p\prim{r})\,,
\label{eqn:S-matrix-Singlechannel-0030}
\end{equation}
The scattering matrix (S-matrix) ${\rm{S}}_{\ell}(p)$ is a function of momentum and independent of $r$; therefore, ${\rm{S}}_{\ell}(p)$ can be defined in terms of a real and momentum dependent phase angle $\delta_{\ell}(p)$,
\begin{equation}
{\rm{S}}_{\ell}(p) = e^{2i\delta_{\ell}(p)}\,.
\label{eqn:S-matrix-0001}
\end{equation}
Then \eref{eqn:SchrodingerEq-0001} in the asymptotic limit, up to a normalization, becomes
\begin{align}
\psi^{+}_{\ell}(pr)  \sim
 \sin\left(pr-\frac{\ell \pi}{2} + \delta_{\ell}(p)\right) \,.
\label{eqn:scattWave-0001}
\end{align}
This scattered wave relative to \eref{eqn:asymp-RiccatiBessel-0001} implies that the impact of the scattering event is to introduce the shift $\delta_{\ell}(p)$ in the phase of the outgoing wave relative to the incoming wave. The radial wave function is usually written in terms of the reaction matrix\footnote{Sometimes it is called the reactance matrix~\cite{R.G.Newton}.} (K-matrix) rather than the S-matrix.  The relation between the reaction and scattering matrix is
\begin{equation}
{\rm{K}}_{\ell}=i\frac{(1-{\rm{S}}_{\ell})}{(1+{\rm{S}}_{\ell})},
\label{chap1:fromKtoS}
\end{equation}
and the radial wave function in terms of the reaction matrix reads, up to a normalization,
\begin{equation}
\psi^{+}_{\ell}(pr)
 \sim \, S_{\ell}(pr) +
{\rm{K}}_{\ell}(p) \, \, C_{\ell}(pr)  \,.
\label{eqn:radialwave-Singlechannel-0030}
\end{equation}

\subsection{Coupled channels}
\label{sec:Coupled-channels-0001}
In this section we make the scattering problem more complicated  by considering that particles have intrinsic spins. This brings some complications into the formalism introduced in the preceding sections due to the fact that the spin is an additional degree of freedom. We consider the scattering of two particles with total spin angular momenta $s = s_{1} + s_{2}$ where $s_{1}$ and $s_{2}$ are the individual particle spins. Here the orbital angular momentum $\ell$ and the spin angular momentum $s$ are coupled to give the total angular momentum $j$. Denote the magnetic quantum number of $j$, $\ell$ and $s$ in the $z$ direction by $M$, $m$ and $m_{s}$, respectively. Since the Hamiltonian of the system commutes with the total angular momentum operator $\bold{J}$ in order to be  rotationally invariant, the conserved quantities of the system are $j$ and $M$. Therefore, depending on the values that $s$ takes, the orbital angular momentum takes different values each of which corresponds to a different radial wave function.

Let us generalize the spherical harmonics for the system of particles with spin and define the following functions
\begin{equation}
\mathcal{Y}_{j\ell s}^{M}(\hat{r})
=
\sum_{m\,m_{s}}
C(\ell s j,m\,m_{s} M) \, \text{Y}_{\ell}^{m}(\hat{r}) \, \chi^{s}_{m_{s}}
\end{equation}
and
\begin{equation}
\slashed{\mathcal{Y}}_{j}^{M}(\ell sm_{s};\hat{z})
=
-i \, \chi^{s\,*}_{m_{s}}\cdot \mathcal{Y}_{j\ell s}^{M}(\hat{z})
\end{equation}
where $\chi^{s}_{m_{s}}$ is the normalized eigenfunction of the total spin, $C(\ell s j,m_{\ell} m_{s} m)$ are the Clebsch-Gordan coefficients, and the dot signifies an inner product with respect to the internal coordinates.  Since the individual particle spins are conserved, we will suppress them in the expressions. The completeness relation of the Clebsch-Gordan coefficients is 
\begin{equation}
\sum_{m_{\ell} m_{s}}
C(\ell s j,m_{\ell} m_{s} m) \,
C(\ell s \prim{j},m_{\ell} m_{s} \prim{m})
=
\delta_{jj^{\prime}}\delta_{m m^{\prime}} \,.
\label{eqn:CG-coeff-completeness}
\end{equation}
Therefore, the total wave function is written as
\begin{equation}
 \Psi^{\pm}(sm_{s};p,\vec{r})
 =
\left(\frac{2\mu}{\pi p}\right)^{1/2} \,
\, \sum_{j M \ell \prim{\ell} \prim{s}}
      \, \frac{\psi_{\prim{\ell} \prim{s},\ell s}^{j(\pm)}(p,r)}{r}  \,
      \mathcal{Y}_{j\prim{\ell} \prim{s}}^{M}(\hat{r})
      \, \slashed{\mathcal{Y}}_{j}^{M \, *}(\ell sm_{s};\hat{z}) \,.
\label{eqn:totalwave-Coupledchannel-0001}
\end{equation}
After inserting \eref{eqn:totalwave-Coupledchannel-0001} into the Lippmann-Schwinger integral equation, we obtain the general form of the radial wave function,
\begin{align}
\psi_{\prim{\ell} \prim{s},\ell s}^{j(\pm)}(p,r)
=
& S_{\ell}(pr) \,
 \delta_{\ell \prim{\ell}} \, \delta_{s \prim{s}} 
\nonumber\\
& 
+
 2 \mu \sum_{\pprim{\ell} \pprim{s}}
  \int_{0}^{\infty}\int_{0}^{\infty} d\vec{r}^{\prime} d\pprim{\vec{r}} G^{\pm}_{\prim{\ell}}(p;r,\prim{r})
\,
W_{\prim{\ell} \prim{s},\pprim{\ell} \pprim{s}}(\prim{r},\pprim{r})
\,
\psi_{\pprim{\ell} \pprim{s},\ell s}^{j(\pm)}(p,r^{\prime})
 \,,
\label{eqn:totalwave-Coupledchannel-0005}
\end{align}
where $G^{\pm}_{\prim{\ell}}$ is given in \eref{eqn:Greens-func-0005}, and $W_{\prim{\ell} \prim{s},\pprim{\ell} \pprim{s}}$ is the spin-dependent interaction potential
\begin{equation}
W_{\prim{\ell} \prim{s},\pprim{\ell} \pprim{s}}(\prim{r},\pprim{r})
= 
\int d {\Omega} \, \mathcal{Y}_{j\prim{\ell} \prim{s}}^{M\,*}(\prim{\hat{r}}) \,
W(\prim{r},\pprim{r}) \,
 \mathcal{Y}_{j\pprim{\ell} \pprim{s}}^{M}(\prim{\hat{r}})
  \,.
\label{eqn:coupled-potential-0005}
\end{equation}
The radial wave function $\psi_{\prim{\ell} \prim{s},\ell s}^{j(\pm)}$ is the solution of the coupled Schr\"{o}dinger equation,
\begin{align}
&\left[
-\frac{d^{2}}{dr^{2}}
+\frac{\prim{\ell}(\prim{\ell}+1)}{r^{2}}
\right]
\psi_{\prim{\ell} \prim{s},\ell s}^{j(+)}
 +2\mu
\sum_{\pprim{\ell} \pprim{s}}
\int_{0}^{\infty} \,dr^{\prime} \,
{W}_{\prim{\ell} \prim{s},\pprim{\ell} \pprim{s}}(r,\prim{r})
\psi_{\pprim{\ell} \pprim{s},{\ell} {s}}^{j(+)} 
=
p^{2}
\psi_{\ell^{\prime} s^{\prime},\ell s}^{j(+)}\,.
\label{eqn:coupled-Schr-0001}
\end{align}
Therefore, the S-matrix for the particles of the total spin $s$ is 
\begin{equation}
{\rm{S}}_{\prim{\ell} \prim{s},\ell s}  =  \delta_{\ell \prim{\ell}} \, \delta_{s \prim{s}} 
 -
  \frac{4i\mu}{p} 
  \sum_{\pprim{\ell} \, \pprim{s}}
  \int_{0}^{\infty}\int_{0}^{\infty} d\prim{r} d\pprim{r} \,
  S_{\prim{\ell}}(p\prim{r}) \,
\,
{W}_{\prim{\ell} \prim{s},\pprim{\ell} \pprim{s}}(\prim{r},\pprim{r})
\,
\psi_{\pprim{\ell} \pprim{s},\ell s}^{j(+)}(p,r^{\prime})
\,.
\label{eqn:S-matrix-Coupledchannel-0010}
\end{equation}

\section{Interaction potential}
\label{chap1:interaction-potential}
In \eref{eqn:S-matrix-0001} we have na\"{i}vely defined the S-matrix without imposing any conditions on the potential. However, in Eqs.~(\ref{eqn:S-matrix-Singlechannel-0030}) and (\ref{eqn:S-matrix-Coupledchannel-0010}) the unitarity condition we have discussed in \sref{sec:ScattThey-Preliminaries} requires some constraints on the interaction potential.

Throughout our analysis we assume that the interaction is sufficiently well-behaved at the origin to admit the regular solution $S_{\ell}(pr)$ . This assumption imposes the restriction that as $r\to0$ the potential is not too singular such that the radial wave function satisfies the regularity condition
\begin{equation}
 \lim_{\rho\rightarrow0}\psi_{\ell}(\rho)\frac{d}{d\rho}\psi_{\ell}(\rho)=0 \,.
\label{eqn:regularity}
\end{equation}
In Ref.~\cite{PhysRevA.30.1279} it is proven that this condition is fulfilled
by a class of potentials $V(r)$ provided that
\begin{equation}
 \int_{0}^{R}
 \, d\prim{r} \,  \prim{r}\left|V(\prim{r})\right|  < \infty \,.
\end{equation}
We also consider only energy independent interactions, and we assume that the interactions
have a finite range $R$. The finite-range condition implies that
\begin{align}
W(r,r^{\prime})=0 \qquad \text{if} \,\,\,\, r>R  \quad \text{or}  \,\,\,\, r^{\prime}>R \,.
\end{align}
These assumptions assure that the interaction can be written as a local potential,
\begin{align}
W(r,\prim{r}) = V(r) \delta\left(r-r^{\prime}\right)\,.
\end{align}

\section{Effective range expansion}

Under our assumptions on the potential in the preceding section we can obtain the scattering information by measuring \eref{eqn:radialwave-Singlechannel-0010}
relative to the free solution in the asymptotic limit,
\begin{equation}
\psi_{\ell}(r) = S_{\ell}(pr) + p f_{\ell}(p) \, h_{\ell}^{+}(pr)\,,
\label{eqn:asymtoticWave-0010}
\end{equation}
where $f_{\ell}(p)$ is the partial wave amplitude
\begin{equation}
f_{\ell}(p)  = \frac{{\rm{S}}_{\ell}(p)-1}{2ip} = \frac{p^{2\ell}}{p^{2\ell+1} \cot\delta_{\ell}(p) -ip^{2\ell+1}} \,.
\label{eqn:partialWave-0001}
\end{equation}
For the system of particles interacting with a finite-range potential, $p^{2\ell+1} \cot\delta_{\ell}(p)$ is described by the well-known power series expansion around $p^{2}$, the so called \emph{effective range expansion}~\cite{RevModPhys.22.77,PhysRev.76.38},
\begin{equation}
p^{2\ell+1}\cot\delta_{\ell}(p)
=-\frac{1}{a_{\ell}}
+\frac{1}{2} r_{\ell} \, p^{2}
+P_{\ell} \, p^{4}
+Q_{\ell} \,  p^{6}
+\mathcal{O}(p^{8}),
\label{eqn:ERE-0001}
\end{equation}
where ${a_{\ell}}$ is the \emph{scattering length}, ${r_{\ell}}$ is the \emph{effective range}, and coefficients in higher order terms of $p^{2}$ are the shape parameters. The effective range formula for the multi channel scattering problem is of the following form~\cite{deSwart1962458,Ross1961147,Blatt-B.I,Blatt-B.II,Kermode1967605},
\begin{equation}
\sum_{m',n'} \textit{\textbf{p}}_{mm'} \, [{\rm{K}}^{-1}]_{m'n'} \, \textit{\textbf{p}}_{n'n}
=-\frac{1}{ \mathbf{a}_{mn}}
+\frac{1}{2} \mathbf{r}_{mn} \, p^{2}
+\mathbf{P}_{mn} \,  p^{4}
+\mathbf{Q}_{mn} \, p^{6}
+O(p^{8}),
\label{eqn:multichannel-ERE-0001}
\end{equation}
where $\mathbf{\hat{K}}$ is the multi channel reaction matrix, $\textit{\textbf{p}}$ is the diagonal momentum matrix
\begin{align}
\textit{\textbf{p}} =
\left(\begin{array}{ccc}
p^{j-s+\frac{1}{2}} &  &  \\ 
 & \ddots &  \\ 
 &  & p^{j+s+\frac{1}{2}} 
\end{array} \right) \,,
\end{align}
$\mathbf{a}_{mn}$ is the scattering length matrix, $\mathbf{r}_{mn}$ is the effective range matrix, and $\mathbf{P}_{mn}$ and $\mathbf{Q}_{mn}$ are the first two lowest shape parameter matrices.

In the cases where particles are interacting via long-range forces the power series or the convergence (analyticity) of \eref{eqn:ERE-0001} is spoiled. Then the function $p^{2\ell+1}\cot\delta_{\ell}(p)$ needs special treatment to modify the expansion to make it an analytic function of $p^{2}$. For instance, for a class of potentials falling off as $e^{-m\, r}$ at large distances, the potential introduces a branch cut starting at $p^{2} = -m^{2}/4$ where $m$ is the mass of the particle which mediates the interaction, see Ref.~\cite{Goldberger:Watson} for the detailed proof. Nevertheless, the expansion is still converges in a circle of radius $m^{2}/4$ around the origin of the complex $p^{2}$ plane.

For the system of charged particles, a modified effective range expansion is used to deal with the Coulomb tails~\cite{PhysRev.75.1637,PhysRev.76.38,PhysRevA.30.1279}. For long-range potentials of the form of $1/r^{\alpha}$ with $\alpha>2$ the irregular behavior of the radial Schr\"{o}dinger equation at short distances limits the determination  of the partial wave amplitude. For example, for $\alpha = 4$ only the scattering length can be well-defined~\cite{jmp1703735}. However, the so called quantum defect formulation for the scattering phase shifts allows one to define the total phase shift for $\alpha>2$ as a sum of a weakly energy-dependent short-range phase shift and strongly energy-dependent long-range phase shift~\cite{PhysRevA.78.012702}. This formalism has been developed to describe atom-atom scattering. In particular, quantum defect theory and the modified effective range expansion for $\alpha= 6$ will be discussed in Section~\ref{chap5:QDT-modifiedERE}.

\section{Scattering solutions}
\label{chap2:scattering-solutions}

\subsection{Neutral particles}
\label{chap1:neutral-particles}

In this section we return to the idealized system of two spinless particles. Here we consider the system of two-particle interacting via a spherically symmetric potential with finite-range $R$. This system is described by the radial Schr\"{o}dinger equation
\begin{equation}
 \left[-\frac{d^{2}}{dr^{2}}+\frac{\ell(\ell+1)}{r^{2}} - p^{2}\right] U_{\ell}^{(p)}(r)
  +
  2\mu \int_{0}^{R} d\prim{r} \,
  W(r,r^{\prime}) \,
  \,U_{\ell}^{(p)}(\prim{r})
 = 0 \,.
\label{eqn:SchrodingerEq-0002}
\end{equation}
where  $\mu$ is the reduced mass, and $W(r,r^{\prime})$ is the finite-range potential, $W(r,r^{\prime})=0$ for
$r>R$ or $r^{\prime}>R$. 
Therefore, for $r>R$  with an arbitrary normalization the solution of \eref{eqn:SchrodingerEq-0002} is written as
\begin{equation}
 U_{\ell}^{(p)}(r)=p^{\ell}\left[\cot\delta_{\ell}(p) 
 \,  S_{\ell}(pr) + C_{\ell}(pr)\right] \,.
 \label{eqn:radialwave-Singlechannel-0040}
\end{equation}
For later convenience we define the rescaled Riccati-Bessel and Riccati-Neumann functions by
\begin{align}
s_{\ell}(p,r) = p^{-\ell-1}S_{\ell}(pr)
\quad
\text{and}
\quad
c_{\ell}(p,r)  &  =p^{\ell}C_{\ell}(pr)\,.
 \label{eqn:rescaled-Riccati-functions}
 \end{align}
Insertion of the rescaled functions and the effective range expansion, \eref{eqn:radialwave-Singlechannel-0040} is rewritten as an expansion in powers of $p^{2}$,
\begin{align}
U_{\ell}^{(p)}(r)=  u_{0,\ell}(r)
+u_{2,\ell}(r) \, p^{2}
+u_{4,\ell}(r) \, p^{4}
+u_{6,\ell}(r) \, p^{6}
+\mathcal{O}(p^{8})\,,
 \label{eqn:radialwave-Singlechannel-0060}
\end{align}
where $u(r)$'s are defined in terms of the effective range expansion parameters by
\begin{align}
u_{0,\ell}(r) = &
\frac{-1}{a_{\ell}} \, s_{0,\ell}(r) + c_{0,\ell}(r),
 \label{eqn:radialwave-u0}
\\
u_{2,\ell}(r)= &
\frac{1}{2}r_{\ell} \, s_{0,\ell}(r)
-\frac{1}{a_{\ell}} \, s_{2,\ell}(r)
+c_{2,\ell}(r),
 \label{eqn:radialwave-u2}
\\
u_{4,\ell}(r)= &
P_{\ell} \, s_{0,\ell}(r)
+\frac{1}{2}r_{\ell} \, s_{2,\ell}(r)
-\frac{1}{a_{\ell}} \, s_{4,\ell}(r)
+c_{4,\ell}(r)\,,
 \label{eqn:radialwave-u4}
 \\
u_{6,\ell}(r)= &
Q_{\ell} \, s_{0,\ell}(r)
+ P_{\ell} \,  s_{2,\ell}(r)
+\frac{1}{2}r_{\ell} \, s_{4,\ell}(r)
-\frac{1}{a_{\ell}} \, s_{6,\ell}(r)
+c_{6,\ell}(r).
 \label{eqn:radialwave-u6}
\end{align}
and the functions $s_{n,\ell}(r)$ and $c_{n,\ell}(r)$ are given in \aref{append:Bessel-functions}.

\subsection{Charged particles}
We consider two particles interacting at long distances in addition to the short-range potential. The example we consider in detail is the system of two particles carrying electric charges $eZ_{1}$ and $eZ_{2}$. The radial Schr\"{o}dinger equation (\ref{eqn:SchrodingerEq-0002}) has a Coulomb potential term $\gamma/r$,
\begin{align}
 \left[-\frac{d^{2}}{dr^{2}}
 +\frac{\ell(\ell+1)}{r^{2}}
 +\frac{\gamma}{r}
 -p^{2}\right] V_{\ell}^{(p)}(r)
  + & 2\mu \int_{0}^{R}dr^{\prime}\,W(r,r^{\prime}) \, V_{\ell}^{(p)}(r^{\prime})
  = 0
 \,,
 \label{uncoup-charg-radialEq}%
\end{align}
where $\gamma = 2 \mu \alpha Z_{1}Z_{2}$. Here $V_{\ell}^{(p)}(r)$ is the rescaled radial
wave function of a two-body system of charged particles,
\begin{equation}
V_{\ell}^{(p)}(r)
=r \cdot R_{\ell}^{(p)}(r)\,.
\label{uncoup-charg-rescaledV}%
\end{equation}
We choose a normalization such that, for $r > R$,  $V_{\ell}^{(p)}(r)$ is
\begin{align}
V_{\ell}^{(p)}(r)= & p^{\ell} \, C_{\eta,\ell} \, \left[
\cot\tilde{\delta}_{\ell}(p) \times F^{(p)}_{\ell}(r)+ G^{(p)} _{\ell}(r)\right] \,,
 \nonumber\\
 = & p^{2\ell+1} \, C_{\eta,\ell}^{2} \, \cot\tilde{\delta}_{\ell}(p) \times f_{\ell}(p,r) +g_{\ell}(p,r)\,,
 \label{uncoup-charg-Vfunc-001}
\end{align}
where $F^{(p)}_{\ell}(r)$ and $G^{(p)} _{\ell}(r)$ are the regular and irregular Coulomb wave functions, and the related functions $f_{\ell}(p,r)$ and $g_{\ell}(p,r)$ are defined for later convenience as
\begin{align}
f_{\ell}(p,r) & = \frac{1}{p^{\ell+1} C_{\eta,\ell}} \, F^{(p)}_{\ell}(r) \,,
\label{uncoup-charg-fLpr-001}
\end{align}
\begin{align}
g_{\ell}(p,r) = p^{\ell} \, C_{\eta,\ell} \,G^{(p)} _{\ell}(r) 
= & \tilde{g}_{\ell}(p,r) 
+ \left[\gamma\,\tilde{h}_{\ell}(p) -ip^{2\ell+1} C_{\eta,\ell}^{2}\right] \, f_{\ell}(p,r)
\,.
 \label{uncoup-charg-gLpr-001}
\end{align}
See Appendix~\ref{append:CoulombSolutions} for the functions $f_{\ell}(p,r)$, $g_{\ell}(p,r)$ and $\tilde{g}_{\ell}(p,r)$.
The factor $C_{\eta,\ell}$ is given by
\begin{align}
C_{\eta,\ell}^{2} = \frac{2^{2\ell}}{\Gamma\left(2\ell+2\right)^{2}}
\prod_{s = 1}^{\ell} (s^{2}+\eta^{2}) \, C_{\eta,0}^{2} \,,
 \label{uncoup-charg-Cofeta-001}
\end{align}
and the function $\tilde{h}_{\ell}(p)$ is defined as
\begin{align}
\tilde{h}_{\ell}(p)
 = \frac{(2p)^{2\ell}}{\Gamma\left(2\ell+2\right)^{2}}
    \frac{|\Gamma\left(\ell+1+i\eta\right)|^{2}}{|\Gamma\left(1+i\eta\right)|^{2}}
     \left[ \psi(i\eta)+\frac{1}{2i\eta}-\log(i\eta)\right] \,,
 \label{uncoup-charg-hofp-001}
\end{align}
where
\begin{align}
\eta = \frac{\gamma}{2p} \,,
 \label{uncoup-charg-eta-001}
\end{align}
\begin{align}
C_{\eta,0}^{2} = \frac{2\pi\eta}{e^{2\pi\eta}-1} \,,
 \label{uncoup-charg-Cofeta-005}
\end{align}
and $\psi(z)$ is the digamma function. Using these new expressions given in Eqs.(\ref{uncoup-charg-fLpr-001})-(\ref{uncoup-charg-Cofeta-005}), we can rewrite Eq.~(\ref{uncoup-charg-Vfunc-001}) as, 
\begin{align}
V_{\ell}^{(p)}(r)=
\left[ p^{2\ell+1} \, C_{\eta,\ell}^{2} \, \left(\cot\tilde{\delta}_{\ell}(p)-i\right) + \gamma \,  \tilde{h}_{\ell}(p) \right] \times f_{\ell}(p,r) + \tilde{g}_{\ell}(p,r)\,,
 \label{uncoup-charg-Vfunc-005}
\end{align}
The expression in square brackets is the modified Coulomb effective range expansion~\cite{PhysRevA.30.1279},
\begin{equation}
 p^{2\ell+1} \, C_{\eta,\ell}^{2} \, \left(\cot\tilde{\delta}_{\ell}(p)-i\right) + \gamma \,  \tilde{h}_{\ell}(p)
  =\frac{-1}{a^{c}_{\ell}}+\frac{1}{2}r^{c}_{\ell} \, p^{2}
+P^{c}_{\ell} \, p^{4}
+Q^{c}_{\ell} \, p^{6}
+\mathcal{O}\left(  p^{8}\right)\,.
\label{uncoup-charg-coulERE}%
\end{equation}
Finally, the Coulomb wave function is written as an expansion in powers of $p^{2}$,
\begin{align}
V_{\ell}^{(p)}(r)=  v_{0,\ell}(r)
+v_{2,\ell}(r) \, p^{2}
+v_{4,\ell}(r) \, p^{4}
+v_{6,\ell}(r) \, p^{6}
+\mathcal{O}(p^{8}),
 \label{uncoup-charg-Vfunc-009}
\end{align}
where $v(r)$'s are written in terms of the modified Coulomb effective range expansion parameters,
\begin{align}
v_{0,\ell}(r) = &
\frac{-1}{a^{c}_{\ell}} \, f_{0,\ell}(r) + g_{0,\ell}(r),
 \label{uncoup-charg-V0}
\\
v_{2,\ell}(r)= &
\frac{1}{2}r^{c}_{\ell} \, f_{0,\ell}(r)
-\frac{1}{a^{c}_{\ell}} \, f_{2,\ell}(r)
+g_{2,\ell}(r),
 \label{uncoup-charg-V2}
\\
v_{4,\ell}(r)= &
P^{c}_{\ell} \, f_{0,\ell}(r)
+\frac{1}{2}r^{c}_{\ell} \, f_{2,\ell}(r)
-\frac{1}{a^{c}_{\ell}} \, f_{4,\ell}(r)
+g_{4,\ell}(r)\,,
 \label{uncoup-charg-V4}
 \\
v_{6,\ell}(r)= &
Q^{c}_{\ell} \, f_{0,\ell}(r)
+P^{c}_{\ell} f_{2,\ell}(r)
+\frac{1}{2}r^{c}_{\ell} \, f_{4,\ell}(r)
-\frac{1}{a^{c}_{\ell}} \, f_{6,\ell}(r)
+g_{6,\ell}(r).
 \label{uncoup-charg-V6}
\end{align}

\subsection{Wronskian integral formula}
\label{chap1:Wronskian-integral formula}
In this section we follow the steps in Ref.~\cite{HammerDean27} and derive the Wronskian integral formula for a two-particle system. We consider two solutions of Eq.~(\ref{eqn:SchrodingerEq-0002}) for momentum $p_{a}$ and $p_{b}$,
\begin{equation}
 \left[-\frac{d^{2}}{dr^{2}}
 +\frac{\ell(\ell+1)}{r^{2}}
 -p_{a}^{2}\right] U_{a}(r)
  + 2\mu \int_{0}^{R}dr^{\prime}\,W(r,r^{\prime}) \,  U_{a}(r')
  = 0
 \,,
 \label{uncoup-generalDifEq-pa}
\end{equation}
\begin{equation}
 \left[-\frac{d^{2}}{dr^{2}}
 +\frac{\ell(\ell+1)}{r^{2}}
 -p_{b}^{2}\right] U_{b}(r)
  + 2\mu \int_{0}^{R}dr^{\prime}\,W(r,r^{\prime}) \,  U_{b}(r')
  = 0
 \,,
 \label{uncoup-generalDifEq-pb}
\end{equation}
where we use the shorthand notation $U_{a}(r)=U_{\ell}^{(p_{a})}(r)$ and $U_{b}(r)=U_{\ell}^{(p_{b})}(r)$. We multiply Eq.~(\ref{uncoup-generalDifEq-pa}) by $V_{b}(r)$ on the left, Eq.~(\ref{uncoup-generalDifEq-pb}) by $V_{a}(r)$, and then subtracting the resulting equations yields
\begin{align}
U_{a}(r) & U'_{b}(r)-U_{b}(r)U'_{a}(r) +
\left(p_{b}^{2}-p_{a}^{2}\right)U_{b}(r)U_{a}(r)
 \nonumber\\
 &= 2\mu\int_{0}^{R}dr^{\prime}\, \left[U_{a}(r) \,W(r,r^{\prime}) \, U_{b}(r^{\prime})-U_{b}(r) \,W(r,r^{\prime}) \, U_{a}(r^{\prime})\right] \,.
 \label{uncoup-generalDifEq-papb}%
\end{align}
Integrating Eq.~(\ref{uncoup-generalDifEq-papb}) from radius $\rho$ to some radius $r\geq R$, we get
\begin{align}
[U_{a} & U''_{b}-U_{b}U''_{a}]|_{\rho}^{r} +
\left(p_{b}^{2}-p_{a}^{2}\right)\int_{0}^{R}dr^{\prime} \, U_{b}(r')U_{a}(r')
 \nonumber\\
 &= 2\mu\int_{\rho}^{R}dr \int_{0}^{R}dr^{\prime}\, \left[U_{a}(r) \,W(r,r^{\prime}) \, U_{b}(r^{\prime})-U_{b}(r) \,W(r,r^{\prime}) \, U_{a}(r^{\prime})\right] \,.
 \label{uncoup-generalDifEq-papb-001}%
\end{align}
 Now for the right hand side of this equation using the condition given in Eq.~(\ref{eqn:regularity}), we finally obtain the Wronskian integral formula,
\begin{equation}
 \frac{W[U_{\ell}^{(p_{b})},U_{\ell}^{(p_{a})}](r)}{p_{b}^{2}-p_{a}^{2}}
 =\int_{0}^{r}
 U_{\ell}^{(p_{a})}(r^{\prime}) \, U_{\ell}^{(p_{b})}(r^{\prime})\,dr^{\prime} \,.
\label{uncoup-wrons-integ-formula-001}
\end{equation}
In the low energy regime, when we set $ p_{a} = p_{b} =p $, Eq.~(\ref{uncoup-wrons-integ-formula-001}) reads
\begin{align}
W[u_{2,\ell}&,u_{0,\ell}](r)
+ 2 p^{2}  \, W[u_{4,\ell},u_{0,\ell}](r)
+p^{4}  \, W[u_{4,\ell},u_{2,\ell}](r)
\nonumber\\
&+ 3 p^{4}  \, W[u_{6,\ell},u_{0,\ell}](r)
 - \int_{0}^{r}
 \left[U_{\ell}^{(p)}(r^{\prime})\right]^{2}\,dr^{\prime}
 +\mathcal{O}(p^{6}) =0
 \ \,.
\label{uncoup-wrons-integ-formula-005}
\end{align}
In Appendix~\ref{append:Wronskians} the Wronskians of the functions $u_{n}(r)$ are given in terms of the scattering parameters.

The integral terms can be rearranged and rewritten as
\begin{align}
\int_{0}^{r}
 \left[U_{\ell}^{(p)}(r^{\prime})\right]^{2}\,dr^{\prime}
=
\int_{0}^{\infty}
 \left[U_{\ell}^{(p)}(r^{\prime})\right]^{2}\,dr^{\prime}
 -
 \int_{r}^{\infty}
  \left[U_{\ell}^{(p)}(r^{\prime})\right]^{2}\,dr^{\prime}
  \,.
\label{uncoup-wrons-integ-formula-009}
\end{align}
Since Eq.~(\ref{eqn:radialwave-Singlechannel-0060}) is the solution of the function $U_{\ell}^{(p)}(r)$ for $r>R$, it can be used in the second integral of the right hand side. Inserting Eq.~(\ref{eqn:radialwave-Singlechannel-0060}) and Eq.~(\ref{eqn:wronsk-005})-(\ref{eqn:wronsk-015}) into Eq.~(\ref{uncoup-wrons-integ-formula-005}) we get\footnote{Eq.~(\ref{uncoup-wrons-integ-formula-001})-(\ref{uncoup-delta4-400}) are valid for Coulomb case. In that case, $U_{\ell}^{(p)}(r)$ and $u_{n,\ell}(r)$  are replaced by $V_{\ell}^{(p)}(r)$ and $v_{n,\ell}(r)$, respectively. In addition, a superscript $c$ is used in the scattering parameters to denote the Coulomb scattering parameters ($a^{c}_{\ell}$ ,$r^{c}_{\ell}$, $P^{c}_{\ell}$ and $Q^{c}_{\ell}$).}
\begin{align}
\int_{0}^{\infty}
\left[U_{\ell}^{(p)}(r^{\prime})\right]^{2}\,dr^{\prime}
 =
 -
 &
 \frac{r_{\ell}}{2}
+\varDelta_{1,\ell}
- 2\left(P_{\ell}-\varDelta_{2,\ell}\right) \, p^{2}
\nonumber\\
&
-
3\left(Q_{\ell}-\varDelta_{3,\ell}-\frac{\varDelta_{4,\ell}}{3}\right)\,p^{4}
+\mathcal{O}(p^{6})
 \,,
\label{uncoup-wrons-integ-formula-011}
\end{align}
where $\varDelta_{n,\ell}$ are integration constants and calculated from the following equations,
\begin{align}
\varDelta_{1,\ell} = \frac{1}{2}\frac{d}{dr} b_{1,\ell}(r)
+
\int_{r}^{\infty} dr^{\prime} \left[u_{0,\ell}(r^{\prime})\right]^{2}
\,,
 \label{uncoup-delta1-100}
\end{align}
\begin{align}
\varDelta_{2,\ell} = \frac{d}{dr} b_{2,\ell}(r)
+
\int_{r}^{\infty} dr^{\prime} \left[u_{2,\ell}(r^{\prime})u_{0,\ell}(r^{\prime})\right]
\,,
 \label{uncoup-delta2-200}
\end{align}
\begin{align}
\varDelta_{3,\ell} = \frac{d}{dr} b_{3,\ell}(r)
+
\int_{r}^{\infty} dr^{\prime} \left[u_{4,\ell}(r^{\prime})u_{0,\ell}(r^{\prime})\right]
\,,
 \label{uncoup-delta3-300}
\end{align}
\begin{align}
\varDelta_{4,\ell} = \frac{d}{dr} b_{4,\ell}(r)
+
&
\int_{r}^{\infty} dr^{\prime} \left[u_{2,\ell}(r^{\prime})\right]^{2}
-
\int_{r}^{\infty} dr^{\prime} \left[u_{4,\ell}(r^{\prime})u_{0,\ell}(r^{\prime})\right]
\,,
 \label{uncoup-delta4-400}
\end{align}
where the  $b_{n,\ell}(r)$ functions are given in Appendix~\ref{append:Wronskians}.

\section{Loosely bound systems}
\subsection{Asymptotic Normalization Coefficients}

\subsubsection{Neutral case}
\label{sec:uncoupled-ANC}

The bound state wave function with momentum $p=i\kappa$ in the asymptotic region is
\begin{align}
\psi_{\ell}(r)
=  i^{\ell} A \, h^{(1)} _{\ell}(i \kappa r)\,
\label{uncoupANC-neut-asymWave-001}
\end{align}
where $A$ is the asymptotic normalization coefficient (ANC) and $h^{(1)}_{\ell}$ is the Riccati Hankel function. The bound state solution is normalized according to
\begin{align}
\int_{0}^{\infty} \left[ \psi_{\ell}(r^{\prime}) \right]^{2} dr^{\prime} = 1
\,.
\label{uncoupANC-neut-asymWave-005}
\end{align}
Furthermore, for the bound state regime, we have
\begin{align}
\cot\delta_{\ell}(i\kappa) = i \,,
\label{uncoupANC-neut-ERE-001}
\end{align}
and the effective range expansion reads
\begin{equation}
(-1)^{\ell} \kappa^{2\ell+1}
=\frac{1}{a_{\ell}}
+\frac{1}{2} r_{\ell} \, \kappa^{2}
-P_{\ell} \kappa^{4}
+Q_{\ell} \, \kappa^{6}
+\mathcal{O}(\kappa^{8}),
\label{eqn:ERE-0002}
\end{equation}
The relation between $\psi_{\ell}(r)$ and  $U_{\ell}^{(i\kappa)}(r)$ can be obtained, for $ r > R $,
\begin{align}
 U_{\ell}^{(i\kappa)}(r)= & (i\kappa)^{\ell} \, \left[
 \cot\delta_{\ell}(i\kappa) \times S_{\ell}(i\kappa r)+ C _{\ell}(i \kappa r)\right]
 \nonumber\\
 = & (i\kappa)^{\ell} \, \left[
 i \times S_{\ell}(i\kappa r)+ C _{\ell}(i \kappa r)\right]
 = (i \kappa)^{\ell} \,  h^{(1)}_{\ell}(i\kappa r)
 =  \frac{\kappa^{\ell}}{A} \,  \psi_{\ell}(r) \,
 \label{eqn:uncoup-ANC-001}
\end{align}
Inserting Eq.~(\ref{eqn:uncoup-ANC-001}) into the integral term  of Eq.~(\ref{uncoup-wrons-integ-formula-011})
we obtain the following expression for the ANC
\begin{align}
A_{\ell}
\approx
\frac{\kappa^{\ell}}{\sqrt{-\frac{r_{\ell}}{2}
+\varDelta_{1,\ell}
+2\left(P_{\ell}-\varDelta_{2,\ell}\right) \, \kappa^{2}
- 3\left(Q_{\ell}-\varDelta_{3,\ell}-\frac{\varDelta_{4,\ell}}{3}\right)\,\kappa^{4}}
}
 \,,
\label{uncoupANC-neut-001}
\end{align}
where integration constants $\Delta_{n,\ell}$ are given in Table~\ref{tab:delta-neutral}.
\begin{table}
\caption{The integration constants $\Delta_{n,\ell}$ of a two-neutral-particles system for $\ell \leq 2$.}%
\label{tab:delta-neutral}
\centering%
\rowcolors{3}{gray!15}{white}
\begin{tabular}{c||c|c|c|c}
\hline\hline
$\ell$& $\varDelta_{1,\ell}$  & $\varDelta_{2,\ell}$ & $\varDelta_{3,\ell}$ & $\varDelta_{4,\ell}$ \\ \hline\hline
0&$\frac{\kappa}{2}$&$0$&$0$ & 0	\\ 
1&$-\frac{3\kappa^{3}}{2}$&$0$&$0$ & 0		\\ 
2&$\frac{5\kappa^{5}}{2}$&$0$&$0$ & 0		\\
\hline\hline	
\end{tabular}
\end{table}
An alternative expression can be obtained by eliminating the effective range $r_{\ell}$ using Eq.~(\ref{eqn:ERE-0002})
\begin{align}
A_{\ell}
\approx
\frac{\kappa^{\ell+1}}{\sqrt{\frac{1}{a_{\ell}}
-(-1)^{\ell} \kappa^{2\ell+1}
+\varDelta_{1,\ell}\kappa^{2}
+\left(P_{\ell}-2\varDelta_{2,\ell}\right) \, \kappa^{4}
-\left(2Q_{\ell}-3\varDelta_{3,\ell}-\varDelta_{4,\ell}\right)\,\kappa^{6}}
}
 \,.
\label{uncoupANC-neut-002}
\end{align}
This expression agrees with Eq.(11) of Ref.~\cite{Konig20121450} up to the given order.

\subsubsection{Coulomb case}
\label{sec:Coulm:uncoupled-ANC}
The bound state Coulomb wave function with binding momentum $\kappa$ in the asymptotic region is
\begin{align}
\psi^{c}_{\ell}(r)
=  A^{c} \, W_{-i\eta,\ell+\frac{1}{2}}(2 \kappa r)\,
 \label{uncoupANC-Coulm-001}
\end{align}
where $A^{c}$ is the Coulomb-ANC and $ W_{-i\eta,\ell+\frac{1}{2}}$ is the Whittaker function. The bound state solution is normalized according to
\begin{align}
\int_{0}^{\infty} \left[ \psi^{c}_{\ell}(r^{\prime}) \right]^{2} dr^{\prime} = 1
\,.
 \label{uncoupANC-Coulm-005}
\end{align}
In the bound state regime, we have
\begin{align}
\cot\tilde{\delta}_{\ell}(i\kappa) = i \,.
 \label{uncoupANC-Coulm-009}
\end{align}
Therefore, the wave function becomes, for $ r > R $,
\begin{align}
V_{\ell}^{(i\kappa)}(r)= & (i\kappa)^{\ell} \, C_{\eta,\ell} \, \left[
 i \, F^{(i\kappa)}_{\ell}(r)+ G^{(i\kappa)} _{\ell}(r)\right]
 \nonumber\\
 =
 &
 (i\kappa)^{\ell} \, C_{\eta,\ell}
 \,
 e^{i\sigma_{\ell}} e^{-i\pi\left(\ell+i\eta\right)}
 \,
 W_{-i\eta,\ell+\frac{1}{2}}(2\kappa r)
 \,,
 \label{uncoupANC-Coulm-011}
\end{align}
where $\sigma_{\ell}$ is the Coulomb phase shift,
\begin{align}
 e^{i\sigma_{\ell}} 
 =
 \frac
 {\Gamma\left(\ell+1+i\eta\right)}
 {\Gamma\left(\ell+1-i\eta\right)}\,.
 \label{uncoupANC-Coulm-015}
\end{align}
The wave function $V_{\ell}^{(i\kappa)}(r)$ in terms of the bound state Coulomb wave function reads
\begin{align}
V_{\ell}^{(i\kappa)}(r)=
 &
 \frac{\kappa^{\ell}}{A^{c}} \, \tilde{C}_{\eta,\ell}
 \,
 \psi^{c}_{\ell}(r)
 \,,
 \label{uncoupANC-Coulm-019}
 \end{align}
where
\begin{align}
\tilde{C}_{\eta,\ell}
=
\frac{2^{\ell} \Gamma\left(\ell+1+i\eta\right)}{\Gamma\left(2\ell+2\right)}
\,.
 \label{uncoupANC-Coulm-025}
\end{align}
Finally, the ANC can be written as
\begin{align}
|A^{c}_{\ell}|
\approx
\frac{ \kappa^{\ell}  \tilde{C}_{\eta,\ell}}{\left[
-\frac{r^{c}_{\ell}}{2}
+\varDelta^{c}_{1,\ell}
+2\left(P^{c}_{\ell}-\varDelta^{c}_{2,\ell}\right) \, \kappa^{2}
- 3\left(Q^{c}_{\ell}-\varDelta^{c}_{3,\ell}-\frac{\varDelta^{c}_{4,\ell}}{3}\right)\,\kappa^{4}
\right]^{1/2}}
 \label{uncoupANC-Coulm-029}\,.
\end{align}
Integration constants $\varDelta^{c}_{n,\ell}$ for $\ell \leq 2$ in the Coulomb case are given in Table~\ref{tab:delta-coulomb}.
\begin{table}[H]
\caption{Integration constants $\varDelta^{c}_{n,\ell}$ calculated from Eqs.~(\ref{uncoup-delta1-100})-(\ref{uncoup-delta4-400}) for $\ell \leq 2$ in Coulomb case .}%
\label{tab:delta-coulomb}
\centering%
\rowcolors{3}{gray!15}{white}
\begin{tabular}{c||c|c|c|c}
\hline\hline
$\ell$& $\varDelta^{c}_{1,\ell}$  & $\varDelta^{c}_{2,\ell}$ & $\varDelta^{c}_{3,\ell}$ & $\varDelta^{c}_{4,\ell}$ \\ \hline\hline
0&$\frac{1}{3\gamma}$&$\frac{2}{15\gamma^{3}}$&$\frac{16}{63\gamma^{5}}$ & 0	\\ 
1&$\frac{\gamma}{108}$&$\frac{11}{270\gamma}$&$\frac{62}{2835\gamma^{3}}$ & 0		\\
2&$\frac{\gamma^{3}}{43200}$&$\frac{17\gamma}{36000}$&$\frac{191}{113400\gamma}$ & 0			
\\
\hline\hline
\end{tabular}
\end{table}
An alternative expression for $A^{c}$ can be found by eliminating the effective range parameter $r_{\ell}^{c}$ in Eq.~(\ref{uncoupANC-Coulm-029}) using the Coulomb modified effective range expansion in the bound state regime,
\begin{equation}
\gamma \, \tilde{h}_{\ell}(i\kappa)
=\frac{-1}{a^{c}_{\ell}}-\frac{1}{2}r^{c}_{\ell}\kappa^{2}
+P^{c}_{\ell}\kappa^{4}
-Q^{c}_{\ell} \kappa^{6}
+\mathcal{O}\left(  k^{8}\right)\,,
 \label{uncoupANC-Coulm-031}
\end{equation}
where $\gamma \, \tilde{h}_{\ell}(i\kappa)$ are given in Table~\ref{tab:gamma-tildeH} for $\ell \leq 2$.
\begin{table}[tbh]
\caption{The function $\gamma \, \tilde{h}_{\ell}(i\kappa)$ in the Coulomb modified effective range expansion for $\ell \leq 2$.}%
\label{tab:gamma-tildeH}
\centering%
\rowcolors{3}{gray!15}{white}
\begin{tabular}{c||c}
\hline\hline
$\ell$& $\gamma \, \tilde{h}_{\ell}(i\kappa)$ \\ \hline\hline
0&$-\frac{\kappa^{2}}{3\gamma}+\frac{2 \kappa^{4}}{15\gamma^{3}}-\frac{16 \kappa^{6}}{63\gamma^{5}} +\mathcal{O}(\kappa^{8})$	\\ 
1&$-\frac{\gamma \kappa^{2}}{108}+\frac{11 \kappa^{4}}{270\gamma}-\frac{62 \kappa^{6}}{2835\gamma^{3}} +\mathcal{O}(\kappa^{8})$	\\
2&$-\frac{\gamma^{3}\kappa^{2}}{43200}
+\frac{17\gamma\kappa^{4}}{36000}
-\frac{191\kappa^{6}}{113400\gamma} +\mathcal{O}(\kappa^{8})$
\\
\hline\hline
\end{tabular}
\end{table}
We find
\begin{align}
|A^{c}_{\ell}|
\approx
\frac{\kappa^{\ell+1}  \tilde{C}_{\eta,\ell}}
{\left[
\frac{1}{a^{c}_{\ell}}
+\gamma \, \tilde{h}_{\ell}(i\kappa)
+\varDelta^{c}_{1,\ell}\kappa^{2}
+\left(P^{c}_{\ell}-2\varDelta^{c}_{2,\ell}\right) \, \kappa^{4}
- \left(2Q^{c}_{\ell}-3\varDelta^{c}_{3,\ell}-\varDelta^{c}_{4,\ell}\right)\,\kappa^{6}
\right]^{1/2}}
 \label{uncoupANC-Coulm-030}
\end{align}
This expression matches with Eq.~(85) of Ref.~\cite{Koenig:2012bv} [cf. Eq.(5.87) of Ref.\cite{konig2013effective}] up to the given order and Eq.~(21) of Ref.~\cite{SparenbergBaye2010} for $\ell=2$. The relation between our convention of the effective range parameters and the convention of  Ref.~\cite{SparenbergBaye2010}
is 
\begin{align}
\tilde{A}^{c}
=
\left(\frac{2^{\ell}\ell!}{\Gamma(2\ell+2)}\right)^{2}
A^{c}\,.
\end{align}

%
%
%
%
%
%
%
%
%
%
%
%
%

%% file: Chapter-3.tex
\chapter{Lattice Effective Field Theory}
\label{chap:LatticeEFT}

\section{Introduction}
\label{chap3:introduction}
In the first theoretical descriptions initiated by Yukawa~\cite{yukawa1935} the strong nuclear forces between nucleons are mediated by massive bosons called mesons. Phenomenological models which were only based on one-boson-exchange (OBE) well described the strong interactions at large distances ~\cite{PhysRev.135.B434,Erkelenz1974191,PhysRevD.17.768}. Later, efforts were made to construct highly sophisticated potential models in order to improve the shape of the potentials in intermediate ranges~\cite{Jackson1975397,PhysRevC.21.861,Stoks1993NijII,PhysRevC.51.38,PhysRevC.63.024001}. For more on potential models and a historical review see Ref.~\cite{Machleidt2001,Machleidt20111}.

At the same time attempts were taken to describe the strong interactions between nucleons within the framework of quantum field theory (QFT), and a breakthrough came with the discovery of quarks~\cite{GellMann1964214}. Quarks are elementary particles in the Standard Model and they are constituents of hadrons. The theory of the strong interactions is called quantum chromodynamics (QCD). The fundamental degrees of freedom  in QCD are gluons as well as quarks. Quarks have six different flavors (up, down, strange, charmed, top, bottom) and three colors (red, green, blue). Colors are the charges of quarks, and the strong interactions are governed by a non-abelian gauge theory with the SU(3)-color group~\cite{FritzschGell-Mann1972,Fritzsch1973365}. The eight generators of the SU(3) group correspond with the eight gluons.

The running coupling constant of the strong interactions makes it possible to carry out calculations of observables perturbatively at higher energies. However, the same feature of the theory causes a breakdown in  perturbative treatments at low energies. This clearly manifests the necessity of non-perturbative methods in order to predict observables from QCD. An elegant method was proposed by Wilson~\cite{PhysRevD.10.2445}. He formulated lattice gauge theory on a discretized space-time lattice, which is commonly known as lattice QCD (LQCD), and this method gave access to study QCD in the low energy limit or at large distances using numerical methods. Therefore, LQCD has become a powerful approach to probe the structure of hadrons using quarks and gluons as degrees of freedom~\cite{SilasEtAl2008,RevModPhys.84.449}. Also, LQCD has been used for studying elastic scattering of meson-mesons~\cite{PhysRevLett.71.2387,PhysRevD.67.014502,Lin2003229,PhysRevD.73.054503,PhysRevD.74.114503,PhysRevD.77.094507}, meson-baryon~\cite{PhysRevD.52.3003}, and baryon-baryon~\cite{Beane2004106,PhysRevLett.97.012001,PhysRevLett.99.022001,Beane200762}.

Another direct approach to access the low-energy regime of QCD is effective field theory (EFT) which is based upon the seminal work of Weinberg~\cite{Weinberg:1978kz}. This idea is rooted in a general concept of separation of scales. Physical processes and observables are well defined in energy scales relevant to the dynamics of the system. In the low energy limit of QCD since quarks and gluons are strongly confined in hadrons by color charge forces, the relevant degrees of freedom at large distance scales are hadrons, instead of quarks and gluons. Therefore, a scale separation is inevitable here and it is indeed the key point of EFT. The spectrum of hadrons shows a visible large gap between the masses of light mesons ($\pi^{0}$,$\pi^{\pm}$) and the masses of  nucleons ($N$) and heavier mesons. The EFT formulation sets a soft scale $Q$ at the mass scale of light mesons and a hard scale $\Lambda$ at the scale of the nucleon mass. Then using EFT one can perform systematic calculations as an expansion in powers of a small parameter $Q/\Lambda$. This formulation is known as chiral effective field theory ($\chi$EFT).

The hard scale $\Lambda$ also corresponds to the spontaneous chiral symmetry breaking scale. Chiral symmetry is a symmetry of QCD associated with the smallness of the light quark masses. In the limit of zero quark mass, one can do independent unitary transformations of the left and the right components of the quarks. Chiral EFT enforces the fact that chiral symmetry must be manifested in the phenomenology of hadrons at low energies. Chiral EFT was first applied to the elastic scattering of $\pi$$\pi$~\cite{Gasser1984142} and $\pi$$N$~\cite{Gasser1988779}. Later, its applicability to the nuclear structure and interactions was derived by Weinberg~\cite{Weinberg1990288,Weinberg19913}. Refs.~\cite{Machleidt20111} and \cite{RevModPhys.81.1773} provide detailed reviews on the subject.

In such systems where momenta is smaller than the pion mass the pionless effective field theory ($\slashed{\pi}$EFT) is a more economic and efficient formulation to use. In the $\slashed{\pi}$EFT  pions are integrated out and the interactions are only local contact interactions between dynamical nucleons~\cite{vanKolck:1998bw,Bedaque:1998kg,Bedaque:1998km,Bedaque:1999ve,Chen:1999tn,Platter:2004zs,
Hammer:2006ct}. For example, the deuteron binding momentum is $\gamma_{d} =45$ MeV which is much smaller than the lightest pion mass $m_{\pi} = 140$ MeV. This clearly suggests here that the pion mass can be considered as the hard scale here since the relevant energy scale is much lower than the pion mass.

Interactions derived from EFT mentioned above have been combined with powerful numerical methods to study low energy nuclear physics from first principles. This growing field is known as lattice effective field theory (lattice EFT). Lattice EFT was formulated on discretized space-time from the chiral EFT~\cite{Borasoy2342007}. Ref.~\cite{Lee:2008fa} provides a detailed review of lattice EFT from the $\slashed{\pi}$EFT and chiral EFT. In the last decade lattice EFT methods have proven its successes with significant contributions made to nuclear structure studies~\cite{Borasoy1312007,Borasoy3352008,PhysRevLett.104.142501,Epelbaum3452010}. Some recent successes of lattice EFT are the \textit{ab initio} calculation of Hoyle state of carbon-12~\cite{PhysRevLett.106.192501}, which is the states that is responsible for the carbon-12 production in the stars, and \textit{ab initio} calculation of the spectrum and structure of $^{16}$O~\cite{PhysRevLett.112.102501}. Also, very recently these calculations have been extended to medium mass nuclei~\cite{Lähde2014110}. Nuclear reaction calculations from lattice EFT were initiated by Ref.~\cite{Rupak:2013aue}.

\section{The path integral}
Following Dirac's pioneer work~\cite{PAMDirac} that suggested that there is a connection between the exponent of the classical action $e^{iS[q(t)]}$ and the transition amplitude of a quantum mechanical particle at two points, Feynman was the first who incorporated classical Lagrangian approaches into quantum mechanics~\cite{RevModPhys.20.367}. With his work, Feynman reformulated quantum mechanics and quantum field theory by the so called \textit{path integral} (PI) method. The PI formulation has brought particular advantages in quantum field theories and become very a important tool for numerical techniques in quantum systems.

Starting with the time evolution operator $e^{-iHt}$ in the Hamiltonian formalism, the transition amplitude of a quantum mechanical particle from an initial point $q_{I}$ to a final point $q_{F}$ is defined by
\begin{align}
\bra{q_{F}}e^{-iHt}\ket{q_{I}}\,,
\label{eqn:transition-amplitude-0001}
\end{align}
where $\ket{q}$ denote the complete set of states in the Dirac bra-ket notation $1 = \int dq \ket{q}\bra{q}$. Now, we want to obtain an expression from \eref{eqn:transition-amplitude-0001} in a path integral form. To achieve this we split the time $t$ into $L_{t}$ equal segments $\alpha_{t} = t/L_{t}$ and rewrite the evolution operator $e^{-iHt}$ as $L_{t}$ products of $e^{-iH \alpha_{t} }$. By insertion of the completeness relations between those segmented operators, \eref{eqn:transition-amplitude-0001} becomes the products of the transition amplitudes (the propagators) at two points over a small time segment $\alpha_{t}$,
\begin{align}
\bra{q_{F}}e^{-iHt}\ket{q_{I}} = 
\int \ldots\int & dq_{1}dq_{2}\ldots dq_{L_{t}-2}dq_{L_{t}-1}
\nonumber\\
&
\times
\bra{q_{F}}e^{-iH\alpha_{t}}\ket{q_{L_{t}-1}}
\bra{q_{L_{t}-1}}e^{-iH\alpha_{t}}\ket{q_{L_{t}-2}} \ldots
\nonumber\\
&
\ldots
\times
\bra{q_{3}}e^{-iH\alpha_{t}}\ket{q_{2}}
\bra{q_{2}}e^{-iH\alpha_{t}}\ket{q_{1}}
\bra{q_{1}}e^{-iH\alpha_{t}}\ket{q_{I}}\,.
\label{eqn:transition-amplitude-0005}
\end{align}
An individual propagator for the Hamiltonian $H = \frac{\hat{p}^{2}}{2 m} + V(\hat{q})$ which describes a particle in a potential $V(\hat{q})$ is of the following form,
\begin{align}
\bra{q_{n+1}}e^{-iH \alpha_{t}}\ket{q_{n}}
=
&
\bra{q_{n+1}}e^{-i\frac{\hat{p}^{2}}{2 m}\alpha_{t}}\ket{q_{n}} e^{ -i V(q_{n})\alpha_{t}}
\label{eqn:Indv-Propagator-0001}
\\
&
=
\sqrt{\frac{-im}{2\pi\alpha_{t}}}
e^{i\frac{m\alpha_{t}}{2}\left(\frac{q_{n+1}-q_{n}}{\alpha_{t}}\right)^{2}}
e^{-i V(q_{n})\alpha_{t}}\,.
\label{eqn:Indv-Propagator-0005}
\end{align}
This is an infinitesimal transition amplitude which describes the particle's evolution from $q_{n+1}$ to $q_{n}$. In \eref{eqn:Indv-Propagator-0001} we use the Baker-Campbell-Hausdorff formula\footnote{$e^{A+B+\frac{1}{2}[A,B]+\ldots} = e^{A}e^{B}$}, and from \eref{eqn:Indv-Propagator-0001} to \eref{eqn:Indv-Propagator-0005} we use the state $\ket{p}$ which is the eigenstate of $\hat{p}$ and whose normalization is such that $\int \frac{dp}{2\pi}\ket{p}\bra{p} = 1$.
Plugging in the individual propagator given by \eref{eqn:Indv-Propagator-0005} into \eref{eqn:transition-amplitude-0005} we get
\begin{align}
\bra{q_{F}}e^{-iHt}\ket{q_{I}} =
\left(\frac{-im}{2\pi \alpha_{t}}\right)^{L_{t}/2}
\int \prod_{n = 1}^{L_{t}-1} dq_{n}
\exp{\left\{i\alpha_{t}\sum_{k=0}^{N-1} \left[
\frac{m}{2}\left(\frac{q_{k+1}-q_{k}}{\alpha_{t}}\right)^{2}-V(q_{k})\right]\right\}}
\,,
\label{eqn:transition-amplitude-0010}
\end{align}
and in the continuum limit $\alpha_{t} \to 0$ \eref{eqn:transition-amplitude-0010} is 
\begin{align}
\bra{q_{F}}e^{-iHt}\ket{q_{I}} = \int\mathcal{D}q(t) \, e^{iS[q(t)]},
\label{eqn:Feynman-PI-0001}
\end{align}
where the short-hand notation for the integral over all paths is
\begin{align}
\int\mathcal{D}q(t) =
\left(\frac{-im}{2\pi \alpha_{t}}\right)^{L_{t}/2}
\left(
\prod_{n = 1}^{L_{t}-1} dq_{n}
\int
\right)
\end{align}
and $S[q(t)]$ is the action defined in terms of the classical Lagrangian $L(q,\dot{q},t) = \frac{m}{2} \dot{q}^{2} -V(q)$,
\begin{align}
\exp{ i S[q(t)]} = \exp{ i \int dt L(q,\dot{q},t) }\,.
\end{align}
In this formalism the classical Lagrangian is the fundamental quantity. \eref{eqn:Feynman-PI-0001} is the Feynman PI formulation which reformulates the quantum mechanical amplitude as the integral over all possible paths weighted by $e^{iS}$.

The weighting function in the PI has  an oscillatory nature. For later convenience and the favor of numerical methods to be employed we want to suppress these oscillations and desire the weighting function to be positive semi-definite and  non-oscillating. Therefore, rotation\footnote{This is the so called Wick rotation. The integrand of the action is rotated from the Re$\,t$ axis to the Im$\,t$ in the complex $t$-plane.} to the Euclidean time direction $it  \to \tau$ is a crucial step to obtain the Euclidean action $S_{E}$,
\begin{align}
S_{E}[q(\tau)] = \int d\tau \, L_{E}(q,\dot{q},\tau)= \int d\tau\ \left[\frac{m}{2} \dot{q}^{2} +V(q)\right]\,,
\end{align}
and the Euclidean time formulation of the PI becomes
\begin{align}
\bra{q_{F}}e^{-iHt}\ket{q_{I}} =
\bra{q_{F}}e^{-H \tau}\ket{q_{I}} = \int\mathcal{D}q(\tau) \, e^{-S_{E}[q(\tau)]}
\,.
\label{eqn:Feynman-PI-0005}
\end{align}

The PI formulations derived above hold for any quantum system as well as  quantum field theory. \eref{eqn:Feynman-PI-0005} can be rearranged according to the following table,
\begin{center}
\rowcolors{1}{gray!15}{white}
\begin{tabular}{l}
\parbox{5cm}{$$q(\tau)  \to \phi(\vec{r})$$}\\
\parbox{5cm}{$$S_{E}[q(\tau)]   \to S_{E}(\phi)$$}\\
\parbox{5cm}{$$\int\mathcal{D}q(\tau)  \to \int\mathcal{D}\phi$$}
\end{tabular}
\end{center}
and the PI formulation for fields becomes
\begin{align}
\bra{\phi_{F}(\vec{r})}e^{-H\tau}\ket{\phi_{I}(\vec{r})} = \int\mathcal{D}\phi \, e^{-S_{E}(\phi)}.
\label{eqn:Feynman-PI-0010}
\end{align}
Here $\phi$ is the field amplitude which is the dynamical variable of quantum field theory, and the euclidean action $S_{E}(\phi)$ is defined in terms of the Lagrangian in the non-relativistic limit of quantum field theory with
\begin{align}
\mathcal{L}_{E} = -\phi^{\dagger}\frac{\partial}{\partial \tau}\phi
&
-\frac{1}{2m}(\nabla \phi^{\dagger})\cdot (\nabla\phi)
\nonumber\\
&- \int d^{3}\vec{r}\,'
\,
\phi^{\dagger}(\vec{r})\phi(\vec{r})
 V(\vec{r}-\vec{r}\,')
 \phi^{\dagger}(\vec{r}\,')\phi(\vec{r}\,')
\end{align}
with the density operator
\begin{align}
\rho(\vec{r}) = \phi^{\dagger}(\vec{r}) \, \phi(\vec{r})\,.
\end{align}
The Hamiltonian of the Schr\"{o}dinger field equation is
\begin{align}
H 
&
= \int \, d^{3} \vec{r} \, \left[ \pi(\vec{r}) \frac{\partial}{\partial \tau}\phi(\vec{r})- \mathcal{L}_{E}\right] = H_{0} + H_{V}
\,,
\label{eqn:Hamiltonian-fields-0001}
\end{align}
where $\pi(\vec{r}) = \frac{\partial \mathcal{L}}{\partial(\partial\phi/\partial \tau)}$ is the momentum density conjugate to $\phi(\vec{r})$, $H_{0}$ is the free Hamiltonian 
\begin{align}
H_{0}
=
\frac{1}{2m}
\int d^{3} \vec{r} \,  \, \nabla \phi^{\dagger}(\vec{r}) \cdot \nabla\phi(\vec{r})
\,,
\label{eqn:Hamiltonian-fields-0002}
\end{align}
and $H_{V}$ is the interaction term
\begin{align}
H_{V}
=
\int\int d^{3}\vec{r} \, d^{3}\vec{r}\,' \, 
\rho(\vec{r})\,
 V(\vec{r}-\vec{r}\,')\,
 \rho({\vec{r}}\,')
\,.
\label{eqn:Hamiltonian-fields-0003}
\end{align}

\section{Grassmann variables}
The quantizations of bosonic fields are based upon the commutation relation, while only the anti-commutation relation yields a consistent theory for the fermionic fields. See Refs.\cite{AZeeQFT,PeskinSchroederQFT} for details and comprehensive discussions on the topic.

The anti-commuting variables that we are in need of are Grassmann variables. In order to study the PI for fermions we necessarily reconsider Eqs.(\ref{eqn:Feynman-PI-0010})-(\ref{eqn:Hamiltonian-fields-0001}) in terms of Grassmann variables. Therefore, we discuss some basic properties of Grassmann variables, and their integration and differentiation are introduced.

Let $\eta_{k}$ for $l=1,2,...,N$ be a set of Grassmann variables which satisfy the anti-commutation relation
\begin{align}
\eta_{k}\eta_{l}+\eta_{l}\eta_{k} \equiv \{\eta_{k},\eta_{l} \} = 0,
\label{eqn:anti-commutation-0001}
\end{align} 
for any $k$ and $l$. \eref{eqn:anti-commutation-0001} imposes that $\eta_{k}^{2}=0$. Assuming that Grassmann variables can be expanded in a Taylor series, the most general function of a Grassmann variable has the form $f(\eta) = a + b \, \eta$ where $a$ and $b$ are ordinary numbers.

The left and right differentiation of Grassmann variables are defined as
\begin{align}
\frac{\overrightarrow{\partial}}{\partial \eta_{k}} \eta_{l}
=
-\eta_{l}
\frac{\overleftarrow{\partial}}{\partial \eta_{k}} =\delta_{kl}\,,
\end{align} 
and using this we can write
\begin{align}
\frac{\overrightarrow{\partial}}{\partial \eta_{k}} \eta_{l} \eta_{k}
=
-\eta_{l}
=
 \eta_{l} \eta_{k}\frac{\overleftarrow{\partial}}{\partial\eta_{k}}\,.
\end{align} 
We use the standard notation for Grassmann variables so that the integration can be written by
\begin{align}
\int \, d\eta_{k} \, 1 = 0 \qquad \text{and} \qquad \int \, d\eta_{k} \, \eta_{k} = -\int \, \eta_{k} \, d\eta_{k} = 1\,.
\end{align}

A complex Grassmann variable can be written as a combination of two real Grassmann variables, $\theta = \eta_{k} + i \eta_{l}$. The properties of the complex Grassmann variables can be defined by using the properties of real Grassmann variables above, 
 \begin{align}
\theta^{2} = {\theta^{*}}^{2} =0 \quad \text{and} \quad \theta^{*} \theta = i[\eta_{k},\eta_{l}],
\end{align} 
where $\theta^{*} = \eta_{k} - i \eta_{l}$. Ref.~\cite{swanson1992path} provides detailed discussions on Grassmann algebra and some interesting physics applications of Grassmann variables.

\section{Lattice formulation}
\label{chap3:lattice-formulation}

In this section by following Ref.~\cite{Lee:2008fa,Lee:2008xs} we introduce a lattice formalism in which spacetime is a discretized periodic cubic lattice with $L^{3}\times L_{t}$ points. In the lattice formalism of our discussion, the lattice spatial spacing is denoted by $a$, and the lattice temporal spacing is $a_{t}$. Also, we introduce $\alpha_{t} = a_{t}/a$ as the ratio of temporal lattice spacing to spacial lattice spacing. Here we use dimensionless parameters and physical quantities in lattice units multiplied by the appropriate power of $a$.

We consider two-component fermions interacting via zero-range potentials, and  we call the two components $\uparrow$ and $\downarrow$ spins. The PI for fermions is defined by the anti-commuting Grassmann variables on lattice,
 \begin{align}
{Z} = \int
 \left[\prod_{n_{t},\vec{n},s=\uparrow,\downarrow}
 d\theta_{s}(n_{t},\vec{n})d\theta_{s}^{*}(n_{t},\vec{n})\right]  \, \, e^{- S[\theta,\theta^{*}]} \,.
 \label{eqn:Grassmann-path-int-100}
 \end{align}
Grassmann variables are periodic along the spatial direction,
\begin{align}
\theta(\vec{n}+L,n_{t})=\theta(\vec{n},n_{t}) 
\qquad
\theta^{*}(\vec{n}+L,n_{t})=\theta^{*}(\vec{n},n_{t})\,,
\end{align}
and anti-periodic in the temporal direction,
\begin{align}
\theta(\vec{n},n_{t}+L_{t})=-\theta(\vec{n},n_{t})
\qquad
\theta^{*}(\vec{n},n_{t}+L_{t})=-\theta^{*}(\vec{n},n_{t})\,.
\end{align}
The non-relativistic lattice action is defined by
\begin{align}
  S[\theta,\theta^{*}]
  =\sum_{n_t} \left\{S_{t}[\theta,\theta^{*},n_{t}]+S_{H_0}[\theta,\theta^{*},n_t]
  +S_{V}[\theta,\theta^{*},n_{t}]\right\} \,,
 \label{eqn:Action-003}
\end{align}
where $S_{t}$ and $S_{H_0}$ contain temporal hopping and spatial hopping terms of the free lattice action respectively,  
\begin{align}
  S_{t}[\theta,\theta^{*},n_{t}] = \sum_{s=\uparrow,\downarrow} &\sum_{\vec{n}}
  \left[ \theta_{s}^{*}(n_{t}+\hat{0},\vec{n}) \, \theta_{s}(n_{t},\vec{n})-\theta_{s}^{*}(n_{t},\vec{n}) \, \theta_{s}(n_{t},\vec{n})
    \right]\,,
  \label{eqn:Lattice-Action-temporal-001}
\end{align}
\begin{align}
  S_{H_0}[\theta,\theta^{*},n_{t}] =\frac{\alpha_{t}}{2m} \sum_{s=\uparrow,\downarrow} &\sum_{_{}\vec{n}}
  \sum_{l=1}^{3}
  \, \theta_{s}^{*}(n_{t},\vec{n}) \,
   \left[2 \, \theta_{s}(n_{t},\vec{n})- \theta_{s}(n_{t},\vec{n}+\hat{l})-
   \theta_{s}(n_{t},\vec{n}-\hat{l}) \right] \,.
  \label{eqn:Lattice-Action-spatial-001}
\end{align}
Here $\hat{0}$ denotes the lattice unit vector in the forward temporal direction. We can also write $S_{H_0}[\theta,\theta^{*},n_{t}]$
\begin{equation}
S_{H_0}[\theta,\theta^{*},n_{t}] =\alpha_{t}H^{\uparrow}_0[\theta_{s},\theta^*_{s},n_t]+\alpha_{t}H^{\downarrow}_0[\theta_{s},\theta^*_{s},n_t],
\end{equation}
and the interaction term $S_{V}[\theta,\theta^{*},n_{t}]$ as
\begin{align}
 S_{V}[\theta,\theta^{*},n_{t}]
 =
  \alpha_{t} H_{V}[\theta,\theta^{*},n_{t}]\,,
 \label{eqn:Lattice-Action-001}
\end{align}
where
\begin{equation}
H_0^s[\theta_{s},\theta^*_{s},n_{t}] = \frac{1}{2m}\sum_{_{}\vec{n}}
 \sum_{l=1}^{3}
 \, \theta_{s}^{*}(n_{t},\vec{n}) \,
  \left[2 \, \theta_{s}(n_{t},\vec{n})- \theta_{s}(n_{t},\vec{n}+\hat{l})-
  \theta_{s}(n_{t},\vec{n}-\hat{l}) \right]\,,
\end{equation}
and
\begin{align}
 H_{V}[\theta,\theta^{*},n_{t}]
 =
  C_{0} \,  \sum_{_{}\vec{n}} \,  \theta_{\uparrow}^{*}(n_{t},\vec{n}) \, \theta_{\uparrow}(n_{t},\vec{n})
   \, \theta_{\downarrow}^{*}(n_{t},\vec{n}) \, \theta_{\downarrow}(n_{t},\vec{n})\,.
 \label{eqn:Lattice-Action-002}
\end{align}
In the last equation $C_{0}$ is the coupling strength of the zero-range potential.

\subsection{The transfer matrix}
\label{chap1:transfer-matrix}

The Grassmann formalism given by Eq.~(\ref{eqn:Grassmann-path-int-001}) is convenient for deriving the lattice Feynman rules. On the other hand,  the operator formalism or so called \textit{transfer matrix} formalism is more  convenient for numerical calculations. Therefore, in this section we review the connection between the PI formulation and the operator formalism in quantum mechanics~\cite{PhysRevD.15.1128}.

As a first step we analyze the connection in quantum mechanics and rewrite Eq.~(\ref{eqn:transition-amplitude-0010}) for the Euclidean time lattice,
\begin{align}
{Z} =
\int\mathcal{D}q(\alpha_{t}) \,
\exp{\left\{-\alpha_{t}\sum_{n=0}^{L_{t}-1} \left[
\frac{m}{2}\left(\frac{q_{n+1}-q_{n}}{\alpha_{t}}\right)^{2}+V(q_{n})\right]\right\}}
\,.
\label{eqn:transfer-matrix-0001}
\end{align}
For finite lattice of $L_{t}$ sites this expression can be rewritten in the form of
\begin{align}
{Z} =
\int \prod_{n=0}^{L_{t}-1}dq_{n} \, M_{n+1,n}
\,.
\label{eqn:transfer-matrix-0005}
\end{align}
where the matrix $M_{n+1,n}$ is the transfer matrix,
\begin{align}
M_{\prim{n},n} = 
\left(\frac{m}{2\pi \alpha_{t}}\right)^{1/2}
\exp{\left[
-\alpha_{t}\frac{m}{2}\left(\frac{q_{\prim{n}}-q_{n}}{\alpha_{t}}\right)^{2}-\alpha_{t}\,V(q_{n})\right]}
\,.
\label{eqn:transfer-matrix-0010}
\end{align}
From Eqs.~(\ref{eqn:Indv-Propagator-0001}) and (\ref{eqn:Indv-Propagator-0005}), we already know that $M_{\prim{n},n}$ describes the evolution of particles over one lattice spacing in the temporal direction
\begin{align}
M_{\prim{n},n} = \bra{q_{n+1}}e^{-\alpha_{t} H}\ket{q_{n}}
\,.
\label{eqn:transfer-matrix-0015}
\end{align}
It is should be noted that the ${{n},{n}'}$ subscripts are not the matrix indices of the transfer matrix, but rather the coordinates of the particle.

Now we turn to the exact correspondence between the Grassmann path integral and the transfer matrix formalism. For the moment let us consider a single component fermion and use $b$ and $b^{\dagger}$ to denote fermion anti-commuting creation and annihilation operators, respectively,
\begin{align}
\{b,b\}= \{b^{\dagger},b^{\dagger}\} = 0, \quad \{b,b^{\dagger}\} = 1\,.
\end{align}
For any function $f(b,b^{\dagger})$ the exact relation between the Grassmann path integral and the transfer matrix formalism is given by~\cite{FoundPhy.30.487},
\begin{align}
\text{Tr}  & \left[ \, : \, f(b,b^{\dagger})  \, : \, \right]
 = \int
 d\theta^{*}d\theta\, e^{2 \theta^{*} \theta}
 f(\theta^{*},\theta) \,,
 \label{eqn:G-path-T-matrix-0001}
\end{align}
where the symbol :\,: signifies normal ordering. Normal ordering rearranges operators betweens the symbol :\,: such that all annihilation operators are on the right and creation operators are on the left. Using anti-periodicity of Grassmann fields in temporal direction, \ie \, $\theta(1) = - \theta(0)$, Eq.~(\ref{eqn:G-path-T-matrix-0001}) can be rewritten as a path integral over a short time interval,
\begin{align}
\text{Tr}  & \left[ \, : \, f(b,b^{\dagger})  \, : \, \right]
 = \int
 d\theta^{*}(0)d\theta(0)\, e^{\theta^{*}(0) \left[\theta(0)-\theta(1)\right]}
 f(\theta^{*}(0),\theta(0)) \,,
 \label{eqn:G-path-T-matrix-0005}
\end{align}
This can be applied to the product of any normal-ordered functions of different component fermion creation and annihilation operators,  which leads  to the following exact correspondence between the PI integral and operator formalism~\cite{PhysRevD.38.1228,FoundPhy.30.487},
\begin{align}
\text{Tr}  & \Big[ \, : \,  f_{L_{t}-1}[b^{\dagger}_{s^{\prime}} (\vec{n}^{\prime}),b_{s} (\vec{n})] \, : \,
 \cdots \, : \, f_{0}[b^{\dagger}_{s^{\prime}} (\vec{n}^{\prime}),b_{s} (\vec{n})] \, : \, \Big]
 \nonumber\\
 &
 = \int
 \left[\prod_{n_{t},\vec{n},s=\uparrow,\downarrow}
 d\theta_{s}^{*}(n_{t},\vec{n})d\theta_{s}(n_{t},\vec{n})\right]  \, e^{- S_{t}[\theta,\theta^{*}]}
 \prod_{n_{t} = 0}^{L_{t}-1}
 f_{n_{t}}[\theta^{*}_{s^{\prime}} (n_{t},\vec{n}^{\prime}),\theta_{s} (n_{t},\vec{n})] \,,
 \label{eqn:G-path-T-matrix-00010}
\end{align}
Therefore the transfer matrix formulations of the path integral given in Eq.~(\ref{eqn:Grassmann-path-int-100}) has the the following form,
\begin{align}
{Z} (L_{t}) = \text{Tr} \, \left[ \hat{M}^{L_{t}} \right] \,,
 \label{eqn:transfer-matrix-0100}
\end{align}
where $\hat{M}$ is the normal-ordered transfer matrix operator,
\begin{align}
 \hat{ M} = \,  : \, \exp
 \left[
 -\alpha_{t} \hat{H}_{0}-\alpha_{t} C_{0}
 \sum_{\vec{n}}
\hat{\rho}_{\uparrow}(\vec{n})
 \hat{\rho}_{\downarrow}(\vec{n})  \right]
 \, : \,.
 \label{eqn:transfer-matrix-0200}
\end{align}
Here $\hat{H}_{0}$ is the free non-relativistic lattice Hamiltonian in terms of anti-commuting creation and annihilation operators
\begin{align}
 \hat{H}_{0} = \hat{H}^{\uparrow}_{0} + \hat{H}^{\downarrow}_{0}  \,,
  \label{eqn:lattice-Hamiltonian-001}
\end{align}
where
\begin{align}
 \hat{H}^{s}_{0} = \frac{1}{2m} \sum_{\hat{l}=1}^{3}
  &
  \sum_{\vec{n}}
  \left[
  2b_{s}^{\dagger}(\vec{n})b_{s}(\vec{n})
  -b_{s}^{\dagger}(\vec{n})b_{s}(\vec{n}+\hat{l})
  -b_{s}^{\dagger}(\vec{n})b_{s}(\vec{n}-\hat{l})
  \right] \,,
  \label{eqn:lattice-Hamiltonian-002}
\end{align}
and $\hat{\rho}_{s}$ are the lattice density operators,
\begin{align}
 \hat{\rho}_{s}(\vec{r}) =  b_{s}^{\dagger}(\vec{r}) b_{s}(\vec{r})\,.
  \label{eqn:Latt-densty-op-001}
\end{align}
$s$ signify the spin component of fermions and $\hat{l}$ denotes the spatial lattice unit vectors.

\subsection{Adiabatic projection methods}
\label{sec:Adiabatic-Projection-Method}
The adiabatic projection method is a general procedure for calculating scattering and reactions on the lattice. The main tools of the method are initial cluster states of the system. By clusters we mean either a single particle or a composite state of several particles. The method constructs a low energy effective theory for clusters, and in the limit of large Euclidean time projection these cluster states will span the low-energy subspace of the Hamiltonian.

The initial cluster states can be parameterized by either the initial spatial separations~\cite{Pine:2013zja} or alternatively the relative momentum between clusters. The latter reduces the number of required initial states, and it is quite advantageous to adopt for improving the efficiency of the calculations. Let us use $\ket{\Psi_{\vec{\rho}}}$ to denote a set of initial cluster states where $\vec{\rho}$ stands for the parameters chosen to define the state, \ie, $\vec{R}$ and $\vec{p}$. The dressed cluster states are formed by projecting the states $\ket{\Psi_{\vec{\rho}}}$ in the Euclidean time,
\begin{align}
\ket{\Psi_{\vec{\rho}}}_{t} = e^{-\hat{H}t}\ket{\Psi_{\vec{\rho}}}
\end{align}
Now the adiabatic projection method uses these dressed cluster states to calculate the matrix elements of observables such as the Hamiltonian and the transfer matrix. 

The dressed cluster states are generally non-orthogonal, and as a result of this the method involves calculating a norm matrix. As an example in the following we consider the calculation of the adiabatic matrix representation of the Hamiltonian operator $\hat{H}$ by following the procedure used in Ref.~\cite{Pine:2013zja}. Let us define the dual state ${}_{t}(\Psi_{\vec{\rho}}|$ written as a linear functional,
\begin{align}
\prescript{}{t}(\Psi_{\vec{\rho}}\ket{u}
=
\sum_{\vec{\rho}\,'}\prescript{}{t}(\Psi_{\vec{\rho}}\ket{\Psi_{\vec{\rho}\,'}}_{t}\prescript{}{t}{\braket{\Psi_{\vec{\rho}\,'}|u}}
\,,
 \label{eqn:dual-state-001}
\end{align}
such that the dual state ${}_{t}(\Psi_{\vec{\rho}}|$ satisfies that
\begin{align}
 \prescript{}{t}{\Braket{\Psi_{\vec{\rho}}|u}}=0  \quad \text{for all $\vec{\rho}$}
\quad
&\Rightarrow
\quad
 \prescript{}{t}(\Psi_{i}\ket{u} = 0 \quad  \text{for all $\vec{\rho}$}
 \\
\prescript{}{t}(\Psi_{\vec{\rho}}\ket{\Psi_{\vec{\rho}\,'}}_{t}
&
=
\delta_{\vec{\rho}\vec{\rho}\,'}
\,.
 \label{eqn:dual-state-005}
\end{align}
The inner product of the propagated initial and final state is the norm matrix,
\begin{align}
[N_t]_{\vec{\rho}\vec{\rho}\,'}
=
\prescript{}{t}{\braket{\Psi_{\vec{\rho}}|\Psi_{\vec{\rho}\,'}}_{t}}
\,.
 \label{eqn:Norm-matrix-elements-001}
\end{align}
and the inner product of the propagated initial and dual state defines
\begin{align}
\prescript{}{t}(\Psi_{\vec{\rho}}\ket{\Psi_{\vec{\rho}\,'}}_{t}
=
[N^{-1}_{t}]_{\vec{\rho}\vec{\rho}\,'}
\,.
 \label{eqn:dual-state-009}
\end{align}
Therefore, the adiabatic matrix of the Hamiltonian $\hat{H}$ projected onto the set of dressed cluster states,
\begin{align}
[\hat{H}^{a}_{t}]_{\vec{\rho}\vec{\rho}\,'}
=
&
\sum_{\vec{\rho}\,''}\prescript{}{t}(\Psi_{\vec{\rho}}\ket{\Psi_{\vec{\rho}\,''}}_{t}\prescript{}{t}{\braket{\Psi_{\vec{\rho}\,''}|\hat{H}|\Psi_{\vec{\rho}\,'}}}_{t}
\nonumber\\
&
=
[N^{-1}_{t}]_{\vec{\rho}\vec{\rho}\,''} \, \, \prescript{}{t}{\braket{\Psi_{\vec{\rho}\,''}|\hat{H}|\Psi_{\vec{\rho}\,'}}}_{t}
\,.
 \label{eqn:adiab-transfer-matrix-001}
\end{align}
By using a similarity transformation, we can define the Hermitian adiabatic Hamiltonian as
\begin{align}
[\hat{H}^{a}_{t}]_{\vec{\rho}\vec{\rho}\,'}
=
[N^{-1/2}_{t}]_{\vec{\rho}\vec{\rho}\,''}
 \, \, \prescript{}{t}{\braket{\Psi_{\vec{\rho}\,''}|\hat{H}|\Psi_{\vec{\rho}\,'''}}}_{t}
 \, \,
 [N^{-1/2}_{t}]_{\vec{\rho}\,'''\vec{\rho}\,'}
\,.
 \label{eqn:adiab-transfer-matrix-002}
\end{align}
$\hat{H}^{a}_{t}$ is the two-body adiabatic Hamiltonian describing the scattering and reactions between interacting clusters, and the calculations become systematically more accurate as the projection time $t$
is increased. See Ref.~\cite{Pine:2013zja} for detailed analysis on an estimate of the residual error as a function of the projection time.

\section{Scattering phase shifts from the lattice}
In the finite-volume calculation, we compute the volume dependent energy spectrum of the system. However, the information about the short-range interaction potentials between two clusters is encoded in the scattering phase shifts. Therefore, in the following we review the mathematical tools and methods that we use in order to determine the scattering phase shift in finite volume calculations.

\subsection{Cubic rotational group}
\label{sec:cubic-rotational-group}

In the discretized lattice the rotational symmetry cannot be explored using an arbitrary rotation angle since the $\mathrm{SO}(3)$ rotational symmetry of continuum space is broken to the finite rotational group $\mathrm{SO}(3,Z)$. The cubic rotational group $\mathrm{SO}(3,Z)$ which is also known as the octahedral group consists of 24 rotations about the $x$, $y$ and $z$ axes. Since a finite rotation can be obtained by a set of infinitesimal rotations about an axis, the rotation operator $R_{\hat{{n}}}(\phi)$ of  the $\mathrm{SO}(3)$ also defines elements of the $\mathrm{SO}(3,Z)$ group for a rotation by $\phi = m\pi/2$ about the $n$ axis, where $m$ is integer and $n$ denotes the axes. Therefore, the angular momentum operators $L_{x}$, $L_{y}$ and $L_{z}$ in the $\mathrm{SO}(3,Z)$ group are defined by
\begin{align}
R_{\hat{{n}}}(\phi) = \exp\left(-i L_{\hat{n} }\phi\right)
\end{align}
It is clear that the eigenvalues of $L_{\hat{n}}$ are integers modulo 4.

The $2\ell+1$ elements of angular momentum transform according to the irreducible representations of the $\mathrm{SO}(3)$ group. Under the $\mathrm{SO}(3,Z)$ group these representations are reducible in most cases and they break up into the five irreducible representations denoted by $A_{1}$, $T_{1}$, $E$, $T_{2}$ and $A_{2}$.
\begin{table}[htp]
\caption{Decomposition of the $\mathrm{SO}(3)$ into the irreducible representations of the $\mathrm{SO}(3,Z)$ for $\ell \leq 6$.}
\label{tab:GroupDecomposition}
\begin{center}
\rowcolors{3}{gray!15}{white}
\begin{tabular}
[c]{c||c}
\hline\hline
$\mathrm{SO}(3)$ & $\mathrm{SO}(3,Z)$\\
\hline\hline
$\ell = 0$ & $A_{1}$
\\
$\ell = 1$ & $T_{1}$ 
\\
$\ell = 2$ & $E\oplus T_{2}$
\\
$\ell = 3$ & $T_{1}\oplus T_{2}\oplus A_{2}$
\\
$\ell = 4$ & $A_{1}\oplus T_{1} \oplus E \oplus T_{2}$ 
\\
$\ell = 5$ & $ T_{1} \oplus T_{1} \oplus E\oplus T_{2}$ 
\\
$\ell = 6$ & $A_{1}\oplus T_{1} \oplus E \oplus T_{2} \oplus T_{2} \oplus A_{2}$ 
\\
\hline\hline
\end{tabular}
\end{center}
\end{table}
 Examples for the decompositions of the orbital angular momentum eigenstates $\ell\leq6$ into the irreducible representation of the $\mathrm{SO}(3,Z)$ group are given in Table~\ref{tab:GroupDecomposition}~\cite{Johnson1982147}

\subsection{L\"{u}scher's finite-volume method}
\label{intro:Luescher-method-001}

L\"{u}scher's method~\cite{Luscher1986105,Luscher1991531} is a well-known tool used to determine elastic phase shifts for two-body scattering from the volume dependence of two-body scattering states in a periodic cubic box. The method has been extended to higher partial waves, two-body systems in moving frames, multi-channel scattering cases, and scattering of particles with spin~\cite{Rummukainen1995397,PhysRevD.83.114508,PhysRevD.85.014506,PhysRevD.85.114507,PhysRevD.88.034502,PhysRevD.88.094507,PhysRevD.88.114507,PhysRevD.89.074507}. L\"{u}scher's framework has also been successfully applied to the determination of resonance parameters~\cite{Bernard0806.4495}, and recently this technique has been applied to moving frame calculations \cite{Eur.Phys.J.A48.114,PhysRevD.86.094513}. See Ref.~\cite{Bernard1010.6018,PhysRevD.85.014027,Doring1111.0616,PhysRevD.87.014502,Doring2013185} for further studies on the extraction of resonance properties at finite volume.
  We note also recent work on improving lattice interactions in effective field theories using L\"{u}scher's
method \cite{Endres:2011er}.

In the following we summarize how L\"{u}scher's method relates the $s$-wave scattering phase shift to two-body energy levels in a periodic cubic box. Later we come back to this discussion for higher angular momentum in Section~\ref{1407.2784-Luescher-method}.

We consider a two-body system in a periodic box of length $L$. The relation between the $s$-wave scattering phase shift and the two-particle energy levels in the center of mass frame is defined by
\begin{equation}
p \cot\delta_{0}(p)
= \frac{2}{\sqrt{\pi}L}
 \mathcal{Z}_{0,0}(1;\eta)
\label{eqn:Luescher-001}
\end{equation}
where
\begin{align}
\eta = \left(\frac{L p}{2\pi}\right)^{2}\,,
\end{align}
and $\mathcal{Z}_{0,0}(1;\eta)$ is the three-dimensional zeta function,
\begin{align}
\mathcal{Z}_{0,0}(1;\eta)
=
\frac{1}{\sqrt{4\pi}}
\lim_{\Lambda\to\infty}
\left[
\sum_{\vec{n}}
\frac
{ \theta(\Lambda^{2}-\vec{n}^{2})}
{|\vec{n}|^{2} - \eta}
-4\pi\Lambda
\right]
\,.
\label{eqn:Luescher-005}
\end{align}
Alternatively, we can evaluate the zeta function using exponentially-accelerated expression~\cite{PhysRevD.83.114508}
\begin{align}
\mathcal{Z}_{0,0}(1;\eta)
=
\pi e^{\eta}(2\eta -1) 
&+ \frac{e^{\eta}}{2\sqrt{\pi}}\sum_{\vec{n}}
\frac{e^{-|\vec{n}|^{2}}}{|\vec{n}|^{2}-\eta}
\nonumber\\
&-\frac{\pi}{2}\int_{0}^{1} d\lambda\frac{e^{\lambda \eta}}{\lambda^{3/2}}
\left(4\lambda^{2}\eta^{2} -
 \sum_{\vec{n}}e^{-\pi^{2}|\vec{n}|^{2}/\lambda}\right)
\,.
\label{eqn:Luescher-009}
\end{align}

\subsection{Spherical wall method}
While L\"uscher's method is very powerful at low energies, the method is limited to calculations of scattering phase shifts below the inelastic threshold. Also, the phase shifts obtained using L\"{u}scher's method depend crucially on an accurate calculation and analysis of finite-volume energy levels.  Furthermore, since the eigenstates of the angular momentum for $\ell>1$ decompose into the irreducible representations of the $\mathrm{SO}(3,Z)$, there is no one-to-one correspondence between the finite-volume energy spectrum and the phase shifts~\cite{PhysRevD.83.114508}. Therefore, the scattering phase shift calculations become more and more difficult at higher angular momentum.

Borasoy at al.~\cite{Borasoy2342007} proposed how to compute phase shifts for non-relativistic fundamental particles on the lattice without encountering some of the difficulties mentioned above. This method uses a spherical boundary condition on the lattice. A hard wall boundary of radius $R_{\rm{wall}}$ is imposed on the relative separation of the two particles which removes the periodic lattice effects between particles.

For two particles interacting via a finite-range $R$ potential, the radial part of the solution at values $r>R$ is described by Eq.~(\ref{eqn:radialwave-Singlechannel-0040}) which vanishes at $r=R_{\rm{wall}}$,

\begin{equation}
\cot\delta_{\ell}(p) 
 \,  S_{\ell}(pR_{\rm{wall}}) + C_{\ell}(pR_{\rm{wall}}) = 0 \,.
 \label{eqn:Rwall-001}
\end{equation}
Here $p$ is the relative momentum of the particles, and it can be determined from the energy $E = p^{2}/2\mu$ with the reduced mass $\mu$. Therefore, the scattering phase shifts can be computed by the following expression,
\begin{equation}
\cot\delta_{\ell}(p) =
\cot^{-1}\left[- \frac{C_{\ell}(pR_{\rm{wall}})}{S_{\ell}(pR_{\rm{wall}})} \right]\,.
 \label{eqn:Rwall-005}
\end{equation}
See Ref.~\cite{Borasoy2342007,Lee:2008fa} for the detailed discussion on the topic and for the case of partial-wave mixing.

\section{Numerical methods}

\subsection{Monte Carlo methods}
In the preceding sections we have derived two elegant approaches which describe evolutions of particles between two space-time points in quantum systems. In principle numerical solutions for Eqs.~(\ref{eqn:Grassmann-path-int-001}) or (\ref{eqn:transfer-matrix-0100}) are not impossible, but it is computationally very expensive, perhaps not practical, due to the multi-dimensional integral in the PI formula and massive matrix operations in the operator formalism. In the presence of such difficulties Monte Carlo methods are the most powerful and commonly used techniques to approximate physical observables. 

Monte Carlo techniques are based on the idea of simulating particle configurations by using random numbers as dynamics of the method. The fact that makes Monte Carlo techniques very advantageous is that among all states of the physical system only a small fraction are chosen at random according to some weighting function $P(c)$ in order to  estimate mean values of physical quantities. Suppose $\mathcal{O}$ be an observable that we want to estimate from Monte Carlo simulations, then the approximated expectation value (the estimator) of $\mathcal{O}$ is the average over some subset states $M$ of the complete states of the system\footnote{The path integrals in Euclidean time are computed using a larger samples weighted by the exponential Boltzmann factor, $P(c) =  e^{-\beta E(c_{i})}$.},
\begin{align}
\braket{\mathcal{O}}_M =\frac{\sum_{i=1}^{M} \mathcal{O}(c_{i}) \,  P(c_{i})}
{\sum_{i=1}^{M} \,  P(c_{i}) }\,,
\label{eqn:estimator-0001}
\end{align}
where $\mathcal{O}(c_{i})$ is the value of the observable for the configuration $c_{i}$. The neglected subset states introduce some statistical errors in Monte Carlo simulations which are suppressed by the size of the sample.

\subsection{Importance sampling}
The chosen subset states in Eq.~(\ref{eqn:estimator-0001}) determines the accuracy of the estimator and the performance of Monte Carlo simulations. One powerful technique that improves the accuracy of the simulations is \textit{importance sampling}. 

Importance sampling is implemented by selecting states with the probability function $P(c)$ in Eq.~(\ref{eqn:estimator-0001}). Then \eref{eqn:estimator-0001} takes the following simpler form\footnote{The estimator is simply the average of the selected configurations.}
\begin{align}
\braket{\mathcal{O}}_M = \frac{1}{M}\sum_{i=1}^{M} \mathcal{O}(c_{i}) \,,
\label{eqn:estimator-0002}
\end{align}
and the statistical error is determined from the variance of the estimator by
\begin{align}
\sigma =\sqrt{\frac{\sum_{i=0}^{N} \left[ \mathcal{O}(c_{i}) -  \braket{\mathcal{O}}_{N} \right]^{2}}{N-1}}\,.
\end{align}
This expression indicates that the statistical error can be easily kept under control by the size of the sample used in simulations, and the estimator can be made as accurate as desired by increasing the number of samples.

\subsection{Markov process}
Importance sampling has the key role of reducing the variance. However, if we randomly generate states, then most of them would be rejected and the computation time would be mostly wasted in such simulations. In order to make computations more efficient, we need a technique that generates some random set of states according to the distribution $P(c)$.  This is done using a \textit{Markov process}, which we now explain.

Suppose the probability distribution is time dependent and at time $t$ the system is in a state $A$ with probability $P(c_{A},t)$. Let $\Omega(A\to B)$ be the probability that generates a state $B$ at a later time $t+\Delta t$, then the evolution of $P(c_{A},t)$ is
\begin{align}
P(c_{A},t)-P(c_{A},t+\Delta t) = \sum_{B\neq A}\left[P({c_{B}},t) \, \Omega(B\to A) - P(c_{A},t) \, \Omega(A\to B)\right]\,,
\label{eqn:master-equation-0001}
\end{align}
where $\Omega(B\to A)$ is the transition probability for selecting the state $A$ from the state $B$. The most important features of the transition probability for a Markov process are that;
\begin{itemize}
\item it is independent of time,
\item it does not depend on any prior state that the system in before the state $B$,
\item $\Omega(A\to B)\geq 0$,
\item $\sum_{B} \Omega(A\to B) = 1$.
\end{itemize}
It should be stressed that every state in a simulation must be accessible despite the fact that the transition probability from one particular state to another could be zero. This requires a Markov process to reach any state of the system from any other state, and it is know as \textit{ergodicity}.

After the simulation runs for many steps, the system comes to an equilibrium,
\begin{align}
\lim_{t\to\infty}P(c,t)\simeq P(c)\,,
\label{eqn:equilibrium-0001}
\end{align}
and the Markov process eventually generates successive states with the probability $P(c)$. If we repeat this Markov process in the simulation to generate successive states, we construct a \textit{Markov chain} which is a set of states each of which is selected with the probability $P(c)$.

A sufficient condition which ensures that in equilibrium all states are generated according to the probability distribution $P(c)$ is the condition of \textit{detailed balance}. The detailed balance requires that each term in the sum in \eref{eqn:master-equation-0001} must be zero\footnote{See Ref.\cite{newman1999monte} for rigorous proofs and detailed discussions.}. This will then satisfy Eq.~\ref{eqn:equilibrium-0001},
\begin{align}
P({c_{B}},t) \, \Omega(B\to A) =
P(c_{A},t) \, \Omega(A\to B)\,.
\label{eqn:detailed-balance-0001}
\end{align}
This condition also ensures that the transition rate into any state is equal to the transition rate out of the same state.

\subsection{Metropolis algorithm}

The detailed balance equation (\ref{eqn:detailed-balance-0001}) imposes a constraint on the transition probability from a state $A$ to another state $B$ by
\begin{align}
\frac{ \Omega(A\to B)}{\Omega(B\to A)} =
\frac{P(c_{B},t)}{P({c_{A}},t)} = e^{-\beta\left[E(c_{B})-E(c_{A})\right]} \,.
\label{eqn:detailed-balance-0005}
\end{align}
where the equilibrium distribution is chosen to be the Boltzmann distribution, $P(c) = P_{0} \, e^{-\beta E(c)}$. Therefore, the states of the desired Markov chain are distributed according to the probability distribution $P_{0} \, e^{-\beta E(c)}$. The most popular method for generating such successive states by respecting the detailed balance condition is the Metropolis algorithm~\cite{Metropolis1953}. In the following we give a simple recipe for the Metropolis algorithm.
\begin{description}
\item[Step 1:] Initially start with an arbitrarily configuration $C_{i}$.
\item[Step 2:] Generate a proposed configuration $C_{p}$ from the configuration $C_{i}$. 
\item[Step 3:] Select a random number $r \in [0,1)$.
\item[Step 4:] Accept the proposed configuration if
$r < e^{-\beta\left[E(C_{p}) - E(C_{i})\right]}\,$ 
and set $C_{i} = C_{p}$. Otherwise leave $C_{i}$ the same.
\item[Step 5:] Repeat \textbf{Step 2-4}.
\end{description}
If the proposed configuration has a lower energy, $E(C_{p}) - E(C_{i})\leq0$, then the Metropolis algorithm always accept that configuration. If a configuration with higher energy is proposed, then it is only accepted with some probability given in \textbf{Step 4}.

\subsection{Sign problem}

Before ending this chapter we would like to discuss the sign problem that Monte Carlo methods suffer from simulating fermions. The main difficulty of simulations with fermions is the sign cancellation due to the identical particle permutations. When we perform calculations by sampling configurations, the fluctuation in the signs associated with Fermi-Dirac statistics results in significant cancellations. In order for rigorous expression let us consider the calculation of the expectation value of a physical observable $\mathcal{O}$,
\begin{align}
\braket{\mathcal{O}} = \frac{\sum_{c} s(c) P(c) \,  \mathcal{O}(c)}{\sum_{c} s(c) P(c)}
\equiv
\frac{\braket{ s(c) \, \mathcal{O}} }{\braket{s(c)}} ,
\label{eqn:sign-001}
\end{align}
where $c$ is the number of configurations, $s(c)=\pm1$ is the sign, and $P(c)$ is the magnitude of the weight. Since the average sign appears as the denominator in Eq.~(\ref{eqn:sign-001}), it is crucial that the cancellation is tolerable.  The average sign over $c$ configurations is
\begin{align}
\braket{s(c)} = \frac{\sum_{c} s(c) P(c)}{\sum_{c} P(c)} = \exp(-t \, \Delta E)\,,
\label{eqn:sign-002}
\end{align}
where $\Delta E = E_{phy}-E_{bos}$ is the difference between the physical ground state energy ($E_{phy}$) and the ground state energy ($E_{bos}$) due to the bosonic ensemble. Since $\Delta E > 0$ and scales with the system size, the average sign is exponentially small in the size of the system which makes the simulations exponentially difficult for large systems.

%% file: Chapter-4.tex
\chapter{Neutron-proton scattering}
\label{chap:Neutron-proton-scattering}

\section{Introduction}

\label{chap4:introduction}

Chiral effective field theory describes the
low-energy interactions of protons and neutrons. \ If one neglects
electromagnetic effects, the long range behavior of the nuclear interactions
is determined by pion exchange processes. \ See
Ref.~\cite{vanKolck:1999mw,Bedaque:2002mn,Epelbaum:2005pn,Epelbaum:2008ga} for
reviews on chiral effective field theory. \ But there are also systems of
interest where momenta smaller than the pion mass are relevant. \ In such
cases it is more economical to use pionless effective field theory with only
local contact interactions involving the nucleons. \ The pionless formulation
is theoretically elegant since the theory at leading order is renormalizable
and the momentum cutoff scale can be arbitrarily large
\cite{vanKolck:1998bw,Bedaque:1998kg,Bedaque:1998km,Bedaque:1999ve,Chen:1999tn,Platter:2004zs,
Hammer:2006ct}. \ This allows an elegant connection with the universal
low-energy physics of fermions at large scattering length and other systems
such as ultracold atoms \cite{Braaten:2004a,Giorgini:2007a}.

For local contact interactions the range of the interactions are set by the
momentum cutoff scale for the effective theory. \ There are rigorous
constraints for strictly finite-range interactions set by causality and
unitarity. \ Some violations of unitarity can relax these constraints if one
works at finite order in perturbation theory or includes unphysical
propagating modes with negative norm. \ However at some point one must
accurately reproduce the underlying unitary quantum system by going to
sufficiently high order in perturbation theory or decoupling the effects of
propagating unphysical modes.

The time evolution of any quantum mechanical system obeys causality and
unitarity. \ Causality requires that the cause of an event must occur before
any resulting consequences are produced, and unitarity requires that the sum
of all outcome probabilities equals one. \ In the case of non-relativistic
scattering, these constraints mean that the outgoing wave may depart only
after the incoming wave reaches the scattering object and must preserve the
normalization of the incoming wave. \ In this chapter we discuss the constraints
of causality and unitarity for finite range interactions. \ Specifically we
consider neutron-proton scattering in all spin channels up to $j=3$.

The constraints of causality and unitarity
for two-body scattering with finite-range interactions were first investigated by Wigner~\cite{Wigner1954}. The time delay between an incoming wave packet and the scattered outgoing wave packet is equal to the energy derivative of the elastic phase shift,
\begin{equation}
 \Delta t=2\frac{d\delta}{dE} \,.
\end{equation}
If $d\delta/dE$ is negative, the outgoing wave is produced earlier than that for
the non-interacting system. \ However the incoming wave must first arrive in
the interacting region before the outgoing wave can be produced. \ For each
partial wave, $\ell$, this puts an upper bound on the effective range parameter,
$r_{\ell}$, in the effective range expansion,%
\begin{equation}
p^{2\ell+1}\cot\delta_{\ell}(p)=-\frac{1}{a_{\ell}}+\frac{1}{2}r_{\ell}p^{2}+O(p^{4})\,.
\label{chap4:EFE}
\end{equation}
We introduce the effective range expansion in Eq.~(\ref{eqn:ERE-0001}). In this chapter we truncate the series at $p^{2}$.

Phillips and Cohen \cite{Phillips1997} derived the causality bound for the
$s$-wave effective range parameter for finite-range interactions in three
dimensions. Constraints on nucleon-nucleon scattering and the chiral two-pion exchange
potential was considered in Ref.~\cite{PavonValderrama:2005wv}, and correlations
between the scattering length and effective range have been explored for
one-boson exchange potentials~\cite{Cordon:2009pj}.  Same authors studied the relationship between the scattering length and
effective range for the van der Waals
interaction~\cite{Cordon:2010a,RuizArriola:2011a}

In Refs.~\cite{Hammer-Lee9,HammerDean27} the causality and unitarity bounds
for finite-range interactions were extended to an arbitrary number of space-time
dimensions or value of angular momentum.  A complementary discussion based
upon conformal symmetry and scaling dimensions can be found in
Ref.~\cite{Nishida:2010a}. Also the interactions with
attractive and repulsive Coulomb tails were first considered in
Ref.~\cite{Koenig:2012bv}.

Let $R$
be the range of the interaction. \ For the case $d=3$, it was found that the
effective range parameter must satisfy the upper bound~\cite{Hammer-Lee9,HammerDean27}
\begin{align}
r_{\ell}  &  \leq b_{\ell}(r)=-\frac{2\Gamma(\ell-\frac{1}{2})\Gamma(\ell+\frac{1}{2}%
)}{\pi}\left(  \frac{r}{2}\right)  ^{-2\ell+1}\nonumber\\
&  \qquad-\frac{4}{\ell+\frac{1}{2}}\frac{1}{a_{\ell}}\left(  \frac{r}{2}\right)
^{2}+\frac{2\pi}{\Gamma(\ell+\frac{3}{2})\Gamma(\ell+\frac{5}{2})}\frac{1}{a_{\ell}%
^{2}}\left(  \frac{r}{2}\right)  ^{2\ell+3}, \label{bL}%
\end{align}
for any $r\geq R$. \ This inequality can be used to determine a length scale,
$R^{b}$, which we call the causal range,%
\begin{equation}
r_{\ell}=b_{\ell}(R^{b}).
\end{equation}
The physical meaning of $R^{b}$ is that any set of interactions with strictly
finite range that reproduces the physical scattering data must have a range
greater than or equal to $R^{b}$.

In this chapter we extend the causality bound to the case
of two coupled partial-wave channels. \ For applications to nucleon-nucleon
scattering the relevant coupled channels are $^{3}S_{1}$-$^{3}D_{1}$,
$^{3}P_{2}$-$^{3}F_{2}$, $^{3}D_{3}$-$^{3}G_{3},$ etc. \ As we will show,
there is some modification of the effective range bound in \eref{bL} due
to mixing. \ For total spin $j$ we show that the lower partial-wave channel
$\ell=j-1$ satisfies the new causality bound,%
\begin{equation}
r_{j-1}\leq b_{j-1}(r)-2q_{0}^{2}\frac{\Gamma(j+\frac{1}{2})\Gamma(j+\frac
{3}{2})}{\pi}\left(  \frac{r}{2}\right)  ^{-2j-1}, \label{preview_r_J-1}%
\end{equation}
where $q_{0}$ is the first term in the expansion of the mixing angle
$\varepsilon_{j}$ in the Blatt-Biedenharn eigenphase convention
\cite{Blatt-B.I},%
\begin{equation}
\tan\varepsilon_{j}(p)=q_{0}p^{2}+q_{1}p^{4}+O(p^{6}). \label{taneps}%
\end{equation}
We note that the last term in \eref{preview_r_J-1} is negative
semi-definite and diverges as $r\rightarrow0$. \ From this observation we make
the general statement that non-vanishing partial-wave mixing is inconsistent
with zero-range interactions. \ We will explore in detail the consequences of
this result as it applies to nuclear effective field theory.

We also derive a new causality bound associated with the mixing angle itself.
\ Using the Cauchy-Schwarz inequality we derive a bound for the parameter
$q_{1}$ in the expansion \eref{taneps}. \ This leads to another minimum
interaction length scale, which we call the Cauchy-Schwarz range,
$R^{\text{C-S}}$. We use the new causality bounds to determine the minimum
causal and Cauchy-Schwarz ranges for each $^{2s+1}\ell_{j}$ channel in
neutron-proton scattering up to $j=3$. \ Since the long range behavior of the
nuclear interactions is determined by pion exchange processes, one expects
$R^{b}\sim R^{\text{C-S}}\sim m_{\pi}^{-1}=1.5$ fm. \ However in some higher
partial-wave channels we find that these length scales are as large as $5$~fm.
\ We show these large ranges are generated by the one-pion exchange tail.
\ Using a potential model we show that the causal range and Cauchy-Schwarz
range are both significantly reduced when the one-pion exchange tail is
chopped off at distances beyond $5$~fm. \ We discuss the impact of this
finding on the choice of momentum cutoff scales in effective field theory.

In the limit of isospin symmetry our analysis of the isospin triplet channels
can also be applied to neutron-neutron scattering and therefore has relevance
to dilute neutron matter. \ The physics of dilute neutron matter is important
for describing the crust of neutron stars as well as connections to the
universal physics of fermions near the unitarity limit. \ Our analysis of the
causality and unitarity bounds show that there are constraints on the
universal character of neutron-neutron interactions in channels with
partial-wave mixing as well as higher uncoupled partial-wave channels. \ In
other words some low-energy phenomenology cannot be cleanly separated from
microscopic details such the range of the interaction. \ Reviews of the theory
of ultracold Fermi gases close to the unitarity limit and their numerical
simulations are given in Ref.~\cite{Giorgini:2007a,Lee:2008fa}. A general
overview of universality at large scattering length can be found in
Ref.~\cite{Braaten:2004a}. \ See Ref.~\cite{Koehler:2006A, Regal:2006thesis}
for reviews of recent cold atom experiments at unitarity.

\section{Uncoupled Channels}

We analyze in this section the channels with only one partial wave, $\ell$. \ We
summarize the results obtained Section~\ref{chap2:scattering-solutions} [cf. in Ref.~\cite{HammerDean27}]. \ For simplicity,
we will assume throughout the calculations that the interaction has finite
range $R$, and we use units where $\hbar=1$. \ For the two-body system the rescaled radial wave function $U_{\ell}^{(p)}(r)$
satisfies the radial Schr\"{o}dinger equation,
\begin{equation}
\left[  -\frac{d^{2}}{dr^{2}}
+\frac{\ell(\ell+1)}{r^{2}}-p^{2}\right]  U_{\ell}^{(p)}(r)+2\mu\int_{0}^{R}%
W(r,r^{\prime})U_{\ell}^{(p)}(r^{\prime})dr^{\prime}=0.
\label{eqn:uncoup:radialeqR}%
\end{equation}
We write $W(r,r^{\prime})$ for the non-local
interaction potential as a real symmetric integral operator. As we discuss in Section~\ref{chap1:interaction-potential} we assume that the potential has finite range $R$ which requires that $W(r,r^{\prime})=0$ for
$r>R$ or $r^{\prime}>R$.

In Eq.~(\ref{uncoup-wrons-integ-formula-005}) taking the limit $p\to 0$, we obtain that for any $r>R$ the effective range satisfies the following relation,
\begin{equation}
r_{\ell}=b_{\ell}(r)-2\int_{0}^{r}\left[  U_{\ell}^{(0)}(r^{\prime})\right]^{2}
dr^{\prime}, \label{eqn:uncoup:effectrang1}%
\end{equation}
where $b_{\ell}(r)$ is
\begin{align}
b_{\ell}(r)=  &  \frac{1}{a_{\ell}^{2}}\frac{2\pi}{\Gamma\left(  \ell+\frac{3}%
{2}\right)  \Gamma\left(  \ell+\frac{5}{2}\right)  }\left(  \frac{r}{2}\right)
^{2\ell+3}\nonumber\\
&  -\frac{1}{a_{\ell}}\frac{4}{\ell+\frac{1}{2}}\left(  \frac{r}{2}\right)
^{2}-\frac{2\Gamma\left(  \ell-\frac{1}{2}\right)  \Gamma\left(  \ell+\frac{1}%
{2}\right)  }{\pi}\left(  \frac{r}{2}\right)  ^{-2\ell+1}.
\label{eqn:uncoup:generalb}%
\end{align}
Since the wave function is real and the integral term in \eref{eqn:uncoup:effectrang1} is positive semi-definite, this equation puts an upper bound on the effective range, $r_{\ell}\leq b_{\ell}(r)$. This relation and causality bound are analyzed in Ref.~\cite{HammerDean27} for arbitrary dimension or angular momentum $\ell$.

\section{Coupled Channels}

\label{sec:coupledChannel}

In this section we derive the general wave functions for spin-triplet
scattering with mixing between orbital angular momentum $\ell=j-1$ and $\ell=j+1$.
The coupled-channel wave functions satisfy the following coupled radial
Schr\"{o}dinger equations,%
\begin{align}
&  \left[  -\frac{d^{2}}{dr^{2}}-p^{2}+\frac{j(j-1)}{r^{2}}\right]
U_{j-1}^{(p)}(r)\nonumber\\
&  +2\mu\int_{0}^{R}[W_{11}(r,r^{\prime})U_{j-1}^{(p)}(r^{\prime}%
)+W_{12}(r,r^{\prime})V_{j+1}^{(p)}(r^{\prime})]\,dr=0,
\label{eqn:couplediff1}%
\end{align}%
\begin{align}
&  \left[  -\frac{d^{2}}{dr^{2}}-p^{2}+\frac{(j+1)(j+2)}{r^{2}}\right]
V_{j+1}^{(p)}(r)\nonumber\\
&  +2\mu\int_{0}^{R}[W_{21}(r,r^{\prime})U_{j-1}^{(p)}(r^{\prime}%
)+W_{22}(r,r^{\prime})V_{j+1}^{(p)}(r^{\prime})]\,dr^{\prime}=0.
\label{eqn:couplediff2}%
\end{align}
Here the non-local interaction potentials are represented by a real symmetric
$2\times2$ matrix $W(r,r^{\prime})$,
\begin{equation}
W(r,r^{\prime})=\left(
\begin{array}
[c]{cc}%
W_{11}(r,r^{\prime}) & W_{12}(r,r^{\prime})\\
W_{12}(r,r^{\prime}) & W_{22}(r,r^{\prime})
\end{array}
\right)  . \label{eqn:PotentialMatrix}%
\end{equation}
In \eref{eqn:couplediff1}-\eref{eqn:couplediff2} the $U_{j-1}^{(p)}(r)$
corresponds with the spin-triplet $\ell=j-1$ channel and the $V_{j+1}^{(p)}(r)$
is for the spin-triplet $\ell=j+1$. These wave functions are the rescaled form of
the radial wave functions. \ In the non-interacting region $r\geq R$ the
coupled radial Schr\"{o}dinger equations reduce to the free radial
Schr\"{o}dinger equations
\begin{equation}
\left[-\frac{d^{2}}{dr^{2}}-p^{2}+\frac{j(j-1)}{r^{2}}\right]  U_{j-1}%
^{(p)}(r)=0, \label{eqn:freediff1}%
\end{equation}%
\begin{equation}
\left[ -\frac{d^{2}}{dr^{2}}-p^{2}+\frac{(j+1)(j+2)}{r^{2}}\right]
V_{j+1}^{(p)}(r)=0. \label{eqn:freediff2}%
\end{equation}
The solutions of these differential equations are the Riccati-Bessel
functions,
\begin{equation}
U_{j-1}^{(p)}(r)=A_{1}S_{j-1}(pr)+B_{1}C_{j-1}(pr),
\label{eqn:uwave1}%
\end{equation}%
\begin{equation}
V_{j+1}^{(p)}(r)=A_{2}S_{j+1}(pr)+B_{2}C_{j+1}(pr).
\label{eqn:vwave1}%
\end{equation}
where $A_{1,2}$ and $B_{1,2}$ are amplitudes associated with incoming and
outgoing waves, respectively. More details regarding the Riccati-Bessel functions are given in Appendix
\ref{append:Bessel-functions}. \ The relation between incoming and outgoing wave
amplitudes is
\begin{equation}
B={{\rm{K}}} \, A, \label{eqn:matr1}%
\end{equation}
${\rm{K}}$ is the reaction matrix and is defined in terms of the unitary scattering matrix ${\rm{S}}$ by Eq.~(\ref{chap1:fromKtoS}). Therefore, the \eref{eqn:matr1} is written as
\begin{equation}
\tilde{B}={\rm{S}} \,  \tilde{A}, \label{eqn:matr2}%
\end{equation}
where $\tilde{A}_{1,2}$ and $\tilde{B}_{1,2}$ are rescaled amplitudes associated with incoming and
outgoing waves. \ For two coupled channels the
$2\times2$ scattering matrix can also be made symmetric. \ It is possible to
write several different $2\times2$ $\rm{S}$-matrices which satisfy the unitarity
and symmetry properties. In the literature, there are two conventionally used
$2\times2$ $\rm{S}$-matrices \cite{Stapp1956, Blatt-B.I}. In this study we adopt
the \textquotedblleft eigenphase" parameterizations of Blatt and Biedernharn
\cite{Blatt-B.I}, and the relations between the eigenphase and nuclear bar
\cite{Stapp1956} parameterizations are shown in Appendix~\ref{append:B}.

The $\rm{S}$-matrix can be diagonalized by an orthogonal matrix $U$
\begin{equation}
{\rm{S}}_{d}=U\,{\rm{S}}\,U^{-1}=\left(
\begin{array}
[c]{cc}%
e^{2i\delta_{\alpha}} & 0\\
0 & e^{2i\delta_{\beta}}%
\end{array}
\right)  ,\label{eqn:sdiag}%
\end{equation}
that contains one real parameter $\varepsilon,$%
\begin{equation}
U=\left(
\begin{array}
[c]{cc}%
\cos\varepsilon & \sin\varepsilon\\
-\sin\varepsilon & \cos\varepsilon
\end{array}
\right)  .\label{eqn:unitmatr}%
\end{equation}
$\delta_{\alpha}(p)$ and $\delta_{\beta}(p)$ are the two phase shifts, and
$\varepsilon(p)$ is the mixing angle. The $\rm{S}$-matrix explicitly is%
\begin{equation}
{\rm{S}}=\left(
\begin{array}
[c]{cc}%
e^{2i\delta_{\alpha}}\cos^{2}\varepsilon+e^{2i\delta_{\beta}}\sin
^{2}\varepsilon & \cos\varepsilon\sin\varepsilon\left(  e^{2i\delta_{\alpha}%
}-e^{2i\delta_{\beta}}\right)  \\
\cos\varepsilon\sin\varepsilon\left(  e^{2i\delta_{\alpha}}-e^{2i\delta
_{\beta}}\right)   & e^{2i\delta_{\alpha}}\sin^{2}\varepsilon+e^{2i\delta
_{\beta}}\cos^{2}\varepsilon
\end{array}
\right)  .\label{eqn:scattmat}%
\end{equation}

The eigenvalue equation ${\rm{S}}\Ket{X}=\lambda\Ket{X}$ results in eigenvalues
$\lambda_{1}=e^{2i\delta_{\alpha}}$ and $\lambda_{2}=e^{2i\delta_{\beta}}$,
with corresponding eigenstates,
\begin{equation}
\Ket{X_1}=\left(
\begin{array}
[c]{c}%
\cos\varepsilon\\
\sin\varepsilon
\end{array}
\right)  \ \ \text{and}\ \ \Ket{X_2}=\left(
\begin{array}
[c]{c}%
-\sin\varepsilon\\
\cos\varepsilon
\end{array}
\right)  ,
\end{equation}
which satisfy the orthogonality condition
\begin{equation}
\Braket{X_1 | X_2}=0. \label{eqn:orthg}%
\end{equation}
We can write \eref{eqn:matr2} as
\begin{equation}
\left(
\begin{array}
[c]{cc}%
\tilde{B}_{1\alpha} & \tilde{B}_{1\beta}\\
\tilde{B}_{2\alpha} & \tilde{B}_{2\beta}%
\end{array}
\right)  =\left(
\begin{array}
[c]{cc}%
S_{11} & S_{12}\\
S_{12} & S_{22}%
\end{array}
\right)  \left(
\begin{array}
[c]{cc}%
\tilde{A}_{1\alpha} & \tilde{A}_{1\beta}\\
\tilde{A}_{2\alpha} & \tilde{A}_{2\beta}%
\end{array}
\right)  , \label{eqn:matr3}%
\end{equation}
where the matrices $\tilde{A}$ and $\tilde{B}$ are
\begin{equation}
\tilde{A}=\left(
\begin{array}
[c]{cc}%
e^{-i\delta_{\alpha}}\cos\varepsilon & -e^{-i\delta_{\beta}}\sin\varepsilon\\
e^{-i\delta_{\alpha}}\sin\varepsilon & e^{-i\delta_{\beta}}\cos\varepsilon
\end{array}
\right)  , \label{eqn:Amatrix}%
\end{equation}%
\begin{equation}
\tilde{B}=\left(
\begin{array}
[c]{cc}%
e^{i\delta_{\alpha}}\cos\varepsilon & -e^{i\delta_{\beta}}\sin\varepsilon\\
e^{i\delta_{\alpha}}\sin\varepsilon & e^{i\delta_{\beta}}\cos\varepsilon
\end{array}
\right)  .\hspace{0.4cm} \label{eqn:Bmatrix}%
\end{equation}

We now define some additional notation. \ We write all $\alpha$-state
phaseshifts $\delta_{\alpha}(p)$ as $\delta_{j-1}(p)$ and all $\beta$-state
phaseshifts $\delta_{\beta}(p)$ as $\delta_{j+1}(p)$. \ The notation is
appropriate since in the $p\rightarrow0$ limit the $\alpha$-state is purely
$\ell=j-1$ and the $\beta$-state is purely $\ell=j+1$. \ We also drop the
superscript $p$ in the wave functions. \ We choose the normalization of the
wave function to be well-behaved in the zero-energy limit. \ Using the
relations%
\begin{equation}
S_{j\pm1}(pr)\underset{\text{as}\ p\rightarrow0}{\longrightarrow}\sqrt{\pi
}(pr)^{j\pm1+1}\frac{2^{-j\mp1-1}}{\Gamma(j\pm1+3/2)},
\label{eqn:firstricbes2}%
\end{equation}%
\begin{equation}
C_{j\pm1}(pr)\underset{\text{as}\ p\rightarrow0}{\longrightarrow}%
\frac{(pr)^{-j\mp1}}{\sqrt{\pi}}2^{j\pm1}\Gamma(j\pm1+1/2),
\label{eqn:secondricbes2}%
\end{equation}
and removing an overall phase factor, we get wave functions of the form%
\begin{align}
&  U_{\alpha}(r)=\cos\varepsilon_{j}(p)\ p^{j-1}[\cot\delta_{j-1}%
(p)S_{j-1}(pr)+C_{j-1}(pr)],\label{eqn:ualpwave3}\\
&  V_{\alpha}(r)=\sin\varepsilon_{j}(p)\ p^{j-1}[\cot\delta_{j-1}%
(p)S_{j+1}(pr)+C_{j+1}(pr)],\\
&  U_{\beta}(r)=-\sin\varepsilon_{j}(p)\ p^{j+1}[\cot\delta_{j+1}%
(p)S_{j-1}(pr)+C_{j-1}(pr)],\\
&  V_{\beta}(r)=\cos\varepsilon_{j}(p)\ p^{j+1}[\cot\delta_{j+1}%
(p)S_{j+1}(pr)+C_{j+1}(pr)]. \label{eqn:vbetwave3}%
\end{align}

For later convenience we define
\begin{align}
s_{\ell}(p,r)  &  =p^{-\ell-1}S_{\ell}(pr),\label{eqn:expfuncS}\\
c_{\ell}(p,r)  &  =p^{\ell}C_{\ell}(pr).\hspace{0.4cm} \label{eqn:expfuncC}%
\end{align}
$s_{\ell}(p,r)$ and $c_{\ell}(p,r)$
are given in Appendix~\ref{append:Bessel-functions}.
Plugging in these function into \eref{eqn:ualpwave3}-\eref{eqn:vbetwave3} we obtain
\begin{align}
&  U_{\alpha}(r)=\cos\varepsilon_{j}(p)[p^{2j-1}\cot\delta_{j-1}%
(p)s_{j-1}(p,r)+c_{j-1}(p,r)],\label{eqn:ualpwave4}\\
&  V_{\alpha}(r)=\sin\varepsilon_{j}(p)[p^{2j+1}\cot\delta_{j-1}%
(p)s_{j+1}(p,r)+p^{-2}c_{j+1}(p,r)], \label{eqn:valphwave4}%
\end{align}%
\begin{align}
&  U_{\beta}(r)=-\sin\varepsilon_{j}(p)[p^{2j+1}\cot\delta_{j+1}%
(p)s_{j-1}(p,r)+p^{2}c_{j-1}(p,r)],\label{eqn:ubetwave4}\\
&  V_{\beta}(r)=\cos\varepsilon_{j}(p)[p^{2j+3}\cot\delta_{j+1}(p)s_{j+1}%
(p,r)+c_{j+1}(p,r)]. \label{eqn:vbetwave4}%
\end{align}

The multi channel effective range expansion formula is given in Eq.~(\ref{eqn:multichannel-ERE-0001}), and here we truncate the expansion at $p^{2}$ and write the two-channel effective range expansion  for the total spin angular momentum $s = 1$ in the following form,
\begin{equation}
\sum_{m',n'} \textit{\textbf{p}}_{mm'} \, [{\rm{K}}^{-1}]_{m'n'} \, \textit{\textbf{p}}_{n'n}
=-\frac{1}{ \mathbf{a}_{mn}}
+\frac{1}{2} \mathbf{r}_{mn} \, p^{2}
+O(p^{4}),
\label{chap4:twochannel-ERE}
\end{equation}
where $\mathbf{a}_{{mn}}$ is the scattering length matrix, $\mathbf{r}_{{mn}%
}$ is the effective range matrix, and $\textit{\textbf{p}}_{mn}$ is the diagonal momentum
matrix $\text{diag}(p^{j-1/2},p^{j+3/2})$. The two-channel effective range
expansion in the Blatt and Biedernharn parameterization is%
\begin{equation}
\sum_{m'm''n''n'} \textit{\textbf{p}}_{mm'} \,
U_{m'm''}
\,
 [{\rm{K}}^{-1}]_{m''n''} \,
 [U^{-1}]_{n''n'}
 \,
 \textit{\textbf{p}}_{n'n}
=-\frac{1}{a_{{mn}}}+\frac{1}{2}r_{{mn}}p^{2}+\mathcal{O}(p^{4})
\label{chap4:ERE}
\end{equation}
where $a_{{mn}}=\text{diag}(a_{j-1},a_{j+1})$, $r_{{mn}}%
=\text{diag}(r_{j-1},r_{j+1})$, and we drop the indices the reaction and unitary matrices for the sake of simplicity. In addition we get an analytic expansion for
the tangent mixing angle \cite{Blatt-B.II}
\begin{equation}
\tan\varepsilon_{j}(p)=q_{0}p^{2}+q_{1}p^{4}+\mathcal{O}\left(  p^{6}\right)
\label{eqn:mixingangleexpans}%
\end{equation}
with mixing parameters $q_{0}$ and $q_{1}$. \ Now using
\eref{eqn:mixingangleexpans} we obtain the following final forms of wave
functions for $r\geq R$,%
\begin{align}
U_{\alpha}(r)  &  =\frac{-1}{a_{j-1}}s_{0,j-1}(r)+c_{0,j-1}(r)\nonumber\\
&  +p^{2}\Big\{\frac{1}{2}r_{j-1}s_{0,j-1}(r)-\frac{1}{a_{j-1}}s_{2,j-1}%
(r)+c_{2,j-1}(r)\Big\}+\mathcal{O}(p^{4}),
\end{align}%
\begin{equation}
V_{\alpha}(r)=q_{0}c_{0,j+1}(r)+p^{2}\Big\{q_{1}c_{0,j+1}(r)+q_{0}%
c_{2,j+1}(r)\Big\}+\mathcal{O}(p^{4}),
\end{equation}%
\begin{align}
U_{\beta}(r)  &  =q_{0}\frac{1}{a_{j+1}}s_{0,j-1}(r)\nonumber\\
&  +p^{2}\Big\{q_{0}\frac{1}{a_{j+1}}s_{2,j-1}(r)-q_{0}\frac{r_{j+1}}%
{2}s_{0,j-1}(r)+q_{1}\frac{1}{a_{j+1}}s_{0,j-1}(r)\Big\}+\mathcal{O}(p^{4}),
\end{align}%
\begin{align}
V_{\beta}(r)  &  =\frac{-1}{a_{j+1}}s_{0,j+1}(r)+c_{0,j+1}(r)\nonumber\\
&  +p^{2}\Big\{\frac{1}{2}r_{j+1}s_{0,j+1}(r)-\frac{1}{a_{j+1}}s_{2,j+1}%
(r)+c_{2,j+1}(r)\Big\}+\mathcal{O}(p^{4}). \label{eqn:vbetwavefinal}%
\end{align}

As in the single channel case, the tool that we use to derive the causality
bound is the Wronskian identity. \ Through the derivation we recall assumption on the potential in Section~\ref{chap1:interaction-potential}. We assume that the potential is not singular at the origin and regular
solutions of the Schr\"{o}dinger equations $U(r)$ and $V(r)$ for two different values of momenta, $p_{a}$ and $p_{b}$, satisfy
\begin{equation}
\lim_{\rho\rightarrow0^{+}}U_{b}(\rho)U_{a}^{\prime}(\rho)=\lim_{\rho
\rightarrow0^{+}}U_{a}(\rho)U_{b}^{\prime}(\rho) = 0,
\end{equation}
\begin{equation}
\lim_{\rho\rightarrow0^{+}}V_{b}(\rho)V_{a}^{\prime}(\rho)=\lim_{\rho
\rightarrow0^{+}}V_{a}(\rho)V_{b}^{\prime}(\rho) = 0.
\end{equation}
Following the procedure given in Section.~\ref{chap1:Wronskian-integral formula}, for $\gamma=\alpha,\beta$ states we obtain
\begin{align}
(p_{a}^{2}-p_{b}^{2})  &  \int_{0}^{r}[U_{a\gamma}(r^{\prime})U_{b\gamma
}(r^{\prime})+V_{a\gamma}(r^{\prime})V_{b\gamma}(r^{\prime})]\,dr^{\prime
}\nonumber\label{eqn:sum2}\\
&  =W[U_{a\gamma}(r),U_{b\gamma}(r)]+W[V_{a\gamma}(r),V_{b\gamma}(r)],
\end{align}
and for the combination of $\alpha$ and $\beta$ states, we get
\begin{align}
(p_{a}^{2}-p_{b}^{2})  &  \int_{0}^{r}[U_{a\alpha}(r^{\prime})U_{b\beta
}(r^{\prime})+V_{a\alpha}(r^{\prime})V_{b\beta}(r^{\prime})+U_{b\alpha
}(r^{\prime})U_{a\beta}(r^{\prime})+V_{b\alpha}(r^{\prime})V_{a\beta
}(r^{\prime})]\,dr^{\prime}\nonumber\\
&  =W[U_{a\alpha}(r),U_{b\beta}(r)]+W[U_{a\beta}(r),U_{b\alpha}%
(r)]+W[V_{a\alpha}(r),V_{b\beta}(r)]+W[V_{a\beta}(r),V_{b\alpha}(r)].
\label{eqn:sum5}%
\end{align}
The Wronskian of the $\alpha$-state wave functions and the $\beta$-state wave
functions for the non-interacting region $r\geq R$ are given in Appendix
\ref{append:Wronskiansof functions}.

In \eref{eqn:sum2}, we set $p_{a}=0$ and take the limit $p=p_{b}%
\rightarrow0$. In the region $r\geq R$ we obtain the following relations for
the effective range parameters,
\begin{align}
&  r_{j-1}=b_{j-1}(r)+2q_{0}^{2}W[c_{2}(r),c_{0}(r)]_{j+1}-2\int_{0}%
^{r}\left(  \left[  U_{\alpha}^{(0)}(r^{\prime})\right]  ^{2}+\left[
V_{\alpha}^{(0)}(r^{\prime})\right]  ^{2}\right)  dr^{\prime}%
,\label{eqn:effectrangAlph2}\\
&  r_{j+1}=b_{j+1}(r)+2q_{0}^{2}\frac{1}{a_{j+1}^{2}}W[s_{2}(r),s_{0}%
(r)]_{j-1}-2\int_{0}^{r}\left(  \left[  U_{\beta}^{(0)}(r^{\prime}){}\right]
^{2}+\left[  V_{\beta}^{(0)}(r^{\prime})\right]  ^{2}\right)  dr^{\prime}.
\label{eqn:effectrangBet2}%
\end{align}
Here $b_{j\mp1}$ are
\begin{align}
b_{j\mp1}(r)  &  =\frac{2}{a_{j\mp1}^{2}}W[s_{2}(r),s_{0}(r)]_{j\mp1}+\frac
{2}{a_{j\mp1}}W[c_{0}(r),s_{2}(r)]_{j\mp1}\nonumber\\
&  +\frac{2}{a_{j\mp1}}W[s_{0}(r),c_{2}(r)]_{j\mp1}+2W[c_{2}(r),c_{0}%
(r)]_{j\mp1}, \label{eqn:bAlph1}%
\end{align}
which reduce to the form
\begin{align}
b_{j\mp1}(r)  &  =\frac{1}{a_{j\mp1}^{2}}\frac{2\pi}{\Gamma\left(  j\mp
1+\frac{3}{2}\right)  \Gamma\left(  j\mp1+\frac{5}{2}\right)  }\left(
\frac{r}{2}\right)  ^{2(j\mp1)+3}\nonumber\\
&  -\frac{1}{a_{j\mp1}}\frac{4}{j\mp1+\frac{1}{2}}\left(  \frac{r}{2}\right)
^{2}-\frac{2\Gamma\left(  j\mp1-\frac{1}{2}\right)  \Gamma\left(  j\mp
1+\frac{1}{2}\right)  }{\pi}\left(  \frac{r}{2}\right)  ^{-2(j\mp1)+1}.
\label{eqn:generalb}%
\end{align}

In \eref{eqn:sum5}, we set $p_{a}=0$ and take the same limit,
$p=p_{b}\rightarrow0$. In the region $r\geq R$ we obtain
\begin{equation}
q_{1}\frac{2}{a_{j+1}}=d_{j}(r)-2\int_{0}^{r}\left[  U_{\alpha}^{(0)}%
(r^{\prime})U_{\beta}^{(0)}(r^{\prime})+V_{\alpha}^{(0)}(r^{\prime})V_{\beta
}^{(0)}(r^{\prime})\right]  \,dr^{\prime}. \label{eqn:sum8}%
\end{equation}
Here $d_{j}(r)$ is
\begin{align}
d_{j}(r)=  &  -q_{0}\frac{2}{a_{j-1}a_{j+1}}W[s_{2}(r),s_{0}(r)]_{j-1}%
+2q_{0}W[c_{2}(r),c_{0}(r)]_{j+1}\nonumber\\
&  +q_{0}\frac{2}{a_{j+1}}\Big\{W[c_{2}(r),s_{0}(r)]_{j-1}-W[c_{2}%
(r),s_{0}(r)]_{j+1}\Big\}, \label{eqn:sum9}%
\end{align}
and this can be written as
\begin{align}
d_{j}(r)=  &  \frac{-q_{0}}{a_{j-1}a_{j+1}}\frac{2\pi}{\Gamma\left(  \frac
{1}{2}+j\right)  \Gamma\left(  \frac{3}{2}+j\right)  }\left(  \frac{r}%
{2}\right)  ^{2j+1}\nonumber\\
&  +\frac{q_{0}}{a_{j+1}}\frac{4}{(2j-1)(2j+3)}r^{2}-2q_{0}\frac{\Gamma\left(
j+\frac{1}{2}\right)  \Gamma\left(  j+\frac{3}{2}\right)  }{\pi}\left(
\frac{r}{2}\right)  ^{-2j-1}. \label{eqn:sum10}%
\end{align}

All of equations derived here have been numerically checked using a simple
potential model.\ The numerical calculations using
delta-function shell potentials with partial-wave mixing have been performed, and details are
given in Appendix.~\ref{append:C}.

\section{Causality Bounds}

\label{sec:causalityBounds}

The terms in the integrals in \eref{eqn:effectrangAlph2} and
\eref{eqn:effectrangBet2} are positive semi-definite since the wave
functions are real. Therefore \eref{eqn:effectrangAlph2} and
\eref{eqn:effectrangBet2} place upper bounds for the effective range
$r_{j-1}$ and $r_{j+1}$ respectively. As noted in the introduction, these
upper bounds result from the causality and unitarity in the quantum scattering
problem. \ Our results are extensions of single-channel results in
Ref.~\cite{Phillips1997} for the $s$-wave in three dimensions and in
Ref.~\cite{Hammer-Lee9} for arbitrary angular momentum and arbitrary dimensions.

The causality bounds for the lower and higher partial-wave effective ranges
are
\begin{equation}
r_{j-1}\leq b_{j-1}(r)-2q_{0}^{2}\frac{\Gamma(j+\frac{1}{2})\Gamma(j+\frac
{3}{2})}{\pi}\left(  \frac{r}{2}\right)  ^{-2j-1}, \label{eqn:rJ-1bound}%
\end{equation}%
\begin{equation}
r_{j+1}\leq b_{j+1}(r)+\frac{2q_{0}^{2}}{a_{j+1}^{2}}\frac{\pi}{\Gamma
(j+\frac{1}{2})\Gamma(j+\frac{3}{2})}\left(  \frac{r}{2}\right)  ^{2j+1}.
\label{eqn:rJ+1bound}%
\end{equation}
We note that the effective range bounds are modified due to partial-wave
mixing. \ The causality upper bound for $r_{j-1}$ is lowered by the negative
term on the right hand side of \eref{eqn:rJ-1bound}, while the causality
upper bound for the higher partial-wave is increased by the term on the right
hand side of \eref{eqn:rJ+1bound}. \ When $q_{0}$ is nonzero and we take
the limit of zero range interactions, \eref{eqn:rJ-1bound} tells us that
$r_{j-1}$ is driven to negative infinity for any $j$. \ We conclude that the
physics of partial-wave mixing requires a non-zero range for the interactions
in order to comply with the constraints of causality and unitarity. \ In Ref.
\cite{Hammer-Lee9} a similar negative divergence in the effective range
parameter was found for single-channel partial waves with $\ell>0$. \ What is
interesting here is that the negative divergence of the effective range occurs
already in the $^{3}S_{1}$ channel due to partial-wave mixing.

We note that the integral terms in \eref{eqn:effectrangAlph2},
\eref{eqn:effectrangBet2} and \eref{eqn:sum8} are closely related.
Analysis of these equations using the Cauchy-Schwarz inequality provides
another useful relation for the coupled-channel wave functions. For real
functions $f_{1}(r)$, $f_{2}(r)$, $g_{1}(r)$ and $g_{2}(r)$, the
Cauchy-Schwarz inequality is%
\begin{align}
\Big(\int[f_{1}(r)\ f_{2}(r)]\left[
\begin{array}
[c]{c}%
f_{1}(r)\\
f_{2}(r)
\end{array}
\right]  \,dr\Big)\Big(\int[g_{1}(r)\ g_{2}(r)]  &  \left[
\begin{array}
[c]{c}%
g_{1}(r)\\
g_{2}(r)
\end{array}
\right]  \,dr\Big)\nonumber\\
\geq &  \Big|\int[f_{1}(r)g_{1}(r)+f_{2}(r)g_{2}(r)]\,dr\Big|^{2}.
\label{eqn:Cauch-Sch_Eq}%
\end{align}
When we apply the inequality to our coupled wave functions, we get
\begin{equation}
f_{j-1}(r)g_{j+1}(r)\geq\left[  h_{j}(r)\right]  ^{2},
\label{eqn:the Cauchy-Schwarz ineqality}%
\end{equation}
where
\begin{equation}
f_{j-1}(r)=b_{j-1}(r)-2q_{0}^{2}\frac{\Gamma(j+\frac{1}{2})\Gamma(j+\frac
{3}{2})}{\pi}\left(  \frac{r}{2}\right)  ^{-2j-1}-r_{j-1},
\label{eqn:f(r)CoupledChannel}%
\end{equation}%
\begin{equation}
g_{j+1}(r)=b_{j+1}(r)+\frac{2q_{0}^{2}}{a_{j+1}^{2}}\frac{\pi}{\Gamma
(j+\frac{1}{2})\Gamma(j+\frac{3}{2})}\left(  \frac{r}{2}\right)
^{2j+1}-r_{j+1}, \label{eqn:g(r)CoupledChannel}%
\end{equation}
and
\begin{equation}
h_{j}(r)=d_{j}(r)-q_{1}\frac{2}{a_{j+1}}. \label{eqn:h(r)CoupledChannel}%
\end{equation}
This inequality is used to define a Cauchy-Schwarz range, $R^{\text{C-S}}$, as
the minimum $r$ for each coupled channel where \eref{eqn:rJ-1bound},
\eref{eqn:rJ+1bound} and \eref{eqn:the Cauchy-Schwarz ineqality} hold.

\section{Neutron-Proton Scattering}

\label{sec:Proton-Neutron-by-Nijmegen}

We now apply our causality bounds to physical neutron-proton data. \ In this
study, we use the low energy neutron-proton scattering data (0-350 MeV) from
the NN data base by the Nijmegen Group \cite{Nijmegen1993}. Tables~\ref{table:tableforOnechannelWaves}- \ref{table:tableforquad} show the
low-energy threshold parameters in the eigenphase parameterization for the
NijmII and the Reid93 potentials. These parameters are calculated using the
results obtained in Ref. \cite{ValderramaArriola} for the low-energy threshold
parameters of the nuclear bar parameterization and relations between
eigenphase and nuclear bar parameterizations given in Appendix \ref{append:B}.
Using these numbers we analyze \eref{eqn:effectrangAlph2},
\eref{eqn:effectrangBet2} and \eref{eqn:sum8}, as well as causality
bounds for the uncoupled channels.

\begin{table}[tbh]
\caption{The eigenphase low energy parameters of uncoupled channels for
neutron-proton scattering by the NijmII and the Reid93 interaction
potentials.}%
\label{table:tableforOnechannelWaves}
\begin{center}
\rowcolors{3}{gray!15}{white}
\begin{tabular}
[c]{||c|c|c||}\hline\hline
Channel & $a_{\ell}$ [$\text{fm}^{2\ell+1}$] & $r_{\ell}$ [$\text{fm}^{-2\ell+1}$]\\
& NijmII (Reid93) & NijmII (Reid93)\\\hline\hline
$^{1}S_{0}$ & -23.727 (-23.735) & 2.670 (2.753)\\\hline
$^{1}P_{1}$ & 2.797 (2.736) & -6.399 (-6.606)\\\hline
$^{3}P_{0}$ & -2.468 (-2.469) & 3.914 (3.870)\\\hline
$^{3}P_{1}$ & 1.529 (1.530) & -8.580 (-8.556)\\\hline
$^{1}D_{2}$ & -1.389 (-1.377) & 14.87 (15.04)\\\hline
$^{3}D_{2}$ & -7.405 (-7.411) & 2.858 (2.851)\\\hline
$^{1}F_{3}$ & 8.383 (8.365) & -3.924 (-3.936)\\\hline
$^{3}F_{3}$ & 2.703 (2.686) & -9.932 (-9.994)\\\hline\hline
\end{tabular}
\end{center}
\end{table}\begin{table}[tbh]
\caption{The eigenphase low energy parameters of coupled channels for
neutron-proton scattering by the NijmII and the Reid93 interaction
potentials.}%
\label{table:tableforTwochannelWaves}
\begin{center}
\rowcolors{3}{gray!15}{white}
\begin{tabular}
[c]{||c|c|c||}\hline\hline
Channel & $a_{\ell}$ [$\text{fm}^{2\ell+1}$] & $r_{\ell}$ [$\text{fm}^{-2\ell+1}$]\\
& NijmII (Reid93) & NijmII (Reid93)\\\hline\hline
$^{3}S_{1}$ & 5.418 (5.422) & 1.7531 (1.7554)\\\hline
$^{3}D_{1}$ & 6.0043 (5.9539) & -3.523 (-3.566)\\\hline
$^{3}P_{2}$ & -0.2844 (-0.2892) & -11.1465 (-10.7127)\\\hline
$^{3}F_{2}$ & 8.126 (7.882) & -5.640 (-5.821)\\\hline
$^{3}D_{3}$ & -0.1449 (-0.177) & 288.428 (198.528)\\\hline
$^{3}G_{3}$ & 648.813 (534.594) & -0.03306 (-0.0529)\\\hline\hline
\end{tabular}
\end{center}
\end{table}\begin{table}[tbh]
\caption{The eigenphase low energy mixing parameters of coupled channels for
neutron-proton scattering by the NijmII and the Reid93 interaction
potentials.}%
\label{table:tableforquad}
\begin{center}\rowcolors{3}{gray!15}{white}
\begin{tabular}
[c]{||c|c|c||}\hline\hline
Mixing angle & $q_{0}$ [$\text{fm}^{2}$] & $q_{1}$ [$\text{fm}^{4}$]\\
& NijmII (Reid93) & NijmII (Reid93)\\\hline\hline
$\varepsilon_{1}$ & 0.303987 (0.303394) & -2.00228 (-1.99129)\\\hline
$\varepsilon_{2}$ & -5.65752 (-5.5325) & 65.8602 (64.2979)\\\hline
$\varepsilon_{3}$ & 66.6632 (54.7062) & 340.988 (94.9015)\\\hline\hline
\end{tabular}
\end{center}
\end{table}

\subsection{Uncoupled Channels}
\label{sec:Cases without Mixing}

We start with channels of a single uncoupled
partial wave. Since there is no mixing between different partial-waves, we
evaluate \eref{eqn:effectrangAlph2} and \eref{eqn:effectrangBet2}
with zero mixing angle, and we obtain the following equation for the effective
range
\begin{equation}
r_{\ell}=b_{\ell}(r)-2\int_{0}^{r}\Big[U_{\ell}^{(0)}(r^{\prime})\Big]^{2}\,dr^{\prime},
\label{eqn:EffecRangNoMix}%
\end{equation}
where $b_{\ell}$ is given in \eref{eqn:uncoup:generalb}. These solutions
were derived by Hammer and Lee~\cite{HammerDean27} for arbitrary dimension and
angular momentum.

\begin{figure}[t]
\begin{center}
\resizebox{120mm}{!}
{\includegraphics{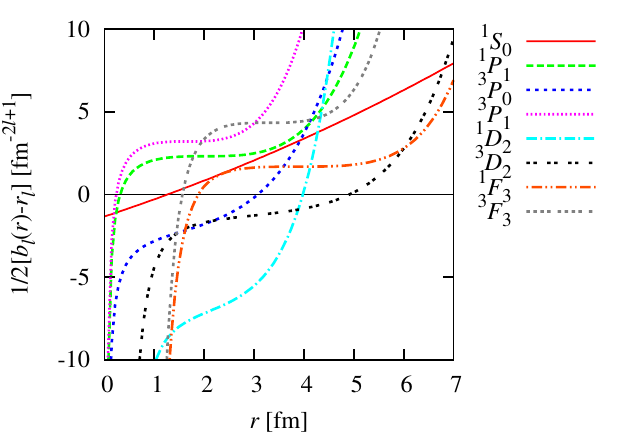}}
\end{center}
\caption{The plot of $[b_{\ell}(r)-r_{\ell}]/2$ as a function of $r$ for
neutron-proton scattering via the NijmII potential in the $^{2s+1}\ell_{j}$
channel.}%
\label{figure:SingletALL}%
\end{figure}

Here, we analyze the causality bound of the effective range for $\ell\leq3$ using
the scattering parameters in \tref{table:tableforOnechannelWaves}. In
\fref{figure:SingletALL}, we plot $\frac{1}{2}[b_{\ell}(r)-r_{\ell}]$ for all
of uncoupled channels with $\ell\leq3$. \ The physical region corresponds with
$\frac{1}{2}[b_{\ell}(r)-r_{\ell}]\geq0$.

For $s$-wave scattering
\begin{equation}
b_{0}(r)=\frac{2}{3a_{0}^{2}}r^{3}-\frac{2}{a_{0}}r^{2}+2r,
\label{eqn:S_wave_b}%
\end{equation}
for $p$-wave,
\begin{equation}
b_{1}(r)=\frac{2r^{5}}{45a_{1}^{2}}-\frac{2r^{2}}{3a_{1}}-\frac{2}{r},
\label{eqn:P_wave_b}%
\end{equation}
for $d$-wave,
\begin{equation}
b_{2}(r)=\frac{2}{1575a_{2}^{2}}r^{7}-\frac{2}{5a_{2}}r^{2}-\frac{6}{r^{3}},
\label{eqn:D_wave_b}%
\end{equation}
for $f$-wave,
\begin{equation}
b_{3}(r)=\frac{2r^{9}}{99225a_{3}^{2}}-\frac{2r^{2}}{7a_{3}}-\frac{90}{r^{5}},
\label{eqn:F_wave_b}%
\end{equation}
and for $g$-wave,
\begin{equation}
b_{4}(r)=\frac{2r^{11}}{9823275a_{4}^{2}}-\frac{2r^{2}}{9a_{4}}-\frac
{3150}{r^{7}}. \label{eqn:G_wave_b}%
\end{equation}

\subsection{Coupled Channels}

\label{sec:triplet Channel} We now analyze channels with coupled partial
waves. \ We plot \eref{eqn:f(r)CoupledChannel} and
\eref{eqn:g(r)CoupledChannel} for all coupled channels with $j\leq3$.
\ The physical region correspond both $f_{j-1}(r)\geq0$ and $g_{j+1}(r)\geq0$.
\begin{figure}[hptb]
\begin{center}
\resizebox{120mm}{!}{\includegraphics{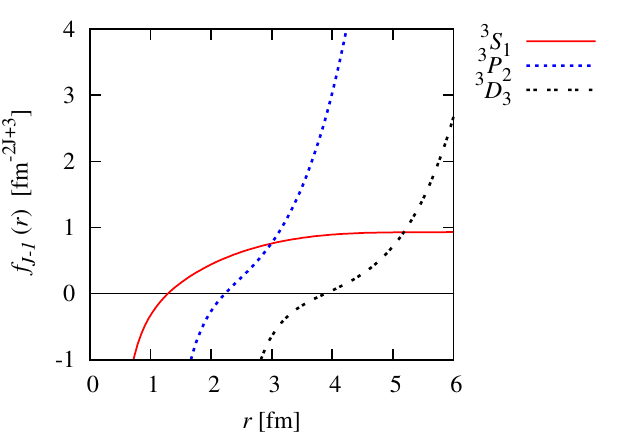}}
\end{center}
\caption{The plot of $f_{j-1}(r)$ as a function of $r$ for neutron-proton
scattering via the NijmII potential for $j\leq3$. Here $f_{1}(r)$ is rescaled
by a factor of $0.01$ and $f_{2}(r)$ is rescaled by a factor of $10^{-4}$.}%
\label{figure:AllTripletLOWER}%
\end{figure}
\begin{figure}[hptb]
\begin{center}
\resizebox{120mm}{!}{\includegraphics{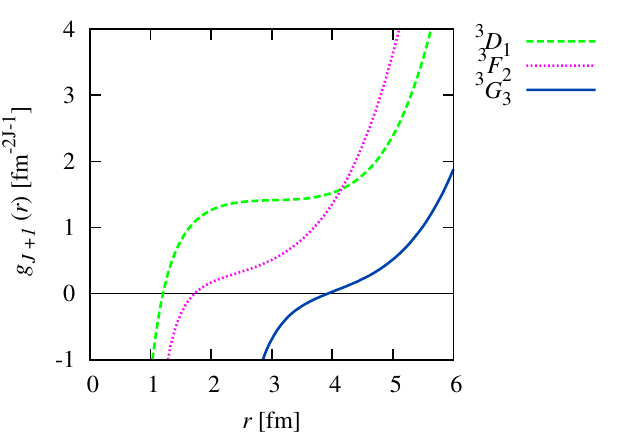}}
\end{center}
\caption{The plot of $g_{j+1}(r)$ as a function of $r$ for neutron-proton
scattering via the NijmII potential for $j\leq3$. Here $g_{3}(r)$ is rescaled
by a factor of $0.1$.}%
\label{figure:AllTripletHIGHER}%
\end{figure}

\subsubsection{\textbf{$\, ^{3}$S$_{1}$-$\, ^{3}$D$_{1}$ Coupling.}}

\label{sec:S-D Proton-Neutron}

We consider \eref{eqn:effectrangAlph2} - \eref{eqn:sum9} for the
$^{3}S_{1}$-$^{3}D_{1}$ coupled channel. We evaluate the Wronskians for $j=1$
and get
\begin{equation}
b_{0}(r)-q_{0}^{2}\frac{6}{r^{3}}-r_{0}=2\int_{0}^{r}\left(  \left[
U_{\alpha}^{(0)}(r^{\prime})\right]  ^{2}+ \left[  V_{\alpha}^{(0)}(r^{\prime
})\right]  ^{2} \right)  dr^{\prime}, \label{eqn:r3S1}%
\end{equation}%
\begin{equation}
b_{2}(r)+q_{0}^{2}\frac{2r^{3}}{3a_{2}^{2}}-r_{2}=2\int_{0}^{r}\left(  \left[
U_{\beta}^{(0)}(r^{\prime})\right]  ^{2} + \left[  V_{\beta}^{(0)}(r^{\prime
})\right]  ^{2} \right)  dr^{\prime}, \label{eqn:r3D1}%
\end{equation}%
\begin{equation}
d_{1}(r)-q_{1}\frac{2}{a_{2}}=2\int_{0}^{r}[U_{0\alpha}(r^{\prime})U_{0\beta
}(r^{\prime})+V_{0\alpha}(r^{\prime})V_{0\beta}(r^{\prime})]\,dr^{\prime}.
\label{eqn:q1 boundE1}%
\end{equation}
$b_{0}(r)$ and $b_{2}(r)$ are given in \eref{eqn:S_wave_b} and in
\eref{eqn:D_wave_b}, respectively, and $d_{1}(r)$ is
\begin{equation}
d_{1}(r)=-q_{0}\frac{1}{a_{0}a_{2}}\frac{2r^{3}}{3}+q_{0}\frac{1}{a_{2}}%
\frac{4r^{2}}{5}-q_{0}\frac{6}{r^{3}}. \label{eqn:d1 bound}%
\end{equation}

Using the scattering parameters in Tables \ref{table:tableforTwochannelWaves} - \ref{table:tableforquad}, we plot \eref{eqn:r3S1}, \eref{eqn:r3D1}
and \eref{eqn:q1 boundE1} as functions of $r$.
\begin{figure}[hptb]
\begin{center}
\resizebox{120mm}{!}{\includegraphics{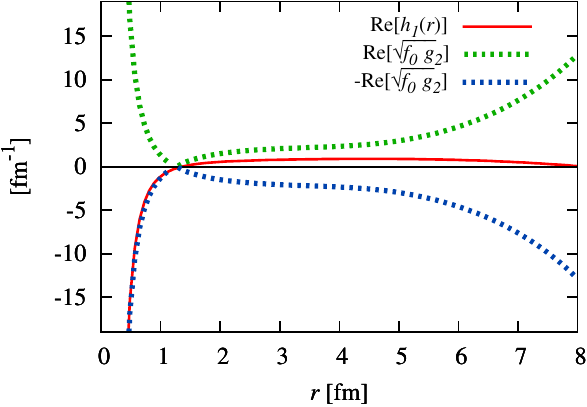}}
\end{center}
\caption{We plot Re$\left[  \sqrt{f_{0}(r)g_{2}(r)}\right]  $, $-$Re$\left[
\sqrt{f_{0}(r)g_{2}(r)}\right]  $, and Re$\left[  h_{1}(r)\right]  $ as
functions of $r$ for neutron-proton scattering in $^{3}S_{1}$-$^{3}D_{1}$
coupled channel.}%
\label{fig:figureS_D_MixingNijmII}%
\end{figure}
In
\fref{fig:figureS_D_MixingNijmII} we show the physical region where the
causality bounds $f_{0}(r)\geq0$, $g_{2}(r)\geq0$, and $f_{0}(r)g_{2}(r)\geq
h_{1}^{2}(r)$, are satisfied. \ Here we have%
\begin{equation}
f_{0}(r)=\frac{2}{3a_{0}^{2}}r^{3}-\frac{2}{a_{0}}r^{2}+2r-q_{0}^{2}\frac
{6}{r^{3}}-r_{0}, \label{eqn:f_0(r)}%
\end{equation}%
\begin{equation}
g_{2}(r)=\frac{2}{1575a_{2}^{2}}r^{7}-\frac{2}{5a_{2}}r^{2}-\frac{6}{r^{3}%
}+q_{0}^{2}\frac{2r^{3}}{3a_{2}^{2}}-r_{2}, \label{eqn:g_2(r)}%
\end{equation}%
\begin{equation}
h_{1}(r)=-q_{0}\frac{1}{a_{0}a_{2}}\frac{2r^{3}}{3}+q_{0}\frac{1}{a_{2}}%
\frac{4r^{2}}{5}-q_{0}\frac{6}{r^{3}}-q_{1}\frac{2}{a_{2}}. \label{eqn:h_1(r)}%
\end{equation}

\subsubsection{\textbf{$\, ^{3}$P$_{2} -\, ^{3}$F$_{2}$ Coupling.}}

\label{sec:P-F Proton-Neutron}

In the $^{3}P_{2}$-$^{3}F_{2}$ coupled channel \eref{eqn:effectrangAlph2}
- \eref{eqn:sum9} take the following forms,
\begin{equation}
b_{1}(r)-q_{0}^{2}\frac{90}{r^{5}}-r_{1}=2\int_{0}^{r}\left(  \left[
U_{\alpha}^{(0)}(r^{\prime})\right]  ^{2}+ \left[  V_{\alpha}^{(0)}(r^{\prime
})\right]  ^{2} \right)  dr^{\prime}, \label{eqn:r3P2}%
\end{equation}%
\begin{equation}
b_{3}(r)+q_{0}^{2}\frac{1}{a_{3}^{2}}\frac{2r^{5}}{45}-r_{3}=2\int_{0}%
^{r}\left(  \left[  U_{\beta}^{(0)}(r^{\prime})\right]  ^{2} + \left[
V_{\beta}^{(0)}(r^{\prime})\right]  ^{2} \right)  dr^{\prime},
\label{eqn:r3F2}%
\end{equation}%
\begin{equation}
d_{2}(r)-q_{1}\frac{2}{a_{3}}=2\int_{0}^{r}[U_{0\alpha}(r^{\prime})U_{0\beta
}(r^{\prime})+V_{0\alpha}(r^{\prime})V_{0\beta}(r^{\prime})]\,dr^{\prime}.
\label{eqn:q1 boundE2}%
\end{equation}
$b_{1}(r)$ and $b_{3}(r)$ are defined in \eref{eqn:P_wave_b} and
\eref{eqn:F_wave_b}, respectively, and $d_{2}(r)$ is
\begin{equation}
d_{2}(r)=-q_{0}\frac{1}{a_{1}a_{3}}\frac{2r^{5}}{45}+q_{0}\frac{1}{a_{3}}%
\frac{4r^{2}}{21}-q_{0}\frac{90}{r^{5}}. \label{eqn:eqn d2}%
\end{equation}
The causality bounds are $f_{1}(r)\geq0$, $g_{3}(r)\geq0$, and $f_{1}%
(r)g_{3}(r)\geq h_{2}^{2}(r)$, where
\begin{equation}
f_{1}(r)=\frac{2r^{5}}{45a_{1}^{2}}-\frac{2r^{2}}{3a_{1}}-\frac{2}{r}%
-q_{0}^{2}\frac{90}{r^{5}}-r_{1}, \label{eqn:f_1(r)}%
\end{equation}%
\begin{equation}
g_{3}(r)=\frac{2r^{9}}{99225a_{3}^{2}}-\frac{2r^{2}}{7a_{3}}-\frac{90}{r^{5}%
}+q_{0}^{2}\frac{1}{a_{3}^{2}}\frac{2r^{5}}{45}-r_{3}, \label{eqn:g_3(r)}%
\end{equation}%
\begin{equation}
h_{2}(r)=-q_{0}\frac{1}{a_{1}a_{3}}\frac{2r^{5}}{45}+q_{0}\frac{1}{a_{3}}%
\frac{4r^{2}}{21}-q_{0}\frac{90}{r^{5}}-q_{1}\frac{2}{a_{3}}.
\label{eqn:h_2(r)}%
\end{equation}
\begin{figure}[hptb]
\begin{center}
\resizebox{120mm}{!}{\includegraphics{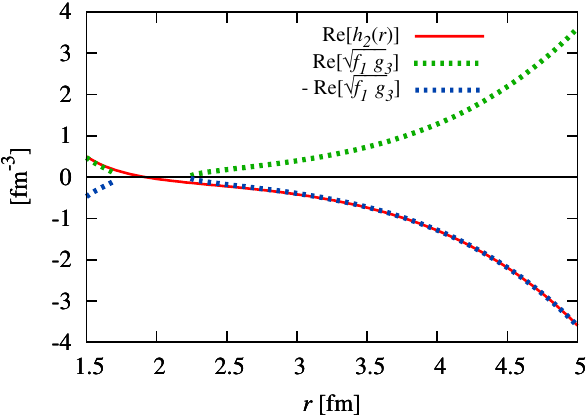}}
\end{center}
\caption{We plot Re$\left[  \sqrt{f_{1}(r)g_{3}(r)}\right]  $, $-$Re$\left[
\sqrt{f_{1}(r)g_{3}(r)}\right]  $, and Re$\left[  h_{2}(r)\right]  $ as
functions of $r$ for neutron-proton scattering in $^{3}P_{2}$-$^{3}F_{2}$
coupled channel. The functions are rescaled by a factor of $0.01$. }%
\label{fig:figureP_F_MixingNijmII}%
\end{figure}
In \fref{fig:figureP_F_MixingNijmII} we show the physical region for the
$^{3}P_{2}$-$^{3}F_{2}$ coupled channel wave functions.

\subsubsection{\textbf{$\, ^{3}$D$_{3} -\, ^{3}$G$_{3}$ Coupling.}}

\label{sec:D-G Proton-Neutron}

For $j=3$, the $^{3}D_{3}$ and $^{3}G_{3}$ channels are coupled. In this case
\eref{eqn:effectrangAlph2}-\eref{eqn:sum9} read
\begin{equation}
b_{2}(r)-q_{0}^{2}\frac{3150}{r^{7}}-r_{2}=2\int_{0}^{r}\left(  \left[
U_{\alpha}^{(0)}(r^{\prime})\right]  ^{2}+ \left[  V_{\alpha}^{(0)}(r^{\prime
})\right]  ^{2} \right)  dr^{\prime}, \label{eqn:r3D3}%
\end{equation}%
\begin{equation}
b_{4}(r)+\frac{q_{0}^{2}}{a_{4}^{2}}\frac{2r^{7}}{1575}-r_{4}=2\int_{0}%
^{r}\left(  \left[  U_{\beta}^{(0)}(r^{\prime})\right]  ^{2} + \left[
V_{\beta}^{(0)}(r^{\prime})\right]  ^{2} \right)  dr^{\prime},
\label{eqn:r3G3}%
\end{equation}%
\begin{equation}
d_{3}(r)-q_{1}\frac{2}{a_{4}}=2\int_{0}^{r}[U_{0\alpha}(r^{\prime})U_{0\beta
}(r^{\prime})+V_{0\alpha}(r^{\prime})V_{0\beta}(r^{\prime})]\,dr^{\prime}.
\label{eqn:q1 boundE3}%
\end{equation}
Here $d_{3}(r)$ is
\begin{equation}
d_{3}(r)=-\frac{q_{0}}{a_{2}a_{4}}\frac{2r^{7}}{1575}+q_{0}\frac{1}{a_{4}%
}\frac{4r^{2}}{45}-q_{0}\frac{3150}{r^{7}}. \label{eqn:eqn d3}%
\end{equation}
The causality bounds are again $f_{2}(r)\geq0$, $g_{4}(r)\geq0$, and
$f_{2}(r)g_{4}(r)\geq h_{3}^{2}(r)$, where
\begin{equation}
f_{2}(r)=\frac{2r^{7}}{1575a_{2}^{2}}-\frac{2r^{2}}{5a_{2}}-\frac{6}{r^{3}%
}-q_{0}^{2}\frac{3150}{r^{7}}-r_{2}, \label{eqn:f_2(r)}%
\end{equation}%
\begin{equation}
g_{4}(r)=\frac{2r^{11}}{9823275a_{4}^{2}}-\frac{2r^{2}}{9a_{4}}-\frac
{3150}{r^{7}}+\frac{q_{0}^{2}}{a_{4}^{2}}\frac{2r^{7}}{1575}-r_{4},
\label{eqn:g_4(r)}%
\end{equation}%
\begin{equation}
h_{3}(r)=-\frac{q_{0}}{a_{2}a_{4}}\frac{2r^{7}}{1575}+q_{0}\frac{1}{a_{4}%
}\frac{4r^{2}}{45}-q_{0}\frac{3150}{r^{7}}-q_{1}\frac{2}{a_{4}}.
\label{eqn:h_3(r)}%
\end{equation}
\begin{figure}[hptb]
\begin{center}
\resizebox{120mm}{!}{\includegraphics{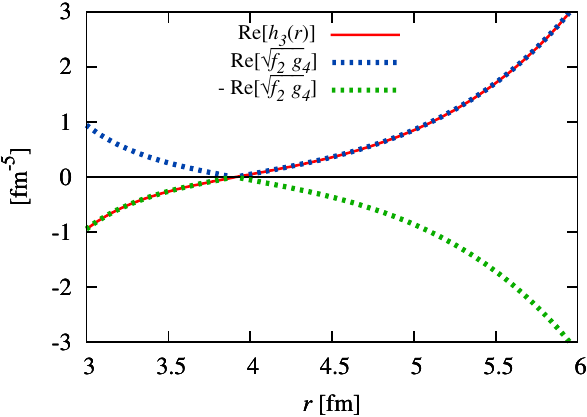}}
\end{center}
\caption{We plot Re$\left[  \sqrt{f_{2}(r)g_{4}(r)}\right]  $, $-$Re$\left[
\sqrt{f_{2}(r)g_{4}(r)}\right]  $, and Re$\left[  h_{3}(r)\right]  $ as
functions of $r$ for neutron-proton scattering in $^{3}D_{3}$-$^{3}G_{3}$
coupled channel. The functions are rescaled by a factor of $0.01$. }%
\label{fig:figureD_G_MixingNijmII}%
\end{figure}
We show plots for the $^{3}D_{3}$-$^{3}G_{3}$ channel in
\fref{fig:figureD_G_MixingNijmII}.

\section{Results and Discussion}

In this section we present the results for the causal and Cauchy-Schwarz ranges, $R^{b}$, and $R^{\text{C-S}}$. We use the NijmII\ scattering data for neutron-proton scattering presented above. In Table \ref{table:RcausalSingleChannel} we show results for the causal range for all uncoupled channels by setting
\begin{equation}
r_{\ell}=b_{\ell}(r).
\end{equation}
In \tref{table:RcausalCoupledChannel} we determine the causal range for all coupled channels using Eqs.~\ref{eqn:f(r)CoupledChannel} - \ref{eqn:g(r)CoupledChannel}. Also, we find the Cauchy-Schwarz ranges shown in \tref{table:R_cauchy_schwarz} using \eref{eqn:the Cauchy-Schwarz ineqality}.
\begin{table}[htbh]
\caption{The causal ranges for uncoupled channels.}%
\label{table:RcausalSingleChannel}
\begin{center}
\begin{tabular}
[c]{||c|c|c|c|c|c|c|c|c||}\hline\hline
Channels & $^{1}S_{0}$ & $^{1}P_{1}$ & $^{3}P_{0}$ & $^{3}P_{1}$ & $^{1}D_{2}
$ & $^{3}D_{2}$ & $^{1}F_{3}$ & $^{3}F_{3}$\\\hline\hline
$R^{b}$ [fm] & $\ 1.27\ $ & $\ 0.31\ $ & $\ 3.07\ $ & $\ 0.23\ $ & $\ 3.98\ $
& $\ 4.91\ $ & $\ 1.88\ $ & $\ 1.56$\\\hline\hline
\end{tabular}
\end{center}
\end{table}\begin{table}[tbh]
\caption{The causal ranges for coupled channels.}%
\label{table:RcausalCoupledChannel}
\begin{center}
\begin{tabular}
[c]{||c|c|c|c|c|c|c||}\hline\hline
Channels & $^{3}S_{1}$ & $^{3}D_{1}$ & $^{3}P_{2}$ & $^{3}F_{2}
$ & $^{3}D_{3}$ & $^{3}G_{3}$\\\hline\hline
$R^{b}$ [fm] & $\ 1.29\ $ & $\ 1.20\ $ & $\ 2.23\ $& $\ 1.73\ $ & $\ 4.03\ $
& $\ 3.92$\\\hline\hline
\end{tabular}
\end{center}
\end{table}
\begin{table}[tbh]
\caption{The Cauchy-Schwarz ranges for coupled channels.}%
\label{table:R_cauchy_schwarz}
\begin{center}
\begin{tabular}
[c]{||c|c|c|c||}\hline\hline
Channels & \ $^{3}S_{1}$-$^{3}D_{1}$ \  & \ $^{3}P_{2}$-$^{3}F_{2}$ \  &
\ $^{3}D_{3}$-$^{3}G_{3}$ \ \\\hline\hline
$R^{\text{C-S}}$ [fm] & $\ 1.29\ $ & $\ 4.65\ $ & $\ 5.68\ $\\\hline\hline
\end{tabular}
\end{center}
\end{table}

We find that in some channels the causal and Cauchy-Schwarzranges are surprisingly large, and it is worthwhile to probe the origin ofthese large ranges. It is convenient to collect together some of the key formulas derived above. The Cauchy-Schwarz inequality has the form
\begin{equation}
f_{j-1}(r)g_{j+1}(r)\geq\left[  h_{j}(r)\right]  ^{2}, \label{CauchySchwarz2}%
\end{equation}
where%
\begin{equation}
f_{j-1}(r)=b_{j-1}(r)-2q_{0}^{2}\frac{\Gamma(j+\frac{1}{2})\Gamma(j+\frac
{3}{2})}{\pi}\left(  \frac{r}{2}\right)  ^{-2j-1}-r_{j-1},
\end{equation}%
\begin{equation}
g_{j+1}(r)=b_{j+1}(r)+\frac{2q_{0}^{2}}{a_{j+1}^{2}}\frac{\pi}{\Gamma
(j+\frac{1}{2})\Gamma(j+\frac{3}{2})}\left(  \frac{r}{2}\right)
^{2j+1}-r_{j+1},
\end{equation}%
\begin{equation}
h_{j}(r)=d_{j}(r)-q_{1}\frac{2}{a_{j+1}},
\end{equation}%
\begin{align}
b_{j\mp1}(r)  &  =\frac{1}{a_{j\mp1}^{2}}\frac{2\pi}{\Gamma\left(  j\mp
1+\frac{3}{2}\right)  \Gamma\left(  j\mp1+\frac{5}{2}\right)  }\left(
\frac{r}{2}\right)  ^{2(j\mp1)+3}\nonumber\\
&  -\frac{1}{a_{j\mp1}}\frac{4}{j\mp1+\frac{1}{2}}\left(  \frac{r}{2}\right)
^{2}-\frac{2\Gamma\left(  j\mp1-\frac{1}{2}\right)  \Gamma\left(  j\mp
1+\frac{1}{2}\right)  }{\pi}\left(  \frac{r}{2}\right)  ^{-2(j\mp1)+1},
\end{align}%
\begin{align}
d_{j}(r)=  &  \frac{-q_{0}}{a_{j-1}a_{j+1}}\frac{2\pi}{\Gamma\left(  \frac
{1}{2}+j\right)  \Gamma\left(  \frac{3}{2}+j\right)  }\left(  \frac{r}%
{2}\right)  ^{2j+1}\nonumber\\
&  +\frac{q_{0}}{a_{j+1}}\frac{4}{(2j-1)(2j+3)}r^{2}-2q_{0}\frac{\Gamma\left(
j+\frac{1}{2}\right)  \Gamma\left(  j+\frac{3}{2}\right)  }{\pi}\left(
\frac{r}{2}\right)  ^{-2j-1}.
\end{align}
\bigskip
We note that the leading power of $r$ in $g_{j+1}(r)$ is%
\begin{equation}
\frac{1}{a_{j+1}^{2}}\frac{2\pi}{\Gamma\left(  j+1+\frac{3}{2}\right)
\Gamma\left(  j+1+\frac{5}{2}\right)  }\left(  \frac{r}{2}\right)  ^{2j+5}.
\end{equation}
This has a very small numerical prefactor multiplying $a_{j+1}^{-2}r^{2j+5}$.
\ For $j=1$ the factor is $2/1575$, for $j=2$ it is $2/99225$, and for $j=3 $
it is $2/9823275$. \ Therefore the term is negligible unless $r$ is large
compared with $(a_{j+1})^{1/(2j+3)}$. \ If we neglect this term, then the term
with the leading power of $r$ on the left hand side of
\eref{CauchySchwarz2} is the same as that on the right hand side,%
\begin{align}
&  \frac{1}{a_{j-1}^{2}}\frac{2\pi}{\Gamma\left(  j-1+\frac{3}{2}\right)
\Gamma\left(  j-1+\frac{5}{2}\right)  }\left(  \frac{r}{2}\right)
^{2j+1}\cdot\frac{2q_{0}^{2}}{a_{j+1}^{2}}\frac{\pi}{\Gamma(j+\frac{1}%
{2})\Gamma(j+\frac{3}{2})}\left(  \frac{r}{2}\right)  ^{2j+1}\nonumber\\
&  =\left[  \frac{-q_{0}}{a_{j-1}a_{j+1}}\frac{2\pi}{\Gamma\left(  \frac{1}%
{2}+j\right)  \Gamma\left(  \frac{3}{2}+j\right)  }\left(  \frac{r}{2}\right)
^{2j+1}\right]  ^{2}.
\end{align}
As a result the curves for $f_{j-1}(r)g_{j+1}(r)$ and $\left[  h_{j}%
(r)\right]  ^{2}$ are approximately parallel for large $r$ until the term that
we have neglected becomes significant. \ These nearly parallel trajectories
inflate the value of the Cauchy-Schwarz range $r=R^{\text{C-S}}$ where the two
curves cross.

For the $^{3}S_{1}$-$^{3}D_{1}$ coupled channel we find $R^{\text{C-S}}$ is
about the same size as the Compton wavelength of the pion, $m_{\pi}^{-1}=1.5 $
fm. \ This is also comparable to what one expects for the range of the
nucleon-nucleon interaction. \ However the results are more interesting for
$j>1$. \ In the $^{3}P_{2}$-$^{3}F_{2}$ channel we have $R^{\text{C-S}}=4.65$
fm. \ And for the $^{3}D_{3}$-$^{3}G_{3}$ coupled channel we find
$R^{\text{C-S}}$ $=$ $5.68$ fm. \ These values are surprisingly large in
comparison with $m_{\pi}^{-1}$.

\section{One-Pion Exchange Potential}

We note that there are some channels where the causal range $R^{b}$ is also
quite large. \ By definition $R^{\text{C-S}}\geq R^{b}$ and so the
Cauchy-Schwarz range will then also be large. \ The causal range is the
minimum value for $r$ such that%
\begin{equation}
f_{j-1}(r)=b_{j-1}(r)-2q_{0}^{2}\frac{\Gamma(j+\frac{1}{2})\Gamma(j+\frac
{3}{2})}{\pi}\left(  \frac{r}{2}\right)  ^{-2j-1}-r_{j-1}\geq0
\end{equation}
for the lower partial wave, or%
\begin{equation}
g_{j+1}(r)=b_{j+1}(r)+\frac{2q_{0}^{2}}{a_{j+1}^{2}}\frac{\pi}{\Gamma
(j+\frac{1}{2})\Gamma(j+\frac{3}{2})}\left(  \frac{r}{2}\right)
^{2j+1}-r_{j+1}\geq0
\end{equation}
for the higher partial wave. \ For uncoupled channels we take $q_{0}=0$.

The largest values for $R^{b}$ occur when the effective range parameter is
positive or near zero. \ See for example the causal ranges for the $^{1}D_{2}
$, $^{3}D_{2}$, $^{3}D_{3}$, and $^{3}G_{3}$ channels. What happens is that
the function $f_{j-1}(r)$ or $g_{j+1}(r)$ remains negative with a rather small
slope until $r$ becomes quite large. \ The small slope is again associated
with the fact that the term with the highest power of $r$ has a small
numerical prefactor.

The range of the interaction plays the dominant role in setting the causal range. \ In the language of local potentials, this is the radius at which the magnitude of the potential is numerically very small. \ However there is also some influence of the exponential tail of the potential upon the causal range.

In all channels where the causal range is unusually large, $^{1}D_{2}$, $^{3}D_{2}$, $^{3}D_{3}$, and $^{3}G_{3}$, we find that the tail of the one-pion exchange potential is attractive. \ At smaller radii, the potential crosses over at some classical turning point to become
repulsive. \ See for example Figures~2 - 4 in Ref.~\cite{Stoks1993NijII}.

The detailed mechanism requires further study, but it appears that this geometry can cause a near-threshold wavepacket to reflect before reaching the classical turning point, thus mimicking a longer range potential. \ However some fine tuning is needed to produce a large causal range, as there is no enhancement in the $^{1}S_{0}$ and $^{3}S_{1}$ channels and a smaller amount of enhancement in the $^{3}P_{0}$ channel.

There seems to be no such enhancement of the causal range
in the $^{1}P_{1}$, $^{3}P_{1}$, and $^{3}D_{1}$ channels where the tail of the potential is repulsive. \ In fact, the causal range for the $^{1}P_{1}$ and $^{3}P_{1}$ channels are unusually small. \ This appears be related to
quantum tunneling into the inner region where the potential is attractive.

In the following analysis we will investigate the
importance of the tail of the one-pion exchange potential plays in setting the
causal range, $R^{b}$. \ We show that even though the one-pion exchange
potential is numerically small at distances larger than $5\,\text{fm}$,
chopping off the one-pion exchange tail at such distances produces a
non-negligible effect. \ The one-pion exchange potential tail appears to be
the source of the large values for $R^{b}$ in higher partial waves where the
central one-pion exchange tail is attractive.

\begin{figure}[hptb]
\begin{center}
\resizebox{120mm}{!}{\includegraphics{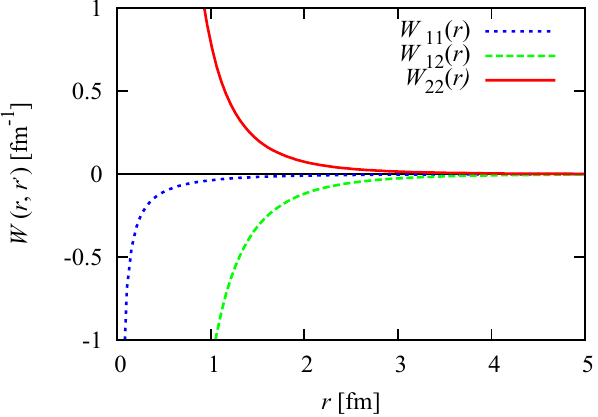}}
\end{center}
\caption{Plot of the potential matrix elements $W_{11}(r)=V_{C}(r)$,
$W_{12}(r)=W_{21}(r)=\sqrt{8}V_{T}(r)$ and $W_{22}(r)=V_{C}(r)-2V_{T}(r)$ as a
function of $r$ in the $^{3}S_{1}$-$^{3}D_{1}$ coupled channel.}%
\label{fig:WrR}%
\end{figure}

If we neglect electromagnetic effects, then the neutron-proton interaction
potential at large distances is governed by the one-pion exchange (OPE)
potential, which in configuration space is
\begin{equation}
V_{OPE}(r)=V_{C}(r)+S_{12}V_{T}(r).
\end{equation}
Here $V_{C}(r)$ is the central potential,
\begin{equation}
V_{C}(r)=\frac{g_{\pi N}^{2}}{12\pi}\left(  \frac{m_{\pi}}{2M_{N}}\right)
^{2}(\vec{\tau}_{1}\cdot\vec{\tau}_{2})(\vec{\sigma}_{1}\cdot\vec{\sigma}%
_{2})\frac{e^{-m_{\pi}r}}{r},\label{central potential}%
\end{equation}
$V_{T}(r)$ is the tensor potential,
\begin{equation}
V_{T}(r)=\frac{g_{\pi N}^{2}}{12\pi}\left(  \frac{m_{\pi}}{2M_{N}}\right)
^{2}\ (\vec{\tau}_{1}\cdot\vec{\tau}_{2})\ \Big(1+\frac{3}{m_{\pi}r}+\frac
{3}{(m_{\pi}r)^{2}}\Big)\frac{e^{-m_{\pi}r}}{r},
\end{equation}
and $S_{12}$ is the tensor operator,
\begin{equation}
S_{12}=3(\vec{\sigma_{1}}\cdot\hat{r})(\vec{\sigma_{2}}\cdot\hat{r}%
)-\vec{\sigma_{1}}\cdot\vec{\sigma_{2}}.\label{eqn:S12}%
\end{equation}
Here $m_{\pi}$ is the pion mass, $M_{N}$ is the nucleon mass, and $g_{\pi
N}=13.0$ is the pion-nucleon coupling constant. \ The one-pion exchange
potential is local in space and the interaction matrix in
Eq.~(\ref{eqn:PotentialMatrix}) takes the following form for $j=1$,%
\begin{equation}
W(r,r^{\prime})=\left(
\begin{array}
[c]{cc}%
V_{C}(r) & \sqrt{8}V_{T}(r)\\
\sqrt{8}V_{T}(r) & V_{C}(r)-2V_{T}(r)
\end{array}
\right)  \delta(r-r^{\prime}).\label{eqn:RealPotential}%
\end{equation}
In Figure~(\ref{fig:WrR}) we plot $W_{11}(r)=V_{C}(r)$, $W_{12}(r)=W_{21}%
(r)=\sqrt{8}V_{T}(r),$ and $W_{22}(r)=V_{C}(r)-2V_{T}(r)$ in the $^{3}S_{1}%
$-$^{3}D_{1}$ coupled channel. 

To demonstrate the origin of large causal ranges found in Table
\ref{table:RcausalSingleChannel} and Table \ref{table:RcausalCoupledChannel},
we will present some simple but illustrative numerical examples. \ For each
channel we add a short range potential to the one-pion exchange potential in
order to reproduce the physical low-energy scattering parameters. The specific
model we use for the short range potential is not important to our general
analysis nor is it the most economical. \ We choose a simple scheme which
consists of three well-defined functions in three different regions and which
is continuously differentiable everywhere. \ The potential has the form%
\begin{align}
V(r)  &  =V_{\text{Gauss}}(r)\theta(R_{\text{Gauss}}-r)+V_{\text{Spline}%
}(r)\theta(r-R_{\text{Gauss}})\theta(R_{\text{Exch.}}-r)\nonumber\\
&  +V_{\text{Exch.}}(r)\theta(r-R_{\text{Exch.}}),
\end{align}
where $\theta$ is a unit step function.

\begin{figure}[hptb]
\begin{center}
\resizebox{120mm}{!}{\includegraphics{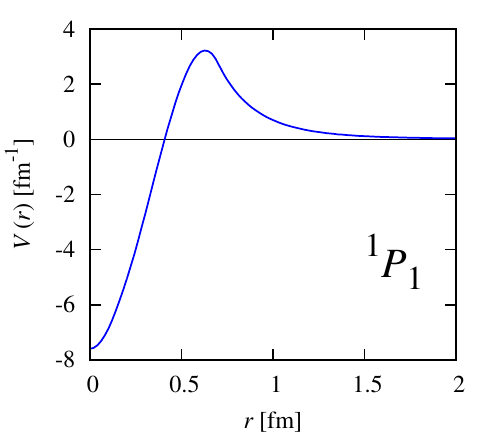}}
\end{center}
\caption{Plot of the model potential in the $^{1}P_{1}$ channel. In this
channel $S=0$, $T=0$, $C_{\pi}=9$, $C_{A}=405.9$, $C_{B}=769.5$, $C_{\Pi}=0$,
$C_{D}=0$, $C_{F}=0$, $C_{G}=-7.6$, $m_{A}=10.0m_{\pi}$, $m_{B}=5.45m_{\pi}$,
$m_{G}=4.46 m_{\pi}$, $C_{1}=-7.464$, $C_{2}=0.179$, $C_{3}=73.933$ and
$C_{4}=-78.011$. We use $R_{\text{Gauss}}=0.2$ fm and $R_{\text{Exch.}}=0.7$
fm.}%
\label{fig:OPE1P1}%
\end{figure}

\begin{figure}[hptb]
\begin{center}
\resizebox{120mm}{!}{\includegraphics{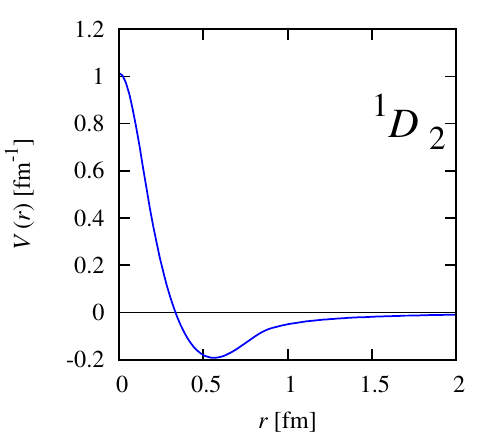}}
\end{center}
\caption{Plot of the model potential in the $^{1}D_{2}$ channel. In this
channel $S=0$, $T=1$, $C_{\pi}=-3$, $C_{A}=-60.5$, $C_{B}=-30.0$, $C_{\Pi}=0$,
$C_{D}=0$, $C_{F}=0$, $C_{G}=1.01$, $m_{A}=9m_{\pi}$, $m_{B}=6m_{\pi}$,
$m_{G}=7.02 m_{\pi}$, $C_{1}=1.463$, $C_{2}=-7.332$, $C_{3}=10.384$ and
$C_{4}=-4.585$. We use $R_{\text{Gauss}}=0.2$ fm and $R_{\text{Exch.}}=0.9$
fm.}%
\label{fig:OPE1D2}%
\end{figure}

\begin{figure}[hptb]
\begin{center}
\resizebox{120mm}{!}{\includegraphics{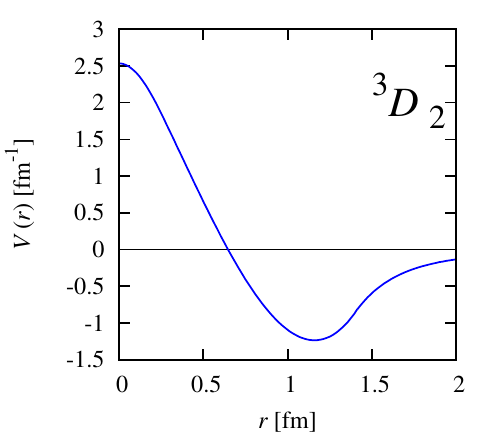}}
\end{center}
\caption{Plot of the model potential in the $^{3}D_{2}$ channel. In this
channel $S=1$, $T=0$, $C_{\pi}=-3$, $C_{A}=-660$, $C_{B}=-1140$, $C_{\Pi}=-3$,
$C_{D}=-930$, $C_{F}=-927$, $C_{G}=2.53$, $m_{A}=m_{D}=8m_{\pi}$, $m_{B}%
=m_{F}=5m_{\pi}$, $m_{G}=3.25 m_{\pi}$, $C_{1}=2.979$, $C_{2}=-4.026$,
$C_{3}=-2.455$ and $C_{4}=2.409$. \ We use $R_{\text{Gauss}}=0.3$ fm and
$R_{\text{Exch.}}=1.4$ fm.}%
\label{fig:OPE3D2}%
\end{figure}

The short-range part is a Gaussian function
\begin{equation}
V_{\text{Gauss}}(r)=C_{G}e^{-m_{G}^{2} r^{2}}. \label{PotentialModelGaussian}%
\end{equation}
The intermediate-range part of the potential is a cubic spline use to connect
the short- and long-range regions,%
\begin{equation}
V_{\text{Spline}}(r)=C_{1}+C_{2}r+C_{3}r^{2}+C_{4}r^{3}.
\label{PotentialModelIntermediate}%
\end{equation}
The long-range part consists of the usual one-pion exchange potential together
with two additional heavy meson exchange terms,%
\begin{equation}
V_{\text{Exch.}}(r)=V_{C}^{\pi,A,B}(r)+S_{12}V_{T}^{\pi,D,F}(r).
\label{PotentialModelOPE}%
\end{equation}
The central part of the potential is composed of Yukawa functions
\begin{equation}
V_{C}^{\pi,A,B}(r)=\frac{g_{\pi N}^{2}}{12\pi}\left(  \frac{m_{\pi}}{2M_{N}%
}\right)  ^{2}\left\{  C_{\pi}\frac{e^{-m_{\pi}r}}{r}+C_{A}\frac{e^{-m_{A}r}%
}{r}+C_{B}\frac{e^{-m_{B}r}}{r}\right\}  , \label{PotentialModel}%
\end{equation}
and the tensor part of the potential has the form%
\begin{align}
V_{T}^{\pi,D,F}(r)  &  =\frac{g_{\pi N}^{2}}{12\pi}\left(  \frac{m_{\pi}%
}{2M_{N}}\right)  ^{2}\left\{  C_{\Pi}\left[  1+\frac{3}{m_{\pi}r}+\frac
{3}{(m_{\pi}r)^{2}}\right]  \frac{e^{-m_{\pi}r}}{r}\right.  \hspace
{5.5cm}\nonumber\\
&  \left.  +C_{D}\left[  1+\frac{3}{m_{D}r}+\frac{3}{(m_{D}r)^{2}}\right]
\frac{e^{-m_{D}r}}{r}+C_{F}\left[  1+\frac{3}{m_{F}r}+\frac{3}{(m_{F}r)^{2}%
}\right]  \frac{e^{-m_{F}r}}{r}\right\}  .
\end{align}
Here $C_{\pi}=(2\vec{S}^{2}-3)(2\vec{T}^{2}-3)$, $C_{\Pi}=2\vec{T}^{2}-3$,
$g_{\pi N}=13.0$, $m_{\pi}=140\,\text{MeV}$, and $M_{N}=938.0$ MeV. The
coefficients which are not parts of the one-pion exchange potential are used as free
parameters to reproduce the physical low-energy scattering parameters. Due
to the abundance of free parameters, the fit process is not unique. However, in
each case we attempt to qualitatively reproduce the shape of the NijmegenII
potentials \cite{Stoks1993NijII}. In each case the heavy meson masses
are kept significantly larger than the pion mass. \ In Figure~\ref{fig:OPE1P1}
we show the potential in the $^{1}P_{1}$ channel. Figure~\ref{fig:OPE1D2}
shows the potential in the $^{1}D_{2}$ channel, and Figure~\ref{fig:OPE3D2}
shows the potential in the $^{3}D_{2}$ channel.


After having recovered the physical low-energy scattering parameters, we now
multiply an additional step function to the potential,%
\begin{equation}
V(r)\rightarrow V(r)\theta(R-r),
\end{equation}
which removes the tail of the potential beyond range $R$. \ We then
recalculate the low-energy scattering parameters with this modification. \ The
results are shown in Table \ref{table:CausalityBounds} for the $^{1}P_{1}$,
$^{1}D_{2}$, and $^{3}D_{2}$ channels. \ The causal ranges for the $^{1}D_{2}$
and $^{3}D_{2}$ channels are quite large for the physical scattering data,
$4.0$ and $4.9$ fm,  respectively. However, if we remove the tail of the model
potential at $R=5$ fm, the causal ranges drop to $2.4$ and $2.7$ fm
respectively. \begin{table}[tbh]
\caption{The potential range dependence of the causal range in various
channels.}%
\label{table:CausalityBounds}
\begin{center}
\begin{tabular}
[c]{||c|c|c|c|c|c||}\hline\hline
\backslashbox{\footnotesize{Causal range}}{\footnotesize{Potential range}} & $2 \ \text{fm}$ &
$5\ \text{fm}$ & $12\ \text{fm}$ & $15\ \text{fm}$ & $50\ \text{fm}%
$\\\hline\hline
$R_{^{1}P_{1}}^{b}$ & 0.4 & 0.4 & 0.3 & 0.3 & 0.3\\\hline
$R_{^{1}D_{2}}^{b}$ & 2.0 & 2.4 & 3.8 & 4.0 & 4.0\\\hline
$R_{^{3}D_{2}}^{b}$ & 1.3 & 2.7 & 4.7 & 4.9 & 4.9\\\hline\hline
\end{tabular}
\end{center}\end{table}

The tail of the model potential is dominated by the one-pion exchange
potential. \ Even though the numerical size of the one-pion exchange potential
is small at distances of $5$ fm, these numerical results show clearly that the
one-pion exchange tail is controlling the size of the causal range. \ The
one-pion exchange potential tail appears to be the source of the large values
for $R^{b}$ in higher partial waves where the central one-pion exchange tail
is attractive.

\section{Summary and Conclusions}

In this study we have derived the constraints of causality and unitarity for
neutron-proton scattering for all spin channels up to $j=3$. \ We have defined
and calculated interaction length scales which we call the causal range,
$R^{b}$, and the Cauchy-Schwarz range, $R^{\text{C-S}}$. \ The causal range is
the minimum value for $r$ such that the causal bounds,%
\begin{equation}
f_{j-1}(r)=b_{j-1}(r)-2q_{0}^{2}\frac{\Gamma(j+\frac{1}{2})\Gamma(j+\frac
{3}{2})}{\pi}\left(  \frac{r}{2}\right)  ^{-2j-1}-r_{j-1}\geq0,
\end{equation}%
\begin{equation}
g_{j+1}(r)=b_{j+1}(r)+\frac{2q_{0}^{2}}{a_{j+1}^{2}}\frac{\pi}{\Gamma
(j+\frac{1}{2})\Gamma(j+\frac{3}{2})}\left(  \frac{r}{2}\right)
^{2j+1}-r_{j+1}\geq0,
\end{equation}
are satisfied. \ For uncoupled channels these bounds simplify to the form%
\begin{equation}
f_{\ell}(r)=g_{\ell}(r)=b_{\ell}(r)-r_{\ell} \geq 0.
\end{equation}
For coupled channels the Cauchy-Schwarz range is the minimum value for $r$
satisfying the causal bounds as well as the Cauchy-Schwarz inequality,%
\begin{equation}
f_{j-1}(r)g_{j+1}(r)\geq\left[  h_{j}(r)\right]  ^{2}.
\end{equation}
If one reproduces the physical scattering data using strictly finite range
interactions, then the range of these interactions must be larger than $R^{b}$
and $R^{\text{C-S}}$. \ From these bounds we have derived the general result
that non-vanishing partial-wave mixing cannot be reproduced with zero-range
interactions. \ As the range of the interaction goes to the zero, the
effective range for the lower partial-wave channel is driven to negative infinity.

This finding has consequences for pionless effective theory where the range of
the interactions is set entirely by the value of the cutoff momentum. \ If the
cutoff momentum is too high, then it is impossible to obtain the correct
threshold physics in coupled channels without violating causality or
unitarity. \ In some channels we find that the causal range and Cauchy-Schwarz
range are as large $5$ fm. \ We have shown that these large values are driven
by the tail of the one-pion exchange potential. \ In these channels the
problems will be even more severe, and the cutoff momentum will need to be
rather low in order to reproduce the physical scattering data in pionless
effective field theory. \ How low this cutoff momentum must be depends on the
particular regularization scheme.

We should note that all of these mixing observables are non-vanishing only
when one reaches higher orders in the power counting expansion, and there is
no direct impact on pionless effective field theory calculations at lower
orders. \ See, for example, Ref.~\cite{Chen:1999tn} for details on power
counting in pionless effective field theory. \ In the zero-range limit, the
term which drives the negative divergence of the effective range parameter
$r_{j-1}$ is%
\begin{equation}
-2q_{0}^{2}\frac{\Gamma(j+\frac{1}{2})\Gamma(j+\frac{3}{2})}{\pi}\left(
\frac{r}{2}\right)  ^{-2j-1}. \label{q_0^2}%
\end{equation}
At leading order there is no divergence since there is no partial wave mixing
and $q_{0}=0$. \ If higher-order terms are iterated non-perturbatively as in
Ref.~\cite{vanKolck:1998bw}, then the divergence appears at order $Q^{2}$, the
first order at which $q_{0}$ is non-vanishing. \ If higher-order terms are
iterated order-by-order in perturbation theory, then the term in
Eq.~(\ref{q_0^2}) appears at order $Q^{4}$. \ This is one order higher than
the analysis presented in Ref.~\cite{Chen:1999tn}, and we predict that
zero-range divergences in $r_{j-1}$ will first appear at this order.

It important to note that if one works order-by-order in perturbation
theory, then the constraints of causality and unitarity always appear
somewhat hidden.  At every order in the effective field theory
calculation there are new operator coefficients which appear and
are determined by matching to physical data.  There are no
obstructions to setting these operator coefficients to reproduce
physical values.

It is only when one iterates the new interactions, i.e., by solving
the Schr{\"o}dinger equation, that non-linear dependencies on
the operator coefficients appear.  In this case one finds that the
constraints of causality and unitarity give necessary conditions
for keeping the operator coefficients real.  Once we fix the
regularization, the bound corresponds with branch cuts of the
effective theory when viewed as a function of physical scattering
parameters.

These branch cuts cannot be seen at any finite order in
perturbation theory.  However a nearby branch point may
spoil the convergence of the perturbative expansion.  In
this context, our causality and unitarity bounds can be
viewed as setting physical constraints for the convergence of
perturbative calculations in pionless effective field theory.

If the cutoff is taken too high, a branch cut develops which
jeopardizes the convergence of the perturbative calculation.
Similarly if one does calculations using dimensional
regularization, then the renormalization scale sets the scale
at which the infrared and ultraviolet physics are regulated
\cite{Georgi:1992xg}.  Similar problems with perturbative
convergence would arise if the renormalization scale is
taken too high.

There is much theoretical interest in the connection between dilute neutron
matter and the universal physics of fermions in the unitarity limit
\cite{Gezerlis:2007fs,Borasoy:2007vk,Epelbaum:2008vj,Lee:2008xs,Gezerlis:2009iw,Wlazlowski:2009yi,Forbes:2012a}%
. \ In the limit of isospin symmetry our analysis of the isospin triplet
channels can be applied to neutron-neutron scattering in dilute neutron
matter. \ In this study we have shown there are intrinsic length scales
associated with the causal range and the Cauchy-Schwarz range. \ When the
average separation between neutrons is smaller than these length scales, one
expects non-universal behavior controlled by the details of the
neutron-neutron interactions. For the $^{1}S_{0}$ channel, $R^{b}=1.3$~fm.
\ For the $^{3}P_{2}$ channel, $R^{b}=2.2$~fm, and for the $^{3}F_{2}$
channel, $R^{b}=1.7$~fm. \ For $^{3}P_{2}$-$^{3}F_{2}$ mixing, we find
$R^{\text{C-S}}=4.7$~fm. \ We see that the physics of $^{3}P_{2}$-$^{3}F_{2}$
mixing will become non-universal at lower densities than the $^{1}S_{0}$
interactions. \ In particular the densities where $^{3}P_{2}$ superfluidity is
expected to occur will be well beyond this universal regime.

%% file: Chapter-5.tex
\chapter{van der Waals interactions}
\label{chap:van-der-Waals}

\section{Introduction}

\label{chap5:introduction}

Low-energy universality appears when there is a large separation between the
short-distance scale of the interaction and the physically relevant
long-distance scales.  Some well-known examples include the unitarity limit
of two-component
fermions~\cite{O'Hara:2002,Kinast:2004,Zwierlein:2004,Bartenstein:2004,Ku:2012a}
and the Efimov effect in three-body and four-body
systems~\cite{Efimov70,Efimov73,Bedaque:1998kg,Platter:2004pra,Hammer:2006ct,vonStecher:2008a,PhysRevA.82.040701,Hadizadeh:2011qj,PhysRevA.84.052503,Kraemer:2006Nat}.
See Refs.~\cite{Braaten:2004a,Giorgini:2007a} for reviews of the subject and
literature.  There have been many theoretical studies of low-energy phenomena
and universality for interactions with finite range.  These studies have direct
applications to nuclear physics systems such as cold dilute neutron matter or
light nuclei such as the triton and alpha particle.  To a good approximation,
the van der Waals interactions between alkali-metal atoms can also be treated as a
finite-range interaction.

However, there are some differences.  For potentials
with an attractive $1/r^{\alpha}$ tail and $\alpha > 2$, the $s$-wave scattering
phase shift near threshold has been formulated in Ref.~\cite{PhysRevA.84.032701}.
For $\alpha > 3$, the modified scattering parameters for an $s$-wave Feshbach
resonance were derived in Ref.~\cite{PhysRevA.85.052703} using coupled-channel
calculations. Analytical expressions for the $s$-wave scattering length and effective
range for two neutral atoms and $\alpha = 6$ have been derived in Ref.~\cite{PhysRevA.59.1998}. However, the applicability of the effective range theory is limited for interactions with
attractive tails. In order to define the scattering length for angular momentum
$\ell\geq2$ and the effective range for $\ell\geq1$, a modified version of effective range theory known as quantum-defect theory is needed~\cite{PhysRevA.58.4222,PhysRevA.80.012702}.
Furthermore, scattering parameters of magnetically tunable multichannel systems
have been studied in the context of multichannel quantum-defect
theory~\cite{PhysRevA.70.012710, PhysRevA.72.042719}. See Ref.~\cite{PhysRevA.87.032706}
for a very recent development of multichannel quantum-defect theory for higher partial waves.
There is also growing empirical evidence that there exists a new type of low-energy universality
that ties together all interactions with an attractive $1/r^{6}$ tail. This
might seem surprising since there is no such analogous behavior for
interactions with a Coulomb tail. In this chapter we derive the theoretical
foundations for this van der Waals universality at low energies by studying
the near-threshold behavior and the constraints of causality.  We also show
that this universality extends to any power-law interaction $1/r^{\alpha}$
with $\alpha\geq2$ in any number of dimensions.  Our analysis applies to
energy-independent interactions.  We first consider a single
scattering channel but then also consider multichannel systems near a
magnetic Feshbach
resonance.

In our analysis we assume that the two-body potential has a long-distance
attractive tail of the form $-C_{6}/r^{6}$.  We define the van der Waals
length scale, $\beta_{6}$, as
\begin{equation}
 \beta_{6}=(2\mu C_{6})^{\frac{1}{4}} \,,
\label{vanderWaalslength}
\end{equation}
where $\mu$ is the reduced mass of the scattering particles.  For simplicity
we use atomic units (a.u.) throughout our discussion.  So, in particular,
we set $\hbar=1$.  In Refs.~\cite{Cordon:2010a,RuizArriola:2011a} it was
noticed that an approximate universal relationship exists between the
effective range and inverse scattering length for $s$-wave scattering in many
different pairs of scattering alkali-metal atoms.  If we write $A_{0}$ as the
scattering length and $R_{0}$ as the effective range, the relation is
\begin{equation}
 R_{0}\approx\frac{\beta_{6}\Gamma(1/4)^{2}}{3\pi}-\frac{4\beta_{6}^{2}}
 {3A_{0}}+\frac{8\pi\beta_{6}^{3}}{3\Gamma(1/4)^{2}A_{0}^{2}} \,.
\label{arriola}
\end{equation}
This approximate relation becomes exact for a pure $-C_{6}/r^{6}$ potential.
What is surprising about \eref{arriola} is that the van der Waals
length $\beta_{6}$ dominates over other length scales which characterize the
short-distance repulsive force between alkali-metal atoms.  This approximate
universality suggests there is some separation of scales between the van der
Waals length $\beta_{6}$ and the length scales of the short-range forces.
This separation of scales will become more transparent later in our analysis
when we determine the coefficients of the short-range $\hat{\rm{K}}$-matrix.  It would
be useful to exploit
the separation of scales as an effective field theory with an explicit
van der Waals tail plus contact interactions.  In this chapter, we discuss the
constraints on such a van der Waals effective field theory.

We note that a similar dominance of the van der Waals length $\beta_{6}$ has
been discovered for the three-body parameter in the Efimov
effect~\cite{Berninger:2011a,Wang:2012a,Wang:2012b}.  In the analysis here we focus
only on two-body systems.  However, our analysis should be useful in developing
the foundations for van der Waals effective field theory.  This in turn could
be used to investigate the Efimov effect and other low-energy phenomena in a
model-independent way. An extension of our analysis may be useful to
understand the recently observed universality of
the three-body parameter for narrow Feshbach resonances
\cite{Roy2013}.

The organization of this chapter is as follows.  We first discuss the connection
between causality bounds and effective field theory.  Next we consider
asymptotic solutions of the Schr\"{o}dinger equation.  After that, we derive
causality bounds for the short-range $\hat{\rm{K}}$-matrix and consider the impact of
these results on van der Waals effective field theory.  Then we discuss
quantum-defect theory and calculate causal ranges for several examples of
single-channel $s$-wave scattering in alkali-metal atoms.  We also consider the
constraints of causality near magnetic Feshbach resonances.  We then
conclude with a summary and discussion.

\section{Causality bounds and effective field theory}

For an effective field theory with local contact interactions, the range of
the interactions is controlled by the momentum cutoff scale.  Problems with
convergence can occur if the cutoff scale is set higher than the scale of the
new physics not described by the effective theory.  It is useful to have a
quantitative measure of when problems may or may not appear, and this is where
the causality bound provides a useful diagnostic tool.  For each scattering
channel we use the physical scattering parameters to compute the causal range, $R^{b}$, which is the minimum range for the interactions
consistent with the requirements of causality and unitarity and discussed in details in Chapter~\ref{chap:Neutron-proton-scattering}.  For any fixed
cutoff scale, the causality bound marks a branch cut of the effective theory
when viewed as a function of physical scattering
parameters, see Chapter~\ref{chap:Neutron-proton-scattering} and Ref.~\cite{Koenig:2012bv}.  The coupling constants of
the effective theory become complex for scattering parameters violating the
causality bound.  These branch cuts do not appear in perturbation theory, but
they can spoil the convergence pattern of the perturbative expansion.

Wigner was the first to recognize the constraints of causality and unitarity
for two-body scattering with finite-range interactions~\cite{Wigner1954}.
The time delay of a scattered wave packet is given by the energy derivative of
the phase shift, $\Delta t=2 d\delta/dE$. 
It is clear that the incoming wave packet must first reach the interacting
region before the outgoing wave packet can leave.  So the causality bound can
be viewed as a lower bound on the time delay, $\Delta t$.  When applied to
wave packets near threshold, the causality bound becomes an upper bound on
the effective range parameter.

A brief historical review on the analysis of the constraints of causality and universality is given in the introduction of Chapter~\ref{chap:Neutron-proton-scattering}. Also in that chapter we have presented the first study of coupled-channel systems with partial-wave mixing

\section{Asymptotic solutions of the Schr\"{o}dinger equation}

We consider a system of two spinless particles interacting via a spherically
symmetric finite range potential in the center-of-mass frame. In addition to the non-singular finite-range
interactions parameterized by $W(r,r^{\prime})$, we assume that there is a long-range local  van der Waals potential $-C_{6}/r^{6}$ for $r>R$.  The van der Waals
length scale $\beta_{6}$ was defined in \eref{vanderWaalslength}. As noted in the
Introduction, we use atomic units where $\hbar=1$.  The
radial Schr{\"{o}}dinger equation is
\begin{equation}
 \left[\frac{d^{2}}{dr^{2}}-\frac{\ell(\ell+1)}{r^{2}}+\frac{\beta_{6}^{4}}{r^{6}}
 \theta(r-R)+p^{2}\right] U_{\ell}^{(p)}(r)
 = 2\mu\int_{0}^{R}dr^{\prime}\,W(r,r^{\prime})U_{\ell}^{(p)}(r^{\prime}) \,.
\label{Schrodinger_van_der_Waals}
\end{equation}
The step function $\theta(r-R)$ cuts off the long-range potential at distances
less than $R$.  This ensures that we satisfy the regularity condition discussed in \eref{eqn:regularity} and avoids mathematical problems~associated with
unregulated singular potentials \cite{RevModPhys.43.36}.  The general form of
the solutions for \eref{Schrodinger_van_der_Waals} has been discussed by
Gao in Ref.~\cite{PhysRevA.78.012702}.

In order to simplify some of the more lengthy expressions to follow, we
introduce dimensionless rescaled variables $r_{s}={r}/{\beta_{6}}$,
$p_{s}=\beta_{6}p$, and $\rho_{s}=1/(2r_{s}^{2})$.  In the outer region,
$r>R$, the Schr{\"{o}}dinger equation reduces to
\begin{equation}
 \left[\frac{d^{2}}{dr^{2}}-\frac{\ell(\ell+1)}{r^{2}}
 + \frac{\beta_{6}^{4}}{r^{6}}+p^{2}\right] U_{\ell}^{(p)}(r)=0 \,
\end{equation}
or
\begin{equation}
 \left[\frac{d^{2}}{dr_{s}^{2}}-\frac{\ell(\ell+1)}{r_{s}^{2}}+\frac{1}{r_{s}^{6}}
 + p_{s}^{2}\right]  U_{\ell}^{(p)}(r)=0 \,.
\label{pure_vdW}
\end{equation}
The exact solutions for \eref{pure_vdW} have been studied in detail in
Ref.~\cite{PhysRevA.58.1728} using the formalism of quantum-defect
theory~\cite{RepProgPhys.46.167,PhysRevA.19.1485,PhysRevA.26.2441}.

The van der Waals wave functions $F_{\ell}$ and $G_{\ell}$ are linearly independent
solutions of \eref{pure_vdW}.  In order to write these out we first
need several functions defined in
Appendix~\ref{append:analytic_solution_vdWaals_eq}.  The van der Waals
wave functions $F_{\ell}$ and $G_{\ell}$ can be written as summations of Bessel and Neumann
functions,
\begin{multline}
 {F}_{\ell}(p,r)=\frac{r_{s}^{1/2}}{x_{\ell}^{2}(p_{s})+y_{\ell}^{2}(p_{s})}\Bigg[
x_{\ell}(p_{s})\sum_{m=-\infty}^{\infty}b_{m}(p_{s})\,J_{\nu+m}
 \left(\rho_{s}\right) \\ - y_{\ell}(p_{s})\sum_{m=-\infty}^{\infty}b_{m}(p_{s})
 \,N_{\nu+m}\left(\rho_{s}\right)\Bigg] \,,
\label{F_L}
\end{multline}
\begin{multline}
 {G}_{\ell}(p,r)=\frac{r_{s}^{1/2}}{x_{\ell}^{2}(p_{s})+y_{\ell}^{2}(p_{s})}\Bigg[
x_{\ell}(p_{s})\sum_{m=-\infty}^{\infty}b_{m}(p_{s})\,N_{\nu+m}
 \left(\rho_{s}\right) \\ + y_{\ell}(p)\sum_{m=-\infty}^{\infty}b_{m}(p_{s})
 \,J_{\nu+m}\left(\rho_{s}\right)\Bigg] \,.
\label{G_L}
\end{multline}
The function $x_{\ell}$ is defined in \eref{X_L}, and $y_{\ell}$ is defined
in \eref{Y_L}.  For $m\geq0$ the function $b_{m}$ is given in
\eref{bm}, while $b_{-m}$ is given in \eref{bmm}.  The offset
$\nu$ appearing in the order of the Bessel functions is given by the solution
of \eref{v} in Appendix~\ref{append:analytic_solution_vdWaals_eq}.
For notational convenience, however, we omit writing the explicit $p_{s}$
dependence of $\nu$.  Let us define $\delta_{\ell}^{(\text{short})}(p)$ to be
the phase shift of the van der Waals wave functions due to the scattering from
the short-range interaction.  The normalization of $U_{\ell}^{(p)}(r)$ is chosen
so that, for $r>R$,
\begin{equation}
 U_{\ell}^{(p)}(r)=F_{\ell}(p,r)-\tan\delta_{\ell}^{(\text{short})}(p)\,G_{\ell}(p,r) \,.
\label{U_L}
\end{equation}
Our van der Waals wave functions are related to the functions $f_{\ell}^{c0}$ and
$g_{\ell}^{c0}$ defined of Ref.~\cite{PhysRevA.78.012702} by the normalization
factors $F_{\ell}=f_{\ell}^{c0}/\sqrt{2}$ and $G_{\ell}=-g_{\ell}^{c0}/\sqrt{2}$.
Henceforth, we write all expressions in terms of the short-range reaction
matrix
\begin{equation}
{\rm{K}}_{\ell}(p)=\tan\delta_{\ell}^{(\text{short})}(p) \,,
\end{equation}
which is related to
the short-range scattering matrix via
\begin{equation}
{\rm{S}}_{\ell}=e^{2i\delta_{\ell}^{(\text{short})}}
 = \frac{i-{\rm{K}}_{\ell}}{i+{\rm{K}}_{\ell}} \,.
\end{equation}

For any finite-range interaction, ${\rm{K}}_{\ell}$ is analytic in $p^{2}$ and can
be calculated by matching solutions for $r\leq R$ and $r>R$ at the boundary.
It can be written in compact form as
\begin{equation}
{\rm{K}}_{\ell}
 =\left.\frac{W(U_{\ell}^{(p)},F_{\ell}^{(p)})}{W(U_{\ell}^{(p)},
 G_{\ell}^{(p)})}\right\vert _{r=R},
\end{equation}
where $U_{\ell}^{(p)}$ is the solution of \eref{Schrodinger_van_der_Waals}
that is regular at the origin, and $W$ denotes the Wronskian of two functions,
\begin{equation*}
 W(f,g)=fg^{\prime}-f^{\prime}g \,.
\end{equation*}

\section{Causality bounds for short-range $\rm{K}$-matrix ${\rm{K}}_{\ell}$}

In this section we derive causality bounds for the short-range $\rm{K}$-matrix.  For this we need to expand the wave function $U_{\ell}^{(p)}(r)$
in powers of $p^{2}$.  The steps we follow are analogous to those used in Chapter~\ref{chap:Neutron-proton-scattering} [cf. in Refs.~\cite{Hammer-Lee9,HammerDean27,Koenig:2012bv}].  We
first expand ${\rm{K}}_{\ell}$,
\begin{equation}
{\rm{K}}_{\ell}
=\tan\delta_{\ell}^{(\text{short})}(p)
 =\sum_{n=0}^{\infty}K_{\ell,2n}\,p^{2n} \,.
\label{eq:K_L_expansion}
\end{equation}
The first two terms $K_{\ell,0}$ and $K_{\ell,2}$ are analogous to the inverse
scattering length and effective range parameters in the usual effective range
expansion.  The higher-order terms can be regarded as analogs of the shape
parameters.  Next we expand the van der Waals wave functions in powers of
$p^{2}$,
\begin{equation}
 {F}_{\ell}(p,r)=f_{\ell,0}(r)+f_{\ell,2}(r)\,p^{2}+{O}(p^{4}) \,,
\label{F_L_expansion}
\end{equation}
\begin{equation}
 {G}_{\ell}(p,r)=g_{\ell,0}(r)+g_{\ell,2}(r)\,p^{2}+{O}(p^{4}) \,.
\label{G_L_expansion}
\end{equation}
In the following, we define
\begin{equation*}
 \nu_{0}=\frac{1}{4}(2\ell+1),
\end{equation*}
which corresponds to the value of $\nu$ at threshold.  Using the low-energy
expansions in Appendix~\ref{low_energy_expansion}, we find that the
coefficients in \eref{F_L_expansion} are
\begin{equation}
f_{\ell,0}(r)=r_{s}^{1/2}J_{\nu_{0}}\left(\rho_{s}\right) \,
\end{equation}
and
\begin{multline}
 f_{\ell,2}(r)=\frac{\Gamma(\nu_{0})\Gamma(2\nu_{0}-1)}{\Gamma(\nu_{0}+1)
 \Gamma(2\nu_{0})}\frac{\beta_{6}^{2}}{16}r_{s}^{1/2}\big[J_{\nu_{0}-1}
 \left(\rho_{s}\right)+N_{\nu_{0}}\left(\rho_{s}\right)\big] \\
 -\frac{\Gamma(\nu_{0})\Gamma(2\nu_{0}+1)}{\Gamma(\nu_{0}+1)\Gamma(2\nu_{0}+2)}
 \frac{\beta_{6}^{2}}{16}r_{s}^{1/2}\big[J_{\nu_{0}+1}\left(\rho_{s}\right)
 -N_{\nu_{0}}\left(\rho_{s}\right)\big] \,.
\label{eqn:second_term_pair_func_F_1a}
\end{multline}
Similarly, the coefficients in \eref{G_L_expansion} are
\begin{equation}
 g_{\ell,0}(r)=r_{s}^{1/2}N_{\nu_{0}}\left(\rho_{s}\right) \,
\end{equation}
and
\begin{multline}
 g_{\ell,2}(r)=\frac{\Gamma(\nu_{0})\Gamma(2\nu_{0}-1)}{\Gamma(\nu_{0}+1)
 \Gamma(2\nu_{0})}\frac{\beta_{6}^{2}}{16}r_{s}^{1/2}\big[N_{\nu_{0}-1}
 \left(\rho_{s}\right)-J_{\nu_{0}}\left(\rho_{s}\right)\big] \\
 -\frac{\Gamma(\nu_{0})\Gamma(2\nu_{0}+1)}{\Gamma(\nu_{0}+1)\Gamma(2\nu_{0}+2)}
 \frac{\beta_{6}^{2}}{16}r_{s}^{1/2}\big[N_{\nu_{0}+1}\left(\rho_{s}\right)
 +J_{\nu_{0}}\left(\rho_{s}\right)\big] \,.
\label{eqn:second_terms_pair_func_G_1a}
\end{multline}
Using \eref{U_L}, we can now express $U_{\ell}^{(p)}(r)$ as an expansion in
powers of $p^{2}$.  For $r>R$, we have
\begin{align}
 U_{\ell}^{(p)}(r)=f_{\ell,0}(r)- & K_{\ell,0} \,  g_{\ell,0}(r) \nonumber\\
 & +p^{2}\big[f_{\ell,2}(r)-K_{\ell,0} \,  g_{\ell,2}(r)-K_{\ell,2} \,  g_{\ell,0}(r)\big]
 +{O}(p^{4}) \,.
\label{eqn:wave_function_vdWaals_2a}
\end{align}

We now consider two solutions of the Schr{\"{o}}dinger equation
$U_{\ell}^{(p_{a})}(r)$ and $U_{\ell}^{(p_{b})}(r)$ with momenta $p_{a}$ and $p_{b}$,
respectively.  We have
\begin{equation}
 \left[\frac{d^{2}}{dr^{2}}-\frac{\ell(\ell+1)}{r^{2}}+\frac{\beta_{6}^{4}}{r^{6}}
 \theta(r-R)+p_{a}^{2}\right]  U_{\ell}^{(p_{a})}(r)=2\mu\int_{0}^{R}dr^{\prime}
 \,W(r,r^{\prime})U_{\ell}^{(p_{a})}(r^{\prime}) \,,
\end{equation}
\begin{equation}
 \left[\frac{d^{2}}{dr^{2}}-\frac{\ell(\ell+1)}{r^{2}}+\frac{\beta_{6}^{4}}{r^{6}}
 \theta(r-R)+p_{b}^{2}\right]  U_{\ell}^{(p_{b})}(r)=2\mu\int_{0}^{R}dr^{\prime}
 \,W(r,r^{\prime})U_{\ell}^{(p_{b})}(r^{\prime}) \,.
\end{equation}
Following the same steps as in Section~\ref{chap1:Wronskian-integral formula}, we
obtain the Wronskian integral formula
\begin{equation}
 \frac{W[U_{\ell}^{(p_{b})},U_{\ell}^{(p_{a})}](r)}{p_{b}^{2}-p_{a}^{2}}=\int_{0}^{r}
 U_{\ell}^{(p_{a})}(r^{\prime})U_{\ell}^{(p_{b})}(r^{\prime})\,dr^{\prime} \,,
\label{eqn:wronskian_integral_relation_1a}
\end{equation}
for any $r>R$.  Using \eref{eqn:wave_function_vdWaals_2a} for momenta
$p_{a}$ and $p_{b}$ we find
\begin{equation}
\begin{split}
 \frac{W[U_{\ell}^{(b)},U_{\ell}^{(a)}](r)}{p_{b}^{2}-p_{a}^{2}}
 &=W[f_{\ell,2},f_{\ell,0}](r)-K_{\ell,0}\big\{
 W[g_{\ell,2},f_{\ell,0}](r)+W[f_{\ell,2},g_{\ell,0}](r)\big\} \\
 &\hspace{2em}+K_{\ell,0}^{2}W[g_{\ell,2},g_{\ell,0}](r)-K_{\ell,2}W[g_{\ell,0},f_{\ell,0}](r)
 +{O}(p_{a}^{2},p_{b}^{2}) \,.
\end{split}
\label{eqn:wronskian_of_wave_functions_1a}
\end{equation}
In the Wronskian integral formula~\eref{eqn:wronskian_integral_relation_1a},
we set $p_{a}=0$ and take the limit $p_{b}\rightarrow0$.  With the
wave function at zero energy written as $U_{\ell}^{(0)}$, the result is
\begin{equation}
 K_{\ell,2}=b^{\rm{vdW}}_{\ell}(r)-\frac{\pi}{4}\int_{0}^{r}
 \left[U_{\ell}^{(0)}(r^{\prime})\right]^{2}\,dr^{\prime} \,,
\label{eqn:causality_bound_1a}
\end{equation}
where
\begin{equation}
\begin{split}
 b^{\rm{vdW}}_{\ell}(r) = &\frac{\pi}{4}W[f_{\ell,2},f_{\ell,0}](r)+\frac{\pi}{4}
 K_{\ell,0}^{2}W[g_{\ell,2},g_{\ell,0}](r) \\
 &-\frac{\pi}{4}K_{\ell,0}
 \big\{W[g_{\ell,2},f_{\ell,0}](r)+W[f_{\ell,2},g_{\ell,0}](r)\big\} \,.
\end{split}
\label{eqn:causality_bound_b_function_1a}
\end{equation}
The Wronskians appearing in \eref{eqn:causality_bound_b_function_1a} can
be written out explicitly as
\begin{align}
 W[f_{\ell,2},f_{\ell,0}](r) &=\frac{\beta_{6}\rho_{s}}{16\nu_{0}(2\nu_{0}-1)}
 \left[J_{\nu_{0}-2}(\rho_{s})\ J_{\nu_{0}}(\rho_{s})
 -J_{\nu_{0}-1}^{2}(\rho_{s})\right] \nonumber \\
 &+\frac{\beta_{6}\rho_{s}}{16\nu_{0}(2\nu_{0}+1)}\left[J_{\nu_{0}+2}(\rho_{s})
 \ J_{\nu_{0}}(\rho_{s})-J_{\nu_{0}+1}^{2}(\rho_{s})\right] \nonumber \\
 &+\frac{\beta_{6}\rho_{s}}{4(2\nu_{0}-1)(2\nu_{0}+1)}
 \left[J_{\nu_{0}+1}(\rho_{s})
 \ J_{\nu_{0}-1}(\rho_{s})-J_{\nu_{0}}^{2}(\rho_{s})
 +\frac{4}{\pi\rho_{s}}\right] \,,
\label{Wff}
\end{align}
\begin{align}
 W[g_{\ell,2},g_{\ell,0}](r) &= \frac{\beta_{6}\rho_{s}}{16\nu_{0}(2\nu_{0}-1)}
 \left[N_{\nu_{0}-2}(\rho_{s})\ N_{\nu_{0}}(\rho_{s})
 -N_{\nu_{0}-1}^{2}(\rho_{s})\right] \nonumber \\
 &+\frac{\beta_{6}\rho_{s}}{16\nu_{0}(2\nu_{0}+1)}\left[N_{\nu_{0}+2}(\rho_{s})
 \ N_{\nu_{0}}(\rho_{s})-N_{\nu_{0}+1}^{2}(\rho_{s})\right] \nonumber \\
 &+\frac{\beta_{6}\rho_{s}}{4(2\nu_{0}-1)(2\nu_{0}+1)}
 \left[N_{\nu_{0}+1}(\rho_{s})\ N_{\nu_{0}-1}(\rho_{s})
 -N_{\nu_{0}}^{2}(\rho_{s})+\frac{4}{\pi\rho_{s}}\right] \,,
\label{Wgg}
\end{align}
and
\begin{multline}
 W[g_{\ell,2},f_{\ell,0}](r) = W[f_{\ell,2},g_{\ell,0}](r) \\
 =\frac{\beta_{6}\rho_{s}}{16\nu_{0}(2\nu_{0}-1)}\big\{J_{\nu_{0}-1}(\rho_{s})
 \left[N_{\nu_{0}+1}(\rho_{s})-N_{\nu_{0}-1}(\rho_{s})\right]
 -N_{\nu_{0}}(\rho_{s})\left[J_{\nu_{0}}(\rho_{s})
 -J_{\nu_{0}-2}(\rho_{s})\right]\big\} \\
 -\frac{\beta_{6}\rho_{s}}{16\nu_{0}(2\nu_{0}+1)}\big\{J_{\nu_{0}+1}(\rho_{s})
 \left[N_{\nu_{0}+1}(\rho_{s})-N_{\nu_{0}-1}(\rho_{s})\right]
 -N_{\nu_{0}}(\rho_{s})\left[J_{\nu_{0}+2}(\rho_{s})
 -J_{\nu_{0}}(\rho_{s})\right]\big\} \,.
\label{Wgf}
\end{multline}
The fact that the integral on the right-hand side of
\eref{eqn:causality_bound_1a} is positive semidefinite sets an upper
bound on the short-range parameter $K_{\ell,2}$.  We find that
\begin{equation}
 K_{\ell,2}\leq b^{\rm{vdW}}_{\ell}(r)
\label{vdW_inequality}
\end{equation}
for any $r>R$.

\section{Impact on effective field theory}

\begin{figure}[hptb]
\begin{center}
\resizebox{120mm}{!}{\includegraphics{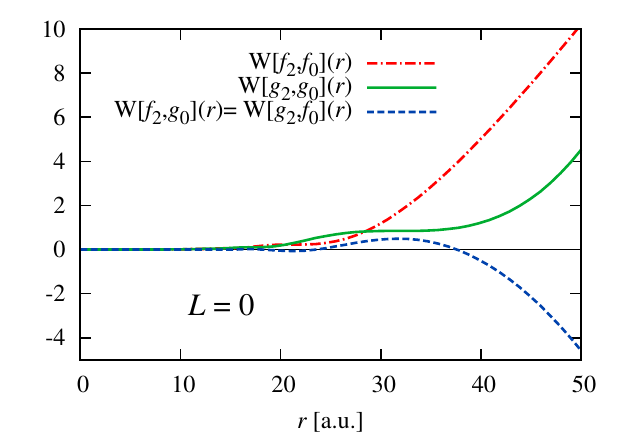}}
\end{center}
\caption{(Color online) Plot of $W[f_{0,2},f_{0,0}](r)$, $W[g_{0,2},g_{0,0}](r)$, and
$W[g_{0,2},f_{0,0}](r)$ as a function of $r$ for $\ell=0$ and $\beta_{6}=50$
(a.u.).}
\label{L0}
\end{figure}

\begin{figure}[hptb]
\begin{center}
\resizebox{120mm}{!}{\includegraphics{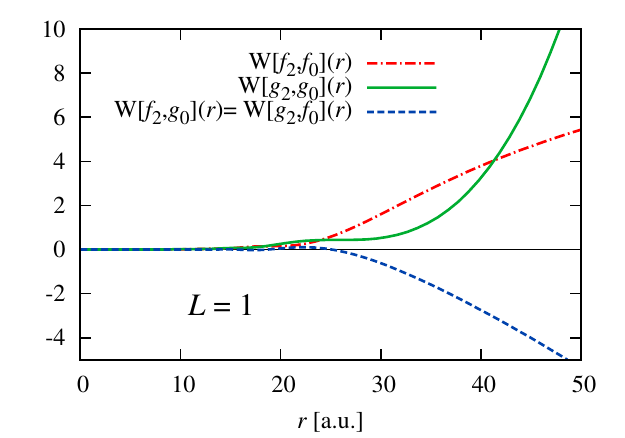}}
\end{center}
\caption{(Color online) Plot of $W[f_{1,2},f_{1,0}](r)$, $W[g_{1,2},g_{1,0}](r)$, and
$W[g_{1,2},f_{1,0}](r)$ as a function of $r$ for $\ell=1$ and $\beta_{6}=50$
(a.u.).}
\label{L1}
\end{figure}

\begin{figure}[hptb]
\begin{center}
\resizebox{120mm}{!}{\includegraphics{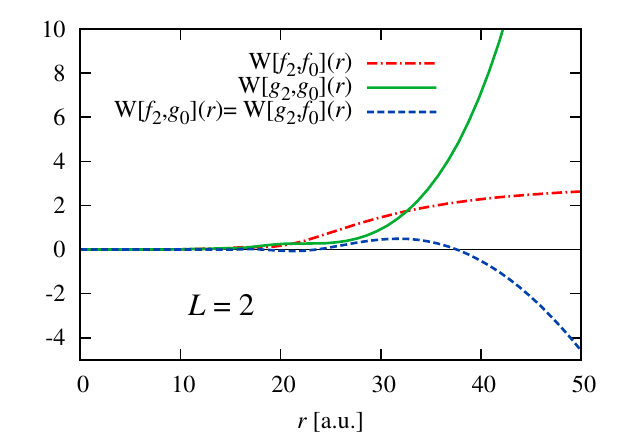}}
\end{center}
\caption{(Color online) Plot of $W[f_{2,2},f_{2,0}](r)$, $W[g_{2,2},g_{2,0}](r)$, and
$W[g_{2,2},f_{2,0}](r)$ as a function of $r$ for $\ell=2$ and $\beta_{6}=50$
(a.u.).}
\label{L2}
\end{figure}

In this section we discuss the impact of our causality bounds for an
effective field theory with short-range interactions and an attractive
$1/r^{6}$ tail.  In \fref{L0} we plot the $\ell=0$ Wronskians
$W[f_{0,2},f_{0,0}]$, $W[g_{0,2},g_{0,0}]$, and $W[g_{0,2},f_{0,0}]$ for
$\beta_{6}=50$ (a.u.).  \fref{L1} and~\fref{L2} show the analogous
plots for $\ell=1$ and $\ell=2$, respectively.


We note that all of the Wronskian functions in Figures~\ref{L0}--\ref{L2} vanish in the limit $r\rightarrow0$.  This stands in clear contrast
to what one finds for purely finite-range
interactions~\cite{Hammer-Lee9,HammerDean27}.  In that case the effective
range parameter, $r_{\ell}$, satisfies the upper bound
\begin{equation}
 r_{\ell}\leq b_{\ell}(r) \,,
\label{rL_bound}
\end{equation}
where the function $b_{\ell}(r)$ is given in Eq.~(\ref{eqn:uncoup:generalb}) [\cf~Eq.~(60) in
Ref.~\cite{HammerDean27}]
\begin{align}
 b_{\ell}(r) = -& \frac{2\Gamma(\ell-\frac{1}{2})\Gamma(\ell+\frac{1}{2})}
 {\pi}
 \left(\frac{r}{2}\right)^{-2\ell+1} 
 \nonumber\\
 & -\frac{4}{\ell+\frac{1}{2}}\frac{1}{a_{\ell}}
 \left(\frac{r}{2}\right)^{2}  +\frac{2\pi}{\Gamma(\ell+\frac{3}{2})\Gamma(\ell+\frac{5}{2})}
 \frac{1}{a_{\ell}^{2}}\left(\frac{r}{2}\right)^{2\ell+3} \,,
\label{2Ld_odd}
\end{align}
and $a_{\ell}$ is the scattering length. As already discussed in Section~\ref{sec:Cases without Mixing} near $r=0$ the behavior of
$b_{\ell}(r)$ is
\begin{equation}
 b_{\ell}(r)=-\frac{2\Gamma(\ell-\frac{1}{2})\Gamma(\ell+\frac{1}{2})}{\pi}
 \left(\frac{r}{2}\right)^{-2\ell+1}+{O}(r^{2}) \,.
\end{equation}
We see that $b_{\ell}(r)$ diverges to negative infinity as
$r\rightarrow0$ for $\ell\geq1$.  The causality bound on $r_{\ell}$ also drives
$r_{\ell}$ to negative infinity for $\ell\geq1,$
\begin{equation}
 r_{\ell}\leq-\frac{2\Gamma(\ell-\frac{1}{2})\Gamma(\ell+\frac{1}{2})}{\pi}
 \left(\frac{r}{2}\right)^{-2\ell+1}+{O}(r^{2}) \,.
\end{equation}

For an effective field theory with local contact interactions, the range of
the interactions are controlled by the momentum cutoff scale.  No matter the
values for $a_{\ell}$ and $r_{\ell}$, it is not possible to take the momentum cutoff
scale arbitrarily high without violating the causality bound for channels with
angular momentum $\ell\geq1$.  For finite-range interactions with an additional
attractive or repulsive Coulomb tail, one finds the same leading
behavior~\cite{Koenig:2012bv}\footnote{To get this analogy, we use here
the normalization of the Coulomb-modified effective range expansion found in
Eq.~(28) of Ref.~\cite{Koenig:2012bv} and insert it in Eqs.~(64), (A.6), and
(A.7) of the same paper, which give the explicit expressions for the
Coulomb-modified causality bound functions for $\ell=0,1,2$.  The statement for
arbitrary $\ell$ then follows by generalization.}
\begin{equation}
 b_{\ell}^{\text{Coulomb}}(r)=-\frac{2\Gamma(\ell-\frac{1}{2})\Gamma(\ell+\frac{1}{2})}
 {\pi} \left(\frac{r}{2}\right)^{-2\ell+1}+{O}(r^{-2\ell}) \,,
\end{equation}
\ie, the only difference in the causality bound relation for the Coulomb-modified effective range is the subleading ${O}(r^{-2\ell})$ pole
term which is absent in the purely finite-range case.  Hence, also for an
effective field theory with contact interactions and long-range Coulomb tail, it
is not possible to take the momentum cutoff scale arbitrarily high for $\ell\geq1$
without violating the causality bound.

There is no such divergence in $b^{\rm{vdW}}_{\ell}(r)$ at $r=0$ for the attractive
$1/r^{6}$ interaction.  For an effective field theory with contact
interactions and van der Waals tail, the causality bound does not impose
convergence problems as long as $K_{\ell,2}$ is less than or equal to zero.  This
holds true for any $\ell$.  There is no constraint from causality and unitarity
preventing one from taking the cutoff momentum to be arbitrarily large.  The key
difference between the van der Waals interaction and the Coulomb interaction is
that, when extended all the way to the origin, the attractive $1/r^{6}$
interaction is singular and the spectrum is unbounded below.  An essential
singularity appears at $r=0$, and both van der Waals wave functions $F_{\ell}$ and
$G_{\ell}$ vanish at the origin.

These exact same features appear in any attractive $1/r^{\alpha}$ interaction
for $\alpha>2$ in any number of spatial dimensions.  The same can be said
about an attractive $1/r^{2}$ interaction when the coupling constant is strong
enough to form bound states.  The key point is that in the zero-range
limit of these attractive singular potentials, the spectrum of bound states
extends to arbitrarily large negative energies.  As a consequence, the scattering
wave functions above threshold must vanish at the origin in order to satisfy
orthogonality with respect to all such bound-state wave functions localized near
the origin.  In all of these cases the function $b^{\rm{vdW}}_{\ell}(r)$
remains finite as $r\rightarrow0$ for any $\ell$.  We conclude that for an
effective field theory with contact interactions and attractive singular power-law interactions, we can take the cutoff momentum arbitrarily large for any
$\ell$ without producing a divergence in the coefficient $K_{\ell,2}$ of the
short-range $\rm{K}$-matrix.

\section{Quantum defect theory and the modified effective range expansion}
\label{chap5:QDT-modifiedERE}

Up to now we have been discussing the short-range phase shift of $\rm{K}$-matrix
for scattering relative to the van der Waals wave functions $F_{\ell}$ and $G_{\ell}%
$.  For power-law interactions $1/r^{\alpha}$ with $\alpha>2$, we also have
the option to define phase shifts relative to the Bessel functions of the free
wave equation.  The problem though is that the usual effective range
expansion given in Eq.~(\ref{eqn:ERE-0001}) is spoiled by nonanalytic terms as a function of $p^{2}$.  For the van der
Waals interactions in the $\ell=0$ channel, the leading nonanalytic term is
proportional to $p^{3}$, and so the scattering parameters $a_{0}$ and $r_{0}$
are well defined, but coefficients in higher order terms of the expansion are not.
For $\ell=1$ the leading nonanalytic term is proportional to $p$ and so only
the scattering length $a_{1}$ is well defined.  For $\ell\geq2$ none of the
low-energy scattering parameters are well defined.  To resolve these
problems, a modified form of the effective range expansion is used which is
known as quantum-defect theory~\cite{RepProgPhys.46.167,PhysRevA.19.1485,PhysRevA.26.2441}.

In quantum-defect theory for attractive $1/r^{6}$ potentials, one defines an
offset for the phase shift~\cite{PhysRevA.80.012702},
\begin{equation}
 \eta_{\ell}=\frac{\pi}{2}(\nu-\nu_{0}) \,.
\end{equation}
The modified effective range expansion is then
\begin{equation}
 p^{2\ell+1}\cot\left({\delta}_{\ell}+2\eta_{\ell}\right) = -\frac{1}{A_{\ell}}
 +\frac{1}{2}R_{\ell}\,p^{2}+{O}\left(p^{4}\ln p\right) \,,
\label{eqn:generalized_ER_expansion_1a}
\end{equation}
where $A_{\ell}$ and $R_{\ell}$ are the generalized scattering length and effective
range parameters.  These definitions coincide with the usual scattering
length $a_{\ell}$ for $\ell=0,1$ and the usual effective range $r_{\ell}$ for $\ell=0$.
The generalized scattering length and effective range can be written in
terms of the short-range $\rm{K}$-matrix parameters as
\begin{equation}
 A_{\ell}=\frac{\pi^{2}\beta_{6}^{2\ell+1}}{2^{4\ell+1}[\Gamma(\frac{\ell}2+\frac14)\Gamma
 (\ell+\frac32)]^{2}}\left[(-1)^{\ell}-\frac{1}{K_{\ell,0}}\right]
\label{eqn:generalized_scattering_length_1a}
\end{equation}
and
\begin{equation}
 R_{\ell}=-\frac{2^{4\ell+2}\Gamma\left(\frac{\ell}{2}+\frac{1}{4}\right)^{2}
 \Gamma\left(\ell+\frac{3}{2}\right)^{2}\beta_{6}^{-2\ell-1}}{\pi^{2}
 \left[K_{\ell,0}(-1)^{\ell}-1\right]^{2}}
 \left[\frac{\beta_{6}^{2}\left(K_{\ell,0}^{2}+1\right)}
 {4\ell^{2}+4\ell-3}-K_{\ell,2}\right] \,.
\label{eqn:generalized_effective_range_1a}
\end{equation}
From these results we see that the short-range parameter $K_{\ell,2}$ appears in
combination with $\beta_{6}^{2}$.  But in nearly all single-channel
scatterings between pairs of alkali-metal atoms, from the following equation,
\begin{align}
 K_{\ell,2} =
 &
 \frac{\beta_{6}^{2}}
 {4\ell^{2}+4\ell-3}
 \left\{
 1
 +\left[(-1)^{\ell}- A_{\ell}\frac{2^{4\ell+1}
  \Gamma\left(\frac{\ell}{2}+\frac{1}{4}\right)^{2}
  \Gamma\left(\ell+\frac{3}{2}\right)^{2}}{\pi^{2} \beta_{6}^{2\ell+1}} \right]^{-2}
 \right\}
\nonumber\\
 + & R_{\ell} A_{\ell}^{2}
  \frac{2^{4\ell}\Gamma\left(\frac{\ell}{2}+\frac{1}{4}\right)^{2}
   \Gamma\left(\ell+\frac{3}{2}\right)^{2}}{\pi^{2}\beta_{6}^{2\ell+1}
 }
 \left[(-1)^{\ell}- A_{\ell}\frac{2^{4\ell+1}
  \Gamma\left(\frac{\ell}{2}+\frac{1}{4}\right)^{2}
  \Gamma\left(\ell+\frac{3}{2}\right)^{2}}{\pi^{2} \beta_{6}^{2\ell+1}} \right]^{-2},
\label{eq:K_2_effective_range}
\end{align}
one quantitatively finds that $K_{\ell,2}$ is at
least one order of magnitude smaller than $\beta_{6}^{2}$.  This separation
of scales is the reason for the approximate universality found in
Refs.~\cite{Cordon:2010a,RuizArriola:2011a}.

The dominance of $\beta_6^{2n}$ over the subleading coefficients $K_{\ell,2n}$
in \eref{eq:K_L_expansion} for $n \geq 1$ holds for nearly all cases of
single-channel scattering between alkali-metal atoms
\cite{PhysRevA.58.4222,PhysRevA.64.010701}.
This phenomenological fact explains the absence of short-distance length scales
in the universality relation in \eref{arriola}.
Furthermore, Gao has shown that when short-range interactions arise from a
repulsive central potential, the fact that the $\rm{K}$-matrix is nearly
independent of energy means that the $\rm{K}$-matrix is also nearly independent of
angular momentum $\ell$ \cite{PhysRevA.64.010701}.  This produces a surprisingly
rich class of universal physics for single-channel van der Waals interactions
where $K_{\ell,2n}$ is negligible compared to $\beta_6^{2n}$ for all $\ell$,
and $K_{\ell,0}$ is approximately the same for all $\ell$.  Therefore $\beta_6$
and the $s$-wave scattering length will determine, to a good approximation,
the threshold scattering behavior for all values of $\ell$.

\section{Causal range for single-channel scattering}

We have shown that for negative $K_{\ell,2}\leq0$, the range $R$ of the
short-range interaction can be taken all the way down to zero.  But when
$K_{\ell,2}$ is positive, there is a constraint on $R$ and we use
\eref{vdW_inequality} to determine the causal range $R^{b}$,
\begin{equation}
 K_{\ell,2}=b^{\rm{vdW}}_{\ell}(R^{b}) \,.
\label{causal_constraint}
\end{equation}
As pointed out in Ref.~\cite{Koenig:2012bv}, one can show \textit{a priori} that
$b^{\rm{vdW}}_{\ell}(r)$ is a monotonically increasing function of $r$.  Therefore, if a
real solution to \eref{causal_constraint} exists, then it is unique.  If,
however, there is no real solution, then there is no constraint on the
interaction range and we define $R^{b}$ to be zero.  For an effective field
theory with contact interactions and van der Waals tail, the cutoff momentum
can be made as large as $\sim1/R^{b}$ before the causality bound is violated.

In the following analysis we extract the single-channel $s$-wave effective range
parameters $a_{0}$ and $r_{0}$ for several different pairs of alkali-metal atoms
$^{7}$Li, $^{23}$Na, and $^{133}$Cs in singlet and triplet channels.  The
data is taken from Refs.~\cite{PhysRevA.50.4827,
PhysRevA.53.234,PhysRevA.50.3177,PhysRevA.59.1998,PhysRevA.58.4222}.  The
reduced masses for $^{7}$Li$_{2}$, $^{23}$Na$_{2}$, and $^{133}$Cs$_{2}$ are
$\mu=6394.7,20954,121100$ (a.u.), respectively.  The van der Waals coupling
constants for $^{7}$Li$_{2}$, $^{23}$Na$_{2}$, and $^{133}$Cs$_{2}$ are
$C_{6}=1388,1472,7020$ (a.u.).  We calculate the corresponding $\rm{K}$-matrix
parameters using \eref{eqn:generalized_scattering_length_1a} and
\eref{eq:K_2_effective_range} and then compute the resulting
causal ranges.  We recall that for $\ell=0$ we simply have $A_{0}=a_{0}$ and
$R_{0}=r_{0}$.
\begin{table}[ptb]
\caption{Scattering parameters and causal ranges for $s$-wave scattering of
$^{7}$Li$,^{23}$Na$,$ and $^{133}$Cs pairs.  The scattering data collection is
taken from Ref.~\cite{PhysRevA.59.1998}. In columns \Rmnum{1} and \Rmnum{4}
the scattering data for $^{7}$Li are from Ref.~\cite{PhysRevA.53.234},
the scattering data for $^{23}$Na are from
Refs.~\cite{PhysRevA.50.4827,PhysRevA.53.234}, and data for $^{133}$Cs are from
Ref.~\cite{PhysRevA.50.3177}. In column \Rmnum{3}  the effective range
parameters, $R_{0}$, are calculated analytically in
Refs.~\cite{PhysRevA.59.1998, PhysRevA.58.4222}. In column \Rmnum{4}, the
$R_{0}$ are obtained from numerical calculations.  The scattering parameters in
columns \Rmnum{2} and \Rmnum{5} are calculated using
\eref{eqn:generalized_scattering_length_1a}
and~\eref{eq:K_2_effective_range}, and the causal ranges in column \Rmnum{6}
are obtained from \eref{causal_constraint}.}
\label{causalranges}
\begin{center}
\rowcolors{3}{gray!15}{white}
\begin{tabular}
[c]{c|c|c|c|c|c|c|c|c}\hline\hline
\multicolumn{3}{c|}{} & \Rmnum{1} & \Rmnum{2}  & \Rmnum{3}  & \Rmnum{4}
& \Rmnum{5}  & \Rmnum{6}\\
\hline\hline
Atoms & State & $\beta_{6}$ & $A_{0}$ & $K_{0,0}$ & $R_{0}$ & $R_{0}$ &
 $K_{0,2}$ & $R^b$
\\
\hline
$^{7}$Li--$^{7}$Li & $^{1}\Sigma_{g}$ & 64.9097 & 36.9 & -5.282 & 66.3 & 66.5 &
 2 $\sim$ 124 & 7 $\sim$ 19
\\
$^{7}$Li--$^{7}$Li & $^{3}\Sigma_{u}$ & 64.9097 & -17.2 & 0.643 & 1006.3
 & 1014.8 & 0 $\sim$ 17 & 3 $\sim$ 25
\\
\hline
$^{23}$Na--$^{23}$Na & $^{1}\Sigma_{g}$ & 88.624 & 34.936 & 5.705 & 187.317
& 187.5 & 0 $\sim$ 86 & 4 $\sim$ 20
\\
$^{23}$Na--$^{23}$Na & $^{3}\Sigma_{u}$ & 88.624 & 77.286 & -1.213
& 62.3756 & 62.5 &  2 $\sim$ 13 & 16 $\sim$ 24
\\
\hline
$^{133}$Cs--$^{133}$Cs & $^{1}\Sigma_{g}$ & 203.62 & 68.216 & 3.365 & 624.013
& 624.55 & 0 $\sim$ 146 & 7 $\sim$ 45\\
\hline\hline
\end{tabular}
\end{center}
\end{table}
The results for the scattering parameters and causal ranges
are given in columns \Rmnum{2}, \Rmnum{5} and \Rmnum{6} of
\tref{causalranges}.  The discrepancies in $R_{0}$ are due to the fact that
in the analytic studies in Refs.~\cite{PhysRevA.59.1998,PhysRevA.58.4222}
$K_{0,2}$ is neglected, while the numerical calculations of
Refs.~\cite{PhysRevA.50.4827,PhysRevA.53.234,PhysRevA.50.3177} include the
short-range contribution from $K_{0,2}$.

In column \Rmnum{5} of \tref{causalranges}, we present an approximate range
for $K_{0,2}$ for each atomic pair using the values for $R_0$ in columns III
and IV.  Since $K_{0,2}$ is positive, we cannot go all the way to the
zero-range limit.  However, in each case $K_{0,2}$ is at least one order of
magnitude smaller than $\beta_{6}^{2}$.\footnote{Note that $K_{\ell,2}$ has the
dimension of an area (in the appropriate atomic units).}  Although we cannot
take the zero-range limit, the causal ranges are small in comparison to
$\beta_{6}$. In each case $R^{b}$ is less than one-third the size of
$\beta_{6}$.  Hence one can probe these interactions in a van der Waals
effective field theory with cutoff momentum up to roughly three times
$1/\beta_{6}$ without violating the causality bound.

\section{Causal range near a magnetic Feshbach resonance}

In Ref.~\cite{Gao:2011a} the multichannel problem of scattering around a
magnetic Feshbach resonance is reduced to a description by an effective
single-channel $\rm{K}$-matrix that depends on the applied magnetic field $B$.
\ The behavior around the resonance is described by several parameters.
$B_{0,\ell}$ is the position of the resonance, while $g_{\text{res}}$
parametrizes the width of the Feshbach resonance. ${\rm{K}}_{\ell}^{\text{bg}}$ is a
background value for the $\rm{K}$-matrix, and the scale $d_{B,\ell}$ is introduced to
define a dimensionless magnetic field.  We write the effective single-channel
$\rm{K}$-matrix as
\begin{equation}
{\rm{K}}_{\ell}^{\text{eff}}(p,B)=-{\rm{K}}_{\ell}^{\text{bg}}
 \left[1+\frac{g_{\text{res}}}{p^{2}\beta_{6}^{2}-g_{\text{res}}
 \left(B_{s}+1\right)}\right] \,,
\label{kL}%
\end{equation}
with
\begin{equation}
 B_{s}=\frac{\left(B-B_{0,\ell}\right)}{d_{B,\ell}} \,.
\label{Bs}
\end{equation}
The parametrization given above corresponds to Eq. (18) in
Ref.~\cite{Gao:2011a}.  Note that we have changed the notation slightly and
are using a different sign convention.

By expanding the right-hand side of \eref{kL} in $p^{2}$, it is
straightforward to determine the $\rm{K}$-matrix expansion parameters $K_{\ell,0}$ and
$K_{\ell,2}$.  A short calculation yields that
\begin{equation}
 K_{\ell,0}^{\text{eff}}=-{\rm{K}}_{\ell}^{\text{bg}}\left(1+\frac{1}{B_{s}+1}\right) \,,
\end{equation}
\begin{equation}
 K_{\ell,2}^{\text{eff}}=\frac{\beta_{6}^{2}{\rm{K}}_{\ell}^{\text{bg}}}
 {g_{\text{res}}(B_{s}+1)^{2}} \,.
\label{KL2}
\end{equation}
As noted in Ref.~\cite{Gao:2011a}, the parameters ${\rm{K}}_{\ell}^{\text{bg}}$ and
$g_{\text{res}}$ are constrained by the condition
\begin{equation}
 {\rm{K}}_{\ell}^{\text{bg}}g_{\text{res}}<0 \,.
\end{equation}
From this we directly see that $K_{\ell,2}^{\text{eff}}$ given by \eref{KL2}
is always negative.  From the causality bound in \eref{vdW_inequality}
it follows that where this effective single-channel description is applicable
and correctly captures the entire energy dependence of the short-range
$\rm{K}$-matrix, the causal range will be zero when the interaction is tuned close
to a Feshbach resonance.

\section{Summary and discussion}

In this chapter we have analyzed two-body scattering with arbitrary short-range
interactions plus an attractive $1/r^{6}$ tail.  We derived the constraints
of causality and unitarity for the short-range $\rm{K}$-matrix,
\begin{equation}
 {\rm{K}}_{\ell}=\tan\delta_{\ell}^{(\text{short})}(p)
 =\sum_{n=0}^{\infty}K_{\ell,2n}\,p^{2n} \,.
\end{equation}
For any $r$ larger than the range of the short-range interactions, $R$, we
find that $K_{\ell,2}$ satisfies the upper bound
\begin{equation}
 K_{\ell,2}\leq b^{\rm{vdW}}_{\ell}(r) \,,
\end{equation}
where $b^{\rm{vdW}}_{\ell}(r)$ is
\begin{align}
 b^{\rm{vdW}}_{\ell}(r) &=\frac{\pi}{4}W[f_{\ell,2},f_{\ell,0}](r)+\frac{\pi}{4}K_{\ell,0}^{2}
 W[g_{\ell,2},g_{\ell,0}](r) \nonumber \\
 &-\frac{\pi}{4}K_{\ell,0}\left\{\vphantom{\frac{1}{2}}W[g_{\ell,2},f_{\ell,0}](r)
 +W[f_{\ell,2},g_{\ell,0}](r)\right\} \,,
\end{align}
and the Wronksians are given in \eref{Wff}, \eref{Wgg}, and \eref{Wgf}.

In clear contrast with the case for only finite-range interactions which was the subject of Chapter~\ref{chap:Neutron-proton-scattering} or with Coulomb
tails~\cite{Koenig:2012bv}, the function $b^{\rm{vdW}}_{\ell}(r)$ does not diverge but rather
vanishes as $r\rightarrow 0$ for all $\ell$.  When $K_{\ell,2}\leq0$, there is no
constraint derived from causality and unitarity that prevents the use of an
effective field theory with zero-range contact interactions plus an attractive
$1/r^{6}$ tail.  This holds true for any angular momentum value $\ell$.  For the
phenomenologically important case of a multichannel system near a
magnetic Feshbach resonance, the effective value for $K_{\ell,2}$ is
negative and so the short-range
interaction can be taken to have zero range.

The van der Waals interaction is qualitatively different from the Coulomb
interaction where $b_{\ell}^{\text{Coulomb}}(r)$ diverges for $\ell\geq1$.  The key difference is
that both van der Waals wave functions $F_{\ell}$ and $G_{\ell}$ vanish at the origin.
This phenomenon also occurs for an attractive $1/r^{\alpha}$ interaction for
$\alpha>2$ in any number of spatial dimensions.  It is also valid for an
attractive $1/r^{2}$ interaction when the coupling constant is strong enough
to form bound states.  For an effective field theory with contact
interactions and attractive singular power-law tail, the cutoff momentum can
be made arbitrarily large for any $\ell$ without producing a divergence in the
coefficient $K_{\ell,2}$ of the short-range ${\rm{K}}$ matrix.

When $K_{\ell,2}$ is positive, there is a lower bound on the range of the
short-range interactions.  We define the causal range $R^{b}$ as this minimum
value for the range, given by the condition%
\begin{equation}
 K_{\ell,2}=b^{\rm{vdW}}_{\ell}(R^{b}) \,.
\end{equation}
We have analyzed several examples of $s$-wave scattering in alkali-metal atoms in
\tref{causalranges}.  We find that the $K_{\ell,2}$ is at least one order
of magnitude smaller than $\beta_{6}^{2}$.  As a result we find that the
causal ranges are small in comparison with $\beta_{6}$.

In summary, we find that $\beta_6$ dominates over distance scales parametrizing
the short-range interactions.  The origin of this van der Waals universality
can be explained by two facts.  The first fact is the phenomenological
observation that, in single-channel scattering between alkali-metal atoms, there
is a significant separation
between the typical length scales of the short-distance physics and $\beta_6$.
This can be seen by the small size of the short-range parameter $K_{\ell,2}$
compared with $\beta_{6}^{2}$.  As Gao has shown, this also leads to
the approximate universal relation that $K_{\ell,0}$ is the same for all $\ell$
\cite{PhysRevA.64.010701}.  Therefore, to a good approximation,
$\beta_6$ and the $s$-wave scattering length will determine the threshold
scattering behavior for all values of $\ell$.  For the multichannel case
near a magnetic Feshbach resonance, we find that the
effective $K_{\ell,2}^{\text{eff}}$ is no
longer negligible.  However, $K_{\ell,2}^{\text{eff}}$ is negative, and this means
that there is no constraint from causality preventing the zero-range limit for
the short-distance interactions.

The second fact underlying the van der Waals universality is that
the zero-range limit of short-distance interactions is
well behaved with regard to scattering near threshold.  We note, however, that
there is still no scale-invariant limit for $\ell \geq 1$ since the effective range
parameter will diverge to negative infinity as $\beta_6$ goes to zero.  This can
be seen from the $\beta_6^{-2\ell+1}$ behavior with negative coefficient for $\ell \geq 1$
in \eref{eqn:generalized_effective_range_1a}.

The analysis in this study should be useful in developing an effective field
theory with an attractive $1/r^{6}$ tail and contact interactions.  Similarly,
one can also construct effective field theories for other attractive singular
potentials $1/r^{\alpha}$ for $\alpha\geq2$.  These effective field theories
could be used to investigate the Efimov effect and other low-energy phenomena in
a model-independent way.

%% file: Chapter-6.tex
\chapter{Impurity Lattice Monte Carlo and the Adiabatic Projection Method}
\label{chap:fermion-dimer-scattering}

\section{Introduction}
\label{sec:introduction}
The adiabatic projection method is a general  framework for calculating scattering and reactions  on the lattice.  The method constructs a low-energy effective theory for clusters which becomes exact in the limit of large Euclidean projection time. Previous studies of this method \cite{Rupak:2013aue,Pine:2013zja} have used exact sparse matrix methods.  In this work we demonstrate the first application using Monte Carlo simulations.  As we will show, the adiabatic projection method significantly improves the  accurate calculation of finite-volume energy levels. As we also will show, the finite-volume energy levels must be calculated with considerable accuracy in order to determine the scattering phase shifts using L\"uscher's method.  We give a short summary of L\"uscher's method later in our discussion.

The goal of this analysis is to benchmark the use of lattice Monte Carlo simulations with the adiabatic projection method.  The example we consider in detail is fermion-dimer scattering for two-component fermions and zero-range interactions.  Our  calculation also corresponds to neutron-deuteron scattering in the spin-quartet channel at leading order in pionless effective field theory. In our interacting system there are two components for the fermions.  We call the two components up and down spins, $\uparrow$ and $\downarrow$.   The bound dimer state is composed of one $\uparrow$ and one $\downarrow$, and our fermion-dimer system consists of two $\uparrow$  and one $\downarrow$.  While $s$-wave scattering has been considered previously \cite{PhysRevD.84.091503,PhysRevC.86.034003,A.Rokash2013.3386,Pine:2013zja}, we will present the first lattice calculations of $p$-wave and $d$-wave fermion-dimer scattering. 

As discussed in Ref.~\cite{Pine:2013zja}, the adiabatic projection method starts with a set of initial cluster states.  By clusters we mean either a single particle or a bound state of several particles.   In our analysis here we consider fermion-dimer elastic scattering where there are two clusters.  In Ref.~\cite{Pine:2013zja}, the initial fermion-dimer states were parameterized by the initial spatial separation between clusters, $\vec{R}$.  The initial cluster states can be written explicitly as
\begin{equation}
\lvert \vec{R} \rangle = \sum_{\vec{n}}b^{\dagger}_{\uparrow}(\vec{n})b^{\dagger}_{\downarrow}(\vec{n})b^{\dagger}_{\uparrow}(\vec{n}+\vec{R}) \lvert 0 \rangle,
\end{equation}
where the spatial volume is a periodic cubic box of length $L$ in lattice units.  The initial states are then projected using Euclidean time to form dressed cluster states,
 \begin{equation}
\lvert \vec{R} \rangle_t = e^{-\hat{H}t}\lvert \vec{R} \rangle. \end{equation}
The adiabatic method uses these dressed cluster states to calculate matrix elements of the Hamiltonian and other observables.  The result is a low-energy effective theory of interacting clusters which becomes systematically more accurate as the projection time $t$ is increased.
 An estimate of the residual error is derived in Ref.~\cite{Pine:2013zja}. 

For our calculations here we follow the same general process except that we build the initial cluster states in a different manner. Instead of working with the relative separation between clusters, we work with the relative momentum between the clusters.  We find that this change improves the efficiency of the Monte Carlo calculation by reducing the number of required initial states.  The new technique involves first constructing  a dimer state with momentum $\vec{p}$ using Euclidean time projection and then   multiplying by a creation operator for a second $\uparrow$ particle with momentum $-\vec{p}$.  For example, we can write the initial fermion-dimer state explicitly as
\begin{equation}
\lvert \vec{p} \rangle = \tilde{b}^{\dagger}_{\uparrow}(-\vec{p}) e^{-\hat{H}t'} \, \tilde{b}^{\dagger}_{\uparrow}(\vec{p})\tilde{b}^{\dagger}_{\downarrow}(\vec{0})\lvert 0 \rangle
\label{eqn:initialState-0001}\,.
\end{equation}
From these states we produce dressed cluster states by Euclidean time projection,
\begin{equation}
\lvert \vec{p} \rangle_t = e^{-\hat{H}t/2} \lvert \vec{p} \rangle
\label{eqn:ClusterState-0001}\,.
\end{equation}
We then proceed in the same manner as in Ref.~\cite{Pine:2013zja} and calculate the matrix elements of the Hamiltonian in the basis of the dressed cluster states.   

For our Monte Carlo simulations we introduce a new algorithm which we call the impurity lattice Monte Carlo 
algorithm.  Credit for developing this algorithm is to be shared with Ref.~\cite{Bour:2014}, where applications to impurities in many-body systems are being investigated using the same method.  It can be viewed as a hybrid algorithm in between worldline and auxiliary-field Monte Carlo simulations.  In worldline algorithms, the quantum amplitude is calculated by sampling particle worldlines in Euclidean spacetime.  In auxiliary-field Monte Carlo simulations, the interactions are recast as single particle interactions, and the  quantum amplitude is computed exactly for each auxiliary field configuration.  In impurity Monte Carlo, we handle the impurities using worldline Monte Carlo simulations while all other particles are treated using the auxiliary-field formalism.  Furthermore, the impurity worldlines themselves are acting as additional auxiliary fields felt by other particles in the system.  We have found that for our system of two $\uparrow$ and one $\downarrow$ particles, impurity lattice Monte Carlo method is computationally superior to other methods such as the auxiliary-field Monte Carlo due to its speed and efficiency as well as control over sign oscillations. We will derive the formalism of impurity Monte Carlo simulations in detail in our discussion here. 

The organization of this chapter is as follows.  We first start with the basic continuum and lattice formulations of our interacting system with zero-range two-component fermions.  We then take a short detour to derive the connection between normal-ordered transfer matrices and lattice Grassmann actions.  Using our dictionary between lattice Grassmann actions and quantum operators, we derive the transfer matrix induced by a given single impurity worldline.  We then describe the implementation of the adiabatic projection method and the details of our Monte Carlo simulations for computing finite-volume energy levels. 

In order to determine scattering phase shifts, we then discuss L\"uscher's finite-volume method.  As part of this discussion we discuss for the first time, the character of topological volume corrections for fermion-dimer scattering in the $p$-wave and $d$-wave channels.  By topological volume corrections, we are specifically referring to  momentum-dependent finite-volume corrections of the  dimer binding energy \cite{PhysRevD.84.091503,Davoudi:2011md}.  Previous studies looking at topological volume corrections had only considered $s$-wave scattering \cite{PhysRevD.84.091503,PhysRevC.86.034003,A.Rokash2013.3386,Pine:2013zja}.  The extension to higher partial waves is given in the appendix.  We then conclude with a comparison of Monte Carlo results as well as exact lattice calculations and continuum calculations.

\section{Lattice Hamiltonian}

We consider a three-body system of two-component fermions with equal mass, $m_{\uparrow}=m_{\downarrow}=m$.
We consider the  limit of large scattering length between the two components  where the the interaction range of the fermions is taken to be negligible. We start with the free non-relativistic Hamiltonian,
\begin{align}
 \hat{H}_{0} = \frac{1}{2m} \sum_{s=\uparrow,\downarrow}
 \int d^{3}\vec{r} \, \,  \vec{\nabla} b_{s}^{\dagger}(\vec{r}) \cdot \vec{\nabla} b_{s}(\vec{r}) \,,
 \label{eqn:free-Hamiltonian-001}
\end{align}
In the low-energy limit the interaction can be simplified as a delta-function interaction between  the two spin components, \begin{align}
 \hat{H} = \frac{1}{2m} \sum_{s=\uparrow,\downarrow}
 \int d^{3}\vec{r} \, \,  \vec{\nabla} b_{s}^{\dagger}(\vec{r}) \cdot \vec{\nabla} b_{s}(\vec{r})
 +
 C_{0} \int d^{3}\vec{r}  \, \,
\hat{\rho}_{\uparrow}(\vec{r}) \,  \hat{\rho}_{\downarrow}(\vec{r}) \,,
 \label{eqn:Hamiltonian-001}
 \end{align}
 where $\hat{\rho}_{\uparrow,\downarrow}(\vec{r})$ are density operators,
 \begin{align}
\hat{\rho}_{\uparrow}(\vec{r}) =  b_{\uparrow}^{\dagger}(\vec{r}) b_{\uparrow}(\vec{r}) \,,
\\
\hat{\rho}_{\downarrow}(\vec{r}) =  b_{\downarrow}^{\dagger}(\vec{r}) b_{\downarrow}(\vec{r})\,.
 \label{eqn:Latt-densty-op-002}
 \end{align}
The ultraviolet physics of this zero-range interaction must be regulated in some manner.  In our case the lattice provides the needed regularization. We denote the spatial lattice spacing as $a$ and the temporal lattice spacing as $a_{t}$. We will write all quantities in lattice units, which are physical units multiplied by the corresponding power of $a$ to render the combination dimensionless. We use the free non-relativistic lattice Hamiltonian defined in Eqs.~(\ref{eqn:lattice-Hamiltonian-001}) and (\ref{eqn:lattice-Hamiltonian-002})
and the contact interaction potential is
\begin{align}
\hat{V} = C_{0}
&\sum_{\vec{n}}
\hat{\rho}_{\uparrow}(\vec{n}) \,
\hat{\rho}_{\downarrow}(\vec{n})\,.
 \label{eqn:lattice-Hamiltonian-005}
\end{align}
Here $\hat{l}$ denotes a lattice unit vector in one of the spatial directions, $\hat{l}=\hat{1},\hat{2},\hat{3}$.  The unknown interaction coefficient $C_0$ is tuned to reproduce the desired binding energy of the dimer at infinite volume.
\section{Lattice path integrals and transfer matrices}
\label{sec:LatticePIandTM}

For our Monte Carlo simulations and exact lattice calculations we use the transfer matrix formalism introduced in Chapter~\ref{chap:LatticeEFT}. It is convenient to collect some of the important formulas given in Chapter~\ref{chap:LatticeEFT}. As it is already discussed, the Grassmann path integral has the form
\begin{align}
 \mathcal{Z} = \int
 \left[\prod_{n_{t},\vec{n},s=\uparrow,\downarrow}
 d\theta_{s}^{*}(n_{t},\vec{n})d\theta_{s}(n_{t},\vec{n})\right]  \, \, e^{- S[\theta,\theta^{*}]} \,,
 \label{eqn:Grassmann-path-int-001}
 \end{align}
where $S[\theta,\theta^{*}]$ is the non-relativistic lattice action and defined in Section~\ref{chap3:lattice-formulation}, and $\theta_{s}^{*}$ and $\theta_{s}$ are anti-commuting Grassmann variables. Our
lattice action can be decomposed into three parts.

While the Grassmann formalism is convenient for deriving the lattice Feynman rules, the transfer matrix formalism is more  convenient for numerical calculations.Therefore, we make the connection between the two formulations, and we use the following exact relation between the Grassmann path integral formula and the transfer matrix formalism~\cite{PhysRevD.38.1228,FoundPhy.30.487}. For any function $f$,
\begin{align}
 \rm{Tr}  & \Big[ \, : \,  f_{L_{t}-1}[a^{\dagger}_{s^{\prime}} (\vec{n^{\prime}}),a_{s} (\vec{n})] \, : \,
 \cdots \, : \, f_{0}[a^{\dagger}_{s^{\prime}} (\vec{n^{\prime}}),a_{s} (\vec{n})] \, : \, \Big]
 \nonumber\\
 &
 = \int
 \left[\prod_{n_{t},\vec{n},s=\uparrow,\downarrow}
 d\theta_{s}^{*}(n_{t},\vec{n})d\theta_{s}(n_{t},\vec{n})\right]  \, e^{-\sum_{n_t} S_{t}[\theta,\theta^{*},n_{t}]}
 \prod_{n_{t} = 0}^{L_{t}-1}
 f_{n_{t}}[\theta^{*}_{s^{\prime}} (n_{t},\vec{n}^{\prime}),\theta_{s} (n_{t},\vec{n})] \,,
 \label{eqn:G-path-T-matrix-002}
\end{align}
where the symbol :\,: signifies normal ordering. Normal ordering rearranges all operators so that all annihilation operators are moved to the right and creation operators are moved to the left with the appropriate number of anticommutation minus signs.  Then the desired transfer matrix formulation of the path integral is
\begin{align}
 \mathcal{Z} = \rm{Tr} \, \left[ \hat{M}^{L_{t}} \right]\,,
 \label{eqn:transfer-matrix-0300}
\end{align}
 where $\hat{M}$ is the normal-ordered transfer matrix operator,
\begin{align}
 \hat{ M} = \,  : \, \exp
 \left[
 -\alpha_{t} \hat{H}_{0}-\alpha_{t} C_{0}
 \sum_{\vec{n}}
\hat{\rho}_{\uparrow}(\vec{n})
 \hat{\rho}_{\downarrow}(\vec{n})  \right]
 \, : \,.
 \label{eqn:transfer-matrix-0200}
\end{align}
Here $\hat{H}_{0}$ is the free lattice Hamiltonian given in Eq.~(\ref{eqn:lattice-Hamiltonian-001}).

\section{Impurity Lattice Monte Carlo:  Single Impurity}
\label{sec:ImpurityLMC}

In this section we derive the formalism for impurity lattice Monte Carlo for a single impurity.  In impurity Monte Carlo the impurities are treated differently from other particles.
The assumption is that there are only a small number of impurities and these can be sampled using worldline Monte Carlo without strong fermion sign oscillation problems from antisymmetrization.  In our case there is exactly one $\downarrow$  particle, and we treat this as a single impurity for our system.

Let us consider the occupation number basis,
\begin{align}
\Ket{\chi^\uparrow_{n_{t}},
\chi^{\downarrow}_{n_{t}}}
=
\prod_{\vec{n}}
\left\{
\left[
b_{\uparrow}^{\dagger}(\vec{n})
\right]^{\chi^{\uparrow}_{n_{t}}(\vec{n})}
\left[
b_{\downarrow}^{\dagger}(\vec{n})
\right]^{\chi^{\downarrow}_{n_{t}}(\vec{n})}
\right\}
 \, \ket{0}
  \label{eqn:transfer-matrix-0311}
\end{align}
where $\chi^{s}_{n_{t}}(\vec{n})$ counts the occupation number on each lattice site at time step $n_t$ and has values which are either 0 or 1. Let us define  the Grassmann functions,
\begin{align}
X(n_{t})
=_{}
\prod_{\vec{n}}
\left[
e^{\theta^{*}_{\uparrow}(n_{t},\vec{n})
   \theta_{\uparrow}(n_{t},\vec{n})}
   \,
e^{\theta^{*}_{\downarrow}(n_{t},\vec{n})
   \theta_{\downarrow}(n_{t},\vec{n})}
\right] \,,
 \label{eqn:transfer-matrix-0313}
\end{align}
and
\begin{equation}
M(n_t) =e^{-S_{H_{0}}[\theta,\theta^{*},n_{t}]}e^{-S_{V}[\theta,\theta^{*},n_{t}]}.
\end{equation}
The transfer matrix element between time steps $n_t$ and $n_t+1$ can be written in terms of these lattice Grassmann functions as
\begin{align}
\bra{\chi^{\uparrow}_{n_{t}+1}
,
\chi^{\downarrow}_{n_{t}+1}}
&
 \hat{M}
\ket{
\chi^{\uparrow}_{n_{t}}
,
\chi^{\downarrow}_{n_{t}}
}
\nonumber\\
&
=
\prod_{\vec{n}}
\left\{
\left[\frac{\overrightarrow{\partial}}{
\partial \theta_{\downarrow}^{*}(n_{t},\vec{n})}
\right]^{\chi^{\downarrow}_{n_{t}+1}(\vec{n})}
\left[\frac{\overrightarrow{\partial}}{
\partial \theta_{\uparrow}^{*}(n_{t},\vec{n})}
\right]^{\chi^{\uparrow}_{n_{t}+1}(\vec{n})}
\right\}
 X(n_{t}) \, M(n_{t})
 \nonumber\\
 &
 \times
\left.
\prod_{\vec{n}^{\prime}}
\left\{
\left[\frac{\overleftarrow{\partial}}{
\partial \theta_{\uparrow}(n_{t},\vec{n}^{\prime})}
\right]^{\chi^{\uparrow}_{n_{t}}(\vec{n}^{\prime})}
\left[\frac{\overleftarrow{\partial}}{
\partial \theta_{\downarrow}(n_{t},\vec{n}^{\prime})}
\right]^{\chi^{\downarrow}_{n_{t}}(\vec{n}^{\prime})}
\right\}\right|_{\tiny{\begin{array}{c}
                  \theta^{\prime}_{\uparrow}
                  =\theta^{\prime}_{\downarrow}=0             \\
                  \theta_{\uparrow}
                   =\theta_{\downarrow}=0 
                \end{array}}}
                \,.
\label{eqn:transfer-matrix-0316}
\end{align}
This result can be verified by checking the different possible combinations for the occupation numbers.  Since we have only one $\downarrow$ particle, the right hand side is nonzero only if
\begin{align}
\sum_{\vec{n}} \chi^{\downarrow}_{n_{t}}(\vec{n})
=
\sum_{\vec{n}} \chi^{\downarrow}_{n_{t}+1}(\vec{n}) = 1\,.
\label{eqn:transfer-matrix-0317}
\end{align}

\begin{figure}[hptb]
\begin{center}
\resizebox{150mm}{!}{\includegraphics{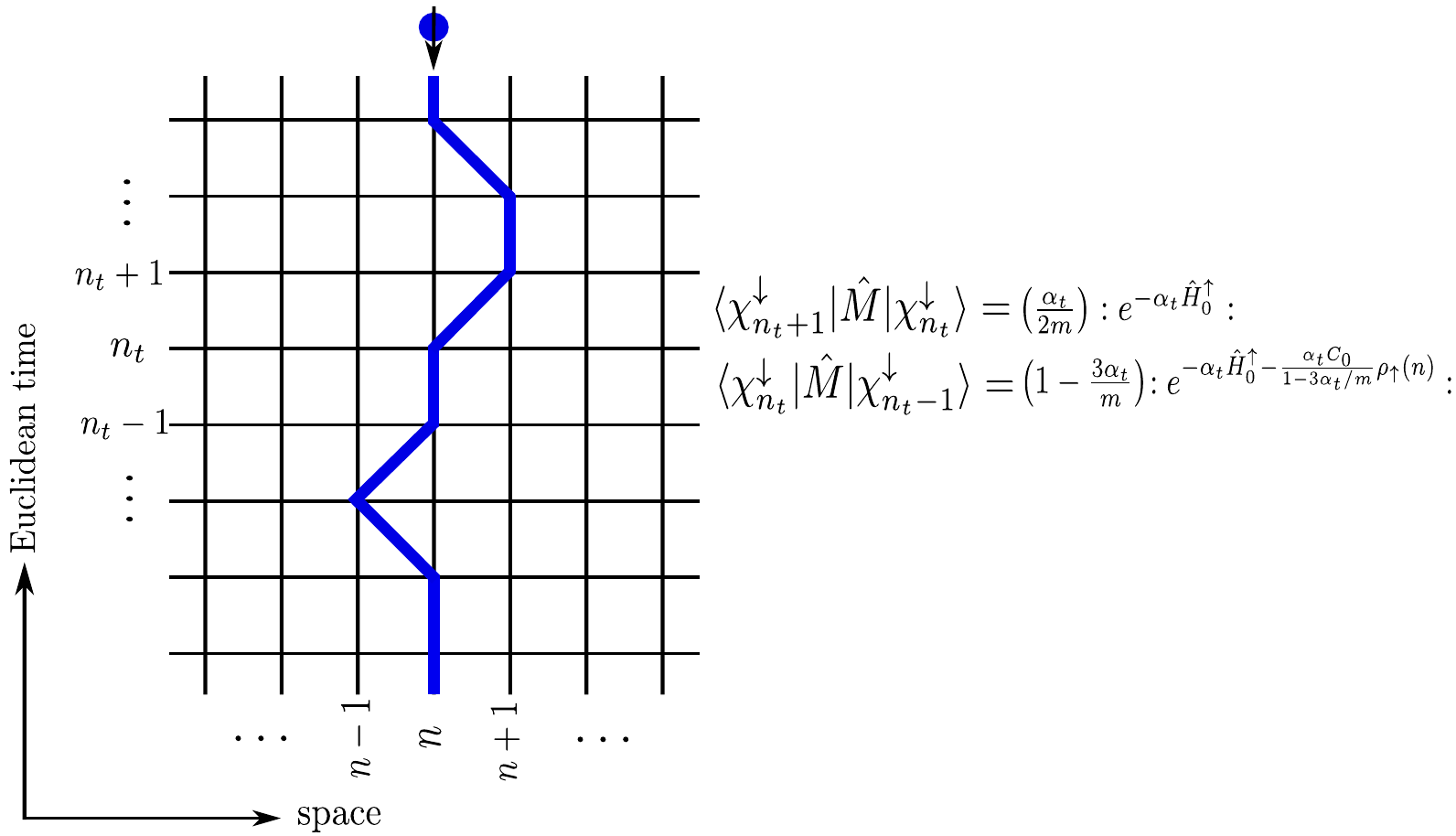}}
\end{center}
\caption{A segment of a worldline configuration on a 1+1 dimensional Euclidean lattice.  See the main text for derivations of the reduced transfer-matrix operators.}%
\label{fig:worldline}
\end{figure}

We now derive the transfer matrix formalism for one spin-$\downarrow$ particle worldline in a medium consisting of an arbitrary number of spin-$\uparrow$ particles.  The impurity worldline is to be considered fixed. To provide  a simple visual representation of the worldine, we draw in Figure~\ref{fig:worldline} an example of a single-particle worldline configuration on a 1+1 dimensional Euclidean lattice.

We now remove or ``integrate out" the impurity particle from the lattice action.  We consider first the case when the $\downarrow$ particle hops from $\vec{n}^{\prime \prime}$ to some nearest neighbor site.  In other words, $\chi^{\downarrow}_{n_{t}}(\vec{n}^{\prime \prime}) =1$ and
$\chi^{\downarrow}_{n_{t}+1}(\vec{n}^{\prime \prime}\pm\hat{l}) =1$ for some unit vector $\hat{l}$. In this case we have
\begin{align}
&\Braket{
\chi^{\uparrow}_{n_{t}+1},
\chi^{\downarrow}_{n_{t}+1}
|\hat{M}|
\chi^{\uparrow}_{n_{t}},
\chi^{\downarrow}_{n_{t}}
}
\nonumber\\
&
=
\prod_{\vec{n}}
\left\{
\left[\frac{\overrightarrow{\partial}}{
\partial \theta_{\uparrow}^{*}(n_{t},\vec{n})}
\right]^{\chi^{\uparrow}_{n_{t}+1}(\vec{n})}
\right\}
\slashed{X}(n_{t}) \,
\slashed{M}_{\vec{n}^{\prime\prime}\pm\hat{l},\vec{n}^{\prime\prime}}(n_{t})
\left.
\prod_{\vec{n}^{\prime}}
\left\{
\left[{\frac{\overleftarrow{\partial}}{
\partial \theta_{\uparrow}(n_{t},\vec{n}^{\prime})}}
\right]^{\chi^{\uparrow}_{n_{t}}(\vec{n}^{\prime})}
\right\}\right|_{\tiny{\begin{array}{c}
                  \theta^{\prime}_{\uparrow}
                  =0 \\
                  \theta_{\uparrow}
                  =0
                \end{array}}}
                \,,
\label{eqn:transfer-matrix-0319}
\end{align}
where
\begin{align}
\slashed{X}(n_{t})
=
\prod_{\vec{n}}
\left[
e^{\theta^{*}_{\uparrow}(n_{t},\vec{n})
   \theta_{\uparrow}(n_{t},\vec{n})}
\right] \,,
 \label{eqn:transfer-matrix-0321}
\end{align}
and\begin{align}
\slashed{M}_{\vec{n}^{\prime \prime}\pm\hat{l},\vec{n}^{\prime \prime}}(n_{t})
=
\left(
\frac{\alpha_{t}}{2 m}
\right)
 \exp
 \left\{
 -\alpha_{t} H^{\uparrow}_0[\theta_{s},\theta^*_{s},n_t]\right\}.
 \label{eqn:transfer-matrix-0323}
\end{align}
Next we consider the case when $\chi^{\downarrow}_{n_{t}}(\vec{n}^{\prime \prime})=1$ and $\chi^{\downarrow}_{n_{t}+1}(\vec{n}^{\prime \prime})=1$ which corresponds to no spatial hopping of the impurity worldline. Then we have
\begin{align}
&\Braket{
\chi^{\uparrow}_{n_{t}+1},
\chi^{\downarrow}_{n_{t}+1}
|\hat{M}|
\chi^{\uparrow}_{n_{t}},
\chi^{\downarrow}_{n_{t}}
}
\nonumber\\
&
=
\prod_{\vec{n}}
\left\{
\left[\frac{\overrightarrow{\partial}}{
\partial \theta_{\uparrow}^{*}(n_{t},\vec{n})}
\right]^{\chi^{\uparrow}_{n_{t}^{\prime}}(\vec{n})}
\right\}
\slashed{X}(n_{t}) \,
\slashed{M}_{\vec{n}^{\prime\prime},\vec{n}^{\prime\prime}}(n_{t})
\left.
\prod_{\vec{n}^{\prime}}
\left\{
\left[\frac{\overleftarrow{\partial}}{
\partial \theta_{\uparrow}(n_{t},\vec{n}^{\prime})}
\right]^{\chi^{\uparrow}_{n_{t}}(\vec{n}^{\prime})}
\right\}\right|_{\tiny{\begin{array}{c}
                  \theta^{\prime}_{\uparrow}
                  =0 \\
                  \theta_{\uparrow}
                  =0
                \end{array}}}
                \,,
 \label{eqn:transfer-matrix-029}
\end{align}
where\begin{align}
\slashed{M}_{\vec{n}^{\prime \prime},\vec{n}^{\prime \prime}}(n_{t})
=
\left(1- \frac{3\alpha_{t}}{m}\right)
&
 \exp
 \left\{
 -\alpha_{t} H^{\uparrow}_0[\theta_{s},\theta^*_{s},n_t]- \frac{\alpha_{t} C_{0}}{1- \frac{3\alpha_{t}}{m_{}}} \,\theta^*_{\uparrow}(n_{t},\vec{n}^{\prime\prime})\theta_{\uparrow}(n_{t},\vec{n}^{\prime\prime})\right\} \,.
 \label{eqn:transfer-matrix-033}
\end{align}

From these Grassmann lattice actions with the impurity integrated out, we can write down the corresponding transfer matrix operators.  When the impurity makes a spatial hop, the reduced transfer-matrix  operator is 
\begin{equation}
\hat{\slashed{M}}_{\vec{n}^{\prime \prime}\pm\hat{l},\vec{n}^{\prime \prime}}=
\left(
\frac{\alpha_{t}}{2 m}
\right ) \,  : \, \exp
 \left[
 -\alpha_{t} \hat{H}^{\uparrow}_{0}  \right]
 \, :.
 \label{eqn:t-matrix-stationary}
\end{equation}
When the impurity worldline remains stationary the reduced transfer-matrix operator is
\begin{equation}
\hat{\slashed{M}}_{\vec{n}^{\prime \prime},\vec{n}^{\prime \prime}}=
\left(1- \frac{3\alpha_{t}}{m}\right ) \,  : \, \exp
 \left[
 -\alpha_{t} \hat{H}^{\uparrow}_{0}-\frac{\alpha_{t} C_{0}}{1-\frac{3\alpha _{t}}{m}}
 \rho_{\uparrow}(\vec{n}^{\prime \prime})  \right]
 \, :.
  \label{eqn:t-matrix-hopping}
\end{equation}We note that these reduced transfer matrices are just one-body operators on the linear space of $\uparrow$ particles.

\section{Adiabatic Projection Method}
\label{sec:Adiabatic-Projection-Method}

In this section we describe our application of the adiabatic projection method using a set of cluster states constructed in momentum space. As already described in Eq.~(\ref{eqn:initialState-0001}) in a simplified notation, we let $\ket{\Psi_{\vec{p}}}$ be the fermion-dimer initial state  with relative momentum $\vec{p}$,
\begin{equation}
\lvert \Psi_{\vec{p}} \rangle = \tilde{b}^{\dagger}_{\uparrow}(-\vec{p})\hat{ M}^{L'_{t}}  \, \tilde{b}^{\dagger}_{\uparrow}(\vec{p})\tilde{b}^{\dagger}_{\downarrow}(\vec{0})\lvert 0 \rangle,
\label{eqn:initialState-0002}\
\end{equation} where we use the transfer matrix operator $\hat{ M}$ given in Eq.~(\ref{eqn:transfer-matrix-0200}) for some number of time steps $L'_{t}$.  The purpose of this time propagation is to allow the dimer to bind its constituents before injecting an additional $\uparrow$ particle.  In this part of the calculation we in fact increase the attractive interactions between the two spins to allow them to form the bound dimer faster.  We find that this trick increases the computational efficiency on large lattice systems. The dressed cluster states are defined as
\begin{align}
\ket{\Psi_{\vec{p}}}_{L_t/2} =  \hat{ M}^{L_t/2}\ket{\Psi_{\vec{p}}},
\end{align}
for some even number $L_t$, and the overlap between dressed cluster states is
\begin{align}
Z_{\vec{p}\vec{p}\,'}(L_t) =\bra{\Psi_{\vec{p}}}\hat{ M}^{L_t}\ket{\Psi_{\vec{p}\,'}}.
 \label{eqn:projection-amplitude-001}
\end{align}

For large $L_t$ we can obtain an accurate representation of the low-energy spectrum of $\hat{M}$ by defining the adiabatic transfer matrix as
\begin{align}
[\hat{M}^{a}(L_t)]_{\vec{p}\vec{p}\,'}
=
\sum_{\vec{p}\,''}Z^{-1}_{\vec{p}\vec{p}\,''}(L_t) \, \, Z_{\vec{p}\,''\vec{p}\,'}(L_t+1).\label{eqn:adibatic-hamitonian-0001} 
\end{align}
Alternatively we can also construct a symmetric version of the adiabatic transfer matrix as
\begin{align}
[\hat{M}^{a}(L_t)]_{\vec{p}\vec{p}\,'}
=
\sum_{\vec{p}\,'',\vec{p}\,'''}Z^{-1/2}_{\vec{p}\vec{p}\,''}(L_t) \, \, Z_{\vec{p}\,''\vec{p}\,'''}(L_t+1)Z^{-1/2}_{\vec{p}\,'''\vec{p}\,'}(L_t).\label{eqn:adibatic-hamitonian-0002}
\end{align}
Either form will produce exactly the same spectrum.  As with any transfer matrix, we interpret the eigenvalues $\lambda_i(L_t)$ of the adiabatic transfer matrix
as energies using the relations
\begin{equation}
e^{-E_i(L_t)\alpha_t} = \lambda_i(L_t), 
\quad  E_{i} (L_t)= -{\alpha_{t}}^{-1}\log \lambda_i(L_t).
\label{eqn:transient_energy} 
\end{equation}
The exact low-energy eigenvalues of the full transfer matrix $\hat{M}$ will be recovered in the limit $L_t \rightarrow \infty$.

As a special case, one can simply restrict the adiabatic projection calculation to  a single initial momentum state, for example, $\vec{p}=0$.  In that case the adiabatic transfer matrix is just the scalar ratio\begin{equation}
Z_{\vec{p}\vec{p}}(L_t+1)/Z_{\vec{p}\vec{p}}(L_t).
\end{equation}
However we find that the energy calculations are significantly more accurate and converge much faster with increasing $L_t$ when using a set of several initial cluster states. 

\section{Impurity Monte Carlo Simulation}
\label{sec:simulation}

The reduced transfer matrices $\hat{\slashed{M}}_{\vec{n},\vec{n}'}$
in Eq.~(\ref{eqn:t-matrix-stationary}) and (\ref{eqn:t-matrix-hopping}) are one-body operators on the linear space of $\uparrow$ particles.
Therefore we can simply multiply the reduced transfer matrices together.  It is perhaps worthwhile to note that the ${\vec{n},\vec{n}'}$ subscripts are not the matrix indices of the reduced transfer matrix, but rather the coordinates of the $\downarrow$ particle that was integrated out.  The matrix indices of $\hat{\slashed{M}}_{\vec{n},\vec{n}'}$ are being left implicit.

The Euclidean time projection can be written  as a sum over worldline configurations of the $\downarrow$ particle. As a convenient shorthand we write \begin{equation}
\hat{\slashed{M}}_{\{\vec{n}_{j}\}}^{[L_t]}=
\hat{\slashed{M}}_{\vec{n}_{L_{t}},\vec{n}_{L_{t}-1}}
\ldots
\hat{\slashed{M}}_{\vec{n}_{1},\vec{n}_{0}},
\end{equation}
where $\vec{n}_{j}$ denotes the spatial position of the spin-$\downarrow$ particle at time step $j$.
 The projection amplitude for cluster states $\ket{\Psi_{\vec{p}}}$ and $\ket{\Psi_{\vec{p}\,'}}$ is then
 \begin{align}
Z_{\vec{p}\vec{p}\,'}(L_t) = \sum_{\vec{n}_{0},\ldots,\vec{n}_{L_{t}}}\bra{\Psi_{\vec{p}}}
\hat{\slashed{M}}_{\{\vec{n}_{j}\}}^{[L_t]}\ket{\Psi_{\vec{p}\,'}}.
 \label{eqn:Projected-matric-elements-010}
\end{align}
The states $\ket{\Psi_{\vec{p}\,}}$
and $\ket{\Psi_{\vec{p}\,'}}$ defined in Eq.~(\ref{eqn:initialState-0002}) are constructed using single particle creation operators, and so the amplitude $Z_{\vec{p}\vec{p}\,'}(L_t)$ is just the determinant of a $2\times 2$ matrix of single-particle amplitudes.  As seen in Eq.~(\ref{eqn:initialState-0002}), there are  an extra $L'_t$ projection steps in between some of the creation operators.  This gives us the following structure, \begin{equation}
Z_{\vec{p}\vec{p}\,'}(L_t)=\sum_{\vec{n}_{0},\ldots,\vec{n}_{L_{t}}}\sum_{\vec{n}'_{0},\ldots,\vec{n}'_{L'_{t}}}\sum_{\vec{n}''_{0},\ldots,\vec{n}''_{L'_{t}}}\det M_{2\times2},
\end{equation} where $\vec{n}''_{0}=\vec{n}_{L_t}$,  $\vec{n}'_{L_t}=\vec{n}_{0}$, and\begin{equation}
M_{2\times2} = \begin{bmatrix}\langle  \vec p \rvert\hat{\slashed{M}}_{\{\vec{n}''_{j}\}}^{[L'_t]}\hat{\slashed{M}}_{\{\vec{n}_{j}\}}^{[L_t]}\hat{\slashed{M}}_{\{\vec{n}'_{j}\}}^{[L'_t]}\lvert \vec{p}\,' \rangle & \langle  \vec p \lvert\hat{\slashed{M}}_{\{\vec{n}''_{j}\}}^{[L'_t]}\hat{\slashed{M}}_{\{\vec{n}_{j}\}}^{[L_t]}\rvert- \vec{p}\,' \rangle \\
\langle-  \vec p \lvert\hat{\slashed{M}}_{\{\vec{n}_{j}\}}^{[L_t]}\hat{\slashed{M}}_{\{\vec{n}'_{j}\}}^{[L'_t]}\rvert \vec{p}\,' \rangle & \langle-  \vec p \lvert\hat{\slashed{M}}_{\{\vec{n}_{j}\}}^{[L_t]}\rvert- \vec{p}\,' \rangle \\
\end{bmatrix}.
\end{equation}The calculation of $Z_{\vec{p}\vec{p}\,'}(L_t)$ has now been recast as a problem of computing the determinant of the matrix $M_{2\times2}$ over all possible impurity worldlines.  We use a Markov chain Monte Carlo process to select worldline configurations.  The Metropolis algorithm is used to accept or reject configurations with importance sampling given by the weight function $\lvert Z_{\vec{p}\vec{p}}(L_t) \rvert$, where $\vec{p}$ is one of the initial momenta.   

We now benchmark our results for the low-energy spectrum calculated using adiabatic projection and the impurity Monte Carlo method.  We compare with exact lattice results computed using the Lanczos iterative eigenvector method with a space of $\sim L^6$ basis states.  Although exact lattice results provide a useful benchmark test for the three-particle system, the extension to larger systems is computationally not viable due to exponential scaling in memory and CPU\ time.  In contrast, the impurity Monte Carlo calculation  does scale well to much larger systems.  In fact, many-body impurity systems are currently being studied in Ref.~\cite{Bour:2014}. 
\begin{table}[h]
\caption{Momentum of the dimer, $\vec{p}_{\rm{d}}$, with $p = 2\pi/L$. The total momentum of the  system is zero.}
\label{table:table-momenta0filling}
\centering
\begin{tabular}{|c|c|}
  \hline
  \, $n$ \, & $\vec{p}_{\rm{d}}$            \\ 
  \hline\hline
  1      & $\Braket{p,0,0}$    \\
  \hline
  2      & $\Braket{0,p,0}$     \\
  \hline
  3      & $\Braket{0,0,p}$     \\
  \hline
  4      & $\Braket{p,-p,0}$    \\
  \hline
  5      & $\Braket{p,0,-p}$    \\
  \hline
  6      & $\Braket{0,p,-p}$    \\
   \hline
\end{tabular}
\end{table}

In our lattice calculations we take the particle mass to be the average nucleon mass, $938.92$ MeV, and the interaction strength $C_{0}$ is tuned to obtain the deuteron energy, $-2.2246$ MeV.   We use an $L^{3}$ periodic cubic volume with spatial lattice spacing $a = 1.97$ fm.  The values of $L$ used will be specified later.  In the temporal direction we use $L_t$ time steps with a temporal lattice spacing $a_{t} = 1.31$ fm/$c$.  

Let $N$ be the number of initial/final states.  We choose  the initial dimer momenta, $\vec{p}_{\rm{d}}$, as shown in Table~\ref{table:table-momenta0filling}.  In all cases the total momentum of the three-particle system is set to zero.  We label  and order the various possible dimer momenta with index $n=1,\cdots,N$.  We then construct the corresponding $N\times N$ adiabatic matrix, $[\hat{M}^{a}(L_t)]_{nn'}$, and obtain the $N$ low-lying energy states of the finite-volume system. There is no restriction on the choice of $N$.  Therefore, so long as the numerical stability of the matrix calculations is under control, it is advantageous to maximize the number $N$. While constructing a large adiabatic matrix requires more computational time, it significantly accelerates the convergence with the number of projection time steps, $L_t$. 
 
\captionsetup[subfigure]{position=bottom, labelfont=normalfont,textfont=normalfont,singlelinecheck=off,margin=115pt}
\begin{figure}[hptb]
\begin{center}
\subfloat[]{\includegraphics[width=75mm]{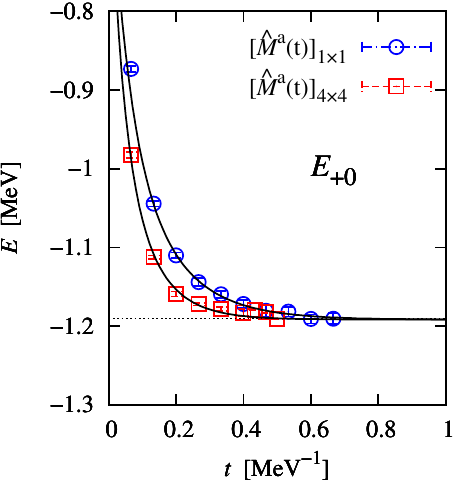}}
\qquad
\subfloat[]{\includegraphics[width=75mm]{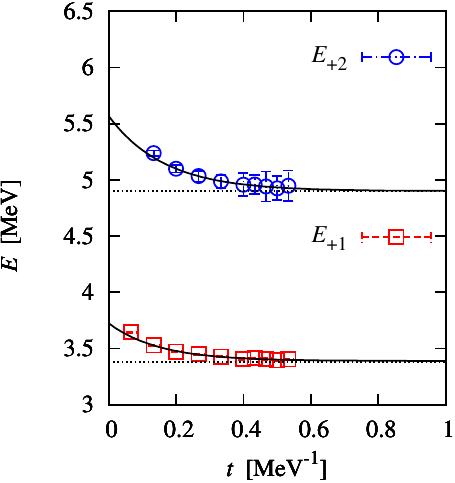}}
\end{center}
\caption{The ground state energy is shown versus projection time $t$ using either one or four initial/final states in Panel (a) and the first two excited state energies with even parity in Panel (b).  For comparison we show the exact lattice energies as dotted horizontal lines. }
\label{fig:Energy1S}
\end{figure}

In Fig.~\ref{fig:Energy1S}(a) we compare the ground state energies  using $[\hat{M}^{a}(t)]_{1\times1}$ and $[\hat{M}^{a}(t)]_{4\times4}$ adiabatic matrices.
We are plotting the energies versus projection time $t=L_ta_t$.  The results shown are obtained using a lattice box of length $L\,a = 13.79$ fm, while the number of time steps is varied over a range of values to extrapolate to  the limit $t \to \infty$.  We use a simple exponential ansatz to extrapolate away the residual contribution from higher-energy states,\begin{equation}
E_i(t)=E_i(\infty)+c_ie^{-\Delta E_it}+\cdots. \label{eqn:error}
\end{equation}As can be seen clearly in the figure, the $[\hat{M}^{a}(t)]_{4\times4}$ results converge with a significantly faster exponential decay than the $[\hat{M}^{a}(t)]_{1\times1}$ results.  This is consistent with the derivation in Ref.~\cite{Pine:2013zja} that the energy gap $\Delta E_i$ in Eq.~(\ref{eqn:error}) is increased by including more initial states. The corresponding extrapolated ground state energies obtained from the $[\hat{M}^{a}(t)]_{1\times1}$ and $[\hat{M}^{a}(t)]_{4\times4}$ adiabatic matrices are $-1.1918(46)$ MeV and  $-1.1916(25)$ MeV, respectively.
\begin{table}[h]
\caption{The exact and Monte Carlo results for the ground state and lowest lying even-parity energies in a periodic box of length $La=13.79$ fm. The Monte Carlo results are obtained from the $[\hat{M}^{a}(t)]_{4\times4}$ adiabatic matrix.}
\label{table:EvenParityEnergies}
\centering
\begin{tabular}{l|l|l|l}
\hline\hline
         & $E_{+0}$ [MeV] & $E_{+1}$ [MeV] & $E_{+2}$ [MeV] \\ 
\hline
${\rm{Exact}}$  & $-1.1904$ &  $3.3828$ & $4.9024$ \\
$\rm{MC}$       &  $-1.1916(25)$& $3.3905(82)$  &  $4.9012(15)$\\
\hline\hline
\end{tabular}
\end{table}

In Figs.~\ref{fig:Energy1S}(a) and (b) we plot the lowest lying even-parity energies as a function of Euclidean projection time $t$.  To be able to calculate the two excited states in  Fig.~\ref{fig:Energy1S}(b) we use seven inital/final states and construct a $7\times7$ adiabatic matrix. We then use symmetry under cubic rotations to reduce the $7\times7$ adiabatic matrix to a $4\times4$ adiabatic matrix. For comparison the horizontal dotted lines in the plots represent the exact lattice energies obtained from the Lanczos iteration method.  The solid lines are exponential fits to the data using the ansatz in Eq.~(\ref{eqn:error}).  As seen from Fig.~\ref{fig:Energy1S} and the corresponding extrapolated energies in Table~\ref{table:EvenParityEnergies}, we find that the calculations using adiabatic projection with impurity Monte Carlo are in excellent agreement with the exact lattice results.

\begin{figure}[hptb]
\begin{center}
{\includegraphics[width=80mm]{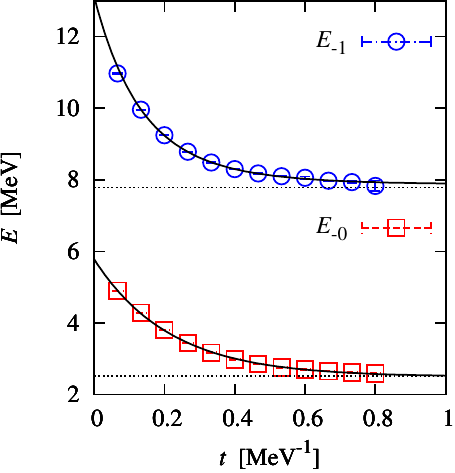}}
\end{center}
\caption{ The lowest two odd parity energies as a function of Euclidean projection time $t$.  For comparison we show the exact lattice energies as dotted horizontal lines. }
\label{fig:EnergyOdd}
\end{figure}
In Fig.~\ref{fig:EnergyOdd} we present the energies for the lowest two states with odd parity. In order to calculate these odd parity energies we use five initial/final states and construct a $5\times5$ adiabatic matrix. The horizontal dotted lines  represent the exact lattice energies obtain from Lanczos iteration, and the solid lines are the exponential extrapolation fits.  We see from Fig.~\ref{fig:EnergyOdd} and Table~\ref{table:OddParityEnergies} again that the calculations using adiabatic projection with impurity Monte Carlo are in excellent agreement with the exact lattice results.   Both energy levels shown in Fig.~\ref{fig:EnergyOdd} have three-fold degeneracy. The degeneracies of these energy levels are not shown here.

\begin{table}[h]
\caption{The exact and Monte Carlo results for the energies of the lowest two odd-parity states in a periodic box of length $La=13.79$ fm. The Monte Carlo results are obtained from the $[\hat{M}^{a}(t)]_{5\times5}$ adiabatic matrix.}
\label{table:OddParityEnergies}
\centering
\begin{tabular}{l|l|l}
\hline\hline
         & $E_{-0}$ [MeV] & $E_{-1}$ [MeV] \\ 
\hline
${\rm{Exact}}$  & $2.509$ &  $7.784$     \\
$\rm{MC}$       &  $2.519(25)$& $7.869(93)$ \\
\hline\hline
\end{tabular}
\end{table}

\section{Composite particles in finite volume}
In this section we present lattice results for the fermion-dimer elastic scattering phase shifts for angular momentum up to $\ell = 2$ using L\"uscher's finite-volume method.  As background for explaining the finite-volume calculations, we first briefly review L\"uscher's method  for the $s$-wave scattering of two particles in Section~\ref{intro:Luescher-method-001}. In the following, we extend the previous discussion to higher partial waves $\ell\leq2$. Since the phase shifts depend crucially on an accurate calculation and analysis of finite-volume energy levels, we also discuss in the appendix some corrections which are due to modifications of the dimer binding energy at finite volume. 

\subsection{L\"{u}scher's finite-volume method}
\label{1407.2784-Luescher-method}

L\"{u}scher~\cite{Luscher1986105,Luscher1991531} is a well-known technique for extracting elastic phase shifts for two-body scattering from the volume dependence of two-body continuum states in a cubic periodic box. L\"{u}scher's relation between scattering phase shifts and two-body energy levels in a cubic periodic box has the following forms~\cite{Luscher1986105,Luscher1991531,PhysRevD.83.114508}
\begin{equation}
p^{2\ell+1} \cot\delta_{\ell}(p)
=
\begin{dcases}
 \frac{2}{\sqrt{\pi}L}
 \mathcal{Z}_{0,0}(1;\eta) & \quad \text{for $\ell = 0$} \,,
\\ 
 \left(\frac{2\pi}{L}\right)^{3} \frac{\eta}{\pi^{3/2}}
 \mathcal{Z}_{0,0}(1;\eta) & \quad \text{for $\ell = 1$} \,,
\\ 
 \left(\frac{2\pi}{L}\right)^{5} \frac{1}{\pi^{3/2}}
 \left[
 \eta^{2}  \mathcal{Z}_{0,0}(1;\eta)
 +
 \frac{6}{7} \mathcal{Z}_{4,0}(1;\eta)
 \right] & \quad \text{for $\ell = 2$}  \,.
\end{dcases} 
\label{eqn:phaseshift-005}
\end{equation}
where
\begin{align}
\eta = \left(\frac{L p}{2\pi}\right)^{2}\,.
\end{align}
Here $\mathcal{Z}_{\ell,m}(1;\eta)$ are the generalized zeta functions~\cite{Luscher1986105,Luscher1991531},
\begin{align}
\mathcal{Z}_{\ell,m}(1;\eta)
=
\sum_{\vec{n}}
\frac
{|\vec{n}|^{\ell} \, \text{Y}_{\ell,m}(\hat{n})}
{|\vec{n}|^{2} - \eta}
\,,
\label{eqn:phaseshift-009}
\end{align}
and $Y_{\ell,m}(\hat{n})$ are the spherical harmonics. We can evaluate the zeta functions using exponentially-accelerated expressions \cite{PhysRevD.83.114508}.  For $\ell,m= 0$ we have
\begin{align}
\mathcal{Z}_{0,0}(1;\eta)
=
\pi e^{\eta}(2\eta -1) 
&+ \frac{e^{\eta}}{2\sqrt{\pi}}\sum_{\vec{n}}
\frac{e^{-|\vec{n}|^{2}}}{|\vec{n}|^{2}-\eta}
\nonumber\\
&-\frac{\pi}{2}\int_{0}^{1} d\lambda\frac{e^{\lambda \eta}}{\lambda^{3/2}}
\left(4\lambda^{2}\eta^{2} -
 \sum_{\vec{n}}e^{-\pi^{2}|\vec{n}|^{2}/\lambda}\right)
\,,
\label{eqn:phaseshift-031}
\end{align}
 and for arbitrary $\ell$ and $m$,
 \begin{align}
 \mathcal{Z}_{\ell,m}(1;\eta)
 =
\sum_{\vec{n}}
&
\frac
{|\vec{n}|^{\ell} \, \text{Y}_{\ell,m}(\hat{n})}
{|\vec{n}|^{2} - \eta}
e^{-\Lambda(|\vec{n}|^{2} - \eta)} 
\nonumber\\
 &+\int_{0}^{\Lambda} d\lambda
 \left(\frac{\pi}{\lambda}\right)^{\ell+3/2}
 e^{\lambda \eta}
\sum_{\vec{n}}
\frac
{|\vec{n}|^{\ell} \, \text{Y}_{\ell,m}(\hat{n})}
{|\vec{n}|^{2} - \eta}
e^{-\pi^{2}|\vec{n}|^{2}/\lambda}
 \,.
 \label{eqn:phaseshift-035}
 \end{align}

\subsection{Results for the elastic phase shifts}

\captionsetup[subfigure]{position=bottom, labelfont=normalfont,textfont=normalfont,singlelinecheck=off,margin=30pt}
\begin{figure}[hptb]
\begin{center}
{\includegraphics[width=120mm]{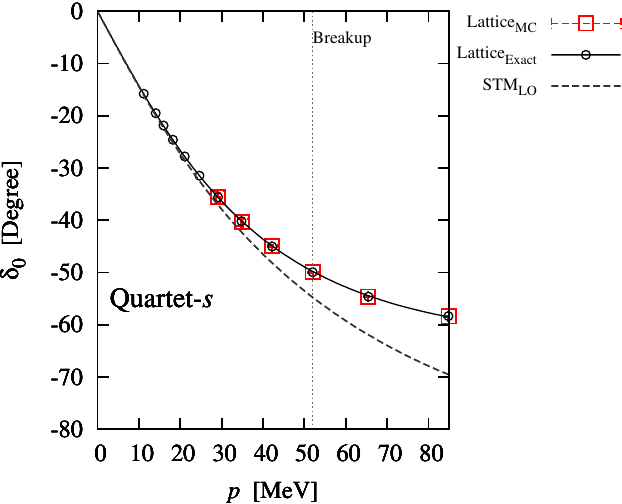}}
\end{center}
\caption{The \textit{s}-wave scattering phase shift versus the relative momentum between fermion and dimer.}
\label{fig:s-wavePhaseshift}
\end{figure}
\captionsetup[subfigure]{position=bottom, labelfont=normalfont,textfont=normalfont,singlelinecheck=off,margin=30pt}
\begin{figure}[hptb]
\begin{center}
{\includegraphics[width=120mm]{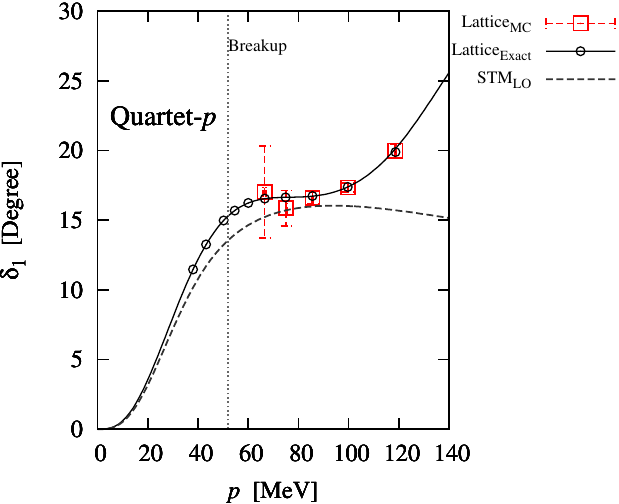}}
\end{center}
\caption{The \textit{p}-wave scattering phase shift versus the relative momentum between fermion and dimer.}
\label{fig:p-wavePhaseshift}
\end{figure}
\captionsetup[subfigure]{position=bottom, labelfont=normalfont,textfont=normalfont,singlelinecheck=off,margin=30pt}
\begin{figure}[hptb]
\begin{center}
{\includegraphics[width=120mm]{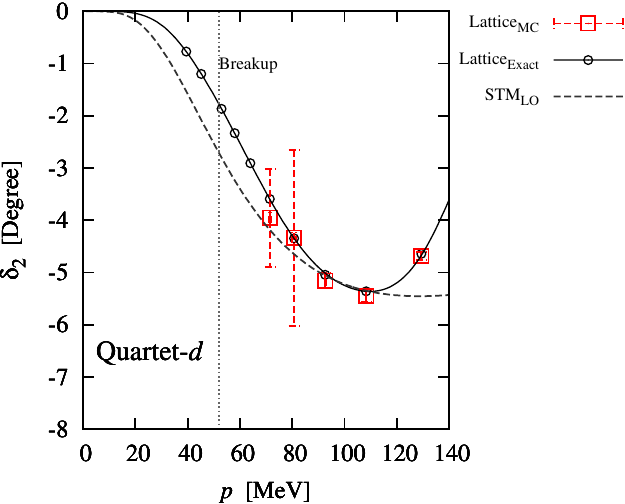}}
\end{center}
\caption{The \textit{d}-wave scattering phase shift versus the relative momentum between fermion and dimer.}
\label{fig:d-wavePhaseshift}
\end{figure}

We now use our lattice results for the finite-volume energies and use Eq.~(\ref{eqn:phaseshift-005})  to determine the elastic phase shifts.  We compute phase shifts using data from the impurity Monte Carlo calculations as well as the exact lattice energies using the Lanczos method.  The fermion-dimer system that we are considering corresponds exactly to neutron-deuteron scattering in the spin-quartet channel at the leading order in pionless effective field theory. Therefore we choose to set the dimer energy to the physical deuteron energy of $-2.2246$ MeV and compare our lattice results to calculations of neutron-deuteron scattering in the continuum and infinite-volume limits at leading order in pionless effective field theory \cite{Bedaque2000357,Gabbiani2000601,Rupak200373}.  The Skorniakov-Ter-Martirosian (STM) integral equation for the $T$-matrix is
\begin{align}
T_{\ell}(k,p) = - & \frac{8 \pi \gamma}{m p k}
\,
Q_{\ell}\left(\frac{p^{2}+k^{2}-mE-i0^{+}}{pk}\right)
\nonumber\\
&
-\frac{2}{\pi}\int_{0}^{\infty} dq \, \frac{q}{p} \frac{T_{\ell}(k,q)}{\sqrt{3q^{2}/4-mE-i0^{+}}-\gamma}
\,
Q_{\ell}\left(\frac{p^{2}+q^{2}-mE-i0^{+}}{pq}\right),
\end{align}
where $\gamma$ is the dimer binding energy, $E = 3p^{2}/(4m)-\gamma^{2}/m$ is the total energy, and $Q_{\ell}$ is the Legendre function of the second kind,
\begin{align}
Q_{\ell}(a) = \frac{1}{2}\int_{-1}^{1}
dx \, \frac{P_{\ell}(x)}{x+a}\,.
\end{align}
The scattering phase shifts can be calculated from the on-shell $T$-matrix formula,
\begin{align}
T_{\ell}(p,p) = \frac{3\pi}{m} \frac{p^{2\ell}}{p^{2\ell+1} \cot\delta_{\ell} - ip^{2\ell+1}}\,.
\end{align}

We show results for the $s$-wave, $p$-wave and $d$-wave phase shifts in Fig.~\ref{fig:s-wavePhaseshift}, \ref{fig:p-wavePhaseshift} and \ref{fig:d-wavePhaseshift} respectively.  The square points  indicate the data from the lattice Monte Carlo simulations, the circular points are the exact lattice calculations, and the solid lines are a fit of the exact lattice data using an effective range expansion, 
\begin{align}
  p^{2\ell+1} \cot\delta_{\ell}(p)
  = -\frac{1}{a_{\ell}}+\frac{1}{2}r_{\ell} \,  p^{2}+\mathcal{O}(p^{4}) \,.
 \label{eqn:append:phaseshift-001}
\end{align}
The dashed lines are leading order results from the STM calculation.  The dotted vertical lines indicate the inelastic breakup threshold of the dimer. The range of lattice box sizes is $L \leq 16$ for the exact lattice and $L \leq 9$ for the Monte Carlo calculations.

Where we have overlapping data, we find excellent agreement between the Monte Carlo and exact lattice phase shifts.  At very low energies we find that Monte Carlo calculations of the phase shifts become impractical due to the high sensitivity of L{\"u}scher's method upon small deviations in the finite-volume energies.  This should be regarded more as a limitation of L{\"u}scher's formalism rather than a deficiency of the adiabatic projection method or impurity Monte Carlo.

We also find quite good agreement between the STM continuum results and the lattice results.  Below the inelastic breakup theshold, the small deviation can be regarded as an estimate of lattice discretization errors.  Above the breakup threshold there are also systematic errors since our analysis using L\"uscher's method does not account for inelastic breakup processes.  Nevertheless we see that the agreement with the STM calculation  for the elastic phase shifts is still quite good, and the STM\ calculation does correctly account for breakup effects.  To our knowledge these results are the first lattice calculations of fermion-dimer scattering in the $p$- and $d$-wave channels. 

\section{Summary and discussion}

In this chapter we have presented the adiabatic projection method and its first application using Monte Carlo methods.  The adiabatic method is a general framework for studying scattering and reactions on the lattice. The method constructs a low-energy effective theory for clusters, and in the limit of large Euclidean projection time the description becomes exact. In previous studies~\cite{Rupak:2013aue,Pine:2013zja} the initial cluster states were parameterized by the initial spatial separations between clusters. In this study we have used a new technique which parameterizes the cluster states according to the relative momentum between clusters. This new approach is crucial for doing calculations with a small number of initial states in order to improve the efficiency of the Monte Carlo calculations.
 The system we have analyzed in detail here is fermion-dimer elastic scattering for two-component fermions interacting via zero-range attractive interactions.

For our calculations we have introduced a new Monte Carlo algorithm which we call impurity lattice Monte Carlo. This can be seen as a hybrid algorithm in between worldline and auxiliary-field Monte Carlo simulations. In impurity Monte Carlo we use worldline Monte Carlo for the impurities, and these impurity worldlines are acting as additional auxiliary fields in the simulation of the other particles.  By using the impurity lattice Monte Carlo algorithm, we have found significant improvement over more standard auxiliary-field Monte Carlo  calculations.  In addition to greater speed and efficiency of the calculations, we also found a reduction of fermonic sign oscillations, and this has greatly improved the resulting accuracy.

We have found that the adiabatic projection method with impurity Monte Carlo enables highly accurate calculations of the finite-volume energy levels of the fermion-dimer system. From these energy levels we have used L\"uscher's method to present the first lattice calculations of $p$-wave and $d$-wave phase shifts for fermion-dimer elastic scattering. In addition to finding excellent agreement between Monte Carlo and exact lattice phase shifts, we have also found good agreement with continuum STM caculations of neutron-deuteron elastic scattering in the spin-quartet channel at leading order in pionless effective field theory. 

Our results show that the adiabatic projection method with Monte Carlo simulations is a viable approach to calculating elastic phase shifts.  The method can be applied in a straightforward manner to other two-cluster scattering systems.  One area where more work is needed is that our application of L\"uscher's method does not account for inelastic breakup processes.  Another area that needs improvement is that L\"uscher's method has too much sensitivity to small changes
in the finite-volume energy levels.  For these reasons we are now working to develop new methods which incorporates more information from the adiabatic projection wavefunction in order to extract scattering information in a more robust manner.

%% file: Appendix-A.tex
\chapter{}

\section{Bessel and related Functions}

\label{append:Bessel-functions}

$S_{\ell}(r)$ and $C_{\ell}(r)$ are Riccati-Bessel and Riccati-Neumann functions, which are defined in terms of the Bessel and Neumann
functions as
\begin{align}
S_{\ell}(x)
=
&
\sqrt{\frac{\pi x}{2}}J_{\ell+\frac{1}{2}%
}(x)
=
\sqrt{\pi}\left(\frac{x}{2}\right)^{\ell+1}\sum_{n=0}^{\infty}\frac{i^{2n}%
}{\Gamma(n+1)\Gamma(n+\ell+\frac{3}{2})}\left(\frac{x}{2}\right)^{2n}, \label{eqn:append-SLpr001}%
\end{align}
\begin{align}
C_{\ell}(x)
=
&
-\sqrt{\frac{\pi r}{2}}N_{\ell+\frac{1}{2}%
}(x)\nonumber\\
=
&
\frac{1}{\sqrt{\pi}}
\left(\frac{x}{2}\right)^{-\ell}
\Gamma(-\ell+\frac{1}{2})
\Gamma(\ell+\frac{1}{2})\sum_{n=0}^{\infty}\frac{i^{2n}}{\Gamma
(n+1)\Gamma(n-\ell+1/2)}\left(\frac{x}{2}\right)^{2n}.
 \label{eqn:append-CLpr001}
 \end{align}
We define the following functions $s_{\ell}(p,r)$ and $c_{\ell}(p,r)$ in terms of Riccati-Bessel and Riccati-Neumann functions,
\begin{align}
s_{\ell}(p,r) = p^{-\ell-1}S_{\ell}(pr)
= \sum_{n = 0}^{\infty}
\frac{\sqrt{\pi} \, i^{2n} \, p^{2n}}
{\Gamma\left(n+1\right)
\Gamma\left(n+\ell+\frac{3}{2}\right)}
\left( \frac{r}{2}\right)^{2n+\ell+1}\,,
\label{eqn:append-SLpr003}
\end{align}
\begin{align}
c_{\ell}(p,r)  &  =p^{\ell}C_{\ell}(pr)
=
\sum_{n=0}^{\infty}
\frac{i^{2n} \, p^{2n}}{\sqrt{\pi}}
\frac{\Gamma(-\ell+\frac{1}{2})
\Gamma(\ell+\frac{1}{2})}{\Gamma
(n+1)\Gamma(n-\ell+1/2)}\left(\frac{r}{2}\right)^{2n-\ell}.
\label{eqn:append-CLpr003}%
\end{align}
The relations in \erefs{eqn:append-SLpr003}{eqn:append-CLpr003} with \erefs{eqn:append-SLpr001}{eqn:append-CLpr001} indicate that $s_{\ell}(p,r)$ and $c_{\ell}(p,r)$ can be written in powers of $p^{2}$,
\begin{align}
s_{\ell}(p,r) =
\sum_{n = 0}^{\infty}
s_{2n,\ell}(r) \, p^{2n+2}
\quad
\text{and}
\quad
c_{\ell}(p,r)   =
\sum_{n = 0}^{\infty}
c_{2n,\ell}(r) \, p^{2n+2} \,,
 \label{uncoup-neut-SCLkr001}
 \end{align}
where $s_{n,\ell}(r)$ and $c_{n,\ell}(r)$ are 
\begin{align}
s_{n,\ell}(r)= 
\frac{\sqrt{\pi} \, i^{2n} }
{\Gamma\left(n+1\right)
\Gamma\left(n+\ell+\frac{3}{2}\right)}
\left( \frac{r}{2}\right)^{2n+\ell+1}\,,
\label{eqn:append-SLpr005}
\end{align}
\begin{align}
c_{n,\ell}(r) =
\frac{i^{2n} }{\sqrt{\pi}}
\frac{\Gamma\left(-\ell+\frac{1}{2}\right)
\Gamma\left(\ell+\frac{1}{2}\right)}{\Gamma
(n+1)\Gamma\left(n-\ell+\frac{1}{2}\right)}\left(\frac{r}{2}\right)^{2n-\ell}.
\label{eqn:append-CLpr005}%
\end{align}

\section{Coulomb wave functions}
\label{append:CoulombSolutions}

We consider radial wave function $V_{\ell}^{(p)}(r)$ that satisfies the radial Schr\"{o}dinger equation for $r > R$,
\begin{align}
 \left[\frac{d^{2}}{dr^{2}}
 -\frac{\ell(\ell+1)}{r^{2}}
 -\frac{\gamma}{r}
 +p^{2}\right] V_{\ell}^{(p)}(r)
  =0
 \,,
 \label{eqn:Coulm:uncoup-radial-010}%
\end{align}
where $\mu$ is the reduced mass, and $\gamma = 2 \mu \alpha Z_{1}Z_{2}$. Defining $ r_{\gamma} =- \gamma r/2 $ and $\epsilon = 4p^{2}/
\gamma^{2}$ Eq.~(\ref{eqn:Coulm:uncoup-radial-010}) can be written as
\begin{align}
 \left[\frac{d^{2}}{dr_{\gamma}^{2}}
 -\frac{\ell(\ell+1)}{r_{\gamma}^{2}}
 +\frac{2}{r_{\gamma}}
 +\epsilon\right] \chi
  =0
 \,,
 \label{eqn:Coulm:uncoup-radial-030}%
\end{align}
The solution to this differential equation has the form~\cite{Seaton2002225} of
\begin{align}
\chi(\epsilon,\ell;r_{\gamma}) = \sum_{q = 0}^{\infty} \epsilon^{q} 
\left[
\sum_{p = 0}^{q} \alpha_{q-p}
\,
\chi_{q} (\ell,r_{\gamma})
\right] \,,
\label{eqn:append:Seaton-XeLr-010}
\end{align}
as linearly independent functions $\chi_{q}(\ell,r_{\gamma})$,
\begin{equation}
\chi_{q}(\ell,r_{\gamma})
=
\begin{dcases}
\sum_{p = 2q}^{3q} C_{q,p} \, \phi_{p}(\ell,r_{\gamma}) & \qquad q > 1\,,
\\
-\frac{(\ell+1)}{4} \phi_{2} + \frac{1}{12} \phi_{3} & \qquad q = 1\,,
\\
\phi_{0}(\ell,r_{\gamma}) & \qquad q = 0\,,
 \end{dcases} 
\label{eqn:append:Seaton-X-020}
\end{equation}
where
\begin{align}
\phi_{p} (\ell,r_{\gamma})
=
a \, P_{p}(\ell,r_{\gamma}) + b \, Q_{p}(\ell,r_{\gamma}),
\label{eqn:append:Seaton-phi-020}
\end{align}
and
\begin{equation}
C_{q,p}
=
\begin{dcases}
 \frac{-(2\ell+p)C_{q-1,p-2} + C_{q-1,p-3}}{4p}
 & \quad \text{for $2q \leq p \leq 3q$}, \\ 
 1 & \quad \text{for $q = p = 0 $}, \\ 
 0 & \quad \text{for $2q > p$ and $p > 3q$.}  
\end{dcases}
\label{eqn:append:Seaton-Cqp-010}
\end{equation}
The coefficient $\alpha_{p}$ in Eq.~(\ref{eqn:append:Seaton-X-020}) and the constants $a$ and $b$ in Eq.~(\ref{eqn:append:Seaton-phi-020}) are to be determined depending on the choice of normalization of the function $\chi(\epsilon,\ell;r_{\gamma})$. The functions $P_{p}(\ell,r_{\gamma})$ and $Q_{p}(\ell,r_{\gamma})$ are defined in terms of Bessel and Modified Bessel functions 
\begin{equation}
P_{p}(\ell,r_{\gamma})
=
\begin{dcases}
 (2r_{\gamma})^{(p+1)/2} J_{2\ell+1+p} (2\sqrt{2r_{\gamma}})
 & \quad \text{for $r_{\gamma} > 0$}, \\ 
  (-1)^{\ell+1+p} (-2r_{\gamma})^{(p+1)/2} I_{2\ell+1+p} (\sqrt{-8r_{\gamma}})
 & \quad \text{for $r_{\gamma} < 0$}, 
\end{dcases}
\label{eqn:append:Seaton-Pp-010}
\end{equation}
and
\begin{equation}
Q_{p}(\ell,r_{\gamma})
=
\begin{dcases}
 (2r_{\gamma})^{(p+1)/2} Y_{2\ell+1+p} (2\sqrt{2r_{\gamma}})
 & \quad \text{for $r_{\gamma} > 0$}, \\ 
  \frac{2}{\pi}(-1)^{\ell+1+p} (-2r_{\gamma})^{(p+1)/2} K_{2\ell+1+p} (\sqrt{-8r_{\gamma}})
 & \quad \text{for $r_{\gamma} < 0$}.
\end{dcases}
\label{eqn:append:Seaton-Qp-010}
\end{equation}

\subsection{Regular solution $\quad$
$f(\epsilon,\ell;r_{\gamma})$}
Here the normalization of the Coulomb wave functions is chosen as the same as that of Ref.~\cite{PhysRevA.30.1279}. From the functions introduced in the previous section, the regular Coulomb wave function is written as a convergent expansion in $\epsilon$, 
\begin{align}
f(\epsilon,\ell;r_{\gamma})
=
 \sum_{q = 0}^{\infty} \epsilon^{q} 
\left[
\sum_{p = 0}^{q} \alpha_{q-p}
\,
\chi^{(f)}_{p} (\ell,r_{\gamma})
\right] \,,
\label{eqn:append:Seaton-fLr-020}
\end{align}
and this function is equivalent to $f_{\ell}(p,r)$ in Eq.~(\ref{uncoup-charg-Vfunc-005}). Comparing to the regular function of Boll\'{e} and Gesztesy~\cite{PhysRevA.30.1279}, we find $a = 1$, $b = 0$ and $\alpha_{q}$ as
\begin{equation}
\alpha_{q}
=
\begin{dcases}
 \frac{(2\ell+1)!}{(-\gamma)^{\ell+1}} 
 & \quad \text{for $q = 0$ and $r_{\gamma} > 0$}\,, \\ 
 (-1)^{-\ell}\frac{(2\ell+1)!}{(-\gamma)^{\ell+1}} 
 & \quad \text{for $q = 0$ and $r_{\gamma} < 0$}\,, \\
 0 & \quad  \text{for $q \geq 1$}\,.
\end{dcases} 
\label{eqn:append:Seaton-a-f-010}
\end{equation}

We find the first few functions of the expansion of $f_{\ell}(p,r)$ in power of $p^{2}$ that we use in Eq.~(\ref{uncoup-charg-V0})--(\ref{uncoup-charg-V6}) are
\begin{align}
f^{\pm}_{0,\ell} (r) 
=
\frac{(2\ell+1)!}{(\pm\gamma)^{\ell+1/2}}
\,
\sqrt{r}
\,
\mathcal{J}^{(\pm)}_{2\ell+1}\left(2\sqrt{\pm\gamma\, r}\right)
 \,,
\label{eqn:append:Seaton-fLr-050}
\end{align}
\begin{align}
f^{\pm}_{2,\ell} (r) 
=-
\frac{(2\ell+1)!}{3}
\frac{\sqrt{r^{3}}}{(\pm\gamma)^{\ell+3/2}}
\left[3(\ell+1)
\mathcal{J}^{(\pm)}_{2\ell+3}\left(2\sqrt{\pm\gamma\, r}\right)
\pm \sqrt{\pm\gamma \, r}\,\mathcal{J}^{(\pm)}_{2\ell+4}\left(2\sqrt{\pm\gamma\, r}\right)
\right]
  \,,
\label{eqn:append:Seaton-fLr-130}
\end{align}
\begin{align}
f^{\pm}_{4,\ell} (r) 
=
\frac{(2\ell+1)!}{90}
\frac{\sqrt{r^{5}}}{(\pm\gamma)^{\ell+5/2}}
&
\left[
5(r\,\gamma+9\ell^{2}+27\ell+18)
\mathcal{J}^{(\pm)}_{2\ell+5}\left(2\sqrt{\pm\gamma\, r}\right)
\right.
\nonumber\\
&
\left.
\pm
(20\ell+18)\sqrt{\pm \gamma\,r}\,\mathcal{J}^{(\pm)}_{2\ell+6}\left(2\sqrt{\pm\gamma\, r}\right)
\right]
  \,,
\label{eqn:append:Seaton-fLr-140}
\end{align}
\begin{align}
f^{\pm}_{6,\ell} (r) 
=
-
&
\frac{(2\ell+1)!}{5670}
\frac{\sqrt{r^{7}}}{(\pm\gamma)^{\ell+7/2}}
\nonumber\\
&
\left\{
7 \left[135 \ell^{3}+810 \ell^{2}+5\ell (7 r\,\gamma+297)+54 (r\,\gamma+15)\right]
\mathcal{J}^{(\pm)}_{2\ell+7}\left(2\sqrt{\pm\gamma\,r}\right)
\right.
\nonumber\\
&
\left.
\pm
\sqrt{\pm\gamma\,r}\left(455\ell^{2}+1253\ell+35r\,\gamma+810\right)
\,\mathcal{J}^{(\pm)}_{2\ell+8}\left(2\sqrt{\pm\gamma\,r}\right)
\right\}
  \,,
\label{eqn:append:Seaton-fLr-160}
\end{align}
where $\mathcal{J}^{(-)}_{n}\left(x\right)$ is the Bessel function of the first kind $J_{n}\left(x\right)$ and $\mathcal{J}^{(+)}_{n}\left(x\right)$ is the modified Bessel fucntion of the first kind $I_{n}\left(x\right)$.

\subsection{Irregular solution.I $\quad$ $h(\epsilon,\ell;r_{\gamma})$}
The first irregular Coulomb wave function is obtained setting $a = 0$ and $b = 1$ in Eq.~(\ref{eqn:append:Seaton-phi-020}),
\begin{align}
h(\epsilon,\ell;r_{\gamma}) 
=
A(\epsilon,\ell)
\left[
\sum_{q = 0}^{M} \epsilon^{q} 
\chi^{(h)}_{q} (\ell,r_{\gamma})
+ \mathcal{O}\left(\epsilon^{M+1}\right)
\right]
 \,,
\label{eqn:append:Seaton-hHLr-010}
\end{align}
where
\begin{equation}
\chi_{q}^{(h)}(\ell,r_{\gamma})
=
\begin{dcases}
\sum_{p = 2q}^{3q} C_{q,p} \, Q_{p}(\ell,r_{\gamma}) & \qquad q > 1\,,
\\
-\frac{(\ell+1)}{4} Q_{2} + \frac{1}{12} Q_{3} & \qquad q = 1\,,
\\
Q_{0}(\ell,r_{\gamma}) & \qquad q = 0\,,
 \end{dcases}
\label{eqn:append:Seaton-hHLr-020}
\end{equation}
and
\begin{align}
A(\epsilon,\ell)
=
\prod_{p = 0}^{\ell} ({1+p^{2} \epsilon})
=
\sum_{n = 0}^{\ell} 
\epsilon^{n} \, \sigma_{n,\ell}
 \,.
\label{eqn:append:Seaton-AeL-020}
\end{align}
However, this irregular solution $h(\epsilon,\ell;r_{\gamma})$ is not a desired function since it is not analytic in $\epsilon$.

\subsection{Irregular solution.II $\quad$ $g(\epsilon,\ell;r_{\gamma})$}
The second irregular Coulomb wave function which is analytic in $\epsilon$ is defined as a linear combination of $h(\epsilon,\ell;r_{\gamma})$ and $f(\epsilon,\ell;r_{\gamma})$,
\begin{align}
g(\epsilon,\ell;r_{\gamma}) 
= 
-h(\epsilon,\ell;r_{\gamma}) 
-A(\epsilon,\ell)B(\epsilon,0) f(\epsilon,\ell;r_{\gamma}) \,,
\label{eqn:append:Seaton-gLr-010}
\end{align}
where
\begin{align}
B(\epsilon,\ell)
&
=
\frac{1}{2\pi}\psi\left(\frac{i}{\sqrt{\epsilon}}+\ell+1\right)
+\frac{1}{2\pi}\psi\left(\frac{i}{\sqrt{\epsilon}}-\ell\right)
-\frac{1}{\pi}\log\left(\frac{i}{\sqrt{\epsilon}}\right)
-\frac{i}{\exp\left(\frac{2\pi}{\sqrt{\epsilon}}\right)-1}
\\
&
=
\frac{\epsilon}{\pi}
\left\{
\sum_{p = 0}^{\ell} 
\frac{p}{1+p^{2} \epsilon}
+\frac{1}{12}
\left(1+\frac{\epsilon}{10}
+\frac{\epsilon^{2}}{21}
+\frac{\epsilon^{3}}{20}
+\mathcal{O}(\epsilon^{4})
\right)
\right\}
 \,.
\label{eqn:append:Seaton-BeL-020}
\end{align}
In Eq.~(\ref{eqn:append:Seaton-gLr-010}) $B(\epsilon,\ell = 0)$ is set according to the normalization which gives 
$\tilde{g}_{\ell}(p,r) = g(\epsilon,\ell;r_{\gamma}) $. Let us define
\begin{align}
A(\epsilon,\ell)B(\epsilon,0) 
=
\sum_{n = 1}^{N} 
 \epsilon^{n} \, \omega_{n,\ell}
 +\mathcal{O}(\epsilon^{N+1}) \,,
\label{eqn:append:Seaton-ABeL-010}
\end{align}
then the convergent expression is written as
\begin{align}
g(\epsilon,\ell;r_{\gamma}) 
= \sum_{q = 0}^{\infty} \epsilon^{q} 
\left[
\sum_{p = 0}^{q} \beta_{q-p}
\,
\chi^{(g)}_{p} (\ell,r_{\gamma})
\right] \,,
\label{eqn:append:Seaton-gLr-030}
\end{align}
where
\begin{equation}
\chi^{(g)}_{p} (\ell,r_{\gamma})
=
\begin{dcases}
 \sigma_{0,\ell} \, \chi^{(h)}_{q} (\ell,r_{\gamma})
 & \quad \text{for $q = 0$}, \\ 
  \sum_{m = 0}^{min(q,\ell)} \sigma_{m,\ell} \, \chi^{(h)}_{q-m} (\ell,r_{\gamma})
   - \sum_{m = 1}^{q} \omega_{m,\ell} \, \chi^{(f)}_{q-m} (\ell,r_{\gamma})
 & \quad \text{for $q>0$}\,,
\end{dcases}
\label{eqn:append:Seaton-gLr-020}
\end{equation}
and
\begin{equation}
\beta_{i}
=
\begin{dcases}
 -\pi \frac{(-\gamma)^{\ell}}{(2\ell+1)!}
 & \quad \text{for $i = 0$ and $r_{\gamma} > 0$}, \\ 
 -\pi \frac{(-\gamma)^{\ell}}{(2\ell+1)!} 
 & \quad \text{for $i = 0$ and $r_{\gamma} < 0$, $R = -r_{\gamma}$}, \\
 0 & \quad  \text{for $i \geq 1$}.
\end{dcases}
\label{eqn:append:Seaton-a-g-010}
\end{equation}

In the following we give the first few functions of the expansion of $\tilde{g}_{\ell}(p,r)$ in power of $p^{2}$ that we use in Eq.~(\ref{uncoup-charg-V0})--(\ref{uncoup-charg-V6},

\begin{align}
g^{(\pm)}_{0,\ell} (r)
= \pm
\frac{2 \, (\pm\gamma)^{\ell+1/2}}{(2\ell+1)!}
\,\sqrt{r}
\, \mathcal{N}^{(\pm)}_{2\ell+1}(2\sqrt{\pm r\,\gamma})
 \,,
\label{eqn:append:Seaton-gLr00-010}
\end{align}
\begin{align}
g^{(\pm)}_{2,0} (r)
= 
\frac{2r^{2}}{3}\,  \mathcal{N}^{(\pm)}_{2}(2\sqrt{\pm r\,\gamma})
\mp
\frac{\sqrt{r}}{3\,(\pm\gamma)^{3/2}}\, \mathcal{J}^{(\pm)}_{1}(2\sqrt{\pm r\,\gamma})
\,,
\label{eqn:append:Seaton-gLr20-000}
\end{align}
\begin{align}
g^{(\pm)}_{2,1} (r)
= 
&
\frac{r}{18}
\Big[
\pm 2 (r\gamma-4)  \mathcal{N}^{(\pm)}_4\left(2 \sqrt{\pm r \gamma }\right)
\nonumber\\
&
\mp \frac{2 (r\gamma  -12)  \mathcal{N}^{(\pm)}_3\left(2 \sqrt{\pm r \gamma }\right)}{\sqrt{\pm r\gamma}}
\mp\frac{ \,  \mathcal{J}^{(\pm)}_3 \left(2 \sqrt{\pm r \gamma }\right)}{\sqrt{\pm r\gamma}}
\Big]
\,,
\label{eqn:append:Seaton-gLr20-010}
\end{align}
\begin{align}
g^{(\pm)}_{2,2} (r_{\gamma})
= 
\frac{\sqrt{\pm r \gamma }}{360}
\Big[
&
\pm 2 \sqrt{\pm r \gamma } (\gamma  r-12)  \mathcal{N}^{(\pm)}_6\left(2 \sqrt{\pm r \gamma }\right)
\nonumber\\
&
\mp 4 (\gamma  r-30)  \mathcal{N}^{(\pm)}_5 \left(2 \sqrt{\pm r \gamma }\right)
\mp  \mathcal{J}^{(\pm)}_5\left(2 \sqrt{\pm r \gamma }\right)
\Big]
\,,
\label{eqn:append:Seaton-gLr20-020}
\end{align}
\begin{align}
g^{(\pm)}_{4,0} (r)
=
\frac{1}{45 \gamma ^4}
&
\Big\{
\mp 18 (-\gamma r)^3 \mathcal{N}^{(\pm)}_6\left(2 \sqrt{\pm r \gamma }\right)
\nonumber\\
&
\pm 5 (\gamma  r+18) (\pm\gamma  r)^{5/2} \mathcal{N}^{(\pm)}_5\left(2 \sqrt{\pm r \gamma }\right)
\nonumber\\
&
\pm 5 \gamma ^2 r^2 \mathcal{J}^{(\pm)}_2\left(2 \sqrt{\pm r \gamma }\right)
\mp
6 \sqrt{\pm\gamma  r} \mathcal{J}^{(\pm)}_1\left(2 \sqrt{\pm r \gamma }\right)\Big\}
\,,
\label{eqn:append:Seaton-gLr40-000}
\end{align}

\begin{align}
g^{(\pm)}_{4,1} (r)
=
\frac{r^{3/2}}{270 (\pm\gamma)^{3/2}}
 &
 \Big\{
 \frac{\left(5 \gamma ^2 r^2-264\right) \mathcal{J}^{(\pm)}_6\left(2 \sqrt{\pm r \gamma }\right)}{(\pm\gamma  r)^{3/2}}
\nonumber\\
  &
 \pm\frac{6 (\gamma  r (5 \gamma  r-11)-220) \mathcal{J}^{(\pm)}_5\left(2 \sqrt{\pm r \gamma }\right)}{(\gamma  r)^2}
 \nonumber\\
 &
 \pm [\gamma  r (5 \gamma  r+4)-720] \mathcal{N}^{(\pm)}_5\left(2 \sqrt{\pm r \gamma }\right)
   \nonumber\\
  &
 \mp 8 \sqrt{\pm \gamma  r} (\gamma  r-18) \mathcal{N}^{(\pm)}_6\left(2 \sqrt{\pm r \gamma }\right)
 \Big\}
 \,,
\label{eqn:append:Seaton-gLr40-010}
\end{align}
\begin{align}
g^{(\pm)}_{4,2} (r)
=
\frac{r^2}{5400 (\pm\gamma  r)^{3/2}} 
&
\Big\{
\pm 2 (5 \gamma  r-153) \mathcal{J}^{(\pm)}_5\left(2 \sqrt{\pm r \gamma }\right)
\nonumber\\
&
\pm  
\sqrt{\pm \gamma  r}
12 (\gamma  r (\gamma  r+8)-96) K_6\left(2 \sqrt{r \gamma }\right)
\nonumber\\
&
+
\sqrt{\pm \gamma  r}
5 (\gamma  r-12) \mathcal{J}^{(\pm)}_6\left(2 \sqrt{\pm r \gamma }\right)
\nonumber\\
&
\pm (5 \gamma ^3 r^3-108 \gamma ^2 r^2-192 \gamma  r+5760) \mathcal{N}^{(\pm)}_5\left(2 \sqrt{\pm r \gamma }\right)
\Big\}
\,,
\label{eqn:append:Seaton-gLr40-020}
\end{align}
\begin{align}
\footnotesize
g^{(\pm)}_{6,0} (r_{\gamma})
=
\frac{r}{5670 \gamma ^5}
\Big\{
&
10 (\pm\gamma r)^3 (7 \gamma  r+162) \mathcal{N}^{(\pm)}_8\left(2 \sqrt{r \gamma }\right)
\nonumber\\
&
-756 (\gamma  r+15) (\pm\gamma  r)^{5/2} \mathcal{N}^{(\pm)}_7\left(2 \sqrt{\pm r \gamma }\right)
\nonumber\\
&
-
\Big(105 \gamma ^6 r^6+8946 \gamma ^5 r^5+75600 \gamma ^4 r^4
-119520 \gamma ^3 r^3
\nonumber\\
&
-462240 \gamma ^2 r^2+2592000 \gamma  r+7257600
\Big)
\frac{ \mathcal{J}^{(\pm)}_7\left(2 \sqrt{r \gamma }\right)}{(\pm\gamma  r)^{7/2}}
\nonumber\\
&
-
\Big(1008 \gamma ^5 r^5+11088 \gamma ^4 r^4-15120 \gamma ^3 r^3
\nonumber\\
&
-
73440 \gamma ^2 r^2+345600 \gamma  r+1036800\Big)\frac{\mathcal{J}^{(\pm)}_8\left(2 \sqrt{\pm r \gamma }\right)}{\gamma ^3 r^3}
\Big\}
\,,
\label{eqn:append:Seaton-gLr60-000}
\end{align}

\begin{align}
g^{(\pm)}_{6,1} (r_{\gamma})
=
\frac{r^2}{34020 \gamma ^3}
&
\Big\{
-\frac{3749760}{(\pm\gamma  r)^{7/2}}\mathcal{J}^{(\pm)}_7\left(2 \sqrt{\pm r \gamma }\right)
\nonumber\\
&
+4 \sqrt{\pm \gamma  r} [\gamma  r (621-154 \gamma  r)+34020] \mathcal{N}^{(\pm)}_7\left(2 \sqrt{r \gamma }\right)
\nonumber\\
&
\pm 2 \gamma  r [\gamma  r (35 \gamma  r+54)-9720] \mathcal{N}^{(\pm)}_8\left(2 \sqrt{\pm r \gamma }\right)
\nonumber\\
&
-\Big(798 \gamma ^5 r^5-2772 \gamma ^4 r^4-99792 \gamma ^3 r^3
\nonumber\\
& \qquad +44640 \gamma ^2 r^2+535680 \gamma  r\Big)\frac{\mathcal{J}^{(\pm)}_8\left(2 \sqrt{\pm r \gamma }\right)}{(\gamma r)^4}
\nonumber\\
&
-(105 \gamma ^4 r^4+5670 \gamma ^3 r^3-36036 \gamma ^2 r^2
\nonumber\\
& \qquad
-694080 \gamma  r+401760)\frac{\mathcal{J}^{(\pm)}_7\left(2 \sqrt{\pm r \gamma }\right)}{(\pm \gamma  r)^{5/2}}
\Big\}
\,,
\label{eqn:append:Seaton-gLr60-010}
\end{align}
\begin{align}
g^{(\pm)}_{6,2} (r_{\gamma})
= 
\frac{r}{680400 \gamma ^3}
\Big\{
&
-
54 \left(7 \gamma ^3 r^3-294 \gamma ^2 r^2+15280\right)
\frac{ \mathcal{J}^{(\pm)}_8\left(2 \sqrt{\pm\gamma r}\right)}{\gamma r}
\nonumber\\
&
-3 \Big(35 \gamma ^4 r^4+126 \gamma ^3 r^3-38556 \gamma ^2 r^2
\nonumber\\
&
\qquad+45840 \gamma  r+1925280\Big) \frac{\mathcal{J}^{(\pm)}_7\left(2 \sqrt{\pm\gamma r}\right)}{(\pm \gamma r)^{3/2}}
\nonumber\\
&
+12\sqrt{\pm\gamma r} \Big(-28 \gamma ^3 r^3+1475 \gamma ^2 r^2
\nonumber\\
&
\qquad
+720 \gamma  r-181440\Big)  \mathcal{N}^{(\pm)}_7\left(2 \sqrt{\pm\gamma r}\right)
\nonumber\\
&
\pm\gamma  r \Big(70 \gamma ^3 r^3-2076 \gamma ^2 r^2-8640 \gamma  r
\nonumber\\
&
\qquad
+311040\Big) \mathcal{N}^{(\pm)}_8\left(2 \sqrt{\pm\gamma  r}\right)
\Big\}
\,,
\label{eqn:append:Seaton-gLr60-020}
\end{align}
where $\mathcal{N}^{(-)}_{n}\left(x\right)$ stands for $\pi/2$ times the Bessel function of the second kind $ N_{n}\left(x\right)$ and $\mathcal{N}^{(+)}_{n}\left(x\right)$ is the modified Bessel function of the second kind $K_{n}\left(x\right)$.

\section{van der Waals wave functions}

\label{append:analytic_solution_vdWaals_eq}

In this section we derive the van der Waals wave functions $F_{\ell}$ and $G_{\ell}$,
following the steps in Ref.~\cite{PhysRevA.58.1728}.  We first redefine the
radial function as $U_{\ell}(r)=\sqrt{r_{s}}Z(\rho_{s})$. This rearrangement
puts \eref{pure_vdW} into the form of an inhomogeneous Bessel equation,
\begin{equation}
 \mathcal{L}_{\nu_{0}}Z(\rho_{s})=\left[\rho_{s}^{2}\frac{d^{2}}{d\rho_{s}^{2}}
 +\rho_{s}\frac{d}{d\rho_{s}}-\nu_{0}^{2}+\rho_{s}^{2}\right]
 Z(\rho_{s})=-\frac{p_{s}^{2}}{8}\frac{Z(\rho_{s})}{\rho_{s}} \,,
\label{eqn:inhomogeneous_Bessel_equ_1a}
\end{equation}
with
\begin{equation*}
\nu_{0}=\frac{1}{4}(2\ell+1).
\end{equation*}
The idea, introduced in Ref.~\cite{PhysRevA.50.2841}, is now to consider
$Z_{\nu}(\rho_{s})$ as a series expansion of solutions,
\begin{equation}
 Z(\rho_{s})=\sum_{n=0}^{\infty}p_{s}^{2n}\,\varphi^{(n)}(\rho_{s}) \,,
\label{eqn:Z_rho_expansion_1a}
\end{equation}
and to use perturbation theory to obtain a solution for $Z_{\nu}(\rho_{s})$.
Substituting \eref{eqn:Z_rho_expansion_1a} into
\eref{eqn:inhomogeneous_Bessel_equ_1a} leads to an infinite number of
differential equations,
\begin{align}
 \mathcal{L}_{\nu_{0}}\,\varphi^{(0)}(\rho_{s})\,+p_{s}^{2}\,
 &\left[\mathcal{L}_{\nu_{0}}\,\varphi^{(1)}(\rho_{s})\,+\frac{1}{8\rho_{s}}
 \,\varphi^{(0)}(\rho_{s})\right] \nonumber \\
 &+p_{s}^{4}\,\left[\mathcal{L}_{\nu_{0}}\,\varphi^{(2)}(\rho_{s})
 \,+\frac{1}{8\rho_{s}}\,\varphi^{(1)}(\rho_{s})\right] +\cdots = 0 \,.
\label{eqn:inhomogeneous_Bessel_equ_2a}
\end{align}
The zeroth-order differential equation is homogenous, while all other orders
are inhomogeneous. This procedure generates a secular perturbation in all
inhomogeneous differential equations as well as driving terms.  The secular
terms here refer to the solutions of the zeroth-order differential equation,
which are Bessel functions.

Following Ref.~\cite{PhysRevA.58.1728}, we introduce a function
$Z_{\nu}(\rho_{s})$ which has an expansion in terms of Bessel functions with
momentum-dependent coefficients,
\begin{equation}
 Z_{\nu}(\rho_{s})=\sum_{m=-\infty}^{\infty}b_{m}(p_{s})
 \,\mathcal{J}_{\nu+m}(\rho_{s}) \,.
\label{Zv}
\end{equation}
We insert this as an ansatz into \eref{eqn:inhomogeneous_Bessel_equ_1a}
with $\nu$ yet to be determined.  Here $\mathcal{J}_{n}$ denotes collectively
the Bessel and Neumann functions, $J_{n}$ and $N_{n}$.
Substitution of \eref{Zv} into \eref{eqn:inhomogeneous_Bessel_equ_1a}
yields a three-term recurrence relation for the $b_{m}$ functions with
$-\infty<m<\infty$,
\begin{equation}
 \lbrack(\nu+m)^{2}-\nu_{0}^{2}]b_{m}(p_{s})+\frac{p_{s}^{2}}{16(m+\nu-1)}
 b_{m-1}(p_{s})+\frac{p_{s}^{2}}{16(m+\nu+1)}b_{m+1}(p_{s})=0 \,.
\label{recurrence}
\end{equation}
Solving these equations for $b_{m}(p_{s})$ yields
\begin{equation}
 b_{m}(p_{s})=\left(-1\right)^{m}\left(\frac{p_{s}}{4}\right)^{2m}
 \frac{\Gamma(\nu)\Gamma(\nu-\nu_{0}+1)\Gamma(\nu+\nu_{0}+1)}
 {\Gamma(\nu+m)\Gamma(\nu-\nu_{0}+m+1)\Gamma(\nu+\nu_{0}+m+1)}\,c_{m}(\nu)
\label{bm}
\end{equation}
and
\begin{equation}
 b_{-m}(p_{s})=\left(-1\right)^{m}\left(\frac{p_{s}}{4}\right)^{2m}
 \frac{\Gamma(\nu-m+1)\Gamma(\nu-\nu_{0}-m)\Gamma(\nu+\nu_{0}-m)}
 {\Gamma(\nu+1)\Gamma(\nu-\nu_{0})\Gamma(\nu+\nu_{0})}\,c_{m}(-\nu)
\label{bmm}
\end{equation}
for $m\geq0$.  The functions $c_{m}(\pm\nu)$ are defined as
\begin{equation}
 c_{m}(\pm\nu)=\prod_{s=0}^{m-1}Q(\pm\nu+s)\,b_{0}(p_{s}) \,,
\end{equation}
where $Q(\nu)$ is given by
\begin{equation}
 Q(\nu)=\cfrac{1}{1-\cfrac{p_{s}^{2} }{16(\nu+1)[(\nu+1)^2-\nu_o^2]
 (\nu+2)[(\nu+2)^2-\nu_o^2]}Q(\nu+1)} \,.
\label{Q}
\end{equation}
The coefficient $b_{0}(p_{s})$ only determines the overall normalization and
is simply set to one in the following.  \eref{recurrence} for $m=0$
determines the shift $\nu$ in the order of the Bessel functions.  We
determine $\nu$ using the constraint
\begin{equation}
 (\nu^{2}-\nu_{0}^{2})-\frac{Q(-\nu)}{16^{2}\nu(\nu-1)[(\nu-1)^{2}-\nu_{0}^{2}]}
 \ p_{s}^{4}-\frac{Q(\nu)}{16^{2}\nu(\nu+1)
 [(\nu+1)^{2}-\nu_{0}^{2}]}\ p_{s}^{4}=0 \,.
\label{v}
\end{equation}
In general there are several roots which become complex beyond a critical
scaled momentum $p_{s}$, and one must be careful to choose the physical
solution.  For a detailed discussion of this point, see
Refs.~\cite{PhysRevA.58.1728,PhysRevA.80.012702}.

Choosing either $\mathcal{J}_{n}=J_{n}$ or $\mathcal{J}_{n}=N_{n}$ already
yields a pair of linearly independent solutions. However, in order to get a
pair with energy-independent normalization as $r_{s}\rightarrow0$ (which
ensures analyticity in the energy), we furthermore define
\begin{equation}
x_{\ell}(p_{s})=\cos\eta_{\ell}\sum_{m=-\infty}^{\infty}(-1)^{m}b_{2m}(p_{s})
 -\sin\eta_{\ell}\sum_{m=-\infty}^{\infty}(-1)^{m}b_{2m+1}(p_{s})
\label{X_L}
\end{equation}
and
\begin{equation}
 y_{\ell}(p_{s})=\sin\eta_{\ell}\sum_{m=-\infty}^{\infty}(-1)^{m}b_{2m}(p_{s})
 +\cos\eta_{\ell}\sum_{m=-\infty}^{\infty}(-1)^{m}b_{2m+1}(p_{s}) \,,
\label{Y_L}
\end{equation}
with
\begin{equation*}
\eta_{\ell}=\frac{\pi}{2}(\nu-\nu_{0}) \,.
\end{equation*}
Combining everything, we arrive at the van der Waals wave functions,
\begin{equation}
 {F}_{\ell}(p,r)=\frac{r_{s}^{1/2}}{x_{\ell}^{2}(p_{s})+y_{\ell}^{2}(p_{s})}
 \left[x_{\ell}(p_{s})\sum_{m=-\infty}^{\infty}b_{m}(p_{s})
 \,J_{\nu+m}\left(\rho_{s}\right)
 -y_{\ell}(p_{s})\sum_{m=-\infty}^{\infty}b_{m}(p_{s})
 \,N_{\nu+m}\left(\rho_{s}\right)\right] \,,
\end{equation}
\begin{equation}
 {G}_{\ell}(p,r)=\frac{r_{s}^{1/2}}{x_{\ell}^{2}(p_{s})+y_{\ell}^{2}(p_{s})}
 \left[x_{\ell}(p_{s})\sum_{m=-\infty}^{\infty}b_{m}(p_{s})
 \,N_{\nu+m}\left(\rho_{s}\right)+y_{\ell}(p)\sum_{m=-\infty}^{\infty}b_{m}(p_{s})
 \,J_{\nu+m}\left(\rho_{s}\right)\right] \,.
\end{equation}

\section{Low-energy expansions of the function terms in van der Waals wave functions}
\label{low_energy_expansion}

In this appendix we expand all functions relating to the van der Waals
wave functions in powers of momentum.  We first consider $\nu$, the shift in
the order of the Bessel functions in \eref{F_L} and \eref{G_L}.
Using \eref{v} in Appendix~\ref{append:analytic_solution_vdWaals_eq},
we find
\begin{equation}
 \nu=\nu_{0}-\frac{3}{2^{8}\nu_{0}(4\nu_{0}^{2}-1)(\nu_{0}^{2}-1)}p_{s}^{4}
 +{O}(p_{s}^{8}) \,,
\end{equation}
where $\nu_{0}=(2\ell+1)/4$.  Using the expansion in \eref{bm},
\eref{bmm}, and~\eref{Q}, we get
\begin{equation}
 b_{m}(p_{s})=\left(-1\right)^{m}\frac{\Gamma(\nu_{0})\Gamma(2\nu_{0}+1)}
 {m!\,\Gamma(\nu_{0}+m)\Gamma(2\nu_{0}+m+1)}\left(\frac{p_{s}}{4}\right)^{2m}
+{O}(p_{s}^{2m+2})
\end{equation}
and
\begin{equation}
 b_{-m}(p_{s})=\frac{\Gamma(\nu_{0}-m+1)\Gamma(2\nu_{0}-m)}{m!\,\Gamma(\nu_{0}
 +1)\Gamma(2\nu_{0})}\left(\frac{p_{s}}{4}\right)^{2m}+{O}(p_{s}^{2m+2})
\end{equation}
for $m\geq0$.  Substituting these expressions into \eref{X_L}
and~\eref{Y_L} we obtain
\begin{equation}
x_{\ell}(p_{s})=1+{O}(p_{s}^{4}) \,,
\end{equation}
\begin{equation}
 y_{\ell}(p_{s})=-\left[  \frac{\Gamma(\nu_{0})\Gamma(2\nu_{0}-1)}{\Gamma(\nu_{0}
 +1)\Gamma(2\nu_{0})}+\frac{\Gamma(\nu_{0})\Gamma(2\nu_{0}+1)}
 {\Gamma(\nu_{0}+1)\Gamma(2\nu_{0}+2)}\right]\left(\frac{p_{s}}{4}\right)^{2}
 +{O}(p_{s}^{4}) \,.
\end{equation}

%% file: Appendix-B.tex
\chapter{}

\section{Wronskians of the wave functions}
\label{append:Wronskians}

\subsection{Single channel}
\label{append:}

Here we calculate Wronskians of the wave function, $U_{\ell}^{(p)}(r)$, and
Wronskians of combinations of the $s_{n,\ell}(r)$ and $c_{n,\ell}(r)$ functions. \ The Wronskian of $U_{\ell}^{(p)}(r)$ for the non-interacting region $r > R$ is
\begin{align}
W[U_{\ell}^{(p_{a})}(r),U_{\ell}^{(p_{b})}(r)]
=
&
(p_{a}^{2}-p_{b}^{2}) \, W[u_{2,\ell},u_{0,\ell}](r)
\nonumber\\
&
+(p_{a}^{4}-p_{b}^{4}) \,  W[u_{4,\ell},u_{0,\ell}](r)
\nonumber\\
&
+(p_{a}^{4}p_{b}^{2}-p_{b}^{4}p_{a}^{2}) \, W[u_{4,\ell},u_{2,\ell}](r)
\nonumber\\
&
+(p_{a}^{6}-p_{b}^{6}) \,  W[u_{6,\ell},u_{0,\ell}](r)
+\mathcal{O}(p_{a}^{8}+p_{b}^{8}) \,,
\label{eqn:wronsk-001}
\end{align}
where
\begin{align}
W[u_{2,\ell},u_{0,\ell}](r)
 =
\frac{1}{2}r_{\ell}W[s_{0,\ell},c_{0,\ell}](r) + \frac{1}{2}b_{1,\ell}(r)
,
\label{eqn:wronsk-005}
\end{align}
\begin{align}
W[u_{4,\ell},u_{0,\ell}](r)
 =
&
P_{\ell}
W[s_{0,\ell},c_{0,\ell}](r)
+
b_{2,\ell}(r)
\,,
 \label{eqn:wronsk-009}
\end{align}
\begin{align}
W[u_{6,\ell},u_{0,\ell}](r)
 =
 &
Q_{\ell} W[s_{0,\ell},c_{0,\ell}](r)
+ b_{3,\ell}(r)
\,,
 \label{eqn:wronsk-011}
\end{align}
\begin{align}
W[u_{4,\ell},u_{2,\ell}](r)
 = b_{4,\ell}(r)
\,.
 \label{eqn:wronsk-015}
\end{align}
It should be noted that $W[f,g]=-W[g,f]$. The functions $b_{n,\ell}(r)$ are defined in terms of Wronskians of combinations of the $s_{n,\ell}(r)$ and $c_{n,\ell}(r)$ functions by
\begin{align}
b_{1,\ell}(r) =
\frac{2}{a_{\ell}^{2}}W[s_{2,\ell},s_{0,\ell}](r)
 &
 -\frac{2}{a_{\ell}}W[s_{2,\ell},c_{0,\ell}](r)
 \nonumber\\
 &
  - \frac{2}{a_{\ell}}W[c_{2,\ell},s_{0,\ell}](r)+2W[c_{2,\ell},c_{0,\ell}](r)
,
\label{eqn:wronsk-019}
\end{align}
\begin{align}
b_{2,\ell}(r)
 =
-\frac{r_{\ell}}{2a_{\ell}}
&
W[s_{2,\ell},s_{0,\ell}](r)
+\frac{1}{2}r_{\ell}
W[s_{2,\ell},c_{0,\ell}](r)
\nonumber\\
&
+\frac{1}{a_{\ell}^{2}}
W[s_{4,\ell},s_{0,\ell}](r)
-\frac{1}{a_{\ell}}W
[s_{4,\ell},c_{0,\ell}](r)
\nonumber\\
&
\qquad
-\frac{1}{a_{\ell}}
W[c_{4,\ell},s_{0,\ell}](r)
+W[c_{4,\ell},c_{0,\ell}](r)
\,,
 \label{eqn:wronsk-021}
\end{align}
\begin{align}
b_{3,\ell}(r)
=
&
W[c_{6,\ell},c_{0,\ell}](r)
+P_{\ell} W[s_{2,\ell},c_{0,\ell}](r)
+\frac{1}{a_{\ell}^{2}}W[s_{6,\ell},s_{0,\ell}](r)
\nonumber\\
&
-\frac{1}{a_{\ell}}W[c_{6,\ell},s_{0,\ell}](r)
-\frac{1}{a_{\ell}}W[s_{6,\ell},c_{0,\ell}](r)
+\frac{r_{\ell}}{2}W[s_{4,\ell},c_{0,\ell}](r)
\nonumber\\
&
-\frac{P_{\ell}}{a_{\ell}} W[s_{2,\ell},s_{0,\ell}](r)
-\frac{r_{\ell}}{2a_{\ell}}W[s_{4,\ell},s_{0,\ell}](r)
\,,
 \label{eqn:wronsk-025}
\end{align}
\begin{align}
b_{4,\ell}(r)
 =
 &
W[c_{4,\ell},c_{2,\ell}](r)
+\left(
\frac{r_{\ell}^{2}}{4}+\frac{P_{\ell}}{a_{\ell}}
\right)
W[s_{2,\ell},s_{0,\ell}](r)
-\frac{r_{\ell}}{2a_{\ell}}W[s_{4,\ell},s_{0,\ell}](r)
\nonumber\\
&
+\frac{r_{\ell}}{2}W[c_{4,\ell},s_{0,\ell}](r)
+\frac{r_{\ell}}{2}W[s_{2,\ell},c_{2,\ell}](r)
+\frac{1}{a_{\ell}^{2}}W[s_{4,\ell},s_{2,\ell}](r)
\nonumber\\
&
+P_{\ell}W[s_{0,\ell},c_{2,\ell}](r)
-\frac{1}{a_{\ell}}W[c_{4,\ell},s_{2,\ell}](r)
-\frac{1}{a_{\ell}}W[s_{4,\ell},c_{2,\ell}](r)
\,.
 \label{eqn:wronsk-029}
\end{align}

We now calculate Wronskians of all possible combinations of $s_{0}(r)$, $s_{2}(r)$, $c_{0}(r)$ and
$c_{2}(r)$ functions. \ We find
\begin{equation}
W[s_{0,\ell},c_{0,\ell}]=-1, \label{eqn:wronskS0C0}%
\end{equation}%
\begin{equation}
W[s_{0,\ell},c_{2,\ell}](r)=-\frac{r^{2}}{2+4\ell}, \label{eqn:wronskS0C2}%
\end{equation}%
\begin{equation}
W[s_{2,\ell},c_{0,\ell}](r)=\frac{r^{2}}{2+4\ell}, \label{eqn:wronsks2c0}%
\end{equation}%
\begin{equation}
W[s_{2,\ell},c_{2,\ell}](r)=\frac{r^{4}}{16\ell(\ell+1)-12}, \label{eqn:wronsks2c2}%
\end{equation}
\begin{equation}
W[s_{0,\ell},s_{2,\ell}](r)-\frac{\pi}{\Gamma\left(  \frac{3}{2}+\ell\right)
\Gamma\left(  \frac{5}{2}+\ell\right)  }\left(  \frac{r}{2}\right)  ^{3+2\ell},
\label{eqn:wronskS0S2}%
\end{equation}%
\begin{equation}
W[c_{0,\ell},c_{2,\ell}](r)=-\frac{\Gamma\left(  -\frac{1}{2}+\ell\right)
\Gamma\left(  \frac{1}{2}+\ell\right)  }{\pi}\left(  \frac{r}{2}\right)  ^{1-2\ell}.
\label{eqn:wronskc0c2}%
\end{equation}

\subsection{Coupled channels}

\label{append:Wronskiansof functions}

Here we calculate Wronskians of the $U(r)$ and $V(r)$ wave functions. \ Wronskians of $U_{\alpha}(r)$ and
$V_{\alpha}(r)$ for the non-interacting region $r\geq R$ are%
\begin{align}
W[U_{a\alpha}(r),U_{b\alpha}(r)]  &  =(p_{a}^{2}-p_{b}^{2})\Big\{\frac{1}%
{2}r_{j-1}W[s_{0}(r),c_{0}(r)]_{j-1}\nonumber\\
&  +\frac{1}{a_{j-1}^{2}}W[s_{2}(r),s_{0}(r)]_{j-1}+\frac{1}{a_{j-1}}%
W[c_{0}(r),s_{2}(r)]_{j-1}\nonumber\\
&  +\frac{1}{a_{j-1}}W[s_{0}(r),c_{2}(r)]_{j-1}+W[c_{2}(r),c_{0}%
(r)]_{j-1}\Big\}\nonumber\\
&  +\mathcal{O}(p_{a}^{4})+\mathcal{O}(p_{b}^{4}), \label{eqn:wronskUalp}%
\end{align}%
\begin{equation}
W[V_{a\alpha}(r),V_{b\alpha}(r)]=(p_{a}^{2}-p_{b}^{2})q_{0}^{2}W[c_{2}%
(r),c_{0}(r)]_{j+1}+\mathcal{O}(p_{a}^{4})+\mathcal{O}(p_{b}^{4}).
\label{eqn:wronskValp}%
\end{equation}
Wronskians of the $\beta$-state wave functions are
\begin{equation}
W[U_{a\beta}(r),U_{b\beta}(r)]=(p_{a}^{2}-p_{b}^{2})q_{0}^{2}\frac{1}{a_{j+1}^{2}%
}W[s_{2}(r),s_{0}(r)]_{j-1}+\mathcal{O}(p_{a}^{4})+\mathcal{O}(p_{b}^{4}),
\label{eqn:wronskUbet}%
\end{equation}%
\begin{align}
W[V_{a\beta}(r),V_{b\beta}(r)]  &  =(p_{a}^{2}-p_{b}^{2})\Big\{\frac{1}%
{2}r_{j+1}W[s_{0}(r),c_{0}(r)]_{j+1}\nonumber\\
&  +\frac{1}{a_{j+1}^{2}}W[s_{2}(r),s_{0}(r)]_{j+1}+\frac{1}{a_{j+1}}%
W[c_{0}(r),s_{2}(r)]_{j+1}\nonumber\\
&  +\frac{1}{a_{j+1}}W[s_{0}(r),c_{2}(r)]_{j+1}+W[c_{2}(r),c_{0}%
(r)]_{j+1}\Big\}\nonumber\\
&  +\mathcal{O}(p_{a}^{4})+\mathcal{O}(p_{b}^{4}). \label{eqn:wronskVbet}%
\end{align}
Wronskian of the combinations of the $\alpha$ and $\beta$-states are
\begin{align}
W[U_{a\alpha}(r),U_{b\beta}(r)]  &  =q_{0}\frac{1}{a_{j+1}}W[c_{0}(r),s_{0}%
(r)]_{j-1}-p_{a}^{2}\Big\{q_{0}\frac{1}{a_{j-1}a_{j+1}}W[s_{2}(r),s_{0}%
(r)]_{j-1}\nonumber\\
&  -q_{0}\frac{1}{a_{j+1}}W[c_{2}(r),s_{0}(r)]_{j-1}\Big\}+p_{b}^{2}%
\Big\{q_{0}\frac{1}{a_{j-1}a_{j+1}}W[s_{2}(r),s_{0}(r)]_{j-1}\nonumber\\
&  -q_{0}\frac{1}{a_{j+1}}W[s_{2}(r),c_{0}(r)]_{j-1}+q_{0}\frac{r_{j+1}}{2}%
W[s_{0}(r),c_{0}(r)]_{j-1}\nonumber\\
&  -q_{1}\frac{1}{a_{j+1}}W[s_{0}(r),c_{0}(r)]_{j-1}\Big\}+\mathcal{O}%
(p^{4}),\hspace{4.5cm} \label{eqn:wronskUaalpUbbet}%
\end{align}%
\begin{align}
W[U_{a\beta}(r),U_{b\alpha}(r)]  &  =-q_{0}\frac{1}{a_{j+1}}W[c_{0}(r),s_{0}%
(r)]_{j-1}+p_{b}^{2}\Big\{q_{0}\frac{1}{a_{j-1}a_{j+1}}W[s_{2}(r),s_{0}%
(r)]_{j-1}\nonumber\\
&  -q_{0}\frac{1}{a_{j+1}}W[c_{2}(r),s_{0}(r)]_{j-1}\Big\}-p_{a}^{2}%
\Big\{q_{0}\frac{1}{a_{j-1}a_{j+1}}W[s_{2}(r),s_{0}(r)]_{j-1}\nonumber\\
&  -q_{0}\frac{1}{a_{j+1}}W[s_{2}(r),c_{0}(r)]_{j-1}+q_{0}\frac{r_{j+1}}{2}%
W[s_{0}(r),c_{0}(r)]_{j-1}\nonumber\\
&  -q_{1}\frac{1}{a_{j+1}}W[s_{0}(r),c_{0}(r)]_{j-1}\Big\}+\mathcal{O}%
(p^{4}),\hspace{4.5cm} \label{eqn:wronskUabetUbalp}%
\end{align}%
\begin{align}
W[V_{a\alpha}(r),V_{b\beta}(r)]  &  =-q_{0}\frac{1}{a_{j+1}}W[c_{0}(r),s_{0}%
(r)]_{j+1}-{p_{a}^{2}}\Big\{q_{1}\frac{1}{a_{j+1}}W[c_{0}(r),s_{0}%
(r)]_{j+1}\nonumber\\
&  +q_{0}\frac{1}{a_{j+1}}W[c_{2}(r),s_{0}(r)]_{j+1}-q_{0}W[c_{2}(r),c_{0}%
(r)]_{j+1}\Big\}\nonumber\\
&  +{p_{b}^{2}}\Big\{q_{0}\frac{r_{j+1}}{2}W[c_{0}(r),s_{0}(r)]_{j+1}-q_{0}\frac
{1}{a_{j+1}}W[c_{0}(r),s_{2}(r)]_{j+1}\nonumber\\
&  +q_{0}W[c_{0}(r),c_{2}(r)]_{j+1}\Big\}+\mathcal{O}(p^{4}),
\label{eqn:wronskVaalpVbbet}%
\end{align}%
\begin{align}
W[V_{a\beta}(r),V_{b\alpha}(r)]  &  =q_{0}\frac{1}{a_{j+1}}W[c_{0}(r),s_{0}%
(r)]_{j+1}+{p_{b}^{2}}\Big\{q_{1}\frac{1}{a_{j+1}}W[c_{0}(r),s_{0}%
(r)]_{j+1}\nonumber\\
&  +q_{0}\frac{1}{a_{j+1}}W[c_{2}(r),s_{0}(r)]_{j+1}-q_{0}W[c_{2}(r),c_{0}%
(r)]_{j+1}\Big\}\nonumber\\
&  -{p_{a}^{2}}\Big\{q_{0}\frac{r_{j+1}}{2}W[c_{0}(r),s_{0}(r)]_{j+1}-q_{0}\frac
{1}{a_{j+1}}W[c_{0}(r),s_{2}(r)]_{j+1}\nonumber\\
&  +q_{0}W[c_{0}(r),c_{2}(r)]_{j+1}\Big\}+\mathcal{O}(p^{4}).
\label{eqn:wronskVabetVbalp}%
\end{align}

%% file: Appendix-C.tex
\chapter{}

\section{Coupled-channel Parameterizations}

\label{append:B}The scattering matrix in terms of the eigenphase parameters
was given in \eref{eqn:scattmat}. \ The scattering matrix in terms of the
nuclear bar parameters is
\begin{equation}
{\rm{S}}=\left(
\begin{array}
[c]{cc}%
e^{2i\bar{\delta}_{\alpha}}\cos2\bar{\varepsilon} & ie^{i\left(  \bar{\delta
}_{\alpha}+\bar{\delta}_{\beta}\right)  }\sin2\bar{\varepsilon}\\
ie^{i\left(  \bar{\delta}_{\alpha}+\bar{\delta}_{\beta}\right)  }\sin
2\bar{\varepsilon} & e^{2i\bar{\delta}_{\beta}}\cos2\bar{\varepsilon}%
\end{array}
\right)  .
\end{equation}
Here $\overline{\delta}_{\alpha}$, $\overline{\delta}_{\beta}$ and
$\overline{\varepsilon}$ are the nuclear bar phase shifts and mixing angle
\cite{Stapp1956}. \ The relations between the eigenphase and the nuclear bar
parameters are
\begin{align}
\sin(\delta_{\alpha}-\delta_{\beta})  &  =\frac{\sin2\overline{\varepsilon}%
}{\sin2{\varepsilon}},\\
\delta_{\alpha}+\delta_{\beta}  &  =\overline{\delta}_{\alpha}+\overline
{\delta}_{\beta},\\
\tan2\varepsilon &  =\frac{\tan2\overline{\varepsilon}}{\sin(\overline{\delta
}_{\alpha}-\overline{\delta}_{\beta})}. \label{eqn:eigen_nuclearbar}%
\end{align}

The two-channel effective range expansion is defined slightly differently in
the eigenphase and the nuclear bar parameterizations. \ In the eigenphase
parameterization,
\begin{equation}
\sum_{m'm''n''n'} \textit{\textbf{p}}_{mm'} \,
U_{m'm''}
\,
 [{\rm{K}}^{-1}]_{m''n''} \,
 [U^{-1}]_{n''n'}
 \,
 \textit{\textbf{p}}_{n'n}
=-\frac{1}{a_{{mn}}}+\frac{1}{2}r_{{mn}}p^{2}+\mathcal{O}(p^{4})\,,
\label{eqn:EFE_BB}%
\end{equation}
and in the nuclear bar parameterization,%
\begin{equation}
\sum_{m'n'} \textit{\textbf{p}}_{mm'}
\,
 [{\rm{K}}^{-1}]_{m''n''}
 \,
 \textit{\textbf{p}}_{n'n}
=-\frac{1}{\bar{a}_{{mn}}}+\frac{1}{2}\bar{r}_{{mn}}p^{2}+\mathcal{O}(p^{4})\,,
\label{eqn:EFE_SYM}%
\end{equation}
where $\textit{\textbf{p}}_{mn}$ is the diagonal momentum
matrix $\text{diag}(p^{j-1/2},p^{j+3/2})$.
Therefore, by straightforward calculations we find the following relations
among the threshold scattering parameters,%
\begin{align}
a_{\alpha}=  &  \bar{a}_{\alpha},\label{eqn:ScattLenAlpha}\\
r_{\alpha}=  &  \bar{r}_{\alpha}+\frac{2\bar{q}_{0}\bar{q}_{1}}{\bar
{a}_{\alpha}}+\frac{\bar{q}^{2}\bar{r}_{\beta}}{\bar{a}_{\alpha}^{2}%
},\label{eqn:EffecRangAlpha}\\
a_{\beta}=  &  \bar{a}_{\beta}-\frac{\bar{q}_{0}^{2}}{\bar{a}_{\alpha}%
},\label{eqn:ScattLenBeta}\\
r_{\beta}=  &  \bar{r}_{\beta},\label{eqn:EffectRangBeta}\\
q_{0}=  &  \frac{\bar{q}_{0}}{\bar{a}_{\alpha}},\label{eqn:MixAngFirst}\\
q_{1}=  &  \frac{\left(  \bar{a}_{\beta}\bar{a}_{\alpha}-\bar{q}_{0}%
^{2}\right)  \left(  \bar{a}_{\alpha}\bar{q}_{1}+\bar{r}_{\beta}\bar{q}%
_{0}\right)  }{2\bar{a}_{\alpha}^{2}}. \label{eqn:MixAngSecond}%
\end{align}
For the uncoupled channels $q_{0}$ and $q_{1}$ are zero, and these relations
become $a_{\alpha}=\bar{a}_{\alpha}$, $r_{\alpha}=\bar{r}_{\alpha}$,
$a_{\beta}=\bar{a}_{\beta}$, and $r_{\beta}=\bar{r}_{\beta}$.

\section{Numerical Test using Delta-Function Shell Potentials in a coupled-channel system}
\label{append:C}

As an example to test the equalities in Eq.~(\ref{eqn:effectrangAlph2}), 
Eq.~(\ref{eqn:effectrangBet2}) and Eq.~(\ref{eqn:sum8}), we consider the
scattering of two spin-$\frac{1}{2}$ particles with a delta-function shell
potential and partial-wave mixing,%
\begin{equation}
W(r,R)=\left(
\begin{array}
[c]{cc}%
C_{11} & C_{12}\\
C_{12} & C_{22}%
\end{array}
\right)  \delta(r-R). \label{eqn:deltfunc22}%
\end{equation}
The coupled radial Schr\"{o}dinger equations become
\begin{equation}
-\frac{d^{2}U(r)}{dr^{2}}-k^{2}U(r)+2\mu C_{11}\delta(r-R)U(r)+2\mu
C_{12}\delta(r-R)V(r)=0, \label{eqn:deltafuncdif01}%
\end{equation}%
\begin{equation}
-\frac{d^{2}V(r)}{dr^{2}}-k^{2}V(r)+\frac{6}{r^{2}}V(r)+2\mu C_{21}%
\delta(r-R)U(r)+2\mu C_{22}\delta(r-R)V(r)=0. \label{eqn:deltafuncdif02}%
\end{equation}
The interaction potentials are non-vansihing only at $r=R$, and everywhere
else the wave functions of particles are free wave solutions. We split the
space in two regions, $r>R$ and $r<R$. Solutions for the region $r>R$ are the
same as functions in Eq.~(\ref{eqn:ualpwave3})-(\ref{eqn:vbetwave3}). For
$r<R$ these functions must satisfy the boundary conditions at the origin.
\ After normalization, the solutions for the region $r>R$ are
\begin{align}
U_{\alpha}^{II}(r)=  &  \cos\varepsilon(k)k^{J-1}\Big[\cot\delta
_{J-1}(k)S_{J-1}(kr)+C_{J-1}(kr)\Big],\label{eqn:deltoutfunc1}\\
V_{\alpha}^{II}(r)=  &  \sin\varepsilon(k)k^{J-1}\Big[\cot\delta
_{J-1}(k)S_{J+1}(kr)+C_{J+1}(kr)\Big],\\
U_{\beta}^{II}(r)=  &  -\sin\varepsilon(k)k^{J+1}\Big[\cot\delta
_{J+1}(k)S_{J-1}(kr)+C_{J-1}(kr)\Big],\\
V_{\beta}^{II}(r)=  &  \cos\varepsilon(k)k^{J+1}\Big[\cot\delta_{J+1}%
(k)S_{J+1}(kr)+C_{J+1}(kr)\Big], \label{eqn:deltoutfunc4}%
\end{align}
and for the region $r<R$,
\begin{align}
U_{\alpha}^{I}(r)=  &  A(k)\cos\varepsilon(k)k^{J-1}S_{J-1}%
(kr),\label{eqn:deltinfunc1}\\
V_{\alpha}^{I}(r)=  &  B(k)\sin\varepsilon(k)k^{J-1}S_{J+1}(kr),\\
U_{\beta}^{I}(r)=  &  -D(k)\sin\varepsilon(k)k^{J+1}S_{J-1}(kr),\\
V_{\beta}^{I}(r)=  &  E(k)\cos\varepsilon(k)k^{J+1}S_{J+1}(kr).
\label{eqn:deltinfunc4}%
\end{align}
Here $A(k)$, $B(k)$, $D(k)$ and $E(k)$ are amplitudes to be determined by
boundary conditions.

At the boundary between two regions we have
\begin{align}
U^{II}(R)=  &  U^{I}(R),\label{eqn:boundcond01}\\
V^{II}(R)=  &  V^{I}(R). \label{eqn:boundcond02}%
\end{align}
In addition, by integrating Eq.~(\ref{eqn:deltafuncdif01}) and 
Eq.~(\ref{eqn:deltafuncdif02}) around $r=R$, we have
\begin{align}
-\left(  \frac{dU(r)}{dr}\right)  _{R-\eta}^{R+\eta}+2\mu C_{11}U(R)+2\mu
C_{12}V(R)  &  =0,\label{eqn:deltafuncdif03}\\
-\left(  \frac{dV(r)}{dr}\right)  _{R-\eta}^{R+\eta}+2\mu C_{22}V(R)+2\mu
C_{21}U(R)  &  =0. \label{eqn:deltafuncdif04}%
\end{align}
Taking $\eta\rightarrow0$, we obtain two more boundary conditions,%
\begin{align}
\lim_{\eta\rightarrow0}\Big(\frac{dU^{II}(r)}{dr}\Big|_{(R+\eta)}-\frac
{dU^{I}(r)}{dr}\Big|_{(R-\eta)}\Big)=  &  2\mu C_{11}U(R)+2\mu C_{12}%
V(R),\label{eqn:boundcond03}\\
\lim_{\eta\rightarrow0}\Big(\frac{dV^{II}(r)}{dr}\Big|_{(R+\eta)}-\frac
{dV^{I}(r)}{dr}\Big|_{(R-\eta)}\Big)=  &  2\mu C_{22}V(R)+2\mu C_{21}U(R).
\label{eqn:boundcond04}%
\end{align}
Next we use these four boundary conditions to find phase shifts and mixing
parameters as well as all unknown amplitudes. After substituting wave
functions in Eq.~(\ref{eqn:deltoutfunc1})-(\ref{eqn:deltinfunc4}) into these
boundary conditions, we get the following equations for the $J-1$ channel,
\begin{align}
A(k)=  &  \cot\delta_{J-1}(k)+\frac{C_{J-1}(kR)}{S_{J-1}(kR)}%
,\label{eqn:BCalph01}\\
B(k)=  &  \cot\delta_{J-1}(k)+\frac{C_{J+1}(kR)}{S_{J+1}(kR)},
\label{eqn:BCalph02}%
\end{align}%
\begin{align}
&  \cot\delta_{J-1}(k)S{^{\prime}}_{J-1}(kR)+C{^{\prime}}_{J-1}%
(kR)-A(k)S{^{\prime}}_{J-1}(kR)\hspace{3cm}\nonumber\\
&  =2\mu C_{11}A(k)S_{J-1}(kR)+2\mu C_{12}\tan\varepsilon(k)B(k)S_{J+1}(kR),
\label{eqn:BCalph03}%
\end{align}%
\begin{align}
&  \tan\varepsilon(k)\Big[\cot\delta_{J-1}(k)S{^{\prime}}_{J+1}(kR)+C{^{\prime
}}_{J+1}(kR)-B(k)S{^{\prime}}_{J+1}(kR)\Big]\hspace{1.2cm}\nonumber\\
&  =2\mu C_{22}\tan\varepsilon(k)B(k)S_{J+1}(kR)+2\mu C_{21}A(k)S_{J-1}(kR),
\label{eqn:BCalph04}%
\end{align}
and following equations for the $J+1$ channel,
\begin{align}
D(k)=  &  \cot\delta_{J+1}(k)+\frac{C_{J-1}(kR)}{S_{J-1}(kR)}%
,\label{eqn:BCbet01}\\
E(k)=  &  \cot\delta_{J+1}(k)+\frac{C_{J+1}(kR)}{S_{J+1}(kR)},
\label{eqn:BCbet02}%
\end{align}%
\begin{align}
&  \tan\varepsilon(k)\Big[\cot\delta_{J+1}(k)S{^{\prime}}_{J-1}(kR)+C{^{\prime
}}_{J-1}(kR)+D(k)S{^{\prime}}_{J-1}(kR)\Big]\hspace{1cm}\nonumber\\
&  =2\mu C_{11}\tan\varepsilon(k)D(k)S_{J-1}(kR)-2\mu C_{12}E(k)S_{J+1}(kR),
\label{eqn:BCbet03}%
\end{align}%
\begin{align}
&  \cot\delta_{J+1}(k)S{^{\prime}}_{J+1}(kR)+C{^{\prime}}_{J+1}%
(kR)-E(k)S{^{\prime}}_{J+1}(kR)\hspace{3cm}\nonumber\\
&  =2\mu C_{22}E(k)S_{J+1}(kR)-2\mu C_{21}\tan\varepsilon(k)D(k)S_{J-1}(kR).
\label{eqn:BCbet04}%
\end{align}

In the following, we present the results from numerical calculations for $j\leq 3$. For numerical solutions we use free parameters $\mu$, $C_{11}$, $C_{12}$, $C_{22}$, $R$ and $r$ to calculate some numerical data for the scattering phase shifts and mixing angle. Then we use some fitting procedure to determine the scattering lengths, effective ranges and mixing parameters. We simply fit the data to Eq.~(\ref{chap4:ERE}) and (\ref{eqn:mixingangleexpans}). It is clear that there is an abundance of free parameters which we can use to determine the scattering parameters. However, we set these free parameters such values that we can obtain very nice fits to the data.

\subsubsection{\textbf{Example 1. $\, ^{3}$S$_{1}$ -$\, ^{3}$D$_{1}$
Coupling.}}

\label{sec:S-D delta}

Our first example is the $^{3}S_{1}$-$^{3}D_{1}$ coupled channel corresponding
with $j=1$. We perform numerical calculations using $2\mu C_{11}=-6.4$ MeV, $2\mu C_{12}=-0.28$ MeV, $2\mu C_{22}=-1.4$ MeV and $R=2.6$ fm. 
\ Results are shown in Table \ref{table:tabledelt1} and Table
\ref{table:tabledelt2}.

\begin{table}[th]
\caption{Numerical results for scattering length and effective range in
two-body interaction by the delta function potentials.}%
\label{table:tabledelt1}
\centering
\rowcolors{3}{gray!15}{white}
\begin{tabular}
[c]{||c|c|c||}\hline\hline
Channel & $a_{L}$ [$\text{fm}^{2L+1}$] & $r_{L}$ [$\text{fm}^{-2L+1}%
$]\\\hline\hline
$^{3}$S$_{1}$ & 2.766 & 1.804\\
$^{3}$D$_{1}$ & -0.326 & -36.876\\
$^{3}$P$_{2}$ & 16.108 & -1.110\\
$^{3}$F$_{2}$ & -0.039 & -287.01\\
$^{3}$D$_{3}$ & 8.832& -0.665\\
$^{3}$G$_{3}$ &0.008 & -1550.6\\\hline\hline
\end{tabular}
\end{table}

\begin{table}[ptb]
\caption{Numerical results for the mixing parameters in two-body interaction by
the delta function potentials.}%
\label{table:tabledelt2}
\centering
\rowcolors{3}{gray!15}{white}
\begin{tabular}
[c]{||c|c|c||}\hline\hline
Mixing angle & $q_{0}$ [$\text{fm}^{2}$] & $q_{1}$ [$\text{fm}^{4}$%
]\\[0.5ex]\hline\hline
$\varepsilon_{1}$ & 0.075 & 0.068\\
$\varepsilon_{2}$ & 0.060 & 0.034\\
$\varepsilon_{3}$ & 0.232 & 0.057\\  \hline\hline
\end{tabular}
\end{table}

For $r=6.5$ fm and $k=0.335$ MeV, we obtain%
\begin{equation}
b_{0}(r)-q_{0}^{2}\frac{6}{r^{3}}-r_{0}=4.577\ \text{fm},
\label{eqn:equalitysdalphleft}%
\end{equation}
and this agrees with the predicted equivalent expression within numerical
precision,%
\begin{equation}
2\int_{0}^{R} \left( \left[U_{\alpha}^{I}(r^{\prime})\right]^2+\left[V_{\alpha}^{I}(r^{\prime}%
) \right]^2 \right) dr^{\prime}+2\int_{R}^{r}\left( \left[ U_{\alpha}^{II}(r^{\prime})\right]^2+\left[V_{\alpha}%
^{II}(r^{\prime})\right]^2 \right) dr^{\prime}=4.575\ \text{fm}.
\label{eqn:equalitysdalphright}%
\end{equation}
We also find%
\begin{equation}
b_{2}(r)+q_{0}^{2}\frac{2r^{3}}{3a_{2}^{2}}-r_{2}=5978.51
\ \text{fm}^{-3}, \label{eqn:equalitysdbetaleft}%
\end{equation}
agrees with%
\begin{align}
&  2\int_{0}^{R}\left( \left[U_{\beta}^{I}(r^{\prime})\right]^2+\left[V_{\beta}^{I}(r^{\prime}%
) \right]^2 \right) dr^{\prime}+2\int_{R}^{r}\left( \left[ U_{\beta}^{II}(r^{\prime})\right]^2+\left[V_{\beta}%
^{II}(r^{\prime})\right]^2 \right) dr^{\prime} = 5934.13 \, \text{fm}^{-3}. \label{eqn:equalitysdbetaright}%
\end{align}

\subsubsection{\textbf{Example 2. $\, ^{3}$P$_{2}$ -$\, ^{3}$F$_{2}$
Coupling.}}\label{sec:P-F delta}

The second example is the coupled channel $^{3}P_{2}$-$^{3}F_{2}$ with $J=2$.
We use $2\mu C_{11}=-1.759$ MeV, $2\mu C_{12}=-0.28$ MeV, $2\mu C_{22}=-1.36$ MeV and $R=2.6$ fm, and the results for $a_{1}$, $a_{3}$,
$r_{3}$, $q_{0}$ and $q_{1}$ shown in Table \ref{table:tabledelt1} and Table
\ref{table:tabledelt2}. \ For $r=8.42$ fm and $k=0.33$ MeV we get%
\begin{equation}
b_{1}(r)-q_{0}^{2}\frac{90}{r^{5}}-r_{1}=5.188\ \text{fm}^{-1},
\label{eqn:equalitypfsigmleft}%
\end{equation}
which agrees within numerical precision with%
\begin{equation}
2\int_{0}^{R} \left( \left[U_{\alpha}^{I}(r^{\prime})\right]^2+\left[V_{\alpha}^{I}(r^{\prime}%
) \right]^2 \right) dr^{\prime}+2\int_{R}^{r}\left( \left[ U_{\alpha}^{II}(r^{\prime})\right]^2+\left[V_{\alpha}%
^{II}(r^{\prime})\right]^2 \right) dr^{\prime} =5.187\ \text{fm}^{-1}.
\label{eqn:equalitypfsigmright}%
\end{equation}
We also find%
\begin{equation}
b_{3}(r)+q_{0}^{2}\frac{2r^{5}}{45a_{3}^{2}}-r_{3}=2.8052\times10^{6}%
\ \text{fm}^{-5}, \label{eqn:equalitypfrholeft}%
\end{equation}
which agrees with%
\begin{align}
&   2\int_{0}^{R}\left( \left[U_{\beta}^{I}(r^{\prime})\right]^2+\left[V_{\beta}^{I}(r^{\prime}%
) \right]^2 \right) dr^{\prime}+2\int_{R}^{r}\left( \left[ U_{\beta}^{II}(r^{\prime})\right]^2+\left[V_{\beta}%
^{II}(r^{\prime})\right]^2 \right) dr^{\prime} =2.8035\times10^{6}\ \text{fm}^{-5}.
\end{align}

\subsubsection{\textbf{Example 3. $\, ^{3}$D$_{3}$ -$\, ^{3}$G$_{3}$
Coupling.}}

\label{sec:D-G delta}

The last example is the $^{3}$D$_{3}$-$^{3}$G$_{3}$ coupled channel with $J=3
$. We use $2\mu C_{11}=-2.255$ MeV, $2\mu C_{12}=-0.28$ MeV, $2\mu C_{22}=-3.27$ MeV and $R=2.6$ fm, and the parameters $a_{2}$, $a_{4}$, $r_{2}$, $a_{4}$, $q_{0}$ and $q_{1}$ are calculated numerically and indicated in Table \ref{table:tabledelt1} and Table
\ref{table:tabledelt2}. \ For $r=9.7$ fm and $k=0.33$ MeV the results for the
$^{3}D_{3}$ channel are
\begin{equation}
b_{2}(r)-q_{0}^{2}\frac{3150}{r^{7}}-r_{2}=120.256\ \text{fm}^{-3},
\label{eqn:equalityDGalphaleft}%
\end{equation}
which agrees within numerical precision with%
\begin{equation}
2\int_{0}^{R} \left( \left[U_{\alpha}^{I}(r^{\prime})\right]^2+\left[V_{\alpha}^{I}(r^{\prime}%
) \right]^2 \right) dr^{\prime}+2\int_{R}^{r}\left( \left[ U_{\alpha}^{II}(r^{\prime})\right]^2+\left[V_{\alpha}%
^{II}(r^{\prime})\right]^2 \right) dr^{\prime}  = 120.88\text{ fm}^{-3}.
\label{eqn:equalityDGalpharight}%
\end{equation}
For $r=2.93$ fm and $k=1.0$ MeV, the results of the $^{3}G_{3}$ channel are
\begin{equation}
b_{4}(r)+\frac{q_{0}^{2}}{a_{4}^{2}}\frac{2r^{7}}{1575}-r_{4}=2.218\times
10^{8}\ \text{fm}^{-7}, \label{eqn:equalityDGbetaleft}%
\end{equation}
which agrees with%
\begin{align}
  2\int_{0}^{R}\left( \left[U_{\beta}^{I}(r^{\prime})\right]^2+\left[V_{\beta}^{I}(r^{\prime}%
) \right]^2 \right) dr^{\prime}+2\int_{R}^{r}\left( \left[ U_{\beta}^{II}(r^{\prime})\right]^2+\left[V_{\beta}%
^{II}(r^{\prime})\right]^2 \right) dr^{\prime} =2.214\times10^{8}\ \text{fm}^{-7}. \label{eqn:equalityDGbetaright}%
\end{align}

%% file: Appendix-D.tex
\chapter{}

\section{Finite-volume binding energy corrections and topological volume corrections for scattering with arbitrary $\ell$}

\label{sec:append:topological-phases}
In order to apply L\"uscher's finite-volume method with maximal accuracy, we consider also finite-volume
corrections to the binding energy of the dimer.  The finite-volume correction to two-body $s$-wave binding energies was derived in Ref.~\cite{Luscher:1985dn} and extended to arbitrary angular momentum in Ref.~\cite{Konig:2011nz,Konig:2011ti}.  There has also been significant work towards understanding three-body binding energy corrections at finite volume \cite{Kreuzer:2010ti,Kreuzer:2012sr}.   

It was noticed in Ref.~\cite{PhysRevD.84.091503} that the finite-volume corrections to the dimer binding energy is dependent on the motion of the dimer.  This fact has been used to cancel out finite-volume corrections to the binding energy \cite{Davoudi:2011md}.  The dimer motion induces phase-twisted boundary conditions on the dimer's relative-coordinate wavefunction.  These effects are called topological volume corrections and were found to have an effect on the finite-volume analysis for scattering of the dimer.  The study of topological volume corrections were carried out for $s$-wave scattering in Ref.~\cite{PhysRevD.84.091503,PhysRevC.86.034003} and further applied in Ref.~\cite{A.Rokash2013.3386,Pine:2013zja}.  In the following we show the extension to general partial wave $\ell$. 

The general solution of the Helmholtz equation has the form of
\begin{align}
\psi_{p}(\vec{r}) = 
  \sum_{\ell,m} c_{\ell,m}(p) \, G_{\ell,m}(\vec{r},p^{2}) \,.
  \label{eqn:topologicalphases-000}
\end{align}
The functions $G_{\ell,m}(\vec{r},p^{2})$ form a linearly independent complete basis set and are defined as
\begin{align}
G_{\ell,m}(\vec{r},p^{2}) = \mathcal{Y}_{\ell,m}(\nabla)
\, G(\vec{r},p^{2})
\,.
\label{eqn:topologicalphases-001}
\end{align}
Here $\mathcal{Y}_{\ell,m}$ are the solid spherical harmonic polynomials and defined in terms of the spherical harmonics as
\begin{align}
\mathcal{Y}^{m}_{\ell}(\vec{r})
=
r^{\ell} \, \text{Y}^{m}_{\ell}(\theta,\phi)
\,,
\label{eqn:topologicalphases-005}
\end{align}
and $G(\vec{r},p^{2})$ is the periodic Green's function solution to the Helmholtz equation for $\ell,m= 0$,
\begin{align}
G_{0,0}(\vec{r},p^{2})
=
G(\vec{r},p^{2})
=
\frac{1}{L^{3}}
\sum_{\vec{k}}
\frac
{e^{\frac{2i\pi}{L}\vec{k}\cdot\vec{r}}}
{\left(\frac{2\pi}{L}\vec{k}\right)^{2} - p^{2}}
\,.
\label{eqn:topologicalphases-011}
\end{align}
Using Eq.~(B1) of Ref.~\cite{Luscher1991531}, we have
\begin{align}
G_{\ell,m}(\vec{r},p^{2}) = r^{\ell} \, \text{Y}^{m}_{\ell}(\theta,\phi)
\left(\frac{1}{r}\frac{\partial}{\partial r}\right)^{\ell}
\, G(\vec{r},p^{2})
\,.
\label{eqn:topologicalphases-009}
\end{align}
Inserting Eq.~(\ref{eqn:topologicalphases-011}) into Eq.~(\ref{eqn:topologicalphases-009}), we write the asymptotic form of the scattering wave function as
\begin{align}
u_{\ell}(r) = 
C
\,
\sum_{\vec{k}}
|\vec{k}|^{\ell}
\frac
{e^{\frac{2i\pi}{L}\vec{k}\cdot\vec{r}}}
{\left(\frac{2\pi}{L}\vec{k}\right)^{2} - p^{2}}
\,,
\label{eqn:topologicalphases-015}
\end{align}
where $C$ is the normalization coefficient. The derivation of the topological volume corrections for the $s$-wave scattering of two composite particles $A$ and $B$ is given in Ref.~\cite{PhysRevD.84.091503,PhysRevC.86.034003,A.Rokash2013.3386}. Here we focus on the fermion-dimer scattering and derive the topological volume corrections for higher partial waves. 

In this analysis we take the continuum limit.  We let the total momentum of the fermion plus dimer system to be zero and let $p$ be the magnitude of the relative momentum between the fermion and dimer.  Let $E_{\text{d},\vec{0}}(\infty)$ be the dimer energy at infinite volume and $m_{\rm{d}}$ be the dimer mass. Then the fermion-dimer energy at infinite volume, $E_{\text{df}}(p,\infty)$, is
\begin{align}
E_{\text{df}}(p,\infty) = \frac{p^{2}}{2 m_{\text{d}}} + \frac{p^2}{2m} + E_{\text{d},\vec{0}}(\infty) \,.
\label{eqn:fermion-dimer-energy-021}
\end{align}
As in previous studies of fermion-dimer scattering on the lattice \cite{PhysRevD.84.091503,PhysRevC.86.034003,A.Rokash2013.3386,Pine:2013zja}, we calculate the effective dimer mass of the dimer on the lattice by computing the dimer dispersion relation. Now we let $E_{\text{df}}(p,L)$ be the finite-volume energy of the fermion-dimer system.  Following  Ref.~\cite{PhysRevD.84.091503,PhysRevC.86.034003,A.Rokash2013.3386}, we can compute the expectation value
\begin{align}
E_{\text{df}}(p,L) = \frac{\int d^3r \, u_{\ell}^{*}(r) \hat{H}  u_{\ell}(r)}{\int d^3r \,|u_{\ell}(r)|^{2}}
= 
\frac{1}{\mathcal{N}_{\ell}} \sum_{\vec{k}}^{{k}_{max}}
|\vec{k}|^{2\ell}
 \frac{\frac{p^{2}}{2m_{\text{d}}}
  +E_{\text{d},\vec{k}}(L)}{\left(\vec{k}^{2} -\eta\right)^{2}}
  \,,
\label{eqn:topologicalphases-019}
\end{align}
where $E_{\text{d},\vec{k}}(L)$ is the finite-volume energy of the dimer with momentum $\vec{k}$,
$\mathcal{N}_{\ell}$ is defined as \begin{align}
\mathcal{N}_{\ell} = \sum_{\vec{k}}^{{k}_{max}}
|\vec{k}|^{2\ell}
\left(\vec{k}^{2} -\eta\right)^{-2},
\end{align} and  $\eta = \left(\frac{Lp}{2 \pi}\right)^{2}$.  For $\ell > 0$ the summations are divergent and we must cutoff the short distance behavior at some momentum scale $\Lambda$ characterizing the range of the fermion-dimer interactions.  The corresponding maximum index value $k_{max}$ scales as $\Lambda L/(2 \pi)$. 

Let $\Delta E_{\text{d},\vec{0}}(L) = E_{\text{d},\vec{0}}(L)-E_{\text{d},\vec{0}}(\infty)$ be the finite-volume energy shift of the dimer energy in its rest frame, and $\Delta E_{\text{d},\vec{k}}(L) = E_{\text{d},\vec{k}}(\infty)- E_{\text{d},\vec{k}}(L)$ be the finite-volume energy shift of the dimer energy with momentum $\vec{k}$.  One can show that \cite{PhysRevD.84.091503},
\begin{align}
\frac{ \Delta E_{\text{d},\vec{k}}(L)}{\Delta E_{\text{d},\vec{0}}(L)} 
 =
 \frac{1}{3} \sum_{i =1}^{3} \cos\left(2\pi k_{i} \, \alpha \right)\, .
\label{eqn:dimer_vol_corr}
\end{align}
Using Eq.~(\ref{eqn:fermion-dimer-energy-021}), (\ref{eqn:topologicalphases-019}), and (\ref{eqn:dimer_vol_corr}), we can now write the fermion-dimer energy correction at finite volume as
\begin{align}
E_{\text{df}}(p,L) -E_{\text{df}}(p,\infty) 
=
\tau_{\ell}(\eta) \, \Delta E_{\text{d},\vec{0}}(L) \,,
\label{eqn:topologicalphases-025}
\end{align}
where $\tau_{\ell}(\eta)$ is the topological factor,
\begin{align}
\tau_{\ell}(\eta)
= 
\frac{1}{\mathcal{N}_{\ell}} \sum_{\vec{k}}^{{k}_{max}}
 \frac{|\vec{k}|^{2\ell} \, 
 \sum_{i =1}^{3} \cos\left(2\pi k_{i} \, \alpha \right)}{3\left(\vec{k}^{2} -\eta\right)^{2}}
\,,
\label{eqn:topologicalphases-029}
\end{align}
with $\alpha = m_{}/(m+m_{\rm{d}})=1/3$. 
Due to the short distance  behavior of the momentum mode summations for $\ell > 0$, we find that the topological phase factor $\tau_{\ell}(\eta)$ is suppressed by the lattice length $L$,
\begin{align}
\tau_{\ell>0}(\eta)
= 
\mathcal{O}\left(L^{-1}\right)\,.
\end{align}
In other words, the topological volume correction for $\ell > 0$ is smaller by a factor of $L$ relative to the $\ell = 0$ correction.  In our analysis of fermion-dimer scattering we have therefore included topological volume corrections as written in Eq.~(\ref{eqn:topologicalphases-029}) for $\ell = 0,$ but neglected the corrections for $\ell > 0$.  We find that this prescription gives good agreement with the continuum infinite-volume STM\ results for partial waves $\ell=0,1,2$.